\documentclass[12pt,a4paper]{article}
\usepackage[english]{babel}
\usepackage[utf8]{inputenc}
\usepackage[T1]{fontenc}
\usepackage[top=1.25in, bottom=1.25in, left=1.05in, right=1.05in]{geometry}

\usepackage{amscd,amsmath,amssymb,amsfonts,amsthm,bbm,bm,latexsym,mathrsfs,mathtools}

\usepackage{natbib}
\usepackage{url}
\usepackage[plain,noend]{algorithm2e}
\usepackage{enumitem}
\usepackage{comment}

\usepackage[section]{placeins}

\usepackage{hyperref}
\hypersetup{bookmarks, unicode, colorlinks, urlcolor=blue, linkcolor=blue, citecolor=blue}
\usepackage{float,array,rotating,booktabs,multirow}
\usepackage{threeparttable}
\usepackage{xcolor,subfigure,epsfig,graphicx}
\usepackage[width=0.80\textwidth]{caption}

\newtheorem{proposition}{Proposition}[section]
\newtheorem{corollary}{Corollary}
\newtheorem{definition}{Definition}[section]
\theoremstyle{remark}
\newtheorem{remark}{Remark}
\theoremstyle{plain}
\newtheorem{example}{Example}

\usepackage[subnum]{cases}	
\allowdisplaybreaks			
\makeatletter
\newcommand*{\distas}[1]{\mathbin{\overset{#1}{\kern\z@\sim}}}		
\makeatother
\usepackage{dsfont}
\def\T {{ \mathrm{\scriptscriptstyle T} }}

\def\Id {\mathbf{I}}
\def\Id {\textnormal{I}}
\def\I  {\mathds{1}}

\def\SL {\mathscr{L}}
\def\CL {\mathcal{L}}

\def\eps {\epsilon}
\def\bpsi {\boldsymbol{\psi}}
\def\bc {\mathbf{c}}
\def\bN {\mathbf{N}}

\def\by {\mathbf{y}}
\def\bw {\mathbf{w}}
\def\bz {\mathbf{z}}
\def\bY {\mathbf{Y}}
\def\bW {\mathbf{W}}
\def\bZ {\mathbf{Z}}
\def\balpha {\boldsymbol{\alpha}}

\def\bxi {\boldsymbol{\xi}}
\def\Y {\mathbb{Y}}

\def\E {\textnormal{E}}
\def\Var {\textnormal{var}}
\def\Cov {\textnormal{cov}}
\def\Cor {\textnormal{corr}}
\def\pr  {\textnormal{pr}}
\def\RE {\mathds{R}}
\def\Rp {\mathds{R}_+}

\def\GP   {\mathcal{G}a\textnormal{P}}
\def\NcGP {\textnormal{Nc}\mathcal{G}a\textnormal{P}}
\def\NcGa {\textnormal{Nc}\mathcal{G}a}
\def\MARG {\textnormal{M-ARG}}
\def\PP   {\textnormal{PP}}

\def\SNMARG {\textnormal{SN-M-ARG}}

\def\BM {\textnormal{BM}}
\newcommand{\coloneqq}{=}

\begin{document}

\title{A Spatiotemporal Gamma Shot Noise Cox Process}

\author{
Federico Bassetti\thanks{University of Milan, Italy. \color{blue}\texttt{federico.bassetti@polimi.it}}
\and
Roberto Casarin\thanks{Ca' Foscari University of Venice, Italy. \color{blue}\texttt{r.casarin@unive.it}}
\and
Matteo Iacopini\thanks{Luiss University of Rome, Italy. \color{blue}\texttt{miacopini@luiss.it}}
\and
Antonio Peruzzi\thanks{Ca' Foscari University of Venice, Italy. \color{blue}\texttt{antonio.peruzzi@unive.it}}
}

\date{\today}


\maketitle

\begin{abstract}
A new discrete-time shot noise Cox process for spatiotemporal data is proposed. The random intensity is driven by a dependent sequence of latent gamma random measures.
Some properties of the latent process are derived, such as an autoregressive representation and the Laplace functional. A simulation method based on the Inverse L\'evy Measure algorithm is provided.
Moreover, these results are used to derive the moment, predictive, and pair correlation measures for the proposed shot noise Cox process. The model is flexible yet tractable, allowing it to capture persistence, global trends, and latent spatial and temporal factors.
A Bayesian inference approach is adopted, and efficient Markov Chain Monte Carlo algorithms based on conditional Sequential Monte Carlo and adaptive Metropolis-Hastings are proposed. An application to georeferenced wildfire data illustrates the properties of the model and inference.
\end{abstract}

\vspace*{10pt}


\section{Introduction}
\label{sec:introduction}

Spatiotemporal random models are extensively used for analysing phenomena characterised by spatial and temporal dependence, allowing for a deeper understanding of local and global dynamics and providing more accurate forecasts in many fields, including environmental science, epidemiology, geoscience, ecology, hydrology, meteorology and oceanography. Cox processes with log-Gaussian random intensity, as introduced by \cite{moller1998log}, are among the most used models in spatial statistics due to their flexibility. They have been studied and extended in different directions, such as the spatiotemporal \citep[e.g., see][]{brix2001spatiotemporal,brix2001space} and the multivariate constructions \citep{waagepetersen2016analysis}. \cite{Diggle2013} provides a review of log-Gaussian Cox Processes (LGCP) for spatial and spatiotemporal data. In this article, we follow another flexible modelling approach based on shot noise Cox processes \citep{Brix99GeneralizedGammaProcess,Wolpert1998PoissonGamma_RandomField}, which has been extended to spatiotemporal \citep{Moeller2010} and multivariate \citep{jalilian2015multivariate} settings. The proposed process is primarily designed to capture stylised facts in spatiotemporal data, such as local and global trends, seasonality, persistence, and unobservable spatiotemporal factors.

A \textit{shot noise Cox process} (SN) is a Poisson point process with random intensity obtained by smoothing a Poisson random measure with a kernel density \citep{moller2003shot}.
Shot noise Cox processes belong to the general class of Cox processes \citep{cox1955some, moller2003statistical,baddeley2007spatial} and, unlike Poisson processes, they allow for complex spatial patterns of point events. The random intensity function can account for the effect of common observable variables and latent factors on the spatial configuration of the Poisson. In this article, we build on the gamma shot noise Cox process of \cite{Wolpert1998PoissonGamma_RandomField} and extend it to a dynamic setting.

We contribute to this literature in several ways.
First,  we propose a new \textit{Measured-valued Autoregressive Gamma} (M-ARG) process $(W_t)_{t \geq 0}$ in discrete time and use it to define a sequence of M-ARG driven SN processes (SN-M-ARG), where the evolution of the dynamic random intensity in a dynamic shot noise Cox process.  As argued by several authors \citep[e.g., see][]{gneiting2010continuous,Diggle2013,richardson2020spatiotemporal}, it is natural to assume a discrete-time setting, since space-time data encountered routinely in meteorological and environmental studies are recorded at regular time intervals, thus forming a spatially indexed time series. Covariates and various forms of non-stationarity can be easily and properly included in discrete-time models with time-varying parameters \citep{Gourieroux06AR_GammaProcess}. For specific parametrizations, our M-ARG construction shares  some properties, such as the finite-dimensional distribution, with the continuous-time gamma process defined in \cite{ethier1993transition} and \cite{etheridge2000introduction} and studied further in \cite{Omiros2016}, \cite{EJS21} and \cite{ascolani2023smoothing}. The proposed M-ARG process also extends the scalar autoregressive gamma process of \cite{pit05b} and \cite{Gourieroux06AR_GammaProcess} to the measure-valued case. In the stationary setting, the M-ARG and the dependent generalised gamma process of \cite{Palla16BNP_Dynamic_SparseGraph} share a common special case when the sparsity (discount) parameter of the generalized gamma Lévy measure is set to zero.

Second, we provide an autoregressive representation of the M-ARG based on a suitable thinning operator and use it to derive properties of these processes, including their unconditional and conditional exponential-affine Laplace functionals and conditional and stationary measures. These properties allow us to obtain explicit formulas for moment, predictive, and pair correlation measures of the M-ARG and the SN-M-ARG. Also, a stochastic simulation procedure is given to generate trajectories from the SN-M-ARG process, based on the Inverse L\'evy Measure (ILM) representation given in \cite{Wolpert1998PoissonGamma_RandomField}.

As a third contribution, we develop a Bayesian approach to inference in this setting, based on Markov Chain Monte Carlo to approximate the posterior distribution of the parameters and the latent intensity process. The procedure relies on adaptive Metropolis-Hastings algorithms \citep{Andrieu08AdaptiveMCMC} and on a particle Gibbs approximation of the likelihood and posterior \citep{Andrieu2010particleMCMC}.

Finally, we illustrate the new model and inference on simulated data and forest fire real data. Nevertheless, the approach can be interesting for several real-world applications where data arise as a spatially indexed time series. We show that the proposed spatiotemporal gamma shot noise Cox process is well suited to capturing spatial patterns and temporal dynamics in forest fires. Besides, it addresses the challenging issue of estimating global trends and seasonality in high-spatial-resolution fire data from a wide region \citep[e.g., see][] {Frontiers22,geo2020}. Moreover, adopting the Bayesian approach allows us to quantify uncertainty in estimates and forecasts, a central issue in climate risk analysis \citep{raftery2017less}.

The remainder of the paper is as follows.
Section~\ref{sec:MARG} defines M-ARG and the spatiotemporal gamma shot noise Cox processes and presents a stochastic simulation algorithm.
Section~\ref{sec:properties} provides some properties of M-ARG and gamma shot noise Cox process.
Section~\ref{sec:Bayesian_inference} describes a Bayesian approach to inference for the proposed process.
Finally, Section~\ref{sec:illustration} illustrates the results of an application of the proposed model to real data on forest fires.  Proofs for the paper results are deferred to the appendices, and further background material and numerical results are included in the Supplement.

\section{Dependent Poisson-Gamma Random Fields}
\label{sec:MARG}

This Section introduces a novel measure-valued autoregressive process based on Poisson and gamma random measures. Then, this process is used to define a new dynamic shot noise Cox process, which is well suited for modelling spatiotemporal phenomena.

\subsection{A Gamma Shot Noise Process}

A \textit{gamma random measure} (process) $W$ on a Polish space $\Theta$, with \textit{base measure} $H$ and inverse scale (i.e., rate) parameter $\beta >0$, denoted $W \sim \GP(W \mid H,\beta)$, is characterised by the Laplace functional 
\begin{equation}
\label{eq:LPgammaP}
\CL_W(f) \coloneqq \E\Big( e^{-\int_{\Theta} f W(\mathrm{d}\theta)} \Big) = \exp\Big( -\int_\Theta \log\Big( 1+ \frac{f}{\beta} \Big) \: H(\mathrm{d}\theta) \Big), \qquad f \in \BM_+(\Theta),
\end{equation}
where $\BM_+(\Theta)$ is the set of measurable positive, bounded functions with bounded support on $\Theta$.
Hereafter, we only consider {\it locally finite measures} on metric spaces, that is, measures that give finite value to any set of bounded diameter.
Recall that the law of any locally finite random measure $W$ is completely characterised by its Laplace functional \citep[e.g., see][ex.10.10.5]{Daley08IntroPointProcesses}.
It is worth recalling that a gamma random measure $W \sim \GP(W \mid H,\beta)$ can be represented as 
\begin{equation}
    \label{eq:propernessNC-gamma}
    W(\mathrm{d}\theta) = \sum_{i\geq 1} w_i \delta_{\theta_i}(\mathrm{d}\theta) = \int_{\RE_+} \! w \: N(\mathrm{d}w, \mathrm{d}\theta), 
\end{equation}
where $N(\mathrm{d}w,\mathrm{d}\theta) = \sum_{i\geq1} \delta_{(\theta_i,w_i)}(\mathrm{d}w,\mathrm{d}\theta)$ is a Poisson point process on $\RE_+ \times \Theta$, with intensity measure $w^{-1} e^{-w\beta} \mathrm{d}w H(\mathrm{d}\theta)$. See Example 15.6 in \cite{LastPenrose2018PoissonProcess_book} for the integrability
condition ensuring that the Poisson integral in \eqref{eq:propernessNC-gamma} is a.s. bounded on bounded sets.

The Poisson-gamma random field introduced in \cite{Wolpert1998PoissonGamma_RandomField} is a shot noise Cox process $N$ with values in a measurable (Polish) space $\Y$ that satisfies the following hierarchical representation:
\begin{equation}
\label{eq:WOLPERT_model}
\begin{split}
     & N \mid \Lambda  \sim \PP(N \mid \Lambda),
    \quad 
     \Lambda(\mathrm{d}y) := \lambda(y) \ell(\mathrm{d}y),  \quad
      \lambda(y) := \int_\Theta K_\phi(y,\theta) \: W(\mathrm{d}\theta)
     \\
     &  W \sim \GP(W \mid H, c), 
    \end{split}
\end{equation}
where $N \sim \PP(N \mid \Lambda)$ denotes a \textit{Poisson random measure} (or Poisson point process) with mean measure (random intensity measure) $\Lambda$, $\ell(\mathrm{d}y)$ is a $\sigma$-finite measure on $\Y$ and $K_\phi$ is a positive density kernel on $\Y\times\Theta$, depending on some parameters $\phi \in \Phi$. 
Using the representation in eq.~\eqref{eq:propernessNC-gamma}, the random intensity $\lambda$ can be written as a kernel mixture  $\lambda(y) = \sum_{i \geq 1} w_i K_\phi(y,\theta_i)$, where weights $\{w_i\}_{i\leq 1}$ need not sum to one.

\begin{example}[Gaussian kernel on  bounded domain]\label{Ex1}
A typical choice in spatiotemporal statistics is that both $\Theta$ and $\Y$ are bounded sets in $\RE^2$ and the kernel is a two-dimensional (un-normalized) Gaussian density with covariance matrix $\phi^2 \Id_2$, that is
\[
K_\phi(y,\theta) =   (\pi \phi^2)^{-1}\exp(- \phi^{-2} \|\theta-y\|^2).
\]
where $||\cdot||$ denotes the Euclidean norm and the range parameter $\phi>0$ controls for the spatial smoothing. In this case $\ell(\mathrm{d}y)$ is the Lebesgue measure on $\Y \subset \RE^2$ and $H$ can be any measure on $\Theta \subset \RE^2$, for example the uniform measure on $\Theta$. 
Note that with this choice $\int_\Y K_\phi(y,\theta) \mathrm{d}y \neq 1$ \citep[e.g., see Section 5.2 in][]{Wolpert1998PoissonGamma_RandomField}.
\end{example}

\begin{example}[Gaussian kernel on $\RE^d$]\label{Ex2}
Another classical choice is $\Y = \Theta = \RE^d$, $H(\mathrm{d}\theta)$ and $\ell(\mathrm{d}y)$ are the Lebesgue measures on $\RE^d$, and $K_\phi(\theta,y) = (\pi \phi^2)^{-d/2} \exp(- \phi^{-2} \|\theta-y\|^2)$ is a Gaussian kernel.
Clearly in this case $\int_\Y K_\phi(y,\theta) \mathrm{d}y = 1$ for every $\theta$. A more general anisotropic specification assumes the kernel then takes the form $K_{\Sigma}(y,\theta) = \pi^{-d/2} |\Sigma|^{-1/2} \exp\big( -(\theta-y)^\top \Sigma^{-1}(\theta-y) \big)$, where the positive definite matrix $\Sigma$ controls the amount of smoothing along different directions. 
\end{example}

\begin{example}[Discrete base measure and continuous kernels]\label{Ex3}
To reduce the model complexity and the parameter space dimension in Example \ref{Ex1}, one can choose a discrete base measure $H$ with finite support $\Theta = \{ \theta_1, \ldots, \theta_{N^g} \}$, that is $H(\mathrm{d}\theta) = \sum_{j=1}^{N^g} \alpha_j \delta_{\theta_{j}}(\mathrm{d}\theta)$.
This can be constructed as a discretisation of a continuous measure $H(\mathrm{d}\theta) = h(\theta) \mathrm{d} \theta$, with $\alpha_j$ the value of $H(A_j)$, $(A_j)_{j=1,\dots, N^g}$ a suitable tessellation of $\Theta$, and $\theta_j$ the ``centre'' of the cell $A_j$. In this way, $H$ is parametrised by a finite $N^g$-dimensional vector $(\alpha_1,\dots,\alpha_{N^g})$. In applications, the cells $A_j$ may correspond to geographic regions or elements of a regular grid. Unlike many spatio-temporal models, where the tessellation is applied directly to the physical domain $\mathbb{Y}$, here the tessellation is applied to the parameter space $\Theta$. Choosing a continuous kernel on the physical space $\mathbb{Y}$ induces smoothing across neighbouring regions. The case of a discrete physical space $\mathbb{Y}$ is discussed in the following example.
\end{example}

\begin{example}[Discrete spaces and discrete kernels]\label{Ex4}On the opposite side with respect to the previous example, one may choose both $\Theta=\{\theta_1,\dots,\theta_{N_1}\}$ and $\Y=\{y_1,\dots,y_{N_2}\}$ be discrete spaces. 
In this case, $H$ is a probability vector on $\Theta$, $\ell(\mathrm{d}y)$ is the counting measure on $\Y$ and the kernel $K_\phi$ is a $N_1 \times N_2$ transition matrix $K_\phi(i,j)=K_\phi(y_i,\theta_j)$. This is the so-called ``discrete model'' described in Section 5.3 of \cite{Wolpert1998PoissonGamma_RandomField}.
\end{example}

The Poisson-gamma random field is an example of a standard shot noise Cox process, and the random density $\lambda$ is referred to as a random intensity function. It belongs to a general family of shot noise processes \citep[][]{Brix99GeneralizedGammaProcess}, where the intensity measure $\Lambda$ satisfies $\Lambda(\mathrm{d}y) = \int_{\Theta\times\RE_+} \mathcal{K}(\mathrm{d}y,w,\theta) N(\mathrm{d}w,\mathrm{d}\theta)$, with $\mathcal{K} : \Y\times\RE_+ \times \Theta\mapsto\RE_+$ a kernel and $N$ a Poisson process.

\subsection{Autoregressive Gamma Random Measures} 
In this Section we introduce a new class of dependent gamma random measures to define a dynamic shot noise Cox process and extend the Poisson-gamma random field model of \cite{Wolpert1998PoissonGamma_RandomField}.
The {\it autoregressive gamma model} has been introduced and studied in \cite{Gourieroux06AR_GammaProcess} to model positive data that feature time-varying complex nonlinear dynamics. 
According to \cite{Gourieroux06AR_GammaProcess}, a positive real-valued process $\{ w_t \}_{t \geq 1}$ is an autoregressive gamma process if the conditional distribution of $w_{t+1}$ given $w_t$ is a noncentral gamma distribution $\NcGa(\delta,\beta_t w_{t},c_t^{-1})$.
It is worth recalling that a random variable $Y$ has a \textit{noncentral gamma distribution} of parameters $\delta>0$, $\beta >0$, and $c^{-1} >0$, denoted $\NcGa(\delta,\beta,c^{-1})$, if it admits the following Poisson mixture of gamma distributions:
\begin{equation}
    Z \sim \mathcal{P}oi(Z \mid \beta), \qquad 
    Y \mid Z \sim \mathcal{G}a(Y \mid \delta+Z,c^{-1}),
\label{eq:noncentralgamma}
\end{equation}
where $\mathcal{P}oi(\cdot \mid \beta)$ denotes the Poisson distribution and $\mathcal{G}a(\cdot \mid a,b)$ the gamma distribution with in the shape-rate parametrization \citep[see][]{Sim90AR_Gamma}.

To define a measure-valued autoregressive gamma random process, we introduce the noncentral gamma random measure.
 
\begin{definition}[$\NcGP$]
\label{def:NCGP}
Given $c >0$ and two base measures $H$ and $M$ on $\Theta$, a random measure $W$ is said to be a \textit{noncentral gamma process} (or noncentral gamma random measure) of parameters $(H,M,c^{-1})$, written $W \sim \NcGP(H,M,c^{-1})$, if its Laplace functional is
\begin{equation*}
\CL_W(f) = \exp\Big( -\! \int_\Theta \log(1+c f) \: H(\mathrm{d}\theta) -\! \int_\Theta \frac{c f}{1+c f} \: M(\mathrm{d}\theta) \Big), \qquad f \in \BM_+(\Theta).
\end{equation*}
\end{definition}

As we shall see later, a non central gamma process has a hierarchical representation analogous to the scalar representation in \eqref{eq:noncentralgamma}. More precisely, if 
\begin{equation}
    V \sim\PP(V \mid M), \qquad 
    W \mid V \sim\GP(W\mid H+W,c^{-1}),
\label{eq:noncentralgammaMEASURE}
\end{equation}
where $V \sim \PP(V \mid M)$ denotes a \textit{Poisson random measure} (or Poisson point process) $V$ with mean measure (intensity measure) $M$, then $W \sim \NcGP(W \mid H,M,c^{-1})$.
We shall prove this in the next Proposition \ref{proposition:Laplace_Wt1_Wt_lag1}. 

The noncentral gamma random measure $W \sim \NcGP(H,M,c^{-1})$ is the sum of two independent homogeneous {\it completely random measures} \citep[see][ Ch. 8]{Kingman93PoissonProcesses_book}. More precisely, $W=W_A+W_B$, where $W_A$ is a gamma random measure, and $W_B$ is a finite activity compound Poisson-gamma random measure. Both these random measures belong to the class of the so-called $G$-random measures \citep[see][]{Brix99GeneralizedGammaProcess}.
In particular, $W \sim \NcGP(H,M,c^{-1})$ turns out to be a completely random measure with L\'evy measure
\begin{equation}
    \mu(\mathrm{d}w \mathrm{d}\theta) = w^{-1}e^{- w/c} \: \mathrm{d}v H(\mathrm{d}\theta) + c^{-1}e^{-w/c} \: \mathrm{d}w M(\mathrm{d}\theta)
\label{eq:muNCGP}
\end{equation}
and hence it can be represented as a Poisson integral as in eq.~\eqref{eq:propernessNC-gamma}, where $N$ is a Poisson random measure on $\RE_+ \times \Theta$ with mean measure $\mu$ given in eq.~\eqref{eq:muNCGP} \citep[e.g., see][Propositon 12.1]{LastPenrose2018PoissonProcess_book}.
Clearly, for $M=0$, a noncentral gamma random measure reduces to a gamma random measure, that is, $\GP(H,\beta) \equiv \NcGP(H,0,\beta)$.

\begin{definition}[$\MARG(1)$]
\label{def:MARG}
A measure-valued stochastic Markov process $(W_t)_{t \geq 1}$ on a Polish space $\Theta$ is called a \textit{measure-valued auto-regressive gamma process} of order $1$, or $(W_t)_{t \geq 1} \sim \MARG(1)$, if the conditional measure governing the transition $(t,t+1)$ is a noncentral gamma process with parameters $(H,\beta_{t+1}W_{t},c_{t+1})$, that is
\begin{equation*}
W_{t+1} \mid W_t \sim \NcGP(W_{t+1} \mid H, \beta_{t+1} W_t, c_{t+1}^{-1}),
\end{equation*}
where $(\beta_t,c_t)_{t\geq 2}$ is a given sequences of positive real numbers  and $H$ is a locally finite measure on $\Theta$. The initial condition $W_1$ is a (possibly random) locally finite measure on $\Theta$. 
\end{definition}

The parameters $\beta_t$ and $c_t$ drive the M-ARG process temporal persistence and dynamic properties, as shown in Section \ref{sec:properties}. The branching representation of the process (see Proposition~\ref{proposition:ARG-AutoregRepre} and Section~\eqref{prob_deth}) shows that large $\beta_t$ values increase the atom survival probability between any two consecutive periods, and large $c_t$ values favour the increase in the number of new atoms introduced each period.

\begin{example}[M-ARG temporal heterogeneity]\label{ExMARGhetero}
A benchmark specification is the \textit{time-homogenous} M-ARG, where $\beta_t = \beta$ and $c_t = c$ for every $t$. This includes a time-stationary regime specification ($\beta c<1$, see Section \ref{SubSec:stationarity}). Temporal heterogeneity in $(W_t)_{t\geq 1}$ can be addressed in several ways:
\begin{itemize}
\item[(i)] A semi-parametric model, \textit{time-varying} M-ARG, can be obtained by   $\beta_t = \beta$ and $c_t$ freely varying over $t$.
\item[(ii)] A M-ARG with \textit{periodic} structure of period $u$, is obtained for $\beta_t = \beta$ and $c_t = c_{t+u} = c_{t+2u} \ldots$ for every $t$, which captures cyclical patterns in several climatic and environmental phenomena.
\item[(iii)] A \textit{predictors and trend} M-ARG with $g(c_t) = \phi_0 + \phi_1 t + \ldots + \phi_p t^p+\boldsymbol\psi' x_t$, where $x_t\in\RE^m$ are observed predictors and $g:\RE_+\to \RE$ is a suitable invertible link function. This model incorporates external information and long-term temporal evolution.
\end{itemize}
\end{example}
See the illustration in Section~\ref{sec:illustration}.

\subsection{A Dynamic Shot Noise Cox Process}

We can now define the SN-M-ARG as a dynamic version of the Poisson-gamma random field model, where the latent intensity is assumed to evolve over time according to a \MARG(1) process, as in Definition \ref{def:MARG}.

\begin{definition}[$\SNMARG(1)$]
\label{def:SNMARG}
A \textit{shot noise measure-valued auto-regressive gamma process} of order 1, or $\SNMARG(1)$, is the time-varying shot noise process defined as
\begin{align*}
    N_{t} \mid \Lambda_{t} \sim \PP(N_t \mid \Lambda_{t}),
\end{align*}
where $\Lambda_{t}(\mathrm{d}y) = \kappa_t \lambda_{t}(y) \ell(\mathrm{d}y)$ is the random intensity measure, $\lambda_{t}(y) = \int_\Theta K_\phi(y,\theta) W_{t}(\mathrm{d}\theta)$ its density,
$\ell(\mathrm{d}y)$ is a $\sigma$-finite measure on $\Y$, $K_\phi$ is a positive density kernel on $\Y\times\Theta$ with range parameter $\phi\in\Rp$, $(\kappa_t)_{t\geq 1}$ a deterministic process and $(W_{t})_{t\geq 1}$ is a \MARG(1) process of parameters $(\beta_t,c_t)_{t\geq 2}$. 
\end{definition}
For $(N_t)_{t\geq 1}$ to be well-defined, one needs some integrability conditions. We already assumed that $H$ is a locally finite measure.
We shall assume that $\bar W_1(\cdot) = \E(W_1(\cdot))$ is also locally finite and that for every bounded measurable $B \subset \Y$ it holds:
\begin{equation}
\label{eq:integrability2}
    \int_B K_\phi(y,\theta) \ell(\mathrm{d}y) < +\infty \;\; \forall \theta, \quad
    \int_B \int_\Theta K_\phi(y,\theta) \: [H(\mathrm{d}\theta) + \bar W_1(\mathrm{d}\theta)] \ell(\mathrm{d}y) < +\infty.
\end{equation}

\begin{proposition}
\label{proposition:finiteN_t}
If eq.~\eqref{eq:integrability2} holds, then $(N_t)_{t\geq 1}$ is a locally finite counting process. 
\end{proposition}

In spatiotemporal applications, observations are made over finite space-time windows, and local finiteness ensures that each $B\subset\mathbb{Y}$ contains only finitely many events for each $t$, yielding a well-defined likelihood function and a meaningful statistical model. Note that
assumption \eqref{eq:integrability2} is satisfied for all in Examples \ref{Ex1}-\ref{Ex4} provided that $\bar W_1(\Theta)<+\infty$.
The intensity $\Lambda_t$ is related to the family of multiplicative intensities that can be written in the general form  $\Lambda_{t}(\mathrm{d}y) = \tilde{\lambda}_1(t) \tilde{\lambda}_2(y) S(y,t) \mathrm{d}y$, where $S(y,t)$ is a spatiotemporal residual process and $\tilde{\lambda}_i$ ($i=1,2$) are deterministic components. In our construction $\tilde{\lambda}_1(t) = \kappa_t$, $\tilde{\lambda}_2(y) \mathrm{d}y = \ell(\mathrm{d}y)$, and the residual process $S(y,t) = \int_{\Theta} K_\phi(y,\theta) W_t(\mathrm{d}\theta)$ has a simple autoregressive structure well-suited for sequential prediction. This structure guarantees analytical tractability and allows for non-stationarity of the residual process and observed counts. Two examples of nonstationary specifications used in our application is given in the following.

\begin{example}[Temporal heterogeneity]\label{Ex5}
A first specification that allows for time dependence includes a trend and an additional regime:
\begin{equation*}
    \kappa_t = \exp( \eta_0 + \eta_{TR} t + \eta_{D,1} d_t ),
\end{equation*}
where the dummy variable $d_t$ takes the value $1$ whenever $t$ belongs to a specific regime (in the application, a month in the dry season), and $0$ otherwise. A second specification augments the trend component with $M$ harmonic terms:
\begin{equation*}
    \kappa_t = \exp\Big( \eta_0 + \eta_{TR} t +\sum_{m=1}^M (\eta_{S,m}\sin(2\pi\omega_1 t) + \eta_{C,m1}\cos(2\pi\omega_m t)) \Big).
\end{equation*}
\end{example}

In conclusion, the $\SNMARG$ parametrization consists of two components, each controlling a distinct aspect of the spatial heterogeneity and temporal dynamics:
\begin{enumerate}
    \item The parameters associated with the shot noise component are $(K_\phi,H,\ell,\kappa_t)$. Several specifications of $(K_\phi,H,\ell)$ are presented in Examples \ref{Ex1}--\ref{Ex4}. The measure $\ell(\mathrm{d}y)$ can also be specified as a function of spatially varying covariates, thereby incorporating additional sources of spatial heterogeneity. The parameter process $(\kappa_t)_{t\geq 0}$ can capture temporal variation via predictors and deterministic components such as trends, cycles, and seasonality. Several specifications of $\kappa_t$ are given in Example \ref{Ex5}.
    \item The autoregressive parameters of the $\MARG$, namely $(\beta_t,c_t)_t$, discussed in Example \ref{ExMARGhetero}. When the $\MARG$ model is used on its own for statistical modelling, predictors can be incorporated through $c_t$ to explain temporal variation in the marginal distribution. When the $\MARG$ is embedded within the $\SNMARG$ specification, however, care must be taken to avoid identifiability issues. Since temporal variation can be captured through the shot noise parameter $\kappa_t$, the same predictors should not be included simultaneously in both $\kappa_t$ and $c_t$. More generally, the roles of $\kappa_t$ and $(\beta_t,c_t)$ should be clearly separated, with the former capturing variation attributable to observed predictors and the latter driving the process $(W_t)_{t\geq 1}$, which accounts for latent temporal and spatial factors.
\end{enumerate}

\subsection{Stochastic Representations and Simulation}
The $\SNMARG(1)$ process $(N_t)_{t\geq 1}$ admits a state-space representation, which is useful for the derivation of some of its properties and for simulation purposes. This representation is based on an auxiliary measure-valued stochastic process $(V_t)_{t\geq2}$ and builds on eq.~\eqref{eq:noncentralgammaMEASURE}. Starting from the initial condition $W_1$, define a process $(W_t,N_t,V_{t+1})_{t \geq 1}$ as follows:
\begin{equation}
\begin{split}
    N_t \mid W_t & \sim \PP(N_t \mid \Lambda_t), \qquad
    \Lambda_t (dy)= \kappa_t\int_{\Theta} K_\phi(y,\theta) W_t(\mathrm{d}\theta)\ell(\mathrm{d}y) \\[3pt]  
    V_{t+1} \mid W_t & \sim \PP(V_{t+1} \mid \beta_{t+1} W_t), \label{eq:MARG1_model} \\[3pt]
    W_{t+1} \mid V_{t+1} & \sim \GP\big( W_{t+1} \mid H + V_{t+1}, c_{t+1}^{-1} \big),
\end{split}
\end{equation}
where $V \sim \PP(V \mid W)$ denotes a \textit{Poisson random measure} $V$ with mean measure $W$.
The (marginal) distribution of the process $W_t$ defined in eq.~\eqref{eq:MARG1_model} turns out to be a $\MARG(1)$ process, as stated in the next proposition.

\begin{proposition}
\label{proposition:Laplace_Wt1_Wt_lag1}
Let $(W_t,N_t)_{t \geq 1}$ be defined by eq.~\eqref{eq:MARG1_model}. 
Then $W_{t+1} \mid W_t \sim \NcGP \big( W_{t+1} \mid H, \beta_{t+1}W_t, c_{t+1}^{-1} \big)$ and $(N_t)_{t\geq 1}$ is a  $\SNMARG(1)$.
\end{proposition}

\begin{figure}[t!h]
\centering
\captionsetup{width=0.92\linewidth}
\includegraphics{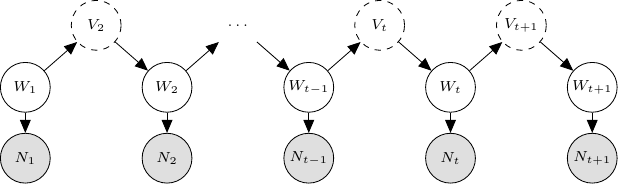}
\caption{Directed acyclic graph of the \SNMARG(1) in Proposition \ref{proposition:Laplace_Wt1_Wt_lag1}. White solid circles represent latent gamma random measures, white dashed circles are auxiliary Poisson random measures, and grey circles represent the observable Poisson point process.}
\label{fig:DAG_model}
\end{figure}

The directed acyclic graph representation of the SN-M-ARG(1) given in  Proposition \ref{proposition:Laplace_Wt1_Wt_lag1} is provided in Figure~\ref{fig:DAG_model}. From Proposition \ref{proposition:Laplace_Wt1_Wt_lag1}, given any measurable set $A$ with $H(A) < +\infty$, the process $\{ W_t(A) \}_{t\geq 1}$ is a scalar-valued autoregressive gamma process in the sense of \cite{Gourieroux06AR_GammaProcess}, that is $W_{t+1}(A) \mid W_{t}(A) \sim \NcGa(H(A),\beta_t,c_t^{-1})$.

The state-space representation in \eqref{eq:MARG1_model} and a suitable choice of the thinning operator allows us to show that the \MARG(1) is an autoregressive measure-valued process. Let us denote with $\NcGa(0,\beta ,c^{-1})$ the limiting case, for $\delta\rightarrow 0$, of the noncentral gamma distribution $\NcGa(\delta,\beta,c^{-1})$, called noncentral gamma-zero distribution \citep[see][]{Monfort17ARG_zeroInflated}. The limiting case $\delta=0$ can be obtained assuming $\mathcal{G}a(0,c^{-1})$ is a Dirac distribution at zero in the representation \eqref{eq:noncentralgamma}.

\begin{proposition}
\label{proposition:ARG-AutoregRepre}
The following autoregressive representation of a $\MARG(1)$ holds:
\begin{align*}
    W_{t+1} \mid W_t \stackrel{\SL}{=} (\beta_{t+1},c_{t+1}) \odot W_t + W^{(I)}_{t+1}
\end{align*}
with $W^{(I)}_{t+1}$ and $W_s$ independent for all $s\leq t$, where $W^{(I)}_{t+1} \sim \GP(W^{(I)}_{t+1} \mid H,c_{t+1}^{-1})$, $W_t=\sum_{i \geq  1} w_{i,t} \delta_{\theta_{i,t}}$ and $\odot$ is a thinning operator defined as
\begin{align*}
    (\beta_{t+1},c_{t+1}) \odot W_t \coloneqq \sum_{i \geq 1 } w_{i,t+1}^{(U)} \delta_{\theta_{i,t}}, \qquad w_{i,t+1}^{(U)} \sim \NcGa(w_{i,t+1}^{(U)} \mid 0,\beta_{t+1} w_{i,t},c_{t+1}^{-1}).
\end{align*}
\end{proposition}

The construction in the Proposition \ref{proposition:ARG-AutoregRepre} exploits a thinning operator since the measure $\NcGP(0,\beta,c^{-1})$ is strictly positive at zero, and the atom $\theta_{i,t}$ in $W_t$ can die with probability
\begin{equation*}\label{prob_deth}
    \pr\{ w_{i,t+1}^{(U)}=0 \mid w_{i,t} \} = 
    e^{-\beta_{t+1} w_{i,t}}. 
\end{equation*}
The transition of the \MARG(1) process decomposes into the sum of finite and infinite activity measures, which are a compound Poisson-gamma random measure, $(\beta_{t+1},c_{t+1}) \odot W_t$, and a gamma random innovation measure, $W_{t+1}^{I}$, respectively.

\begin{remark}\label{rem:kernelconvolution}
Combining the definition of SN-M-ARG(1) in
\eqref{eq:MARG1_model} and Proposition 
\ref{proposition:ARG-AutoregRepre}, one has 
\[
\begin{split}
 \Lambda_{t+1} (dy) & = 
 \kappa_t \sum_{j}
 w_{j,t+1} K_\phi(y,\theta_{j,t+1}) \ell(\mathrm{d}y) \\
 & 
=\kappa_t \Big ( \sum_{j}
 w_{j,t+1}^{(U)} K_\phi(y,\theta_{j,t}) +
 \sum_{j}
 w_{j,t+1}^{(I)} K_\phi(y,\theta_{j,t+1}^{(I)}) 
 \Big) 
 \ell(\mathrm{d}y).
 \\
 \end{split}
\]
Hence, the spatial intensity is a convolution of kernels, where atoms $\theta_{j,t+1}$ and $\theta_{j,t+1}^{(I)}$ indicate latent intensity centers and weights $w_{j,t+1}^{(U)}$ and $w_{j,t+1}^{(I)}$ determine their contribution to the intensity surface. The above decomposition highlights the two mechanisms driving the process dynamics: the first term propagates information from the past through surviving atoms, whereas the second term represents the innovation component generated by newly introduced atoms.
\end{remark}

\begin{remark}[$\MARG(p)$]
Definitions \ref{def:MARG} and \ref{def:SNMARG} and Propositions~\ref{proposition:Laplace_Wt1_Wt_lag1} and~\ref{proposition:ARG-AutoregRepre} can be extended by introducing a more general process M-ARG  of order $p \geq 1$ (M-ARG$(p)$), with the following conditional measure governing the transition
\begin{equation*}
    W_{t+1} \mid W_t,\ldots,W_{t+1-p} \sim \NcGP\Big( W_{t+1} \mid H, \sum_{j=1}^p \beta_{j,t+1} W_{t+1-j}, c_{t+1}^{-1} \Big).
\end{equation*}
\end{remark}

Building on the results in this Section, a simulation method for $(N_t)_{t\geq 1}$ can be derived by extending the ILM algorithm of \cite{Wolpert1998PoissonGamma_RandomField}. 
See the Supplement for a detailed description of the ILM algorithm. 

We describe a stochastic simulation algorithm to generate realisations of an \SNMARG(1) process when $\Theta$ is a bounded subset of $\RE^D$, the base measure is $H(\mathrm{d}\theta) = \alpha \mathrm{d}\theta$ and $\kappa_t = 1$.
Given  an initial measure $W_1(\mathrm{d}\theta) = \sum_{j=1}^M w_{1,j} \delta_{\theta_{j,1}}(\mathrm{d}\theta)$ on $\Theta$, and a truncation parameter $M>0$, we simulate from a \SNMARG(1) process, by iterating the following steps for $t=2,\ldots,T$.
    \begin{enumerate}
    \item Generate $v_{i,t} \mid w_{i,t-1} \sim \mathcal{P}oi(\beta w_{i,t-1})$ and compute the index set of the atoms with nonzero weights, $\mathcal{J}_t = \{i=1,\ldots, M, \, \text{s.t.} \, v_{i,t}>0\}$. Then, set $\theta_{i,t} = \theta_{j_i,t-1}$ for $j_i\in\mathcal{J}_t$ and $i=1,\ldots,|\mathcal{J}_t|$.
    The resulting collection of pairs $\{ (\theta_{i,t},v_{i,t}) \}_{i=1}^{|\mathcal{J}_t|}$ represents a sample from $V_{t} \mid W_{t-1}  \sim \PP(V_{t} \mid \beta_{t} W_{t-1})$.
    
    \item From eq.~\eqref{eq:MARG1_model} and Proposition~\ref{proposition:Laplace_Wt1_Wt_lag1}, the random measure $W_t \mid V_t$ can be decomposed as $W_t \stackrel{\mathscr{L}}{=} W_t^{(U)} + W_t^{(I)}$, where $W_t^{(U)} = \sum_{i=1}^{|\mathcal{J}_t|} w_{i,t} \delta_{\theta_{i,t}}$ is a sample from $\GP(W_t^{(U)} \mid V_t, c^{-1})$ with $w_{i,t} \mid v_{i,t}$ drawn independently from $\mathcal{G}a(\alpha+v_{i,t},c^{-1})$ and $W_t^{(I)}$ is a sample from $\GP(W_t^{(I)} \mid H, c^{-1})$ generated by the ILM algorithm. The ILM generates $M-|\mathcal{J}_t|$ atoms $\theta_{j,t}$ and the associated weights $w_{j,t}$, $j=|\mathcal{J}_t|+1,\ldots,M$.
    The resulting collection of pairs $\{(\theta_i,w_{i,t})\}_{i=1}^M$ represents an approximation to $W_t$.
    
    \item Compute $\Lambda_t(\Y) = \sum_{i=1}^M w_{i,t} \int_\Y K_\phi(y, \theta_{i,t}) \mathrm{d}y$, and generate the number of observations $N^y_t \sim \mathcal{P}oi(\Lambda_t(\Y))$.
    
    \item Compute the normalised weights $\tilde{w}_{i,t} = w_{i,t} \int_\Y K_\phi(y, \theta_{i,t}) \mathrm{d}y / \Lambda_t(\Y)$, $i=1,\ldots,M$, and randomly generate the allocation variable $z_{j,t} \sim \text{Cat}(\mathbf{\tilde{w}}_t)$, for $j=1,\ldots,N_t^y$, where $\text{Cat}(\mathbf{w})$ denotes the categorical distribution with probability vector $\mathbf{w}$.
    \item  Generate independent observations  $y_{j,t} \sim K_\phi(y,\theta_{z_{j,t}})$, for $j=1,\ldots,N_t^y$.
\end{enumerate}

\begin{figure}[tp]
\centering
\hspace{-1cm}
\setlength{\tabcolsep}{2pt}
\renewcommand{\arraystretch}{0.65}
\begin{tabular}{cccc}
& {\footnotesize High persistence, Long Range}  & {\footnotesize Low persistence, Long Range} & {\footnotesize High persistence, Short Range}\\
& {\footnotesize ($\beta c = 0.82$, $\alpha=0.02$, $\phi = 2.2$)}  & {\footnotesize ($\beta c = 0.12$, $\alpha=0.12$, $\phi = 2.2$)} & {\footnotesize ($\beta c = 0.82$, $\alpha=0.02$, $\phi = 1.1$)}\\
\begin{rotate}{90} \hspace{38pt} {\footnotesize Time $t$}\end{rotate} \hspace{-10pt} &
\includegraphics[clip,trim=0 20 75 30, scale=0.31]{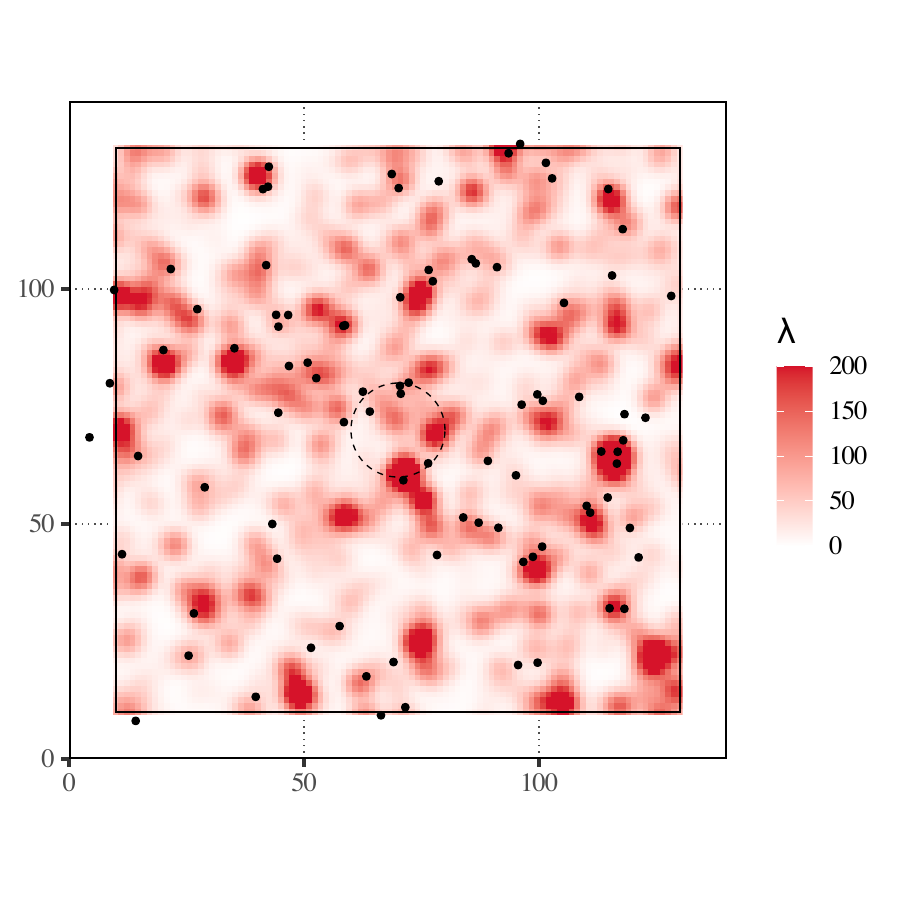} &
\includegraphics[clip,trim=0 20 75 30, scale=0.31]{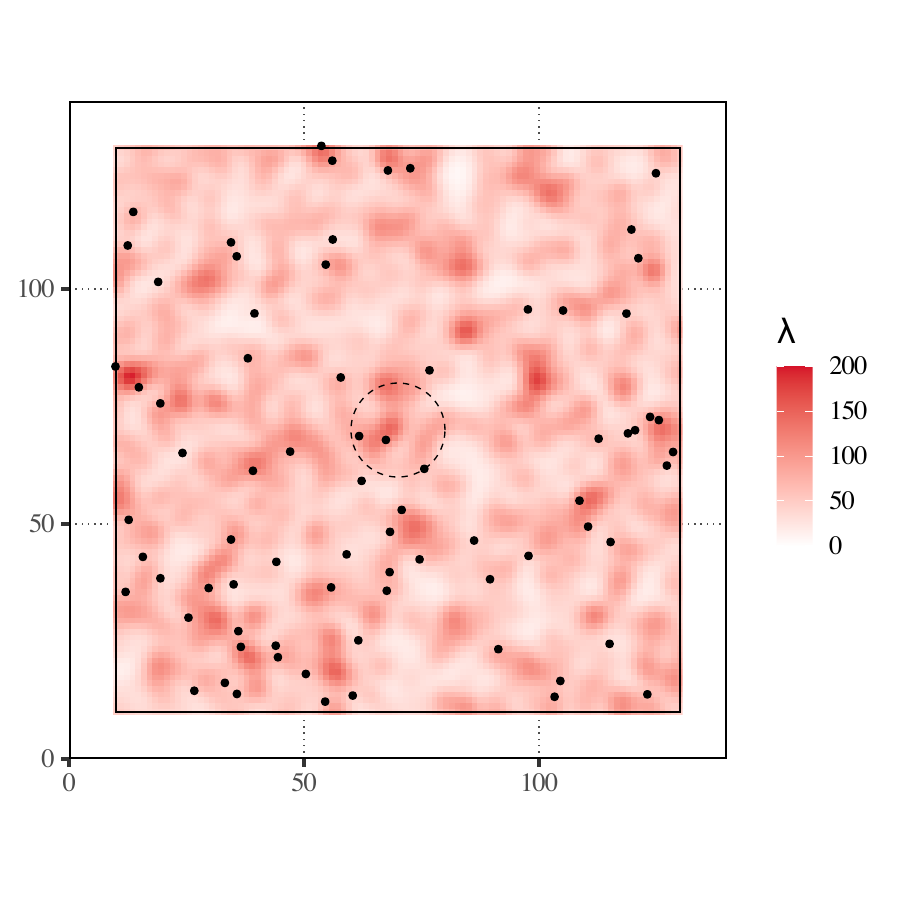}&
\includegraphics[clip,trim=0 20 75 30, scale=0.31]{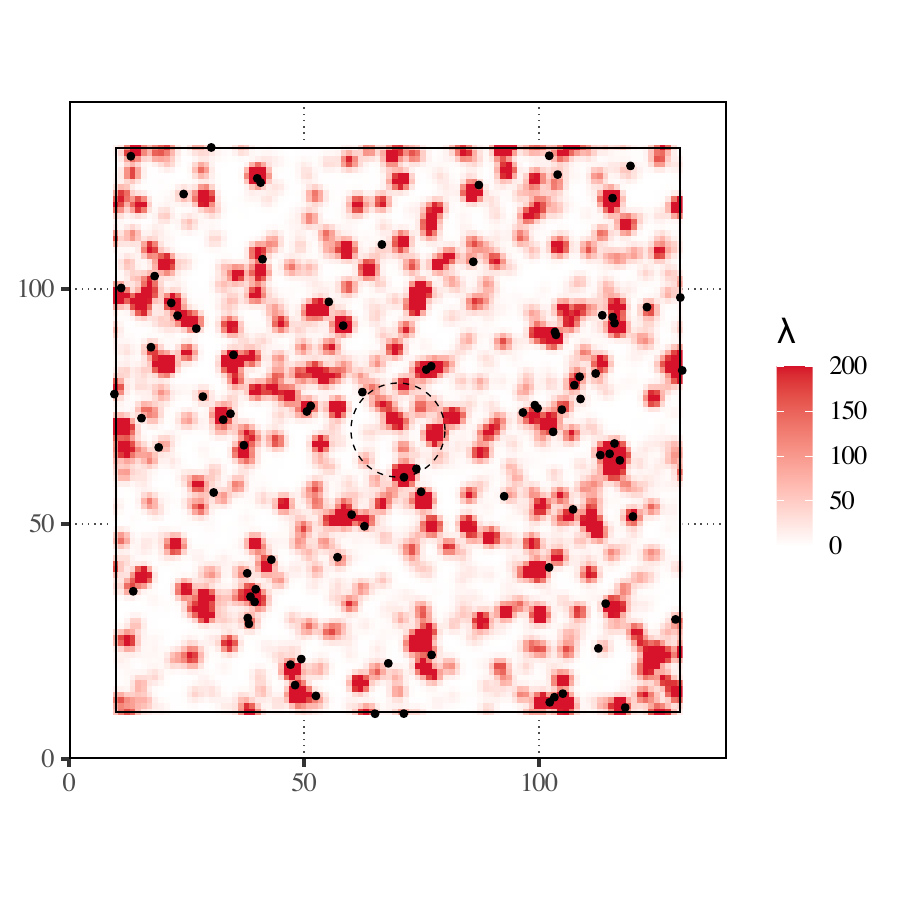} \\[-12pt]
\begin{rotate}{90} \hspace{38pt} {\footnotesize Time $t+1$}\end{rotate} \hspace{-10pt} &
\includegraphics[clip,trim=0 20 75 30, scale=0.31]{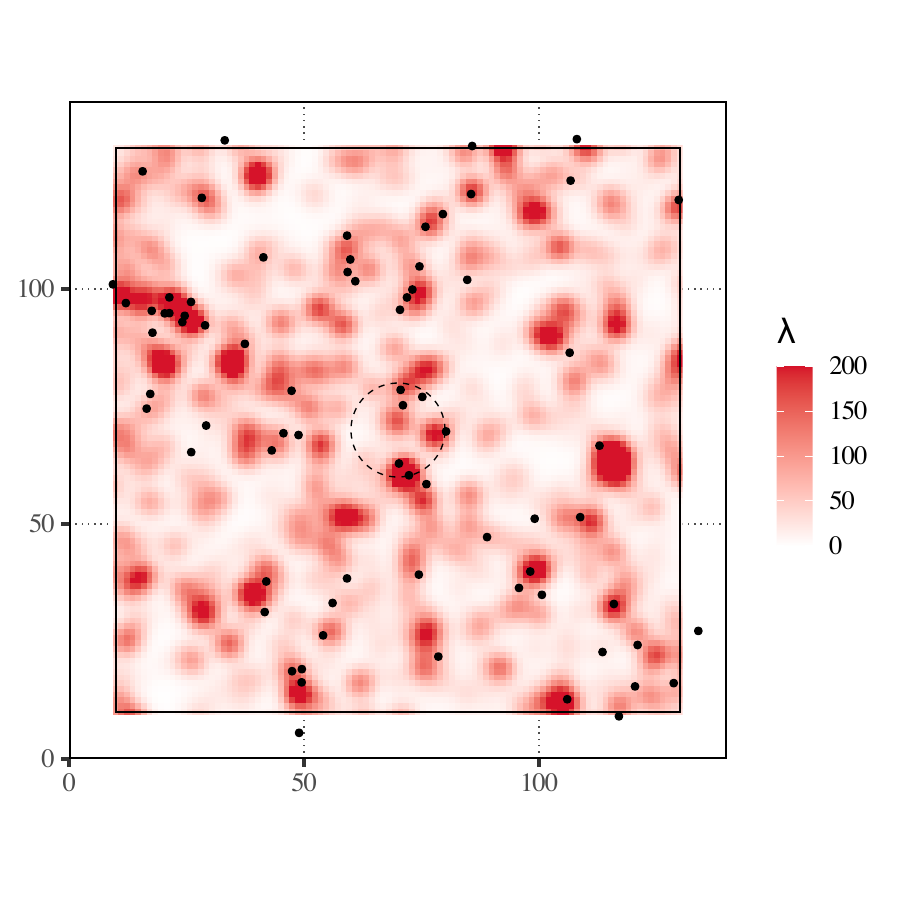} &
\includegraphics[clip,trim=0 20 75 30, scale=0.31]{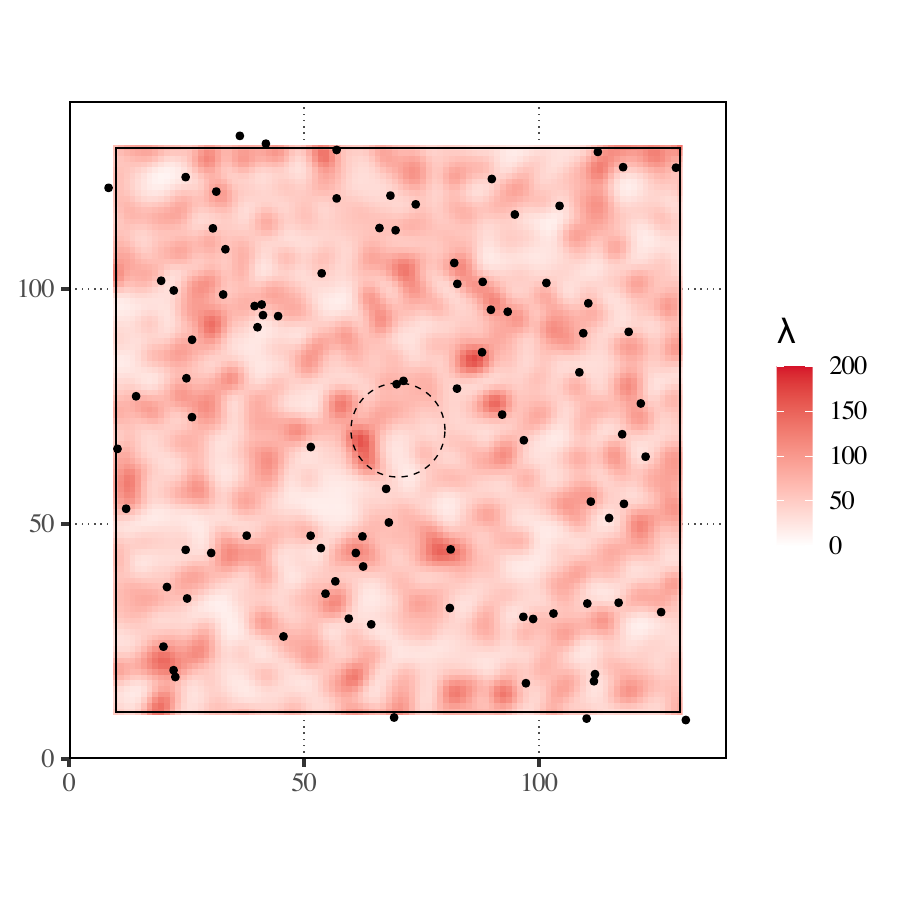} &
\includegraphics[clip,trim=0 20 75 30, scale=0.31]{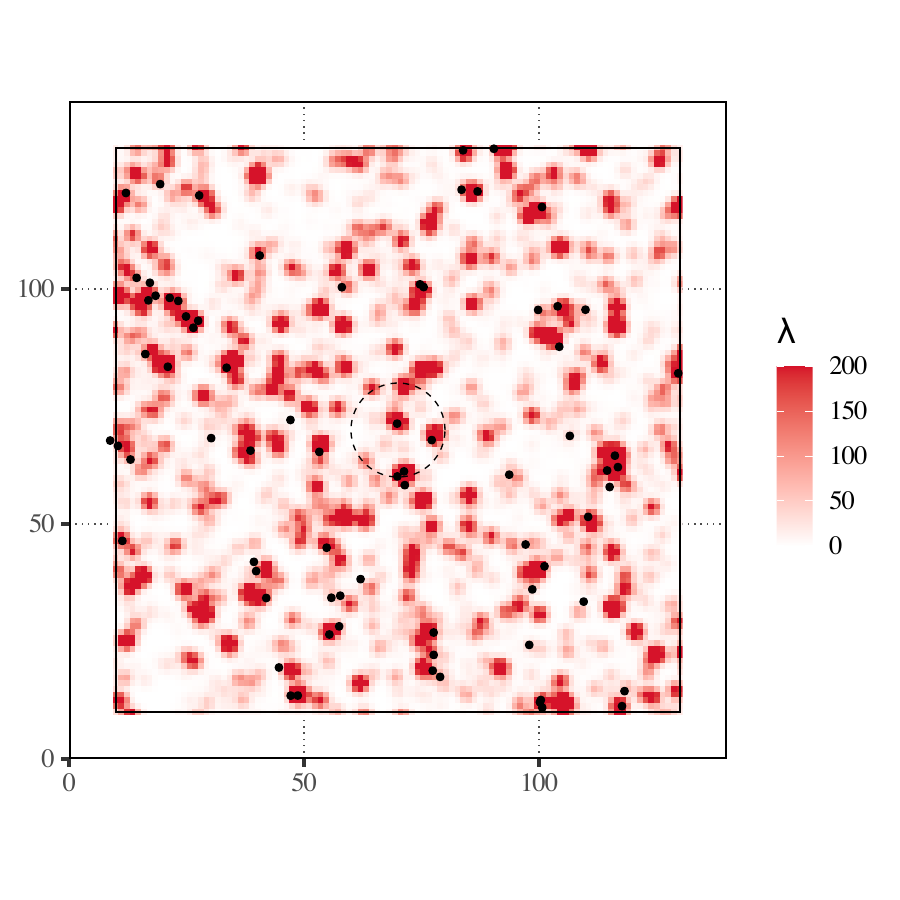} \\[-12pt]
\begin{rotate}{90} \hspace{38pt} {\footnotesize Time $t+2$}\end{rotate} \hspace{-10pt} &
\includegraphics[clip,trim=0 20 75 30, scale=0.31]{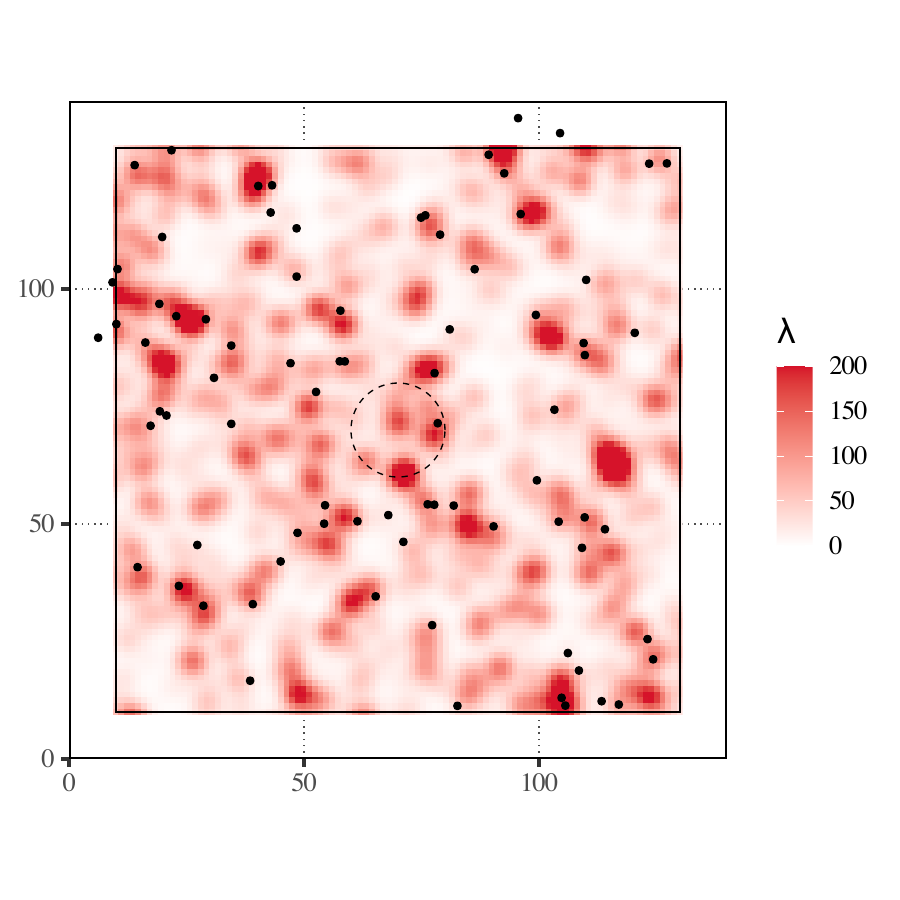}  &
\includegraphics[clip,trim=0 20 75 30, scale=0.31]{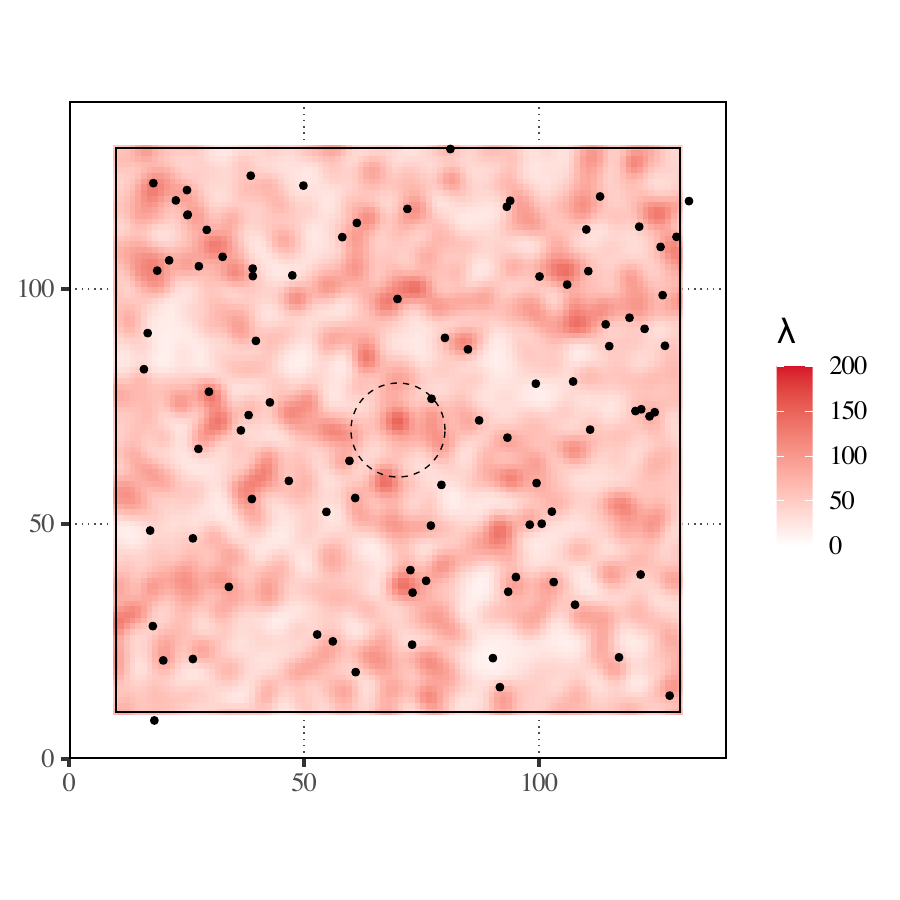}&
\includegraphics[clip,trim=0 20 75 30, scale=0.31]{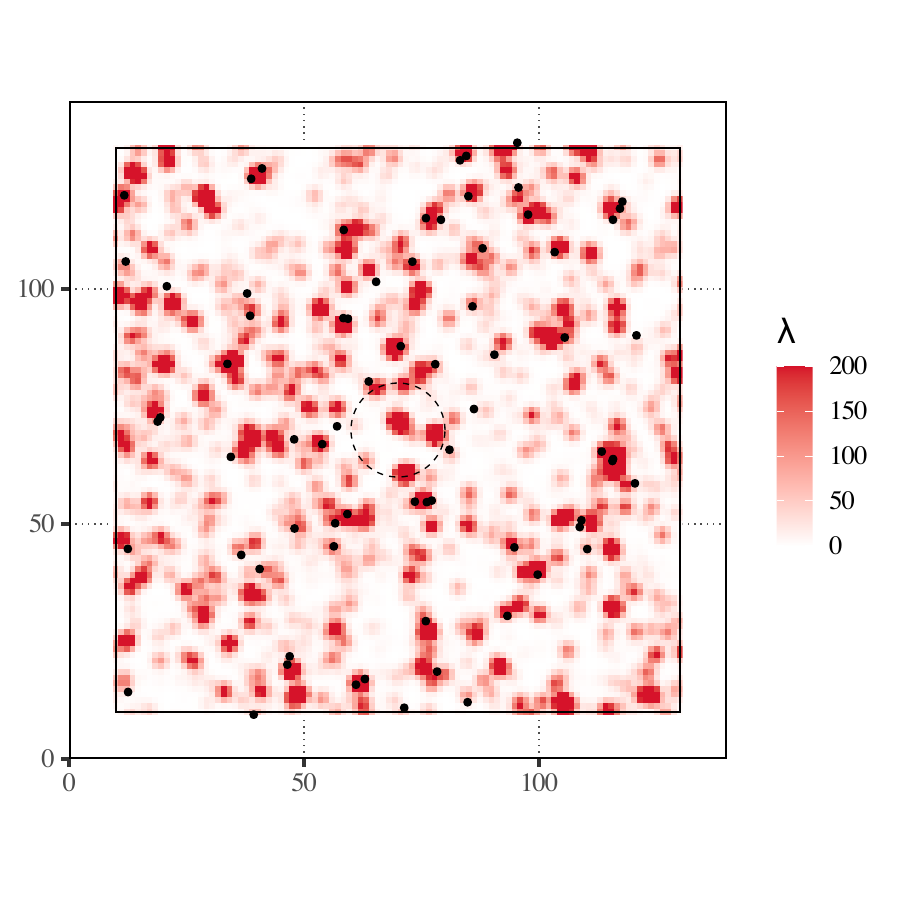} \\[-12pt]
\begin{rotate}{90} \hspace{38pt} {\footnotesize Time $t+3$}\end{rotate} \hspace{-10pt} &
\includegraphics[clip,trim=0 20 75 30, scale=0.31]{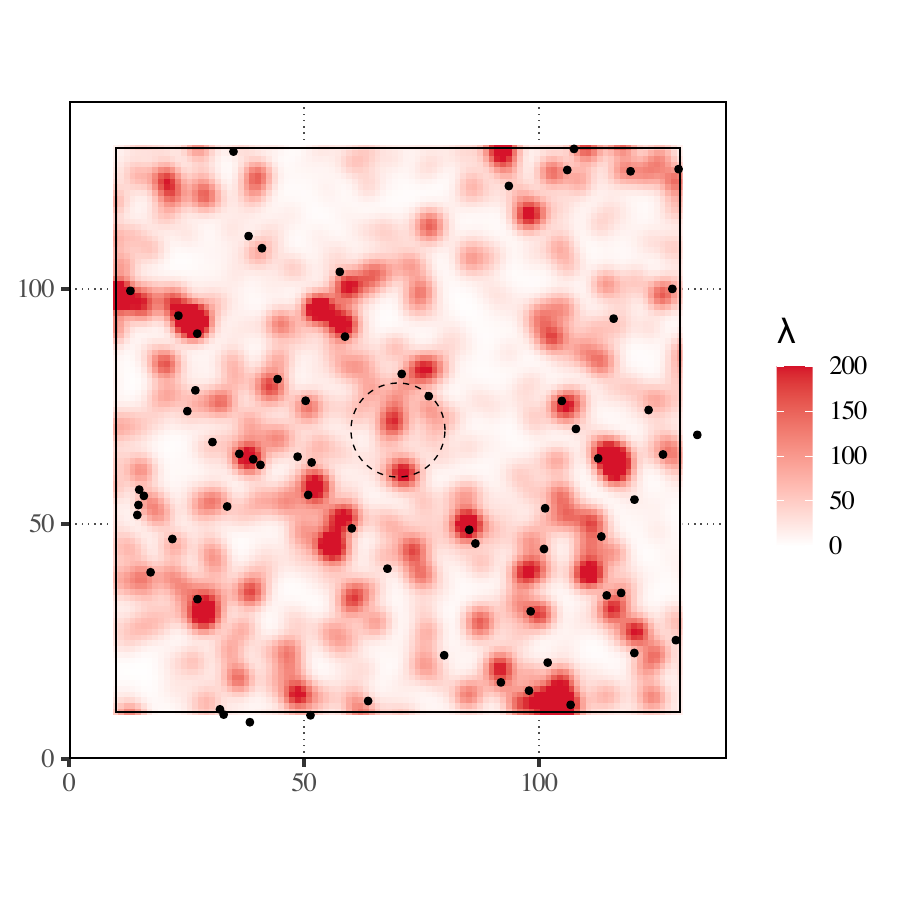} &
\includegraphics[clip,trim=0 20 75 30, scale=0.31]{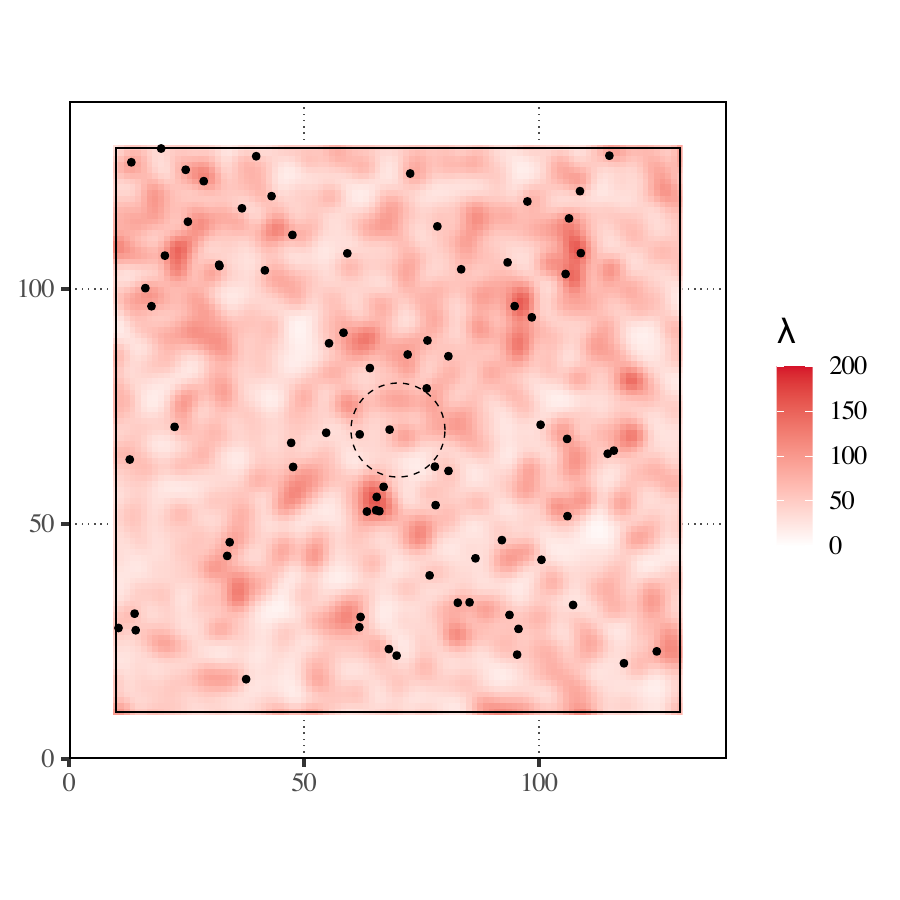}&
\includegraphics[clip,trim=0 20 75 30, scale=0.31]{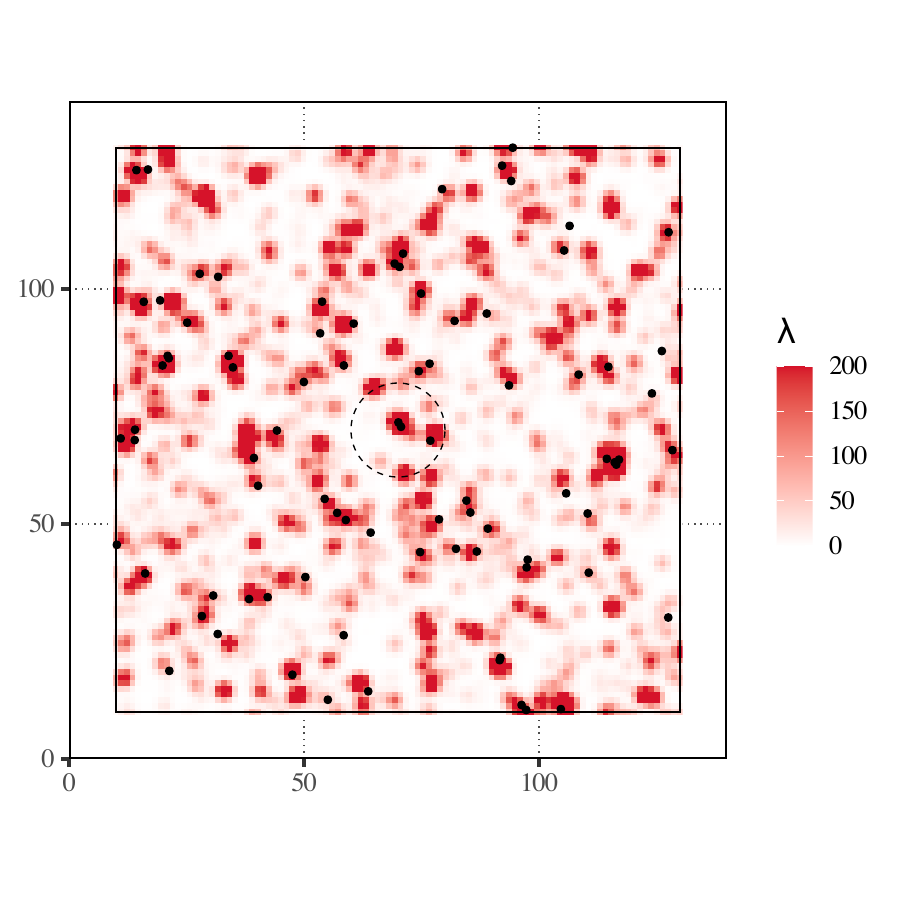} 
\end{tabular}
\captionsetup{width=0.95\textwidth}
\caption{Simulated points ($\bullet$) on $\mathbb{Y}\subset [0,140]^2$ from an homogenous \SNMARG(1) with intensity $\Lambda_t(y)$ (shaded areas), $H(\mathrm{d}\theta)=\alpha \mathrm{d}\theta$, $\Theta=\mathbb{Y}$ in the high-persistence ($\beta c = 0.82$), and 
low-persistence settings ($\beta c = 0.12$) for
low and high global intensity $\alpha \in \{0.02, 0.12\}$, and long and short range $\phi \in \{1.1, 2.2\}$. In each plot: dark red for high intensity, white for null intensity, and a reference dashed circle for comparing the smoothing effects.}
\label{fig:simulation}
\end{figure}

Figure~\ref{fig:simulation} shows the data (dots) and latent intensity (shades) simulated from a \SNMARG(1) model with an isotropic Gaussian kernel and range parameter $\phi\in\{1.1,2.2\}$ for four consecutive periods. The top row stems from a combination of the static parameters, yielding higher process persistence ($\beta c = 0.82$) with long-range spatial dependence ($\phi=2.2$, left column) and short-range dependence ($\phi=1.1$, third column). This results in several peaks (darker shades) in the latent intensity, whose locations remain stable over time due to the atom's high survival probability in the gamma measure process (e.g., see the persistence of dark red shades within the reference circle). Conversely, the low persistence setting with long-range dependence ($\beta c = 0.12$, middle column) is characterised by a latent intensity more uniformly distributed in space and time. Locations of high intensity at two consecutive times are typically different since the atom's survival probability is low and new atoms appear between consecutive time instants. The distribution of intensity over space and time determines the location of the data (black dots), which are more clustered around high-intensity regions (left and right) or randomly scattered across space (middle). The clustering behavior is due to the kernel convolution given in Remark \ref{rem:kernelconvolution}. We refer to the Supplement for additional simulations illustrating the effects of the main parameters.

\subsection{Comparison with Alternative Processes}
The time-dependent family of random measures introduced in \cite{Palla16BNP_Dynamic_SparseGraph} and our \MARG\ process share a common special case. More precisely, the sequence of random measures considered in \cite{Palla16BNP_Dynamic_SparseGraph} is based on a two-component autoregressive construction of the form described in Proposition~\ref{proposition:ARG-AutoregRepre}, with innovations belonging to the Generalized Gamma process (GG) family \citep{Hougaard1986,Brix99GeneralizedGammaProcess,Caron17SparsityPowerLaw_Exchangeableraphs}. The model is parametrized by $(\alpha,\sigma,\tau,\phi)$ and uses innovations distributed as a GG process $\mathcal{GG}(\alpha,\sigma,\tau+\phi)$. When the sparsity parameter satisfies $\sigma=0$, the model reduces to a time-homogeneous \MARG(1) process with
\begin{equation*}
W_1 \sim \GP(W_1 \mid H,\tau),
\qquad
\beta_{t+1}=\phi,
\qquad
c_{t+1}^{-1}=\phi+\tau,
\end{equation*}
for all $t\ge 1$, and base measure $H(\mathrm{d}x) = \mathbb{I}_{[0,\alpha]} \mathrm{d}x$. Since the parameters are constant and satisfy $\beta_t c_t=\frac{\phi}{\phi+\tau}<1$, the process is stationary with marginal distribution $\GP(W_t\mid H,\tau)$ for every $t\ge1$, see next Proposition \ref{proposition:invariant_Wth_Wt_lag1}.

The time-varying GG process is applied by \cite{Palla16BNP_Dynamic_SparseGraph} to modelling sociabilities in sparse time-varying networks, where our \MARG\ is used to define a new time-varying shot noise process. In addition, while \cite{Palla16BNP_Dynamic_SparseGraph} focuses primarily on the construction and inference aspects of the model, the general \MARG\ framework developed here allows for a more detailed probabilistic analysis. In particular, Propositions~\ref{proposition:laplace_Wth_Wt_lag1} and \ref{proposition:invariant_Wth_Wt_lag1} provide the conditional Laplace functional in the general time-inhomogeneous setting and sufficient conditions for the existence of a stationary distribution in the time-homogeneous \MARG(1) case. These results form the basis for deriving higher-order moment properties, including spatial and temporal covariance structures and predictive distributions, which play a central role in spatio-temporal forecasting applications.

A benchmark process in spatiotemporal statistics is the log-Gaussian Cox process. A discrete-time log-Gaussian Cox process (LGCP) can be defined as
\begin{equation}
N_t \mid Y_t \sim \PP(N_t \mid \Lambda_t),
\qquad
\Lambda_t(dy) = \exp(Y_t(y)) \, \mathrm{d}y,
\end{equation}
where $Y_t(y)$ is a latent Gaussian random field, or Gaussian Process (GP), on a continuous spatial domain $\Y$. The latent log-intensity evolves according to the Gaussian autoregressive dynamics
\begin{equation}
    Y_{t}(y) = \mu(y) + \rho\{Y_{t-1}(y)-\mu(y)\} + \varepsilon_{t}(y), \qquad |\rho|<1,
\end{equation}
where $\{\varepsilon_t\}_{t\ge 1}$ is a sequence of independent GPs  with covariance kernel $C_\varepsilon(y,y')$ \citep{Diggle2013}. In an LGCP, both temporal and spatial dependence are induced by the latent GP. In contrast, in the \SNMARG\ model, temporal dependence arises through thinning and innovations, while spatial dependence is determined by the kernel (see Section \ref{sec:properties}). The latent GP makes the LGCP flexible in capturing complex spatial and temporal dependence structures, but inference is often computationally demanding due to large covariance matrices. By contrast, the \SNMARG\ model is computationally more tractable because it relies on independent spatial increments of the \MARG\ and avoids modelling the covariance function. The practical trade-off between flexibility and parsimony, as well as the resulting goodness of fit and computational cost, depends on the application and are examined in the Supplement.

\section{Moments, Covariances, Limit Distributions}
\label{sec:properties}

To investigate how the proposed model accounts for key features of spatiotemporal data, this section presents and discusses the statistical properties of the \MARG\ and \SNMARG\ processes, focusing on spatial and temporal dependence.

\subsection{Conditional Laplace Functionals and Mean for \texorpdfstring{$(W_t)_{t \geq 1}$}{Wt}}

The conditional Laplace functional of a $\MARG(1)$ process at any lag $h \geq 1$ has the appealing feature of log-linear in $W_t$. The exponential-affine form is appealing because conditional moment measures and prediction remain analytically tractable despite the infinite-dimensional and infinite-activity nature of the process. For ease of notation, for every $t\geq 1$ and $h\geq 1$ let $\rho_{t+1} \coloneqq \beta_{t+1} c_{t+1}$ and define
\begin{align}
    \rho_{t+h|t} \coloneqq \prod_{j=t+1}^{t+h} \rho_j, \quad
    c_{t+h|t} \coloneqq c_{t+h} + \sum_{j=t+1}^{t+h-1} c_j \Big( \prod_{i=j+1}^{t+h} \rho_i \Big), \quad
    \beta_{t+h|t} \coloneqq \rho_{t+h|t} c_{t+h|t}^{-1},
\label{eq:parameters_lag_h}
\end{align}
where we use the convention $c_{t+1|t} = c_{t+1}$. Finally, if $(M,N)$ is a vector of random measures, write $\mathcal{L}_{M \mid N}(f) \coloneqq \E\big( e^{-\int f \: \mathrm{d}M} \mid N \big)$ for the conditional Laplace functional.

\begin{proposition}
\label{proposition:laplace_Wth_Wt_lag1}
For any $f \in \BM_+(\Theta)$, $t\geq 1$ and $h\geq 1$
\begin{equation*}
    \CL_{W_{t+h} \mid W_t}(f) = \exp\Big( - \!\! \int_{\Theta} \log(1+c_{t+h|t} f(\theta)) \: H(\mathrm{d}\theta) - \! \int_{\Theta} \dfrac{\rho_{t+h|t} f(\theta)}{1+c_{t+h|t} f(\theta)} \: W_t(\mathrm{d}\theta) \Big).
\end{equation*}
Therefore, the conditional law of $W_{t+h}$ given $W_t$ is the same as a $\NcGP\big( H, \beta_{t+h|t}W_t,c_{t+h|t}^{-1} \big)$.
\end{proposition}

As a corollary of the previous proposition, one can compute the conditional mean measure at any lag $h \geq 1$, which is linear in the past realisations of the process.

\begin{corollary}[Conditional mean measures]
\label{corollary:expected_distribution_Wt1_Wt_lagh}
For $t\geq 1$, $h \geq 1$ and any measurable set $A\subset\Theta$
\begin{equation*}
\E\big( W_{t+h}(A) \mid W_t \big) = c_{t+h|t} H(A) + \rho_{t+h|t} W_t(A).
\end{equation*}
\end{corollary}
The first term represents the expected contribution of newly generated mass. $H$ determines where this new mass is allocated across $\Theta$, while $c_{t+h|t}$ determines its magnitude. In the second term, the coefficient $\rho_{t+h|t}$ determines the temporal persistence of the current mass $W_t(A)$.

The results in this section are reminiscent of those proved for a continuous-time gamma process in \cite{ethier1993transition} and \cite{etheridge2000introduction}.
In particular, the Laplace functional of the transition function given in Lemma 2.2 of \cite{ethier1993transition} coincides with the one of a  $\NcGP$ with suitable parameters. 
However, it is important to note that our model is not a straightforward time-discretization of the continuous-time gamma process in \cite{ethier1993transition}. 
Specifically, the finite-dimensional distributions of the continuous-time process are a special case of our process only for time-independent $\rho_t=\rho$ and a particular choice of $c_t=c$. 

\subsection{Stationary Distributions}\label{SubSec:stationarity}

Provided that the coefficients of the \MARG(1) process do not depend on $t$ and satisfy a suitable condition, the process admits a gamma limiting distribution and a stationary version in the gamma process family.

\begin{proposition}[Stationary and limiting for $(W_t)_{t}$] 
\label{proposition:invariant_Wth_Wt_lag1}
If $c_{t+1} = c$ and $\beta_{t+1} = \beta$ for all $t\geq 1$ and $\rho = \beta c < 1$, then
\begin{equation*}
\lim_{h\to +\infty} \CL_{W_{t+h} \mid W_t}(f)
 = \exp\Big( - \!\! \int_{\Theta} \log\Big( 1+\dfrac{c}{1-\rho} f(\theta) \Big) \: H(\mathrm{d}\theta) \Big), \quad f\in \BM_+(\Theta).
\end{equation*}
Moreover, the limiting distribution is also invariant, that is, if $W_t \sim \GP\big( W_t \mid H, (1-\rho)/c \big)$ then $W_{t+h} \sim \GP\big( W_{t+h} \mid H, (1-\rho)/c \big)$, for every $h \geq 1$.
\end{proposition}

Similarly, one obtains a limiting process and a stationary distribution for the shot noise process $(N_t)_{t \geq 1}$, which is a Poisson-gamma random field \citep{Wolpert1998PoissonGamma_RandomField}.

\begin{proposition}[Stationary and limiting for $(N_t)_t$]
\label{proposition:statN}
Assume eq.~\eqref{eq:integrability2} holds. Let $c_{t+1} = c$, $\beta_{t+1} = \beta$, $\kappa_t=1$ for all $t\geq 1$ and $\rho = \beta c < 1$.
Then the limiting law of $(N_t)_{t\geq 1}$ is the one of a Poisson-gamma random field defined in eq.~\eqref{eq:WOLPERT_model}, with $(1-\rho)/c$ in place of $c$. Equivalently, for every $f$ in $\BM_+(\Y)$ one has $\lim_{t \to +\infty} \CL_{N_t}(f) = \CL_{N_\infty}(f)$, where
\begin{equation*}
    \CL_{N_\infty}(f) = \exp\Big( - \!\! \int_\Theta \log\Big( 1+\dfrac{c}{1-\rho} \int_{\Y} \big( 1-e^{-f(y)} \big) K_\phi(y,\theta) \: \ell(\mathrm{d}y) \Big) \: H(\mathrm{d}\theta) \Big). 
\end{equation*}
Moreover, if $W_1\sim \GP( W_1 \mid H, (1-\rho)/c )$, then the process $(N_t)_{t\geq 1}$ is stationary in the sense that $N_t \stackrel{\SL}{=} N_\infty$ for every $t$, where $N_\infty$ has Laplace functional $\CL_{N_\infty}$.
\end{proposition}

\subsection{Moments and Covariances}
We now study first- and second-order moment measures (both in space and time) of the \MARG(1) and \SNMARG(1) $(W_t)_{t \geq 1}$ and $(N_t)_{t \geq 1}$, respectively, by exploiting the properties of completely random measures and the results in Corollary~\ref{corollary:expected_distribution_Wt1_Wt_lagh}.

\begin{proposition}[Moments of $(W_t)_t$]
\label{proposition:covWtg}
Let $g_1,g_2$ be measurable functions on $\Theta$. Define $\bar W_1(\cdot) \coloneqq \E(W_1(\cdot))$ and $W_t(g_i) = \int_\Theta g_i(\theta) W_t(\mathrm{d}\theta)$, with $i=1,2$. Then for $t\geq 1$
\begin{equation}
\label{eq:meanWg}
    \E(W_t(g_1)) = c_{t|1} \int_\Theta g_1(\theta) H(\mathrm{d}\theta) + \rho_{t|1} \int_\Theta g_1(\theta) \bar W_1(\mathrm{d}\theta),
\end{equation}
\begin{equation}
\label{eq:covWglag0}
\begin{split} 
    \Cov(W_t(g_1),W_t(g_2)) & = c_{t|1}^2 \int_\Theta g_1(\theta)g_2(\theta) H(\mathrm{d}\theta) + 2\rho_{t|1}c_{t|1}\int_\Theta g_1(\theta)g_2(\theta) \bar W_1(\mathrm{d}\theta) \\ 
     & \quad +\rho_{t|1}^2 \Cov(W_1(g_1),W_1(g_2)),
\end{split}
\end{equation}
provided the integrals on the right-hand side are well-defined and 
\begin{equation}
\label{eq:covWglagh}
    \Cov(W_t(g_1),W_{t+h}(g_2)) = \rho_{t+h|t} \Cov(W_t(g_1), W_t(g_2)),
\end{equation}
where $c_{t|1}$, $\rho_{t|1}$ and $\rho_{t+h|t}$ are defined in eq.~\eqref{eq:parameters_lag_h} with the convention $c_{1|1}=0$ and $\rho_{1|1}=1$.
\end{proposition}

\begin{remark}
As a special case of Proposition~\ref{proposition:covWtg}, taking  $g_1(\theta) = \I\{\theta \in A_1\}$ and $g_2(\theta) = \I\{\theta \in A_2\}$, one obtains expressions for $\Cov(W_t(A_1),W_{t+h}(A_2))$ with $h\geq 0$.
Under the assumptions of Proposition~\ref{proposition:invariant_Wth_Wt_lag1}, when the process is time stationary, then $W_t \sim \GP\big( W_t \mid H, (1-\rho)/c \big)$, $\rho_{t+h|t} = \rho^h$, $W_t(A) \sim \mathcal{G}a(H(A),(1-\rho)/c)$ and from eq.~\eqref{eq:covWglagh} it follows
\begin{equation*}
\Cov(W_t(A),W_{t+h}(A)) = \rho^h \Cov(W_t(A),W_{t}(A)) = \rho^h \Var(W_t(A))=\rho^h  H(A) \frac{c^2}{(1-\rho)^2}.
\end{equation*}
In particular, for any measurable $A \subseteq \Theta$, $\Cor(W_t(A),W_{t+h}(A)) = \rho^h$.
\end{remark}

Following Definitions 4.2, 4.3, and 4.8 in \cite{moller2003statistical}, the first- and second-order moment statistics of $(N_t)_{t \geq 1}$ can be derived from the intensity $\mathfrak{D}^{(1)}_{t}(y)$, the second-order product density $\mathfrak{D}_{t}^{(2)}(y_1,y_2)$ and the cross second-order pair intensity $\mathfrak{D}^{(2)}_{t,t+h}(y_1,y_2)$, which are positive functions such that for bounded measurable $B_1,B_2\subset\Y$:
\begin{equation}
\label{eq:densitiestomoments}
\begin{split}
    \E(N_t(B_1)) & = \int_{B_1} \mathfrak{D}^{(1)}_{t}(y) \ell(\mathrm{d}y), \\
    \E(N_t(B_1)N_t(B_2)) & = \int_{B_1 \cap B_2} \!\! \mathfrak{D}^{(1)}_{t}(y) \ell(\mathrm{d}y)+ \int_{B_1 \times B_2} \!\! \mathfrak{D}_{t}^{(2)}(y_1,y_2)\ell(\mathrm{d}y_1)\ell(\mathrm{d}y_2), \\
    \E(N_{t}(B_1)N_{t+h}(B_2)) & = \int_{B_1 \times B_2} \!\! \mathfrak{D}_{t,t+h}^{(2)}(y_1,y_2) \ell(\mathrm{d}y_1)\ell(\mathrm{d}y_2).
\end{split}
\end{equation}
In our shot noise process, conditionally on $(W_t,W_{t+h})$, $(N_t,N_{t+h})$ are independent Poisson random measures with random mean measures $\Lambda_t(\mathrm{d}y) = \kappa_t \lambda_t(y) \ell(\mathrm{d}y)$ and $\Lambda_{t+h}(\mathrm{d}y) = \kappa_{t+h} \lambda_{t+h}(y) \ell(\mathrm{d}y)$, respectively, where $\lambda_{t}(y)$ is given in Definition~\ref{def:SNMARG}.
Combining this observation with the fact that $\E(N(B_1)) = \Lambda(B_1)$ and $\E(N(B_1)N(B_2)) = \Lambda(B_1\cap B_2) + \Lambda(B_1)\Lambda(B_2)$, for $N \sim \PP(N \mid \Lambda)$ \citep[][eq. 4.26]{LastPenrose2018PoissonProcess_book}, it follows: 
\begin{equation}
\label{eq:densitytolambda}
\begin{split}
 & \mathfrak{D}^{(1)}_{t}(y) = \kappa_t \E(\lambda_t(y)), \qquad \mathfrak{D}_{t}^{(2)}(y_1,y_2) = \kappa_t^2 \E(\lambda_t(y_1)\lambda_t(y_2)), \\
 & \mathfrak{D}^{(2)}_{t,t+h}(y_1,y_2) = \kappa_t \kappa_{t+h} \E(\lambda_t(y_1)\lambda_{t+h}(y_2)).
\end{split}
\end{equation}
For dependent sequences of shot noise processes, it is usually difficult to find tractable expressions for the intensity $\mathfrak{D}^{(1)}$ and the correlation densities $\mathfrak{D}^{(2)}_{t}$ and $\mathfrak{D}^{(2)}_{t,t+h}$ \citep[see][for further discussion]{jalilian2015multivariate}. Instead, such expressions are available for our \SNMARG\ model. 

The following integrability conditions are needed to guarantee the convergence of the integrals appearing on the right-hand side of eq.~\eqref{eq:densitiestomoments}:
\begin{enumerate}
    \item[$(H_1)$] For every bounded measurable $B \subset \Y$ condition \eqref{eq:integrability2} is satisfied.
    \item[$(H_2)$] For every couple of bounded measurable sets $B_1,B_2 \subset \Y$ it holds:
    \begin{equation*}
        \int_{B_1 \times B_2} \int_\Theta K_\phi(y_1,\theta) K_\phi(y_2,\theta) \: [ H(\mathrm{d}\theta)+\bar W_1(\mathrm{d}\theta)] \ell(\mathrm{d}y_1)\ell(\mathrm{d}y_2) < +\infty.
    \end{equation*}
    \item[$(H_3)$] For every couple of bounded measurable sets $B_1,B_2 \subset \Y$ it holds:
    \begin{equation*}
        \E\Big( \int_{B_1 \times B_2} \int_\Theta K_\phi(y_1,\theta_1) W_1(\mathrm{d}\theta_2) \int_\Theta K_\phi(y_2,\theta_2) W_1(\mathrm{d}\theta_2) \ell(\mathrm{d}y_1) \ell(\mathrm{d}y_2) \Big) < +\infty.
    \end{equation*}
\end{enumerate}
Conditions $(H_{1})$-$(H_{3})$ are always satisfied when the kernel $K_\phi$ is bounded, and the expected initial measure $\bar W_1$ and the base measure $H$ are locally bounded. 

Simpler forms for eq.~\eqref{eq:densitytolambda} can be deduced in the time-stationary regime guaranteed by Proposition~\ref{proposition:statN} when the following stationarity condition is satisfied:
\begin{enumerate}
\setcounter{enumi}{3}
    \item[$(H_4)$] for all $t\geq1$ $c_t = c$, $\beta_t = \beta$, $\rho = \beta c < 1$ and $W_1 \sim \GP(W_1 \mid H, (1-\rho)/c)$.
\end{enumerate}

\begin{proposition}[First and second order statistics of $(N_t)_{t}$]
\label{proposition:densities}
Let $(N_t)_{t \geq 1}$ be a \SNMARG(1). Assume $(H_{1})$-$(H_{3})$. Then, for every $y$, $y_1$ and $y_2$ in $\Y$ and for every $t$ and $h$ strictly positive integers it holds:
\begin{equation}
    \mathfrak{D}^{(1)}_{t}(y) =\kappa_t c_{t|1} \int_\Theta K_\phi(y,\theta) H(\mathrm{d}\theta) + \kappa_t \rho_{t|1} \int_\Theta K_\phi(y,\theta) \bar W_1(\mathrm{d}\theta),
\label{eq:inensity_t}
\end{equation}
\begin{equation*}
    \mathfrak{D}^{(2)}_{t,t+h}(y_1,y_2) = \kappa_{t+h} \rho_{t+h|t} {\kappa_t}\mathfrak{D}^{(2)}_t(y_1,y_2) 
    +\mathfrak{D}^{(1)}_{t}(y_1)  \kappa_{t+h} c_{t+h|t}  \int_\Theta K_\phi(y_2,\theta) H(\mathrm{d}\theta).
\label{eq:corsdensity2}
\end{equation*}

\begin{equation}
\label{eq:corsdensity1}
\begin{split}
    \mathfrak{D}^{(2)}_t (y_1,y_2) & = \mathfrak{D}^{(1)}_t (y_1)\mathfrak{D}^{(1)}_t (y_2) + \kappa_t^2c_{t|1}^2  \int_\Theta K_\phi(y_1,\theta)K_\phi(y_2,\theta) H(\mathrm{d}\theta) \\
    & \quad + \kappa_t^2\rho_{t|1}^2 \Cov\Big(\int_\Theta K_\phi(y_1,\theta) W_1(\mathrm{d}\theta), \int_\Theta K_\phi(y_2,\theta) W_1(\mathrm{d}\theta)\Big).
\end{split}
\end{equation}
\end{proposition}

The previous proposition can be extended to $h=0$, by using the convention $c_{t|t}=0$ and $\rho_{t|t}=1$, so that $\mathfrak{D}^{(2)}_{t,t} (y_1,y_2) = \mathfrak{D}^{(2)}_t(y_1,y_2)$. 
Expressions of the first and second-order statistics simplify under the stationarity assumption $(H_{4})$, as stated below.

\begin{proposition}[First and second order statistics of stationary $(N_t)_{t}$]
\label{proposition:densities-Stationarity}
Let $(N_t)_{t \geq 1}$ be a \SNMARG(1), assume integrability $(H_{1})$-$(H_{3})$ and time stationarity $(H_{4})$. Then for every $y$, $y_1$, and $y_2$ in $\Y$ and $t$ positive integers, 
\begin{align}
\label{eq:inensity_t:stat}
     & \mathfrak{D}^{(1)}_{t}(y) = \frac{c\kappa_t  }{1-\rho} \int_\Theta K_\phi(y,\theta) H(\mathrm{d}\theta) \\
    \label{eq:pairdens:stat}
     & \mathfrak{D}^{(2)}_{t}(y_1,y_2)- \mathfrak{D}^{(1)}_{t}(y_1) \mathfrak{D}^{(1)}_{t}(y_2) = \frac{c^2 \kappa_t^2 }{ (1-\rho)^2}  \int_\Theta  K_\phi(y_1,\theta)K_\phi(y_2,\theta) H(\mathrm{d}\theta).
\end{align}
\end{proposition}

In spatial point process theory, the intensity measure $\Lambda_t$ describes where points are likely to occur, while the cross-pair correlation function $\mathscr{R}_{t,t+h}$ reveals whether points tend to cluster or repel each other relative to a Poisson process. The cross-pair correlation function \citep[see][def. 4.8]{moller2003statistical} of the point process $(N_t)_{t \geq 1}$ is
\begin{equation*}
    \mathscr{R}_{t_1,t_2}(y_1,y_2) \coloneqq \frac{\mathfrak{D}^{(2)}_{t_1,t_2}(y_1,y_2)}{\mathfrak{D}^{(1)}_{t_1}(y_1)\mathfrak{D}^{(1)}_{t_2}(y_2)}
     = \frac{\E(\lambda_{t_1}(y_1)\lambda_{t_2}(y_2))}{\E(\lambda_{t_1}(y_1))\E(\lambda_{t_2}(y_2))},
\end{equation*}
where the second equality follows by eq.~\eqref{eq:densitytolambda}. A process whose cross-pair correlation function $\mathscr{R}_{t,t+h} \geq 1$ is said to exhibit second-order clustering, or equivalently to be attractive. Intuitively, pairs of points are more likely to occur jointly at locations $y_1$ and $y_2$ than under a Poisson process with the same intensity. In many applications (disease outbreaks, earthquakes, crimes, species occurrences, and financial transactions), spatial patterns are clustered, and ignoring clustering often underestimates uncertainty and leads to overconfident inference. Examples of processes exhibiting such clustering can be found within the shot noise Cox, Thomas, Neyman--Scott, and log-Gaussian Cox process families \cite[see, e.g.][]{moller2003statistical}. The cross-pair correlation function is important for characterizing a process, since different processes can have the same intensity but very different cross-pair correlation functions. It also provides interpretable parameters, as the strength and spatial scale of clustering can often be directly linked to the underlying mechanisms generating the observed pattern.

The general expression of $\mathscr{R}$ for our \SNMARG\ can be easily deduced using Proposition \ref{proposition:densities}. In the time-stationary regime of Proposition~\ref{proposition:densities-Stationarity}, one has
\begin{equation*}
\mathscr{R}_{t,t+h}(y_1,y_2) = \rho^{h} \frac{\int_\Theta K_\phi(y_1,\theta) K_\phi(y_2,\theta)H(\mathrm{d}\theta)}{\int_{\Theta} K_\phi(y_1,\theta_1) H(\mathrm{d}\theta_1) \int_{\Theta} K_\phi(y_2,\theta_2) H(\mathrm{d}\theta_2)}+1
\end{equation*}
for $t\geq 1$ and $h\geq0$. Indeed, in this case $\rho_{t+h|t}=\rho^h$ and $c_{t+h|t}=c(1-\rho^h)/(1-\rho)$, so that combining eq.~\eqref{eq:corsdensity2}, \eqref{eq:inensity_t:stat}, and \eqref{eq:pairdens:stat} the previous expression follows easily. 

In the time-stationary case with $\kappa_t=1$ for every $t$, since $\mathscr{R}_{t,t+h}\geq 1$, the \SNMARG(1) process exhibits second-order clustering. Besides, the clustering effect decreases at larger temporal lags $h$ and increases with $\rho$. In this case, the cross-pair correlation function $\mathscr{R}_{t_1,t_2}$ is time-homogeneous since it depends on $h = |t_1-t_2|$, but it can be either homogeneous or non-homogeneous in space following the choice of $K_\phi$ and $H$ as discussed in the following.

\begin{example}
In the setting of Example \ref{Ex2}, where $\Y = \Theta = \RE^d$, $H$ and $\ell$ are the Lebesgue measure on $\RE^d$ and $K_\phi(\theta,y) = (\pi \phi^2)^{-d/2} \exp( -\phi^{-2} \|\theta-y\|^2 )$ is a Gaussian kernel, then 
\begin{equation*}
    \mathscr{R}_{t,t+h}(y_1,y_2) = (2\pi\phi^2)^{-d/2} \rho^{|h|} \exp( -(2\phi^{2})^{-1} \|y_1-y_2\|^2) +1,
\end{equation*}
which shows the process is spatially stationary. The parameter $\phi$ affects the spatial interaction range, that is, the cluster width.
\end{example}

\begin{example}
In the setting of Example \ref{Ex3}, where $\Y =\RE^d$, $\ell$ is the Lebesgue measure on $\RE^d$, $H(\theta)=\sum_{j=1}^{N^g} \alpha_j \delta(\theta)_{\theta_j}$ and $K_\phi(\theta,y) = (\pi \phi^2)^{-d/2} \exp(- \phi^{-2} \|\theta-y\|^2)$ is a Gaussian kernel, the process is no longer spatially stationary since
\begin{equation*}
    \mathscr{R}_{t,t+h}(y_1,y_2) = \rho^{|h|} \exp(-(2\phi^2)^{-1}\|y_1-y_2\|^2) \frac{r(y_1,y_2)}{r(y_1)r(y_2)}+1,
\end{equation*}
where 
\begin{equation*}
    r(y_1,y_2) = \sum_{j=1}^{N^g}\alpha_j \exp(-2\phi^{-2} \|\theta_j-\bar y\|^2), \,\,\, r(y_k)=\sum_{j=1}^{N^g} \alpha_j \exp(-\phi^{-2} \|y_k-\theta_j\|^2), \quad k=1,2
\end{equation*}
and $\bar y=(y_1+y_2)/2$.
The factor $\rho^{|h|} \exp(-(2\phi^2)^{-1} \|y_1-y_2\|^2)$ is translation-invariant, whereas $r(y_1,y_2)/(r(y_1)r(y_2))$ shows the process is spatial nonstationary. The sequence $(\theta_j,\alpha_j)$, $j=1,\ldots N^g$ determines the form and degree of spatial nonstationarity of the process. The following equivalent representations provide further insight on the process features
\begin{align*}
    \mathscr{R}_{t,t+h}(y_1,y_2) & = \rho^{|h|} \sum_{j=1}^{N^g} \frac{\alpha_j}{r(y_1)r(y_2)} K_{\phi}(y_1,\theta_j)K_{\phi}(y_2,\theta_j)+1 \\
    & = \rho^{|h|} \sum_{j=1}^{N^g} \frac{\alpha_j}{r(y_1)r(y_2)} K_{\phi}^{(2)}(y_1,y_2,\theta_j)+1,
\end{align*}
where 
\begin{equation*}
    K_{\phi}^{(2)}(y_1,y_2,\theta_j) = \exp\left( -\frac{\|y_1-y_2\|^2}{2\phi^2} -\frac{2\|\theta_j-\bar y\|^2}{\phi^2} \right).
\end{equation*}
The first line shows that the dependence is driven by shared centers $\theta_j$ and therefore the process exhibits clustering behaviour reminiscent of the Thomas and Neyman–Scott processes \citep{Daley03IntroPointProcesses}. The parameters  $\alpha_j$ determine the relative importance of each cluster. The parameter $N^g$ specifies the number of clusters and therefore controls the complexity of the spatial structure. The second line shows that the cross-pair correlation is a kernel convolution, which follows from the kernel mixture representation of the intensity given in Remark \ref{rem:kernelconvolution}, and indicates that the SN-M-ARG is a flexible model for spatial nonstationarity. 
\end{example}

Combining Propositions~\ref{proposition:densities} and \ref{proposition:densities-Stationarity} with eq.~\eqref{eq:densitiestomoments}, one obtains closed-form expressions of the second-order moments of $N_t$, given in the next two Corollaries. 
For simplicity, set $K_\phi(A,\theta) = \int_A K_\phi(y,\theta) \ell(\mathrm{d}y)$.

\begin{corollary}
\label{corollary:covariance_Nth}
Assume $(H_{1})$-$(H_{3})$, then for measurable sets $B_1,B_2 \subset \Theta$  and integers $t \geq 1$ and $h \geq 0$, it holds
\begin{equation*}
\begin{split}
    \Cov( & N_t(B_1),N_{t+h}(B_2)) = \kappa_t\kappa_{t+h} \rho_{t+h|t} c_{t|1}^2 \int_\Theta K_\phi(B_1,\theta) K_\phi(B_2,\theta) (H(\mathrm{d}\theta) + 2\beta_{t|1} \bar{W}_1(\mathrm{d}\theta)) \\
     & + \kappa_t\kappa_{t+h} \rho_{t+h|t} \rho_{t|1}^2 \Cov\Big(\int_{\Theta} K_\phi(B_1,\theta) W_1(\mathrm{d}\theta), \int_\Theta K_\phi(B_2,\theta) W_1(\mathrm{d}\theta) \Big) \\
     & + \kappa_t \I\{ h=0 \} \int_{\Theta} c_{t|1} K_\phi(B_1\cap B_2,\theta) \big( H(\mathrm{d}\theta) + \beta_{t|1} \bar{W}_1(\mathrm{d}\theta) \big),
\end{split}
\end{equation*}
provided all the quantities on the right-hand side are well-defined. 
\end{corollary}

\begin{corollary}
\label{corollary:corol_covariance_N}
Assume integrability $(H_{1})$-$(H_{3})$ and time stationarity $(H_{4})$, then
\begin{equation*}
\begin{split}
    \E(N_t(B_1)) & = \frac{c \kappa_t}{1-\rho} \int_\Theta K_\phi(B_1) \: H(\mathrm{d}\theta),\\
    \Cov( N_t(B_1),N_{t+h}(B_2) ) & = \I\{h =0\} \frac{c \kappa_t}{1-\rho} \int_{\Theta} K_\phi(B_1 \cap B_2,\theta)  \: H(\mathrm{d}\theta) \\
     & \quad + \frac{\rho^h c^2 \kappa_t \kappa_{t+h}}{(1-\rho)^2} \int_\Theta K_\phi(B_1,\theta)  K_\phi(B_2,\theta) \: H(\mathrm{d}\theta),
\end{split}
\end{equation*}
provided all the quantities on the right are well-defined. 
\end{corollary}

We remark that, in the time-stationary case, the expression of $\Cov(N_t(B_1),N_t(B_2))$ for each fixed $t=1,\ldots,T$, coincides with the result given in \cite{Wolpert1998PoissonGamma_RandomField} for the Poisson-gamma random field. The analytical expression of the temporal and spatial covariances may be used for method-of-moments estimation.

Using arguments similar to those used to obtain first and second-order statistics, one can deduce closed-form expressions for high-order moments. For simplicity, we provide the high-order moments only in the time-stationary case.

\begin{proposition}
\label{proposition:MomentsHighorders}
Assume time stationarity $(H_{4})$ and 
\begin{equation*}
    J_r(B) \coloneqq \frac{\gamma(r)c^r}{(1-\rho)^r} \int K_\phi(B,\theta)^r \: H(\mathrm{d}\theta) < +\infty,
\end{equation*}
for $r=1,\dots,m$, then 
\begin{equation}
\label{eq:spatialmoments_stationary}
    \E(N_t(B)^m) = \sum_{j=1}^m S_{m,j} \,\, \kappa_t^j\!\! \sum_{\substack{\ell_1,\dots,\ell_j \geq 0 \\ \sum_{r=1}^j r \ell_r=j}}  c(\ell_1,\dots,\ell_j) \prod_{r=1}^j J_r(B)^{\ell_r}
\end{equation}
where $c(\ell_1,\dots,\ell_j) = j! \big( \prod_{r=1}^j (r!)^{\ell_r} \ell_r! \big)^{-1}$ and $S_{m,j}$ are the Stirling numbers of the second kind. 
\end{proposition}

\FloatBarrier
\section{Bayesian Inference}
\label{sec:Bayesian_inference}

To make inferences in a Bayesian setting on the proposed \SNMARG\ process, in this Section, we design a Markov Chain Monte Carlo algorithm for approximating the posterior distribution of parameters and latent processes. We leverage the state-space representation of our shot noise Cox process and apply a particle Gibbs \citep{Andrieu2010particleMCMC} with block updating to draw the trajectories of the latent intensity \citep[e.g., see][]{singh2017blocking,goldman2021spatiotemporal}. We rely on an adaptive Metropolis-Hastings sampler to draw the static parameters \citep{Diggle2013,Andrieu08AdaptiveMCMC}.

\subsection{Data Augmentation}

Given $\Lambda_t$, $N_{t+1}$ is an integer-valued measure which can be represented as the collection of a random number $N_{t}^y \sim \mathcal{P}oi(N_t^y \mid \Lambda_{t}(\Y))$ of unit point masses at not-necessarily-distinct points $y_{i,t}$.  
In the real-data illustration, $N_{t}^y$ is the total number of observed event points (fires) at the time (month) $t$. The points $y_{i,t}$ are drawn conditionally independently from
\begin{equation*}
    y_{i,t} \mid W_{t},N_{t}^y \stackrel{iid}{\sim} Q_{t}^y, \qquad i=1,\ldots,N_{t}^y,
\end{equation*}
where $Q_{t}^y$ is the random probability measure resulting from the mixture
\begin{align}
    Q_{t}^y(\mathrm{d}y) & = \frac{\Lambda_{t}(\mathrm{d}y)}{\Lambda_{t}(\Y)} 
    = \sum_{j=1}^{\infty} w_{j,t}^* K_\phi(y,\theta_{j,t})  \ell(\mathrm{d}y),
\label{eq:Qty_mixture}
\end{align}
where $w_{j,t}^* = \kappa_t w_{j,t}/\Lambda_t(\Y)$, for $j=1,2,\ldots$, are normalised weights.

Following \cite{Wolpert1998PoissonGamma_RandomField}, we consider the data augmentation approach and introduce a collection of latent allocation variables for resolving the mixture. As a result, one obtains a random measure $Z_{t+1}$ on $\Y \times \Theta$ such that
\begin{align*}
    Z_{t+1} \mid W_t, V_{t+1}, W_{t+1} & \sim \PP\big( Z_{t+1} \mid \kappa_{t+1} K_\phi(y,\theta) \ell(\mathrm{d}y) W_{t+1}(\mathrm{d}\theta) \big) \\
    N_{t+1}(\mathrm{d}y) & = Z_{t+1}(\mathrm{d}y \times \Theta).
\end{align*}
The random measure $Z_{t+1}$ assigns unit mass to each pair $(y_{i,t+1},\theta_{i,t+1}^*)$, where $\theta_{i,t+1}^* \in \Theta$ is the atom of the latent measure $W_{t+1}$ to which the observation $y_{i,t+1}$ has been allocated.
We can describe $Z_t$ with the collection of the data points $y_{i,t}$ and a set of allocation variables $z_{i,t}$, is a set of allocation variables which assign the observation $y_{i,t}$ to the the $z_{i,t}$-th element of the collection $\{\theta_{i,t},w_{i,t}\}_{i\in\mathbb{N}}$. 
Therefore, conditioning on $Z_{t}$, the mixture in eq.~\eqref{eq:Qty_mixture} is resolved and one obtains the following conditional distribution, which is used in Section~\ref{sec:posterior} to obtain the full conditional posterior of the parameters and latent variables:
\begin{align*}
    (y_{1,t},\ldots,y_{N_t^y,t}) \mid W_t, N_t^y, Z_t \sim \prod_{i=1}^{N_t^y} w_{i,t}^* K_\phi(y_{i,t}, \theta_{i,t}^*), 
\end{align*}
where $\theta_{i,t}^* = \theta_{z_{i,t}}$,  $w_{i,t}^* = w_{z_{i,t}}$, $\theta_{i,t}$ are points in $\Theta$, $w_{it}$ are positive weights and $z_{i,t}\in\mathbb{N}$, $i=1,\ldots,N_t^y$.

\subsection{Model Specification}

In the application, we assume that both $\Theta$ and $\Y$ are bounded sets in $\RE^2$ and that the kernel is a two-dimensional Gaussian density with covariance matrix $\phi^2 \Id_2$, that is $K_\phi(y,\theta) = \mathcal{N}_2(y \mid \theta, \phi^2 \Id_2)$.
As for the deterministic component, we assume $\kappa_t = \exp(\eta^{\T} x_t)$ where $\eta$ is a parameter vector and the $x_t \in \RE^m$ are time-dependent predictors. See Section~\ref{sec:illustration} for an illustration.

To reduce the complexity of the parameter space, we choose a discrete base measure $H(\mathrm{d}\theta) = \sum_{j=1}^{N^g} \alpha_j \delta_{\theta_{j}}(\mathrm{d}\theta)$ (see discussion in Example \ref{Ex3}).
With this choice, the measure-valued noncentral gamma process $(W_t)_{t \geq 1}$ evaluated at $\Theta$ results in a collection of scalar noncentral gamma processes. In particular, for each $j=1,\ldots,N^g$, define $w_{j,t} \coloneqq W_t( \{ \theta_j \} )$ to obtain:
\begin{equation}
    w_{j,t} \mid w_{j,t-1} \sim \NcGa(w_{j,t} \mid \alpha_j, \beta w_{j,t-1},c_t^{-1}).
\label{eq:latent_ARG}
\end{equation}
Therefore, we obtain $N_{t}^y \sim \mathcal{P}oi(N_t^y \mid \Lambda_{t}(\Y))$ and 
\begin{equation}
\begin{split}
      z_{i,t} \mid N_t^y  & \;\stackrel{iid}{\sim}\; \sum_{j=1}^{N^g} \frac{w_{z_{j,t},t}^*}{\Lambda_{t}(\Y)}\delta_{z_{j,t}}(\mathrm{d}z ), \quad i=1,\dots,N_{t}^y \\
    y_{i,t} \mid z_{i,t}, N_t^y  & \;\stackrel{iid}{\sim}\; K_\phi(y,\theta_{z_{j,t}}) \ell(dy), \quad i=1,\dots,N_{t}^y.
\label{eq:dataaugm}
\end{split}
\end{equation}

\subsection{Posterior Approximation}
\label{sec:posterior}

The collection of unknown parameters is the vector $\bpsi = \{ \balpha,\beta,\bc,\phi,r,\eta \}$, with $\balpha = \{ \alpha_1,\ldots,\alpha_{N^g} \}$ and $\bc = \{ c_1,\ldots,c_T \}$, for which a prior need to be specified.
We assume the following prior for the static parameters:
\begin{align*}
    \alpha_j \mid \gamma & \distas{iid} \mathcal{G}a(\alpha_j \mid \underline{a}_\alpha, \gamma \underline{b}_\alpha), \;\; j=1,\ldots,N^g, \qquad
    \beta  \sim \mathcal{G}a(\beta \mid \underline{a}_\beta, \underline{b}_\beta), \\
    \gamma & \sim \mathcal{G}a(\gamma \mid \underline{a}_\gamma, \underline{b}_\gamma), \qquad
    \phi \sim \mathcal{G}a(\phi \mid \underline{a}_\phi, \underline{b}_\phi).
\end{align*}
We consider two alternative specifications for the scale parameter $\bc$: i) constant, that is, $c_t=c$ for every $t$; ii) time-varying.
In the first setting, we assume $c \sim \mathcal{G}a(c \mid \underline{a}_c, \underline{b}_c)$, whereas in the second one, we assume the hierarchical prior distribution:
\begin{align*}
    c_t \mid r & \distas{iid} \mathcal{G}a(c_t \mid r^2/\underline{\sigma}^2_c, r/\underline{\sigma}^2_c), \;\;\; t=1,\ldots,T, \qquad
    r \sim \mathcal{G}a(r \mid \underline{a}_r, \underline{b}_r),
\end{align*}
such that $\E(c_t \mid r) = r$.
Finally, a multivariate Gaussian prior is assumed for the coefficient vector, $\eta \sim \mathcal{N}_m(\eta \mid \underline{\boldsymbol\mu}_\eta, \underline{\Sigma}_\eta)$.

Let $\by_{1:T} = \{ y_1,\ldots,y_T \}$, $\bw_{1:T} = \{ w_1,\ldots,w_T \}$ and $\bz_{1:T} = \{ z_1,\ldots,z_T \}$, from eq.~\eqref{eq:latent_ARG}-\eqref{eq:dataaugm}, the complete-data likelihood function $L(\by_{1:T}, \bz_{1:T}, \bw_{1:T} \mid \boldsymbol{\psi})$ is
\begin{equation*}
    \prod_{t=1}^T \mathcal{P}oi( N_t^y \mid \Lambda_t(\Y) ) \prod_{j=1}^{N^g} \NcGa(w_{j,t} \mid \alpha_j, \beta w_{j,t-1},c_t^{-1}) \prod_{i=1}^{N_t^y} \frac{w_{z_{i,t},t}\kappa_t \mathcal{N}_2(y_{i,t} \mid \theta_{z_{i,t}}, \Id_2\phi^2)}{\Lambda_t(\Y)}.
\end{equation*}

The joint posterior of the model parameters $\bpsi$ and the latent variables $\bw_{1:T}$ and $\bz_{1:T}$ is approximated using a Markov chain Monte Carlo algorithm.
To cope with the high dimensionality of the latent space, we use a blocking strategy.
We assume the latent states are grouped into $L$ blocks of size $S_\ell$ each, that is $\bw_{1:T} = \{ \bw_{1,1:T},\ldots,\bw_{L,1:T} \}$, where $\bw_{\ell,1:T} = \{ w_{j,t} : j \in S_\ell, \, t=1,\ldots,T \}$ and $S_\ell \subset \{1,\ldots,N^g\}$.
We propose two alternative algorithms that differentiate the estimation of the latent dynamic variables $\bw_{1:T}$, as follows:
\begin{enumerate}
    \item Sample the parameters $\bpsi$ given $\by_{1:T},\bz_{1:T}$, and $\bw_{1:T}$.
    \item Sample the allocation variables $\bz_{1:T}$ given $\by_{1:T},\bw_{1:T}$, and $\bpsi$.
    \item Sample the latent states $\bw_{\ell,1:T}$ sequentially for $\ell=1,\ldots,L$, using either of the following approaches:
    \begin{enumerate}
        \item a single-step update from the full conditional distribution of each $w_{\ell,t}$ given $\bz_t,\by_t$ and $\bpsi$, via an adaptive random walk Metropolis-Hastings step with log-normal proposal distribution.
        \item a conditional Sequential Monte Carlo algorithm \citep[SMC,][]{chopin2013smc2,gerber2015sequential,singh2017blocking,goldman2021spatiotemporal} with target $p(\bw_{\ell,1:T} \mid \by_{1:T},$ $\bz_{1:T},\bw_{-\ell,1:T},\bpsi)$, and sampling $\bw_{\ell,1:T} \sim \widehat{p}(\bw_{\ell,1:T} \mid \by_{1:T},\bz_{1:T},\bw_{-\ell,1:T},\bpsi)$, where the latter is an approximation of the distribution $p(\bw_{\ell,1:T} \mid \by_{1:T}, \bz_{1:T}, \bw_{-\ell,1:T}, \bpsi)$.
    \end{enumerate}
\end{enumerate}
We denote the algorithm consisting of Steps 1, 2, 3(a) \textit{adaptive MCMC}, and the algorithm consisting of Steps 1, 2, 3(b) a \textit{particle MCMC} \citep{Andrieu2010particleMCMC,kantas2015particle}.
In Step 1, we draw from the full conditional distribution of the parameters with an adaptive Metropolis-Hastings step. The allocation variables in Step 2 are drawn exactly.
In Step 3, given the parameters, $\bpsi$, the allocation variables, $\bz_{1:T}$, and the other $L-1$ blocks, $\bw_{-\ell,1:T}$, we propose to use either a single-move Metropolis-Hastings algorithm to sample from the full conditional of $w_{\ell,t}$, or a conditional Sequential Monte Carlo (SMC) for the $\ell$th block $\bw_{\ell,1:T}$. The latter is similar to a standard SMC, but imposes that a prespecified path, $\bw_{\ell,1:T}$, with its ancestral lineage survives all the resampling steps, whereas the remaining $N-1$ particles are generated as usual. Based on a comparison of the two schemes on simulated data, the two schemes have similar performance for small dimension of the grid, $N^g$, but the adaptive MH better scales with $N^g$, making it our suggested choice in the empirical application in Section~\ref{sec:illustration}.
The Supplement provides further details and a comparison of the two approaches.

\FloatBarrier
\section{Illustration}
\label{sec:illustration}

Motivated by the significant increase in wildfires in recent years \citep[e.g., see][]{pontes2021drought}, we apply the \SNMARG\ process to a real dataset of wildfires in South America. This section reports the results and shows that \SNMARG\ captures the common patterns and the heterogeneity of the fire intensity across space and time.

\subsection{Data Description}
Climate change has been a critical factor impacting the risk and extent of wildfires in many areas, such as Australia, California, Canada, Siberia, the Mediterranean coast, and Savannah. Over the past decades, new regions and ecosystems, such as tropical forests, have experienced an increase in the frequency and intensity of large fires, which have severe impacts on ecological systems (e.g., vegetation structure and composition), the economy (e.g., loss of properties), and society (e.g., direct and indirect threats to human health and life). As argued by \cite{Balch20}, the abundance of remote-sensed fire data calls for an effort by the scientific community to investigate changes in fire regimes and the vulnerabilities of society and ecosystems. Measuring the risk and intensity of fires and modelling and predicting their local or global spatiotemporal dynamics can help to support policy decisions in areas affected by future climatic and land-use changes \citep{Hantson16}. 

Our application uses satellite observations with high spatiotemporal resolution and broad spatial coverage to study fires in the Amazon forest. Various types of satellite data are available for investigating fires. Here, we consider NASA's Moderate Resolution Imaging Spectroradiometer (MODIS), the first family of remotely sensed fire datasets \citep{Giglio16}. MODIS provides systematic observations of fires over the entire globe and has been used to answer many scientific questions, such as fire dynamics in forests \citep{Loboda07} and its impact on air quality and Savannah's ecosystems.

\begin{figure}[p]
\setlength{\tabcolsep}{3pt}
\centering
\begin{tabular}{c c c}
& {\footnotesize Constant Scale} & {\footnotesize Time-varying Scale} \\
\begin{rotate}{90} \hspace{35pt} {\scriptsize June $2020$} \end{rotate} \hspace{-14pt} &
\includegraphics[clip,trim=0 10 0 10, scale=0.50]{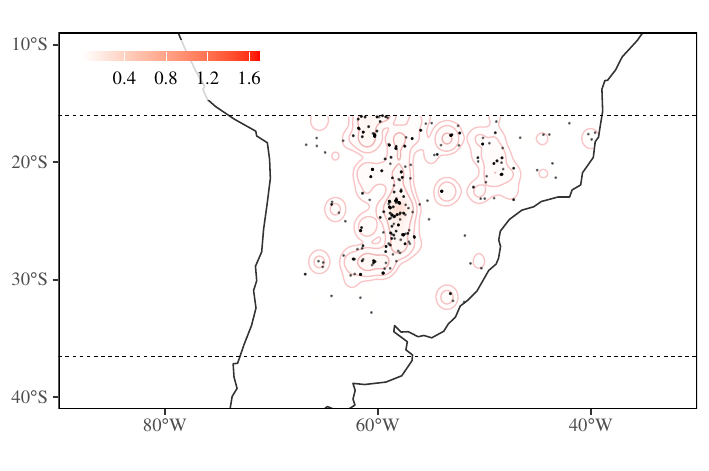}&
\includegraphics[clip,trim=0 10 0 10, scale=0.50]{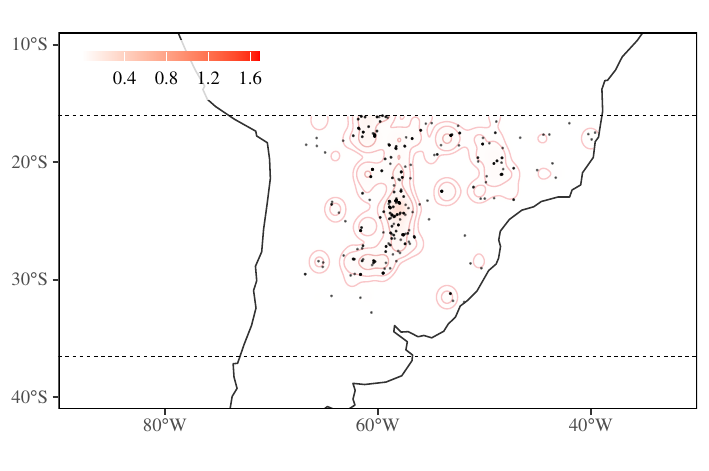}\\
\begin{rotate}{90} \hspace{35pt} {\scriptsize July $2020$} \end{rotate} \hspace{-14pt} &
\includegraphics[clip,trim=0 10 0 10, scale=0.50]{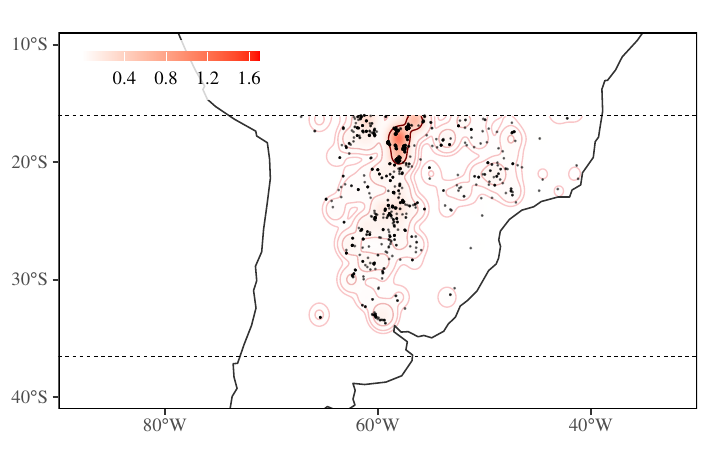}&
\includegraphics[clip,trim=0 10 0 10, scale=0.50]{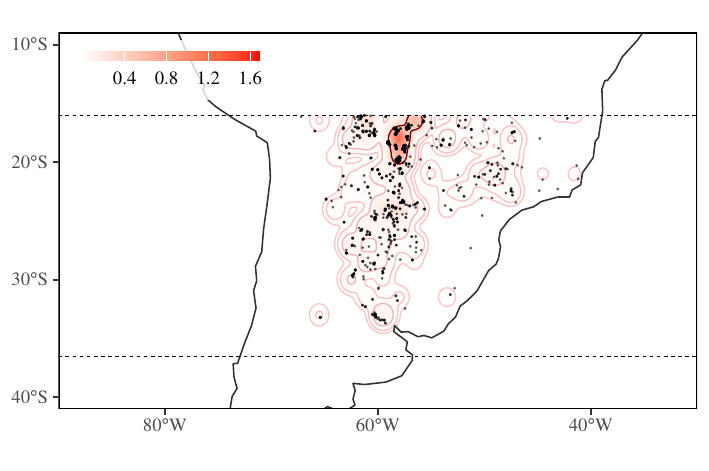}\\
\begin{rotate}{90} \hspace{33pt} {\scriptsize August $2020$} \end{rotate} \hspace{-14pt} &
\includegraphics[clip,trim=0 10 0 10, scale=0.50]{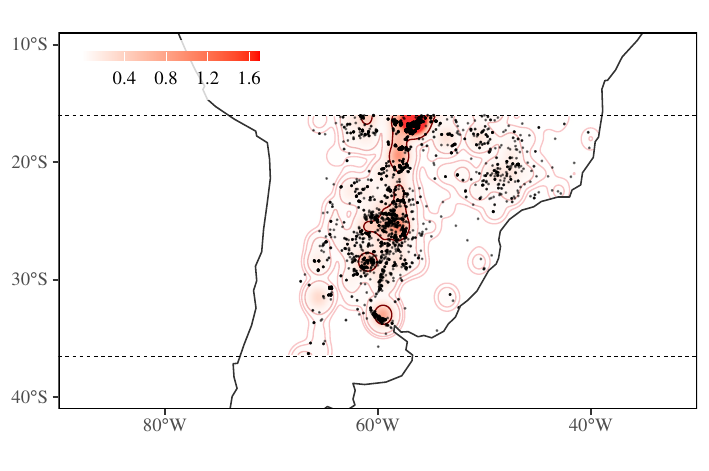}&
\includegraphics[clip,trim=0 10 0 10, scale=0.50]{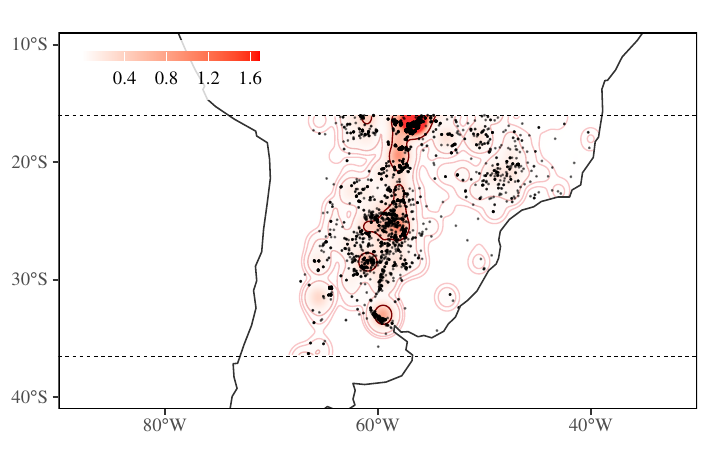}\\
\begin{rotate}{90} \hspace{25pt} {\scriptsize September $2020$} \end{rotate} \hspace{-14pt} &
\includegraphics[clip,trim=0 10 0 10, scale=0.50]{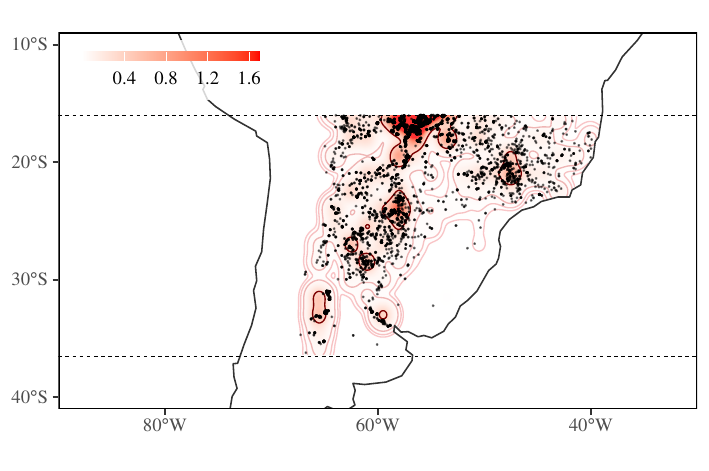}&
\includegraphics[clip,trim=0 10 0 10, scale=0.50]{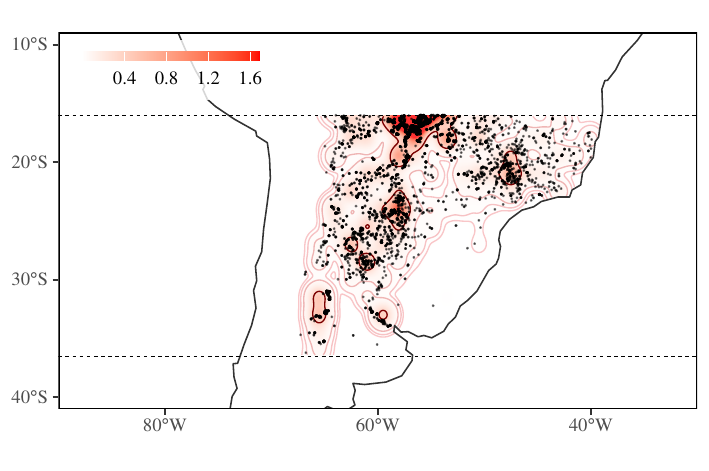}\\
\end{tabular}
\captionsetup{width=0.95\textwidth}
\caption{Monthly fires ($\bullet$) between longitude $80^{o}W$ and $40^{o}W$ and latitude $35^{o}S$ and $10^{o}S$ and estimated fire intensity, $\Lambda_t(y)$ (shaded areas and contour lines) for models with harmonic components and constant (left) and time-varying (right) scale.}
\label{fig:Lambda_mean}
\end{figure}

Fire observations such as fire detection and fire radiative power are collected by MODIS 1-km sensor on Terra and Aqua (see the Supplement for further details). A geographic location, time, and date identify each fire pixel in the dataset. Geodesic coordinates have first been transformed into Euclidean coordinates using the Lambert azimuthal equal-area (LAEA) projection before conducting statistical inference. The study area of this paper covers the territory with a longitude between $82^{o}W$ and $34^{o}W$ and a latitude between $40^{o}S$ and $0^{o}$. The area includes the Amazon forest, the world's largest rainforest, and is central to global warming. The additional pixel-level information allows us to exclude active volcanoes, other static land sources, and offshore fires and select only presumed vegetation fires detected by Terra and Aqua MODIS sensors with detection confidence between 90\% and 100\%. Figure~\ref{fig:Lambda_mean} shows the frequency of observed fires (dots) in Brazil, Bolivia, and Peru, which exhibit spatial and temporal patterns.

\subsection{Results}
We apply our spatiotemporal models with two alternative specifications of the global factor $\kappa_t$ proposed in the literature and three different specifications for the scale parameter $c_t$. As for $\kappa_t$, the first specification (denoted by ``$D$'') has a trend and a dry-season dummy variable,  that is $\kappa_t = \exp(\eta_0 + \eta_{TR} t + \eta_{D,1} d_t )$, where $d_t$ takes the value one if $t$ is a month of the dry season (from July to November), while the second specification (denoted by ``$H$'') has a trend and four harmonic components $\kappa_t = \exp( \eta_0 + \eta_{TR} t + \eta_{S,1} \sin(2\pi \omega_1 t) + \eta_{C,1} \cos(2\pi \omega_1 t) +\ldots+ \eta_{S,4} \sin(2\pi \omega_4 t) + \eta_{C,4} \cos(2\pi \omega_4 t) )$. The frequencies have been estimated from the peaks in the spectral density of the time series of the total number of fires. The estimated frequencies $\hat{\omega}_1 = 0.086$, $\hat{\omega}_2 = 0.168$, $\hat{\omega}_3 = 0.254$, and $\hat{\omega}_4 = 0.336$ correspond to annual, semi-annual, four-month, and three-month periods, respectively. The three scale parameter specifications are constant scale $c_t=c$ for every $t$ (denoted by ``$c$''), a time-varying scale $c_t$ (indicated by ``$v$'') and a monthly specification $c_t = c_{t+12} = c_{t+24} \ldots$ for every $t$ (denoted by ``$v12$''). We obtain three specifications for the harmonic models denoted by $\mathcal{M}^{H,c}$, $\mathcal{M}^{H,v}$, $\mathcal{M}^{H,v12}$ and three other specifications for the model with seasonal dummies marked by $\mathcal{M}^{D,c}$, $\mathcal{M}^{D,v}$, $\mathcal{M}^{D,v12}$.
As a robustness check, we also report in the Supplement the estimated parameters and in-sample performance metrics for a model with an autoregressive scale $\mathcal{M}^{j,AR}$ similar to \cite{gamerman2013non}.

Our framework can provide a globally consistent inference of fire trend and seasonality based on high spatial resolution data from a wide region, which is still an open issue in analysing climate-related risks \citep[e.g., see][]{Frontiers22}.
We note that the Bayesian framework is well-suited to quantifying uncertainty in model estimates and forecasts, which is currently a crucial issue in evaluating model-based estimates and projections of climate-related variables \citep{raftery2017less}.

\begin{table}[p]
\caption{Model estimates and fitting.\label{tab1}}
\def~{\hphantom{0}} 
\centering
\small
\setlength{\tabcolsep}{1pt}
\resizebox{.9\textwidth}{!}{%
\renewcommand{\arraystretch}{0.82}
\begin{tabular*}{\textwidth}{@{\extracolsep\fill}ccccccc@{}}
\toprule
&$\mathcal{M}^{H,c}$&$\mathcal{M}^{H,v}$&$\mathcal{M}^{H,v12}$&$\mathcal{M}^{D,c}$&$\mathcal{M}^{D,v}$&$\mathcal{M}^{D,v12}$\\[4pt]
\multicolumn{7}{c}{(a) Parameter estimates}\\[4pt]
$\eta_{TR}$ & 0.019 & 0.018 & 0.018 & 0.020 & 0.005 & 0.018 \\
 & (0.018,0.022) & (0.017,0.019) & (0.017,0.019) & (0.016,0.022) & (0.003,0.008) & (0.015,0.020) \\
$\eta_{D,1}$ &  &  &  & 1.438 & 0.850 & 0.364 \\
 &  &  &  & (1.410,1.482) & (0.784,0.911) & (0.260,0.489) \\
 
$\eta_{S,1}$ & -0.929 & -0.697 & -0.209 &  &  &  \\
 & (-0.953,-0.898) & (-0.753,-0.646) & (-0.220,-0.198) &  &  &  \\
$\eta_{C,1}$ & 0.686 & -0.052 & -0.126 &  &  &  \\
 & (0.683,0.688) & (-0.054,-0.049) & (-0.128,-0.123) &  &  &  \\
$\eta_{S,2}$ & 0.139 & -0.025 & -0.443 &  &  &  \\
 & (0.130,0.147) & (-0.066,0.052) & (-0.460,-0.420) &  &  &  \\
$\eta_{C,2}$ & -0.513 & -0.260 & 0.077 &  &  &  \\
 & (-0.543,-0.494) & (-0.321,-0.197) & (0.062,0.089) &  &  &  \\
$\eta_{S,3}$ & 0.119 & 0.078 & 0.308 &  &  &  \\
 & (0.117,0.121) & (0.045,0.103) & (0.307,0.310) &  &  &  \\
$\eta_{C,3}$ & 0.025 & -0.058 & 0.054 &  &  &  \\
 & (-0.026,0.050) & (-0.059,-0.055) & (0.052,0.061) &  &  &  \\
$\eta_{S,4}$ & 0.060 & -0.014 & 0.106 &  &  &  \\
 & (0.018,0.145) & (-0.031,0.015) & (0.105,0.107) &  &  &  \\
$\eta_{C,4}$ & 0.073 & -0.054 & 0.215 &  &  &  \\
 & (0.030,0.110) & (-0.058,-0.047) & (0.046,0.304) &  &  &  \\[4pt]
\multicolumn{7}{c}{(b) Model fitting (normalised across models)}\\[4pt]
MSE & 0.259 & 0.034 & 0.130 & 0.377 & 0.046 & 0.154 \\
MAE & 0.210 & 0.073 & 0.169 & 0.288 & 0.072 & 0.187 \\[4pt]
\multicolumn{7}{c}{(c) Model forecasting over horizon $h$ (cumulated and normalised across models)}\\[4pt]
MSE $h = 1$ & 0.0150 & 0.2870 & 0.0045 & 0.2497 & 0.4290 & 0.0148 \\
MSE $h = 3$ & 0.0768 & 0.2591 & 0.0049 & 0.2316 & 0.3997 & 0.0278 \\
MSE $h = 6$ & 0.0717 & 0.2735 & 0.0078 & 0.2297 & 0.3902 & 0.0272 \\[3pt]
MAE $h = 1$ & 0.0612 & 0.2676 & 0.0337 & 0.2496 & 0.3272 & 0.0607 \\
MAE $h = 3$ & 0.1093 & 0.2423 & 0.0324 & 0.2320 & 0.3049 & 0.0792 \\
MAE $h = 6$ & 0.1085 & 0.2481 & 0.0396 & 0.2306 & 0.2944 & 0.0788 \\
\bottomrule
\end{tabular*}}
\label{tab:parest}
\begin{tablenotes}
\item \footnotesize Panel (a) reports the posterior mean and $95\%$ posterior credible intervals (in brackets) of the parameters of the global factor $\kappa_t$ across models. Panels (b) and (c) show the in- and out-of-sample model diagnostics.
For the harmonic ($j=H$) and the dry-season dummy ($j=D$) model three specifications are considered: $\mathcal{M}^{j,c}$, constant scale;
$\mathcal{M}^{j,v}$ time-varying scale; $\mathcal{M}^{j,v12}$ time-varying scale with monthly seasonal dummy.
For model performance, Mean Square Error (MSE) and Mean Absolute Error (MAE) are computed in-sample and normalised across models.
\end{tablenotes}
\end{table}

The parameter estimates of the factor $\kappa_t$ in Panel (a) of Table~\ref{tab:parest} and the factor estimates in the second row of Panel (a) and (b) in Fig. 7 in the Supplement provide strong evidence across specifications of an exponential trend and seasonal effects for the entire area of interest.
The monthly growth rate of the fire intensity varies between 0.005 and 0.020 across models, corresponding to annual percentage growth between 6.184\% and 27.125\%. These findings align with the results reported in other studies on global trends in fire intensity \citep[e.g.][]{geo2020}.
There is strong evidence of temporal persistence regarding the dynamic of the latent random measure, which captures the residual local spatiotemporal variability. The estimated persistence parameter in the constant scale specification is $\hat{\rho}= 0.485$ and $\hat{\rho}=0.532$ in the dummy and harmonic models, respectively. In the time-varying scale model with harmonics, $\hat{\rho}_t$ ranges from 0.136 to 1.636 across seasons with a mean of 0.586, showing high fire persistence in January and July and low persistence in November and December.

\begin{figure}[t]
\centering
\setlength{\tabcolsep}{4pt}
\begin{tabular}{c c c}
& {\footnotesize Constant scale} & {\footnotesize Time-varying Scale} \\
\begin{rotate}{90} \hspace{35pt} {\scriptsize June $2020$} \end{rotate} \hspace{-15pt} &
\includegraphics[clip,trim=0 10 0 10, scale=0.50]{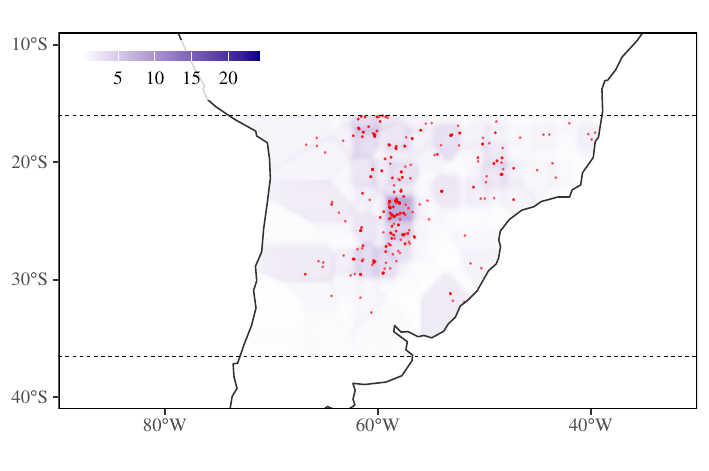} &
\includegraphics[clip,trim=0 10 0 10, scale=0.50]{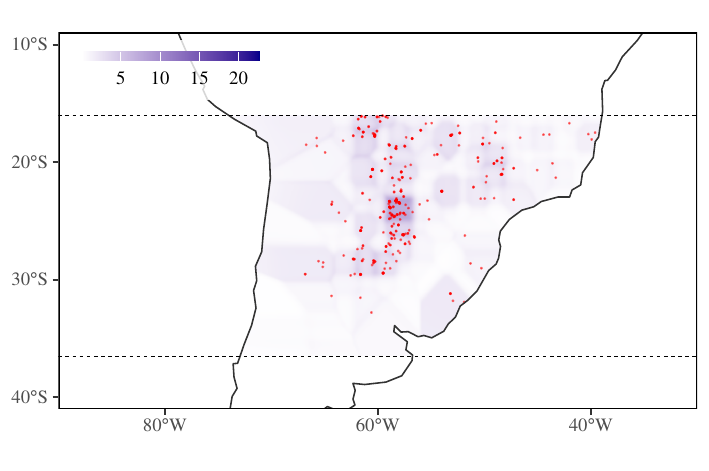} \\
\begin{rotate}{90} \hspace{25pt} {\scriptsize September $2020$} \end{rotate} \hspace{-15pt} &
\includegraphics[clip,trim=0 10 0 10, scale=0.5]{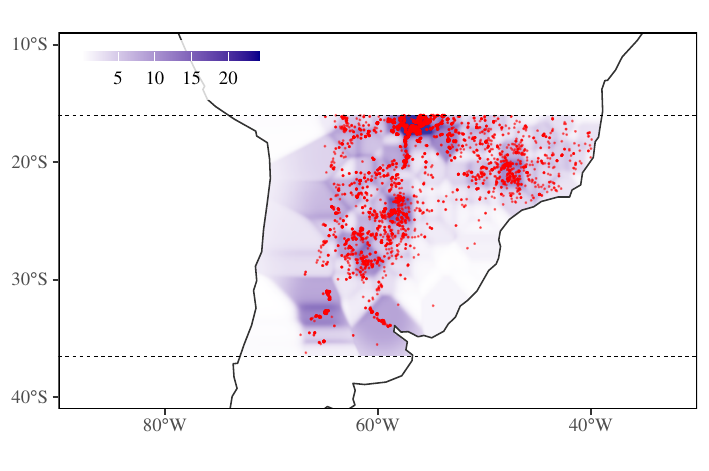} &
\includegraphics[clip,trim=0 10 0 10, scale=0.50]{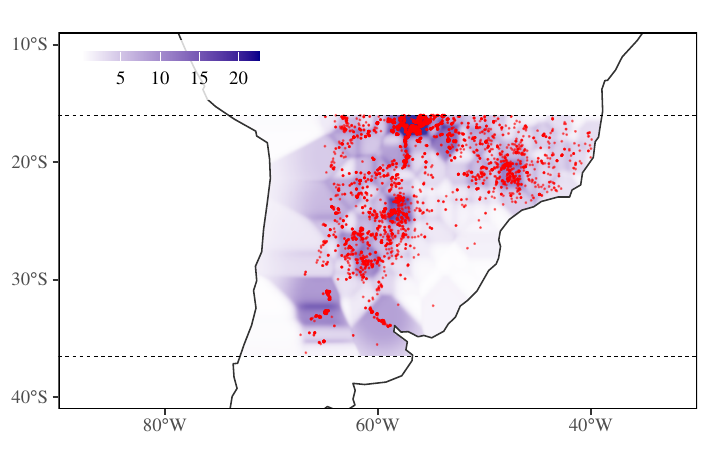}
\end{tabular}
\captionsetup{width=0.95\textwidth}
\caption{Inverse coefficient of variation (iCV) of the fire intensity $\Lambda_t(y)$ (shaded areas and contour lines) for models with harmonic components and constant (left) and time-varying (right) scale, and observed fires (\textcolor{red}{$\bullet$}).}
\label{fig:Lambda_CV}
\end{figure}

Analysing the model performances is crucial in incorporating uncertainty in the prediction, for example, through model combination techniques. We studied the fitting of the time-varying scale model at a global scale by estimating the in-sample mean squared error (MSE) and mean absolute error (MAE) when fitting $N_t$ with the intensity posterior mean $\E(\Lambda_t(\Y) \mid \by_{1:T})$.
Panel (b) of Table~\ref{tab:parest} shows the model relative errors normalised across models to the unit interval. There is evidence of better fitting abilities of models with time-varying scale $c_t$ across different specifications of $\kappa_t$.
Regarding the out-of-sample analysis, Table~\ref{tab:parest} shows that the constant scale model with dry and wet season dummy generally overperforms the other specifications, especially at horizons 1 and 3. However, at larger horizons (e.g. $h=6$), the differences in the model performances reduce significantly.

For illustrative purposes, we report in Fig.~\ref{fig:Lambda_mean} the estimated expected fire intensity $\Lambda_t(y)$ (shaded areas) in four months of 2020 wet (June and July) and dry (August and September) seasons for the harmonic-component specification of $\kappa_t$ assuming alternatively constant scale $c_t=c$ (left) and time-varying scale (right). Similar figures for all the other models are reported in the Supplement. The time-varying scale model seems to better fit the spatial variability compared to the constant scale model and, in some regions, better captures the fire dynamics (e.g., between 50$^\circ$W and 40$^\circ$W).

\begin{figure}[t]
\centering
\setlength{\tabcolsep}{10pt}
\begin{tabular}{cc}
\includegraphics[clip,trim=0 0 0 0, scale=0.42]{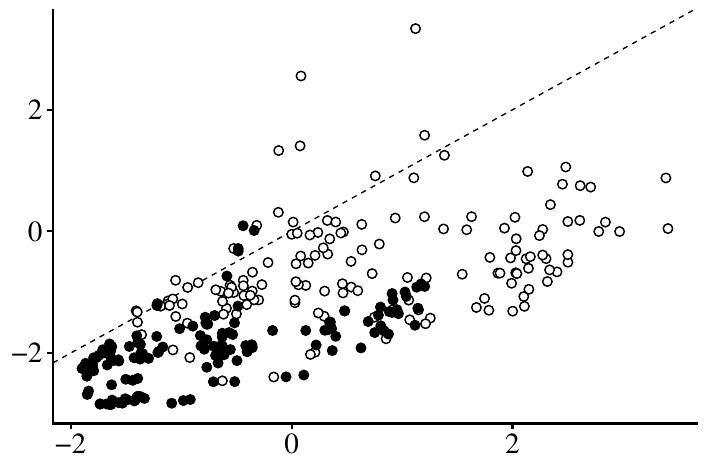} &
\includegraphics[clip,trim=0 0 0 0, scale=0.42]{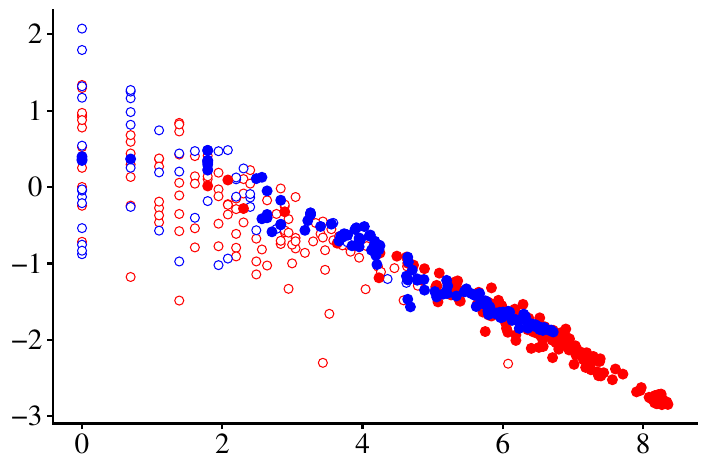}
\end{tabular}
\captionsetup{width=0.95\textwidth}
\caption{Variability of the expected intensity estimates $\E(\Lambda_t(A) | \by_{1:T})$, measured by the normalised interquantile range, $(Q_{97.5}(\Lambda_t(A))-Q_{2.5}(\Lambda_t(A)))/Q_{50}(\Lambda_t(A))$, for 100 different $1^{\circ}\times 1^{\circ}$ and $4^{\circ}\times 4^{\circ}$ areas $A$ ($\circ$ and $\bullet$, respectively).
Left: Dry-season month against wet-season month range. Right: range against the number of fires in the same areas during dry (\textcolor{red}{$\circ$, $\bullet$}) and wet (\textcolor{blue}{$\circ$, $\bullet$}) seasons.}
\label{fig:IQR}
\end{figure}

The inverse of the coefficient of variation (iCV) of the intensity $\Lambda_t$ reported in Fig.~\ref{fig:Lambda_CV} confirms the greater flexibility of the time-varying scale specifications (see also Fig. 5 and 6 in the Supplement).
Since high iCV values are associated with low posterior variance of $\Lambda_t(y)$, this measure is relevant in assessing the uncertainty, a crucial issue in spatial analysis. The iCV of time-varying scale models presents lower spatial heterogeneity when compared to the constant scale models. Their iCV is naturally higher (dark grey) in areas with more fires, which suggests their latent random measures can better capture variability, possibly due to unobserved spatial factors.

To get further insights into the drivers of the uncertainty of $\Lambda_t$ across time and space, we report in Fig.~\ref{fig:IQR} the normalised interquantile range defined as $(Q_{97.5}(\Lambda_t(A))-Q_{2.5}(\Lambda_t(A)))/Q_{50}(\Lambda_t(A))$, for several sub-regions $A \subset \Y$. The left plot shows that the uncertainty is smaller during the dry season than during the wet season since all points lay below the $45^{\circ}$ line.
Also, larger regions (dots) have lower uncertainty than smaller regions (circles) as more observations are available. This is confirmed by comparing the uncertainty and the number of observations per sub-region.
The right plot shows that within each season, the uncertainty decreases when the number of observations increases. Figure 8 in the Supplement reports the results for different sizes of the sub-regions.

Figure 9 in the Supplement reports the value of the function $\mathscr{R}_{t,t+h}(x,y)$, for $t=3$ (March) and $t=11$ (November) at different locations and horizons ($h=0,1,3$). In each plot, there are areas with clustering features, that is, $\mathscr{R}_{t,t+h}(x,y) > 1$, at a distance of $4^{\circ}$ from the centre $x$ (red shades), meaning that fires are likely to occur jointly at location $x$.
Other areas exhibit regularity, that is, $\mathscr{R}_{t,t+h}(x,y) < 1$ (white and blue shades). Overall, there is evidence of spatial heterogeneity and deviation from the standard Poisson process. The local aggregation features decrease as the horizon $h$ increases and do not change across dry and wet season months (e.g., November).

\FloatBarrier
\section{Discussion}
Motivated by the temporal and spatial dependence patterns in several real-world datasets, we have defined a novel dynamic shot noise Cox process. We introduced a measure-valued autoregressive gamma process to account for temporal dependence in latent intensity measures and extended the Poisson-gamma random fields to a dynamic setup.

We investigated some theoretical properties of the proposed process, including spatial and temporal dependence features. Moreover, we proposed a Bayesian inference procedure for parameters and latent intensity and designed an algorithm to perform posterior approximation. The procedure allows for quantifying uncertainty in intensity estimation over space and time. An empirical application of the forest fire time series in South America showed that the \SNMARG\ model can capture non-stationarity intensity features, such as periodic and spatial variability.
An interesting extension left for future work concerns generalising the proposed \MARG\ process to $p>1$ lags within the shot noise Cox process to account for more general dynamics in the latent intensity and observables.
Another interesting line for research is the investigation of alternative computational techniques, like the non-reversible Metropolis of \cite{kamatani2023non}.

\bibliographystyle{chicago}
\bibliography{paper-ref.bib}

\appendix

\renewcommand{\theequation}{A.\arabic{equation}}

\section{Proofs}
\label{sec:proofs}

In this section, we provide proof of the results presented in the main text. 

\begin{proof}[Proof of Proposition~\ref{proposition:finiteN_t}]
It is enough to show that $\E(N_t(B)) < +\infty$ for every bounded $B$, which clearly yields that $\pr\{N_t(B) < +\infty\} = 1$, that is, $N_t$ is a locally finite random measure.
Using eq.~\eqref{eq:meanWg} in Proposition~\ref{proposition:covWtg} with $g_1(\theta) = \kappa_t \int_B K_\phi(y,\theta) \ell(\mathrm{d}y)$, the above expectation is finite thanks to eq.~\eqref{eq:integrability2}.
\end{proof}

\begin{proof}[Proof of Proposition~\ref{proposition:Laplace_Wt1_Wt_lag1}]
By eq.~\eqref{eq:LPgammaP}, 
for any $f \in \BM_+(\Theta)$ one has
\begin{equation}
    \E\Big( e^{-\int_\Theta f(\theta) \: W_{t+1}(\mathrm{d}\theta)} \mid V_{t+1} \Big) = \exp\Big( - \!\! \int_\Theta \log(1+c_{t+1} f(\theta)) \: H(\mathrm{d}\theta) - \!\! \int_\Theta \log(1+c_{t+1} f(\theta)) \: V_{t+1}(\mathrm{d}\theta) \Big).
\label{eq:Laplace_Wt1_Vt1}
\end{equation}
Recall now that a Poisson random measure $V$ on $\Theta$ with intensity measure $H$ has Laplace functional
\begin{equation}
\label{eq:LPoisson}
    \CL_V(f) = \E\Big( e^{-\int_\Theta f(\theta) V(\mathrm{d}\theta)} \Big) = \exp\Big( -\!\! \int_\Theta \big( 1-e^{-f(\theta)} \big) \: H(\mathrm{d}\theta) \Big), \qquad f \in \BM_+(\Theta).
\end{equation}
By iterated expectations and using eq.~\eqref{eq:Laplace_Wt1_Vt1} together with eq.~\eqref{eq:LPoisson}, we have
\begin{align*}
    \E\Big( & e^{-\int_\Theta f(\theta) \: W_{t+1}(\mathrm{d}\theta)} \mid W_t \Big) = \E\Big( \E\Big( \exp\Big( -\int_\Theta f(\theta) \: W_{t+1}(\mathrm{d}\theta) \Big) \mid V_{t+1} \Big) \mid W_t \Big) \\
     & = \exp\Big( - \!\! \int_\Theta \log(1+c_{t+1} f(\theta)) \: H(\mathrm{d}\theta) \Big) \times \E\Big( \exp\Big( -\!\! \int_\Theta \log(1+c_{t+1} f(\theta)) \: V_{t+1}(\mathrm{d}\theta) \Big) \mid W_t \Big) \\
     & = \exp\Big( - \!\! \int_\Theta \log(1+c_{t+1} f(\theta)) \: H(\mathrm{d}\theta) \Big) \times \exp\Big( -\!\beta_{t+1} \!\int_\Theta \Big( 1-e^{-\log(1+c_{t+1} f(\theta))} \Big) \: W_t(\mathrm{d}\theta) \Big) \\
     & = \exp\Big( - \!\! \int_\Theta \log(1+c_{t+1} f(\theta)) \: H(\mathrm{d}\theta) -\!\! \int_\Theta \frac{\beta_{t+1} c_{t+1} f(\theta)}{1+c_{t+1} f(\theta)} \: W_t(\mathrm{d}\theta) \Big).
\end{align*}
By Definition~\ref{def:NCGP}, the thesis follows.
\end{proof}

\begin{proof}[Proof of Proposition~\ref{proposition:ARG-AutoregRepre}]
From eq.~\eqref{eq:Laplace_Wt1_Vt1} one observes that the conditional law of $W_{t+1}$ given $V_{t+1}$ of the state space representation in 
eq.~\eqref{eq:MARG1_model} is equivalent to 
\begin{equation*}
    W_{t+1} \mid V_{t+1}  \stackrel{\SL}{=} W_{t+1}^{(I)} + W_{t+1}^{(U)},
\end{equation*}
where $W_{t+1}^{(I)}$ is independent from $W_{t+1}^{(U)}$ and
\begin{align*}
    W_{t+1}^{(I)} \mid V_{t+1} \sim \GP\big( W_{t+1}^{(I)} \mid H, c_{t+1}^{-1} \big) \qquad W_{t+1}^{(U)} \mid V_{t+1} \sim \GP\big( W_{t+1}^{(U)} \mid V_{t+1}, c_{t+1}^{-1} \big).
\end{align*}

Recalling that $V_{t+1} \mid W_t \sim PP(V_{t+1} \mid \beta_{t+1} W_t)$, conditionally on $W_{t} = \sum_{i \geq 1} w_{i,t} \delta_{\theta_{i,t}}$ one can write  
\begin{equation*}
    V_{t+1} = \sum_{i \geq 1} v_{i,t+1} \delta_{\theta_{i,t}}, \quad \quad
    v_{i,t+1} \sim \mathcal{P}oi( v_{i,t+1} \mid \beta_{t+1} w_{i,t})
\end{equation*}
and 
\begin{equation*}
    W_{t+1}^{(U)} \mid V_{t+1} \stackrel{\SL}{=} \sum_{i \geq 1} w_{i,t+1}^{(U)} \delta_{\theta_{i,t}},
\end{equation*}
where the weights are distributed as
\begin{equation*}
    w_{i,t+1}^{(U)} \sim 
\begin{cases}
    \delta_{0}\big( w_{i,t+1}^{(U)} \big) & \text{ if } v_{i,t+1} = 0\\[5pt]
    \mathcal{G}a\big( w_{i,t+1}^{(U)} \mid v_{i,t+1},c_{t+1}^{-1} \big)  & \text{ if } v_{i,t+1} > 0,
\end{cases}
\end{equation*}
for $i=1,2,\ldots$, which is equivalent to the statement. 
\end{proof}

\begin{proof}[Proof of Proposition~\ref{proposition:laplace_Wth_Wt_lag1}]
Consider the case $h=2$. By iterated use of Proposition~\ref{proposition:Laplace_Wt1_Wt_lag1} and iterated expectations one obtains
\begin{align*}
    \E\Big( & e^{-\int_\Theta f(\theta) \: W_{t+2}(\mathrm{d}\theta)} \mid W_t \Big) = \E\Big( \E\Big( \exp\Big( -\int_\Theta f(\theta) \: W_{t+2}(\mathrm{d}\theta) \Big) \mid W_{t+1} \Big) \mid W_t \Big) \\
     & = \exp\Big( - \!\! \int_\Theta \log\big( 1+c_{t+2} f(\theta) \big) \: H(\mathrm{d}\theta) \Big) \times \E\Big( \exp\Big( - \! \int_\Theta \frac{\rho_{t+2} f(\theta)}{1+c_{t+2} f(\theta)} \: W_{t+1}(\mathrm{d}\theta) \Big) \mid W_t \Big) \\
     & = \exp\Big( - \!\! \int_\Theta \log\big( 1+c_{t+2} f(\theta) \big) \: H(\mathrm{d}\theta) \Big) \times \exp\Big( - \! \int_\Theta \frac{\rho_{t+1} \frac{\rho_{t+2} f(\theta)}{1+c_{t+2} f(\theta)}}{1+c_{t+1} \frac{\rho_{t+2} f(\theta)}{1+c_{t+2} f(\theta)}} \: W_t(\mathrm{d}\theta) \Big) \\
     & \quad \times \exp\Big( - \!\! \int_\Theta \log\Big( 1+c_{t+1} \frac{\rho_{t+2} f(\theta)}{1+c_{t+2} f(\theta)} \Big) \: H(\mathrm{d}\theta) \Big) \\
     & = \exp\Big( - \!\! \int_\Theta \log\big( 1+ (c_{t+2} + c_{t+1}\rho_{t+2})f(\theta) \big) \: H(\mathrm{d}\theta) \\
     & \qquad\quad - \! \int_\Theta \dfrac{\rho_{t+2} \rho_{t+1} f(\theta)}{1+(c_{t+2} +\rho_{t+2} c_{t+1}) f(\theta)} \: W_t(\mathrm{d}\theta) \Big).
\end{align*}
The general case $h > 2$ follows by iterating the same argument.
\end{proof}

\begin{proof}[Proof of Corollary~\ref{corollary:expected_distribution_Wt1_Wt_lagh}]
It can be easily proved that if $M \sim \NcGP(M \mid H,\beta W,c^{-1})$, then
\begin{equation*}
\E(M(A)) = c H(A) +  c \beta W(A).
\end{equation*}
By Proposition~\ref{proposition:laplace_Wth_Wt_lag1}, the thesis follows.
\end{proof}

\begin{proof}[Proof of Proposition~\ref{proposition:invariant_Wth_Wt_lag1}]
By assuming $c_t = c$, $\beta_t = \beta, \forall \, t$, one has $\rho_{t+h|t} = \rho^h$ and $c_{t+h|t} = c(1-\rho^h)/(1-\rho)$ for any $h \geq 1$.
Then, by Proposition~\ref{proposition:laplace_Wth_Wt_lag1}, one gets
\begin{align*}
    \lim_{h\to +\infty} \E\Big( e^{-\int_\Theta f(\theta) \: W_{t+h}(\mathrm{d}\theta)} \mid W_t \Big) & = \lim_{h\to +\infty} \! \exp\Bigg( - \!\! \int_\Theta \log\Big( 1+\frac{c(1-\rho^h)}{1-\rho} f(\theta) \Big) \: H(\mathrm{d}\theta) \\
    & \qquad\qquad - \!\! \int_\Theta \dfrac{\rho^h f(\theta)}{1+\frac{c(1-\rho^h)}{1-\rho} f(\theta)} \: W_t(\mathrm{d}\theta) \Bigg).
\end{align*}
Under the assumption $\rho < 1$, the exponent of the second term converges to $0$ by the dominated convergence theorem. In fact, for any $f \in \BM_+(\Theta)$ and any $h \geq 1$, 
\begin{align*}
    f_h(\theta) \coloneqq \rho^h f(\theta) \Bigg( 1+\frac{c(1-\rho^h)}{1-\rho} f(\theta) \Bigg)^{-1} \leq \rho^h f(\theta) \leq f(\theta), \quad \forall \, \theta \in \Theta, \: \forall \, h\geq 1
\end{align*}
and clearly $\lim_{h \to +\infty}f_h(\theta) = 0$ for any $\theta$.
Analogously, we can establish a bound for the argument of the logarithm in the first exponent as $1+(c(1-\rho^h))(1-\rho)^{-1} f(\theta) \leq 1+c(1-\rho)^{-1} f(\theta)$ and clearly $\lim_{h \to +\infty}  1+(c(1-\rho^h))(1-\rho)^{-1} f(\theta)
=  1+c(1-\rho)^{-1} f(\theta)$. 
Therefore
\begin{equation*}
    \lim_{h\to +\infty} \E\Big( \exp\Big( -\int_\Theta f(\theta) \: W_{t+h}(\mathrm{d}\theta) \Big) \mid W_t \Big) = \exp\Big( - \!\! \int_\Theta \log\Big( 1+\frac{c}{1-\rho} f(\theta) \Big) \: H(\mathrm{d}\theta) \Big).
\end{equation*}
Finally, when $W_t \sim \GP(W_t \mid H, (1-\rho)/c)$, we can prove that the random measure $W_{t+1}$ follows the same law as $W_t$ by using
Proposition~\ref{proposition:Laplace_Wt1_Wt_lag1}, as follows:
\begin{align*}
    \E\Big( \E\Big( & e^{-\int_\Theta f(\theta) \: W_{t+1}(\mathrm{d}\theta)} \mid W_t \Big) \Big) = \exp\Big( - \!\! \int_\Theta \log\big( 1+c f(\theta) \big) \: H(\mathrm{d}\theta) \Big) \\
     & \times \E\Big( \exp\Big( -\int_\Theta \frac{\rho f(\theta)}{1+c f(\theta)} \: W_t(\mathrm{d}\theta) \Big) \Big) \\
     & = \exp\Big( - \!\! \int_\Theta \log\big( 1+c f(\theta) \big) \: H(\mathrm{d}\theta) - \!\! \int_\Theta \log\Big( 1+ \frac{c}{1-\rho} \frac{\rho f(\theta)}{1+c f(\theta)} \Big) \: H(\mathrm{d}\theta) \Big) \\
     & = \exp\Big( - \!\! \int_\Theta \log\Big( 1 + \frac{c}{1-\rho} f(\theta) \Big) \: H(\mathrm{d}\theta) \Big).
\end{align*}
\end{proof}

\begin{proof}[Proof of Proposition~\ref{proposition:statN}]
Since $\Lambda_t(\mathrm{d}y) = \int_{\Theta} K_\phi(y,\theta) \: W_t(\mathrm{d}\theta) \ell(\mathrm{d}y)$, one has:
\begin{equation*}
\begin{split}
    \CL_{N_t}(f) 
     & = \E\Big( \exp\Big( -\!\int_{\Y} \Big( 1-e^{-f(y)} \Big) \: \int_{\Theta} K_\phi(y,\theta) \: W_t(\mathrm{d}\theta) \ell(\mathrm{d}y) \Big) \Big) \\
     & = \E\Big( \exp\Big( -\!\int_{\Theta} g(\theta) \: W_t(\mathrm{d}\theta) \Big) \Big) = \CL_{W_t}(g),
\end{split}
\end{equation*}
where
\begin{equation*}
    g(\theta) \coloneqq  \int_{B_f} \Big( 1-e^{-f(y)} \Big) K_\phi(y,\theta) \: \ell(\mathrm{d}y),
\end{equation*}
and $B_f \subset \Y$ is the support of $f$ which is bounded by assumption. Since $f$ is also bounded $\int_\Theta g(\theta) H(\mathrm{d}\theta) \leq \int_\Theta \int_{B_f} K_\phi(y,\theta) \: \ell(\mathrm{d}y) H(\mathrm{d}\theta) < +\infty$
by eq.~\eqref{eq:integrability2}.
This also yields
\begin{equation*}
    \mathcal{I}(g) \coloneqq \int_\Theta \log\Big( 1+\dfrac{c}{1-\rho} g(\theta) \Big) \: H(\mathrm{d}\theta) < +\infty.
\end{equation*}
Hence $\CL_{N_\infty}(f)$ is well-defined and $\exp( -\mathcal{I}(g) ) = \CL_{N_\infty}(f) = \CL_{W_\infty}(g)$ when $W_\infty \sim \GP( W_\infty \mid H, (1-\rho)/c )$.
At this stage, one gets that if $W_t$ is stationary, then process $N_t$ is stationary in the sense that $N_t \stackrel{\SL}{=} N_\infty$.  
To prove the first part of the statement, choose $A_\eps \subset \Theta$ bounded such that $\mathcal{I}(g\I \{A_\eps^c\})\leq \eps$, which is possible for every $\epsilon$ given $f$ since $\mathcal{I}(g)<+\infty$.
Using eq.~\eqref{eq:meanWg} and again eq.~\eqref{eq:integrability2}, 
one can also choose $A_\eps$ in such a way $\E(\int_{A_\eps^c} W_t(\mathrm{d}\theta)) \leq \epsilon$ for every $t$. 
Now define $g^\eps(\theta) = g(\theta) \I\{\theta \in A_\eps\}$ and write
\begin{equation*}
    |\CL_{N_t}(f) - \CL_{N_\infty}(f)| \leq |\CL_{W_t}(g) - \CL_{W_t}(g^\eps)| + |\CL_{W_t}(g^\eps) - \CL_{W_\infty}(g^\eps)| + |\CL_{W_\infty}(g^\eps) - \CL_{W_\infty}(g)|.
\end{equation*}
When $t\rightarrow +\infty$ one has that $|\CL_{W_t}(g^\eps) -\CL_{W_\infty}(g^\epsilon)|$ goes to zero by Proposition 5,
and using the conditions $\E(\int_{A_\eps^c} W_t(\mathrm{d}\theta)) \leq \epsilon$ and $\mathcal{I}(g\I \{A_\eps^c\})\leq \eps$ one gets $|\CL_{W_t}(g)-\CL_{W_t}(g^\eps)|+|\CL_{W_\infty}(g^\eps)-\CL_{W_\infty}(g)| \leq \eps$.
\end{proof}

\begin{proof}[Proof of Proposition~\ref{proposition:covWtg}]
By Proposition~\ref{proposition:laplace_Wth_Wt_lag1}
one has that $W_t$ given $W_1$ has distribution $\NcGP(H, \beta_{t|1}W_1,c_{t|1}^{-1})$, which is a CRM with L\'evy measure $\mu(\mathrm{d}w,\mathrm{d}\theta) = w^{-1}e^{- w/c_{t|1}} \: \mathrm{d}w H(\mathrm{d}\theta) + \beta_{t|1}/c_{t|1} e^{- w/c_{t|1}} \mathrm{d}w W_1(\mathrm{d}\theta)$,
eq.~\eqref{eq:muNCGP}.
By eq.~\eqref{eq:propernessNC-gamma},
one has the representation $W_t(g_i) = \int_{\RE_+\times \Theta} g_i(\theta) w \: M(\mathrm{d}w \mathrm{d}\theta)$ ($i=1,2$), with $M \sim \PP(M \mid \mu)$.
Now recall that, if $f_1$ and $f_2$ are measurable functions, then
\begin{align}
    & \E\Big( \int_{\RE_+ \times \Theta} f_1(w,\theta) \: M(\mathrm{d}w\mathrm{d}\theta) \Big) = \int_{\Re_+ \times \Theta} f(w,\theta) \: \mu(\mathrm{d}w\mathrm{d}\theta),
    \label{eq:E-PP} \\
    \notag
    & \Cov \Big( \int_{\Re_+ \times \Theta} \! f_1(w,\theta) \: M(\mathrm{d}w \mathrm{d}\theta),  \int_{\RE_+ \times \Theta} \! f_2(w',\theta') \: M(\mathrm{d}w'\mathrm{d}\theta') \Big) = \\
    & \hspace*{35ex} = \!\int_{\RE_+ \times \Theta} \! f_1(w,\theta) f_2(w,\theta) \: \mu(\mathrm{d}w\mathrm{d}\theta),
    \label{eq:cov-PP}
\end{align}
provided the integrals on the right-hand side are well-defined \citep[see, e.g.][Campbell's formula in Proposition 2.7 and eq. 4.26]{LastPenrose2018PoissonProcess_book}. 
The expression \eqref{eq:meanWg} follows from eq.~\eqref{eq:E-PP}. As for eq.~\eqref{eq:covWglag0}, we start with the elementary identity:
\begin{equation*}
    \Cov(W_t(g_1),W_t(g_2)) = \E( \Cov(W_t(g_1),W_t(g_2) \mid W_1) ) + \Cov(\E(W_t(g_1) \mid W_1), \E(W_t(g_2) \mid W_1)).
\end{equation*}
Using Corollary~\ref{corollary:expected_distribution_Wt1_Wt_lagh} 
with $t=1$ and $h=t-1$ one gets $\E(W_t(g_i) \mid W_1) = c_{t|1} H(g_i) + \rho_{t|1} W_1(g_i)$ and hence
\begin{equation*}
    \Cov(\E(W_t(g_1) \mid W_1),\E(W_t(g_2) \mid W_1)) = \rho_{t|1}^2 \Cov(W_1(g_1),W_1(g_2)). 
\end{equation*}
It remains to compute $\Cov(W_t(g_1),W_t(g_2) \mid W_1)$. Using eq.~\eqref{eq:cov-PP} with $f_i(w,\theta) = w g_i(\theta)$ one obtains
\begin{equation*}
\begin{split}
    \Cov(W_t(g_1),W_t(g_2) \mid W_1) & = \int_{\RE_+\times \Theta} g_1(\theta) g_2(\theta) w^2 \: \mu(\mathrm{d}w\mathrm{d}\theta) \\
    & = c_{t|1}^2 \int_{\Theta} g_1(\theta)g_1(\theta) \big( H(\mathrm{d}\theta)+ 2\beta_{t|1} W_1(\mathrm{d}\theta) \big).
\end{split}
\end{equation*}  
Equation~\eqref{eq:covWglag0} follows. Using Corollary~\ref{corollary:expected_distribution_Wt1_Wt_lagh} one gets
\begin{equation}
\begin{split}
    \E(W_t(g_1) W_{t+h}(g_2)) & = \E(W_t(g_1) \E(W_{t+h}(g_2) \mid W_t) ) \\
    & = c_{t+h|t} \E(W_t(g_1)) \int_\Theta g_2(\theta) H(\mathrm{d}\theta) + \rho_{t+h|t} \E(W_t(g_1) W_t(g_2))
\end{split}
\label{eq:Ewt-t+h}
\end{equation}
and 
\begin{equation*}
    \E(W_t(g_1)) \, \E(W_{t+h}(g_2)) = c_{t+h|t} \E(W_t(g_1)) \, \int_\Theta g_2(\theta) H(\mathrm{d}\theta) + \rho_{t+h|t} \E(W_t(g_2)) \E(W_t(g_1)). 
\end{equation*}
Combining these two identities,
eq.~\eqref{eq:covWglagh} follows. 
\end{proof}

\begin{proof}[Proof of Proposition~\ref{proposition:densities}]
Note that the expressions for $\mathfrak{D}^{(1)}_t (y)$, $\mathfrak{D}^{(2)}_t (y_1,y_2)$ and $\mathfrak{D}^{(2)}_{t,t+h} (y_1,y_2)$ are well defined $\ell$-almost everywhere by $(H_{1})$-$(H_3)$.
In addition, integrals in eq.~\eqref{eq:densitiestomoments} for these functions are also well-defined for bounded sets. Starting from eq.~\eqref{eq:densitytolambda}, eq.~\eqref{eq:inensity_t} follows by \eqref{eq:meanWg} with $g(\theta) = \kappa_t K_\phi(y,\theta)$ and \eqref{eq:corsdensity2} follows by \eqref{eq:Ewt-t+h} with $g_1(\theta) = \kappa_t K_\phi(y_1,\theta)$ and $g_2(\theta) = \kappa_{t+h} K_\phi(y_2,\theta)$.
Similarly eq.~\eqref{eq:corsdensity1} follows by eq.~\eqref{eq:meanWg}-\eqref{eq:covWglag0} using the elementary identity $\E(\lambda_t(y_1)\lambda_t(y_2)) = \Cov(\lambda_t(y_1),\lambda_t(y_2)) + \E(\lambda_t(y_1)) \E(\lambda_t(y_2))$.

Note that eq.~\eqref{eq:corsdensity1}
can also be deduced using Proposition 5.1 in \cite{moller2003statistical}, which states that in a Cox process directed by a Poisson process with intensity measure $\mu$ and kernel $k$, the second-order product density $\mathfrak{D}^{(2)}(y_1,y_2)$ satisfies
\begin{equation}
\begin{split}
    \mathfrak{D}^{(2)}_t (y_1,y_2) & =
    \int_{\Re_+\times\Theta}  w K_{\phi}(y_1,\theta)  \mu(\mathrm{d}w \mathrm{d}\theta) \int_{\Re_+\times\Theta} w K_{\phi}(y_2,\theta) \mu(\mathrm{d}w \mathrm{d}\theta) \\
    & +\int_{\Re_+\times\Theta} w^2 K_{\phi}(y_1,\theta) K_{\phi}(y_2,\theta) \mu(\mathrm{d}w, \mathrm{d}\theta).
\end{split}
\label{eq:MoellerProp5.1}
\end{equation}
Since conditionally on $W_1$, $N_t$ is a Cox process directed by a Poisson process with intensity $\mu(\mathrm{d}w,\mathrm{d}\theta) = \kappa_t w^{-1}e^{- w/c_{t|1}} \: \mathrm{d}w H(\mathrm{d}\theta) +\kappa_t \beta_{t|1}/c_{t|1}e^{- w/c_{t|1}} \mathrm{d}w W_1(\mathrm{d}\theta)$, 
eq.~\eqref{eq:corsdensity1}
follows by conditioning and elementary computations starting from eq.~\eqref{eq:MoellerProp5.1}. 
\end{proof}

\begin{proof}[Proof of Proposition~\ref{proposition:densities-Stationarity}]
By Proposition~\ref{proposition:invariant_Wth_Wt_lag1},
$W_t \sim \GP(W_t \mid H, (1-\rho)/c)$. 
Equation~\eqref{eq:inensity_t:stat} follows immediately from Proposition~\ref{proposition:densities}.
As for the second-order product density, one can use the expression recalled in eq.~\eqref{eq:MoellerProp5.1}
where now the L\'evy measure of $W_t$ is $\mu(\mathrm{d}w \mathrm{d}\theta) = w^{-1}e^{-c^*w} \mathrm{d}w H(\mathrm{d}\theta)$, with $c^* = (1-\rho)/c$. 
\end{proof}

\begin{proof}[Proof of Corollary~\ref{corollary:covariance_Nth}]
The proof follows by combining 
eq.~\eqref{eq:densitiestomoments} with Proposition~\ref{proposition:densities}. 
As an alternative proof, recall that if $M \sim PP(M \mid \mu)$ then $E(M(B_1)) = \mu(B_1)$ and $\E(M(B_1)M(B_2)) = \mu(B_1)\mu(B_2) + \mu(B_1 \cap B_1)$ \citep[see, e.g.][Campbell's formula in Proposition 2.7 and eq. 4.26]{LastPenrose2018PoissonProcess_book}.   
Hence, since $N_t \mid W_t \sim PP(N_t \mid \Lambda_t)$ one gets
\begin{equation}
\label{eq:covN_proof_corol1A}
    \Cov(N_t(B_1),N_t(B_2)) = \Cov(\Lambda_t(B_1),\Lambda_t(B_2)) + \E\big( \Lambda_t(B_1 \cap B_2) \big).
\end{equation}
Since for any set $B \subset \Y$, one has $\Lambda_t(B) = \kappa_t \int_\Theta K_\phi(B,\theta) W_t(\mathrm{d}\theta)$, by \eqref{eq:meanWg}
with $g(\theta)=\kappa_tK_\phi(B_1 \cap B_2,\theta)$ it follows
\begin{equation*}
    \E(\Lambda_t(B_1 \cap B_2)) = \kappa_t c_{t|1} \int_{\Theta} K_\phi(B_1 \cap B_2,\theta) (H(\mathrm{d}\theta) + \beta_{t|1} \bar W_1(\mathrm{d}\theta)).
\end{equation*}
Moreover, by Proposition~\ref{proposition:covWtg} 
with $g_1(\theta) = \kappa_t K_\phi(B_1,\theta)$ and $g_2(\theta) = \kappa_{t} K_\phi(B_2,\theta)$ one obtains 
\begin{equation*}
\begin{split}
    \Cov(\Lambda_t(B_1),\Lambda_t(B_2)) & = \kappa_t^2 c_{t|1}^2 \int_\Theta K_\phi(B_1,\theta)  K_\phi(B_2,\theta) (H(\mathrm{d}\theta) + 2\beta_{t|1} \bar{W}_1 (\mathrm{d}\theta)) \\
     & + \kappa_t^2 \rho_{t|1}^2 \Cov\Big( \int_{\Theta} K_\phi(B_1,\theta) W_1(\mathrm{d}\theta), \int_{\Theta} K_\phi(B_2,\theta) W_1(\mathrm{d}\theta) \Big).
\end{split}
\end{equation*}
Collecting all the terms, the statement for $h=0$ follows.
As for $h \geq 1$
\begin{equation*}
\begin{split}
    \E(N_t(B_1)N_{t+h}(B_2)) & = \E\big( \E(N_t(B_1)N_{t+h}(B_2) \mid W_t,W_{t+h}) \big) \\
     & = \E(\Lambda_t(B_1) \Lambda_{t+h}(B_2)).
\end{split}
\end{equation*}
Since $\E(N_t(B_1)) = \E(\Lambda_t(B_1))$ and $\E(N_{t+h}(B_2)) = \E(\Lambda_{t+h}(B_2))$, one gets 
\begin{equation}
\label{eq:covN_proof_corol1B}
    \Cov(N_t(B_1),N_{t+h}(B_2)) = \Cov(\Lambda_t(B_1),\Lambda_{t+h}(B_2)) = \Cov(W_t(g_1),W_{t+h}(g_2)),
\end{equation}
for $g_1(\theta)= \kappa_t K_\phi(B_1,\theta)$ and $g_2(\theta) = \kappa_{t+h} K_\phi(B_2,\theta)$.
By Proposition~\ref{proposition:covWtg},
$\Cov(W_t(g_1),W_{t+h}(g_2)) = \rho_{t+h|t} \Cov(W_t(g_1), W_t(g_2))$ and the thesis follows also for $h \geq 1$. 
\end{proof}

\begin{proof}[Proof of Corollary~\ref{corollary:corol_covariance_N}]
Start by recalling the relationship $\E(N_t(B_1)) = \E(\Lambda_t(B_1)) = \kappa_t \E\big( \int_\Theta K_\phi(B_1,\theta) W_t(\mathrm{d}\theta) \big)$. Under the stated assumptions, it follows that $W_t \sim \GP(W_t \mid H, (1-\rho)/c)$ and $W_t \sim \GP(W_t \mid H, (1-\rho)/c)$, cf. Proposition~\ref{proposition:invariant_Wth_Wt_lag1},
thus proving the first statement. 
If $h=0$, by eq.~\eqref{eq:covN_proof_corol1A} one has $\Cov(N_t(B_1),N_{t+h}(B_2)) = \Cov(W_t(g_1),W_{t}(g_2)) + \E\big( W_t(g_{1,2}) \big)$, where $g_1(\theta) =  \kappa_tK_\phi(B_1,\theta)$, $g_2(\theta) = \kappa_{t}K_\phi(B_2,\theta)$, and $g_{1,2}(\theta) =  \kappa_t K_\phi(B_1 \cap B_2, \theta)$. 
Since $W_t \sim \GP(W_t \mid H, (1-\rho)/c)$, the second part of the statement for $h=0$ follows using eq.~\eqref{eq:cov-PP}.
To conclude, assume $h \geq 1$ and use eq.~\eqref{eq:covN_proof_corol1B} to write $\Cov(N_t(B_1),N_{t+h}(B_2)) = \Cov(W_t(g_1),W_{t+h}(g_2))$, with $g_1(\theta) = \kappa_t K_\phi(B_1,\theta)$ and $g_2(\theta) =  \kappa_{t+h} K_\phi(B_2,\theta)$. 
By Proposition~\ref{proposition:covWtg},
$\Cov(W_t(g_1),W_{t+h}(g_2)) = \rho_{t+h|t} \Cov(W_t(g_1), W_t(g_2))$ and the claim follows using $W_t \sim \GP(W_t \mid H, (1-\rho)/c)$ and eq.~\eqref{eq:cov-PP}.
\end{proof}

\begin{proof}[Proof of Proposition~\ref{proposition:MomentsHighorders}]
The unconditional moments of $N_t(B)$ follow from the moments of $\Lambda_t$.
Since $N_t(B) \mid \Lambda_t \sim \mathcal{P}oi(N_t(B) \mid \Lambda_t(B))$ and the $m$th moment of a Poisson random variable of parameter $\lambda$ is $\sum_{j=1}^m S_{m,j} \lambda^j$, then 
\begin{equation}
    \E\big( N_t(B)^m \big) = E\Big( \E\big( N_t(B)^m \mid \Lambda_t \big)\Big) = \E\Big( \sum_{j=1}^m S_{m,j} \Lambda_t(B)^j \Big).
\label{eq:spatialmomentsA}
\end{equation}
To prove eq.~\eqref{eq:spatialmoments_stationary},
starting from eq.~\eqref{eq:spatialmomentsA}, one needs to compute $\E\big( \Lambda_t(B)^j \big)$.
Since we are assuming that $W_t \sim \GP(W_t \mid H, (1-\rho)/c)$, $W_t$ is a CRM with L\'evy measure $\mu(\mathrm{d}w \mathrm{d}\theta) = w^{-1}e^{-c^*w} \mathrm{d}w H(\mathrm{d}\theta)$, where $c^* = (1-\rho)/c$.
By eq.~\eqref{eq:propernessNC-gamma} one gets
\begin{equation*}
    \Lambda_t(B) = \kappa_t \int_\Theta K_\phi(B,\theta) \: W_{t}(\mathrm{d}\theta) =\kappa_t \int_{\RE_+\times \Theta} K_\phi(B,\theta) w \: M(\mathrm{d}w\mathrm{d}\theta),
\end{equation*}
where $M \sim \PP(M \mid \mu)$.
For $M \sim \PP(M \mid \mu)$, following \cite{bassan1990moments}, one has
\begin{equation*}
    \E\Big( \Big( \int_{\RE_+\times \Theta} K_\phi(B,\theta) w \: M(\mathrm{d}w\mathrm{d}\theta) \Big)^j \Big) = \sum_{\substack{\ell_1,\dots,\ell_j \geq 0 \\ \sum_{r=1}^j r\ell_r=j}} c(\ell_1,\dots,\ell_j) \prod_{r=1}^j J_r(B)^{\ell_r},
\end{equation*}
which concludes the proof. 
\end{proof}


\clearpage

\begin{center}
	{\bf Supplementary Material for: ``A Spatiotemporal Gamma Shot Noise Cox Process''}
\end{center}

\begin{abstract}
Section~\ref{sec:ILM_algorithm} describes the Inverse L\'evy Measure algorithm.
Then, Section~\ref{sec:posterior} provides some details on the posterior approximation methods, including the derivation of the posterior full conditional distributions for the sampler.
Section~\ref{sec:simulation_continuous} reports simulated paths of the \SNMARG\ process illustrating the role of the static parameters.
Then, Section~\ref{sec:simulation_exercise} presents results of the Bayesian inference on synthetic data in the case of a discrete base measure, $H$.
Finally, Section~\ref{Sec:Supp_empirical_appl} contains additional results on the numerical illustration.
\end{abstract}

\renewcommand{\theequation}{S.\arabic{equation}}

\vspace*{5ex}

\appendix
\section{ILM algorithm}
\label{sec:ILM_algorithm}

The inverse L\'{e}vy Measure (ILM) proposed by \cite{Wolpert1998PoissonGamma_RandomField} is an algorithm designed for sampling from inhomogeneous gamma random fields and other random measures assigning independent infinitely divisible random variables to disjoint sets.

For instance, consider the task of sampling from a Gamma random measure, $W \sim \GP(W \mid H,c^{-1})$.
The ILM algorithm to obtain an approximate draw consists of the following steps:
\begin{enumerate}
    \item Fix an integer $M >0$, corresponding to the truncation value, and choose any distribution $\Pi(\mathrm{d}\theta)$ on $\Theta$ which is easy to sample from and such that the shape measure $H(\mathrm{d}\theta)$ has a density $h(\theta) = H(\mathrm{d}\theta)/\Pi(\mathrm{d}\theta)$.
    \item Generate $M$ independent identically distributed atoms $\theta_i \sim H(\mathrm{d}\theta)$, for $i=1,\ldots,M$.
    \item Generate the first $M$ jumps $\tau_i$, for $i=1,\ldots,M$, of a standard Poisson process (for example, by adding successive independent exponential random variables).
    \item Set the weights:
    \begin{equation*}
        w_i = E_1^{-1}\Big( \frac{\tau_i}{h(\theta_i)} \Big) c,
    \end{equation*}
    where $E_1(s) = \int_s^{\infty} e^{-u} u^{-1} \, \mathrm{d}u$, is the exponential integral function.
    \item Define the approximated random measure, $W_M$ as the finite sum:
    \begin{equation*}
        W(\mathrm{d}\theta) \approx W_M(\mathrm{d}\theta) = \sum_{i=1}^M w_i \delta_{\theta_i}.
    \end{equation*}
\end{enumerate}
The approximate measure $W_M$ converges in distribution to $W$ as $M\to \infty$ \citep[see][]{Wolpert1998PoissonGamma_RandomField,Wolpert1998Simulation_Levy_RandomFields}.

\section{Posterior full conditional distributions} \label{sec:posterior}

In this section, we provide a description of the posterior full conditional distributions of the parameters and the algorithms used to draw samples from them.
Let us first define the collection of measures $\bW = \{ W_1,\ldots,W_T \}$, $\bZ = \{ Z_1,\ldots,Z_T \}$, $\bY = \{ y_1,\ldots,y_T \}$, and $\bN^y = \{ N_1^y,\ldots,N_T^y \}$.
Finally, let us recall that $\balpha = \{ \alpha_1,\ldots,\alpha_{N^g} \}$, $\bc = \{ c_1,\ldots,c_T \}$, $\by_{1:T} = \{ y_1,\ldots,y_T \}$, $\bw_{1:T} = \{ w_1,\ldots,w_T \}$ and $\bz_{1:T} = \{ z_1,\ldots,z_T \}$.

\subsection{Sampling the parameters \texorpdfstring{$\psi$}{psi}}
The parameters $\bpsi$ given $\by_{1:T},\bz_{1:T}$, and $\bw_{1:T}$ are drawn from their full conditional distributions by adaptive Metropolis-Hastings. Details are given below. 

The full conditional posterior of the spatial shape parameters $\alpha_j$, $j=1,\ldots,N^g$ is 
\begin{align*}
    \alpha_j \mid \bW, \beta, \bc, \gamma \propto \mathcal{G}a(\alpha_j \mid \underline{a}_\alpha, \gamma \underline{b}_\alpha) \prod_{t=1}^T \NcGa(w_{j,t} \mid \alpha_j, \beta w_{j,t-1}, c_t^{-1}),
\end{align*}
which can be sampled by using an adaptive random walk Metropolis-Hastings (RWMH) step with a Lognormal proposal \citep[e.g., see][]{Andrieu08AdaptiveMCMC,Atchade2005adaptiveMH}.

The full conditional posterior of the shape parameter $\beta$ is 
\begin{align*}
    \beta \mid \bW, \balpha, \bc \propto \mathcal{G}a(\beta \mid \underline{a}_\beta, \underline{b}_\beta) \prod_{t=1}^T \prod_{j=1}^{N^g} \NcGa(w_{j,t} \mid \alpha_j, \beta w_{j,t-1}, c_t^{-1}),
\end{align*}
which can be sampled by using an adaptive RWMH step with a Lognormal proposal.

The full conditional posterior of the constant scale parameter $c$ is
\begin{align*}
    c \mid \bW, \balpha, \beta \propto \mathcal{G}a(c \mid \underline{a}_c, \underline{b}_c) \prod_{t=1}^T \prod_{j=1}^{N^g} \NcGa(w_{j,t} \mid \alpha_j, \beta w_{j,t-1}, c^{-1}),
\end{align*}
which can be sampled by using an adaptive RWMH step with a Lognormal proposal.

The full conditional posterior of the scale parameters $c_t$, $t=1,\ldots,T$ is
\begin{align*}
    c_t \mid \bW, \balpha, \beta, r \propto \mathcal{G}a(c_t \mid r^2/\underline{\sigma}^2_c, r/\underline{\sigma}^2_c) \prod_{j=1}^{N^g} \NcGa(w_{j,t} \mid \alpha_j, \beta w_{j,t-1}, c_t^{-1}),
\end{align*}
which can be sampled by using an adaptive RWMH step with a Lognormal proposal.

The full conditional posterior of the hierarchical scale parameter $r$ is 
\begin{align*}
    r \mid \bc & \propto \mathcal{G}a(r \mid \underline{a}_r, \underline{b}_r) \prod_{t=1}^T \mathcal{G}a\Big( c_t \mid \frac{r^2}{\underline{\sigma}^2_c}, \frac{r}{\underline{\sigma}^2_c} \Big) \\
     & \propto r^{\underline{a}_r-1} e^{-r \underline{b}_r} \prod_{t=1}^T \frac{1}{\Gamma\Big(\frac{r^2}{\underline{\sigma}^2_c}\Big)} \Big(\frac{r}{\underline{\sigma}^2_c} \Big)^{\frac{r^2}{\underline{\sigma}^2_c}} \: c_t^{\frac{r^2}{\underline{\sigma}^2_c}-1} \: \exp\Big( -\frac{r}{\underline{\sigma}^2_c} c_t \Big) \\
     & \propto r^{\underline{a}_r-1} \exp\Bigg\{ -r \Big( \underline{b}_r + \sum_{t=1}^T \frac{c_t}{\underline{\sigma}^2_c} \Big)\Bigg\} \Big( \prod_{t=1}^T c_t \Big)^{\frac{r^2}{\underline{\sigma}^2_c}} \Bigg\{ \frac{1}{\Gamma\Big(\frac{r^2}{\underline{\sigma}^2_c}\Big)} \Big(\frac{r}{\underline{\sigma}^2_c} \Big)^{\frac{r^2}{\underline{\sigma}^2_c}} \Bigg\}^T,
\end{align*}
which can be sampled by using an adaptive RWMH step with a Lognormal proposal.

\vspace*{3ex}

Assuming monthly-specific scales $c_t$, that is
\begin{align*}
    c_t = \begin{cases}
    \xi_k,    & t \bmod 12 = k \wedge k \in \{1,\ldots,11\} \\
    \xi_{12}, & t \bmod 12 = 0.
    \end{cases}
\end{align*}
We choose the hierarchical prior:
\begin{align*}
    \xi_k \mid r & \distas{iid} \mathcal{G}a(\xi_k \mid r^2/\underline{\sigma}_c^2, r/\underline{\sigma}_c^2), \quad k=1,\ldots,12, \\
    r & \sim \mathcal{G}a(r \mid \underline{a}_r,\underline{b}_r).
\end{align*}
Hence, the full conditional posterior of the monthly scale parameters $\xi_k$, $k=1,\ldots,12$ is 
\begin{align*}
    \xi_k \mid r, \balpha, \beta, \bW \propto \mathcal{G}a(\xi_k \mid r^2/\underline{\sigma}^2_c, r/\underline{\sigma}^2_c) \prod_{t \in \mathcal{T}_k} \prod_{j=1}^{N^g} \NcGa(w_{j,t} \mid \alpha_j, \beta w_{j,t-1}, c_t^{-1}),
\end{align*}
where $\mathcal{T}_k = \{ t \in \{1,\ldots,T\} : t \bmod 12 = k \}$, for $k=1,\ldots,11$, and $\mathcal{T}_{12} = \{ t \in \{1,\ldots,T\} : t \bmod 12 = 0 \}$ if $k=12$. We use an adaptive RWMH step to sample from this distribution.

Let $\bxi = \{ \xi_1,\ldots,\xi_{12} \}$, then the full conditional posterior of the hierarchical scale parameter $r$ is 
\begin{align*}
r \mid \bxi & \propto \mathcal{G}a(r \mid \underline{a}_r, \underline{b}_r) \prod_{k=1}^{12} \mathcal{G}a(\xi_k \mid r^2/\underline{\sigma}^2_c, r/\underline{\sigma}^2_c) \\
 & \propto r^{\underline{a}_r-1} e^{-r \underline{b}_r} \prod_{k=1}^{12} \frac{1}{\Gamma\Big(\frac{r^2}{\underline{\sigma}^2_c}\Big)} \Big(\frac{r}{\underline{\sigma}^2_c} \Big)^{\frac{r^2}{\underline{\sigma}^2_c}} \: \xi_k^{\frac{r^2}{\underline{\sigma}^2_c}-1} \: \exp\Big( -\frac{r}{\underline{\sigma}^2_c} \xi_k \Big) \\
 & \propto r^{\underline{a}_r-1} \exp\Bigg\{ -r \Big( \underline{b}_r + \sum_{k=1}^{12} \frac{\xi_k}{\underline{\sigma}^2_c} \Big)\Bigg\} \Big( \prod_{k=1}^{12} \xi_k \Big)^{\frac{r^2}{\underline{\sigma}^2_c}} \Bigg\{ \frac{1}{\Gamma\Big(\frac{r^2}{\underline{\sigma}^2_c}\Big)} \Big(\frac{r}{\underline{\sigma}^2_c} \Big)^{\frac{r^2}{\underline{\sigma}^2_c}} \Bigg\}^{12},
\end{align*}
which can be sampled by using an adaptive RWMH step with a Lognormal proposal.

\vspace*{3ex}

The full conditional posterior of the hierarchical scale parameter $\gamma$ is
\begin{align*}
    \gamma \mid \balpha & \propto \gamma^{\underline{a}_\gamma -1} \exp\big( -\gamma \underline{b}_\gamma \big) \prod_{j=1}^{N^g} \gamma^{\underline{a}_\alpha} \exp\big( -\alpha_j \gamma \underline{b}_\alpha \big) 
    \propto \mathcal{G}a(\gamma \mid \overline{a}_\gamma, \overline{b}_\gamma),
\end{align*}
where $\overline{a}_\gamma = \underline{a}_\gamma + N^g \underline{a}_\alpha$ and $\overline{b}_\gamma = \underline{b}_\gamma + \underline{b}_\alpha \sum_{j=1}^{N^g} \alpha_j$.

The full conditional posterior of the kernel bandwidth parameter $\phi$ is
\begin{align*}
    \phi \mid \bY, \bW, \bZ \propto \mathcal{G}a(\phi \mid \underline{a}_\phi, \underline{b}_\phi) \prod_{j=1}^{N^g} \prod_{(i,t) : z_{i,t}=j} \!\! \frac{\mathcal{N}_2(y_{i,t} \mid \theta_j, \Id_2 \phi^2)}{\mathcal{N}_2(\Y \mid \theta_j, \Id_2 \phi^2)},
\end{align*}
which can be sampled by using an adaptive RWMH step with a Lognormal proposal.

The full conditional posterior of the temporal covariates' coefficients $\eta$ is 
\begin{align*}
    \eta \mid \bW, \bN^y \propto \mathcal{N}_m(\eta \mid \underline{\mu}_\eta, \underline{\Sigma}_\eta) \prod_{t=1}^T \mathcal{P}oi(N_t^y \mid \Lambda_t(\Y)),
\end{align*}
which can be sampled by using an adaptive RWMH step with a multivariate Lognormal proposal.

\subsection{Anisotropic case}
The baseline model can extended to account for anisotropy by replacing the constant scale in the kernel, $\phi$, with a full covariance matrix, $\Sigma$. In this case, we first parametrise the covariance using a Cholesky decomposition to grant positive definiteness of $\Sigma$ at every iteration:
\begin{align*}
    \Sigma & = L L^\T    & 
    L & = \begin{pmatrix}
        e^{\varpi_1} & 0 \\
        \varpi_2 & e^{\varpi_3}
    \end{pmatrix}.
\end{align*}
We assume a multivariate Gaussian prior for $(\varpi_1, \varpi_2, \varpi_3)^\T \sim \mathcal{N}_3(\boldsymbol{0}, \sigma^2_\xi \Id_3)$. An adaptive RWMH with Gaussian proposal is used for each of the three parameters.

\subsection{Sampling the allocation variables \texorpdfstring{$\bz_{1:T}$}{z}}
The posterior distribution of the allocation variables $\bz_{1:T}$ is easily described. 
For $i=1,\ldots,N_t^y$, $t=1,\ldots,T$, one has
\begin{equation*}
    \pr\{ z_{i,t} = m \mid \by_{1:T},\bw_{1:T},\bpsi \} = \frac{w_{m,t} k_\phi(y_{i,t},\theta_{m}) }{{\Lambda}(y_{i,t})}, \qquad m=1,\dots,N^g,
\end{equation*}
where ${\Lambda}(y_{i,t}) = \sum_{m=1}^{N^g} w_{m,t} k_\phi(y_{i,t},\theta_{m})$.

\subsection{Sampling the latent states \texorpdfstring{$\mathbf{w}_{1:T}$}{w}}
\paragraph{Adaptive Metropolis-Hastings} The posterior full conditional of $w_{\ell,t}$ is obtained as
\begin{align*}
    p(w_{\ell,t} & \mid w_{\ell,t-1}, w_{\ell,t+1}, \bw_{-\ell,t}, \by_t, \bz_t, \bpsi)  \propto \mathcal{P}oi( N_t^y \mid \Lambda_t(\Y) ) \prod_{i=1}^{N_t^y} \frac{\kappa_t w_{z_{i,t},t} \mathcal{N}_2(y_{i,t} \mid \theta_{z_{i,t}}, \Id_2\phi^2)}{\Lambda_t(\Y)} \\
    & \quad \times \NcGa(w_{j,t} \mid \alpha_j, \beta w_{j,t-1},c_t^{-1}) \NcGa(w_{j,t+1} \mid \alpha_j, \beta w_{j,t},c_{t+1}^{-1}) \\
    & \propto \Lambda_t(\Y)^{N_t^y} \exp( -\Lambda_t(\Y) ) \Big( \frac{\kappa_t}{\Lambda_t(\Y)} \Big)^{N_t^y} \prod_{i=1}^{N_t^y} \Big[ w_{z_{i,t},t} \mathcal{N}_2(y_{i,t} \mid \theta_{z_{i,t}}, \Id_2\phi^2) \Big] \\
    & \quad \times \NcGa(w_{j,t} \mid \alpha_j, \beta w_{j,t-1},c_t^{-1}) \NcGa(w_{j,t+1} \mid \alpha_j, \beta w_{j,t},c_{t+1}^{-1}) \\
    & \propto \exp( -\Lambda_t(\Y) ) \Big[ \prod_{i=1}^{N_t^y} w_{z_{i,t},t} \Big] \NcGa(w_{j,t} \mid \alpha_j, \beta w_{j,t-1},c_t^{-1}) \NcGa(w_{j,t+1} \mid \alpha_j, \beta w_{j,t},c_{t+1}^{-1}).
\end{align*}
We sample from this distribution using a RWMH with Lognormal proposal and adaptive scale as in \cite{Andrieu08AdaptiveMCMC}.

\paragraph{Conditional Sequential Monte Carlo}
We follow the approach and the notation of the ancestor line as in \cite{Andrieu2010particleMCMC}.
Let $q(\cdot \mid \cdot)$  be an importance density from which the particles are generated (at $t=1$) or extended forward in time (for $t=2,\ldots, T$).
Set $A_{\ell,t-1}^k$ to be the index of the `parent' at time $t-1$ of the particle $\bw_{\ell,1:t}^k$, and let $\mathcal{F}(\cdot)$ be a discrete distribution governing the resampling procedure. For $\mathcal{F}$ equal to the multinomial distribution, one obtains the standard multinomial resampling.
Finally,  $B_{\ell,1:T}^k = (B_{\ell,1}^k,\ldots,B_{\ell,T}^k)$ is the ancestor lineage of the $k$th particle, where $B_{\ell,T}^k \coloneqq k$ and $B_{\ell,t}^k \coloneqq A_{\ell,t}^{B_{\ell,t+1}^k}$ for $t=T-1,\ldots,1$. Note that $\bw_{\ell,1:T}^k = (\bw_{\ell,1}^{B_{\ell,1}^k},\ldots,\bw_{\ell,T}^{B_{\ell,T}^k})$ for all $k$.
The Sequential Monte Carlo algorithm is:
\begin{enumerate}
    \item Let $\bw_{\ell,1:T} = (\bw_{\ell,1}^{B_{\ell,1}},\bw_{\ell,2}^{B_{\ell,2}},\ldots,\bw_{\ell,T}^{B_{\ell,T}})$ be a path associated with the ancestral lineage $B_{\ell,1:T}$.
    \item At time $t=1$, do
    \begin{enumerate}
        \item for $k \neq B_{\ell,1}$, sample $\bw_{\ell,1}^k \sim q(\bw_{\ell,1}^k \mid \by_1, \bz_1, \bw_{-\ell,1}, \bpsi)$
        \item compute the weights $u_{\ell,1}(\bw_{\ell,1}^k)$ and their normalised version, $\tilde{u}_{\ell,1}$, as
        \begin{align*}
            u_{\ell,1}(\bw_{\ell,1}^k) & \coloneqq \frac{p(\bw_{\ell,1}^k, \by_1 \mid \bz_1,\bw_{-\ell,1},\bpsi)}{q(\bw_{\ell,1}^k \mid \by_1, \bz_1, \bw_{-\ell,1}, \bpsi)} \\
            & = \frac{p(\bw_{\ell,1}^k \mid \bpsi) p(\by_1 \mid \bz_1,\bw_{\ell,1}^k,\bw_{-\ell,1},\bpsi)}{q(\bw_{\ell,1}^k \mid \by_1, \bz_1, \bw_{-\ell,1}, \bpsi)},
            \\
            \tilde{u}_{\ell,1}^k & \coloneqq \frac{u_{\ell,1}(\bw_{\ell,1}^k)}{\sum_{j=1}^N u_{\ell,1}(\bw_{\ell,1}^j)}.
        \end{align*}
    \end{enumerate}
    \item For time $t=2,\ldots,T$, do
    \begin{enumerate}
        \item for $k \neq B_{\ell,t}$, sample $A_{\ell,t-1}^k \sim \mathcal{F}(A_{\ell,t-1}^k \mid \tilde{u}_{\ell,1}^k,\ldots,\tilde{u}_{\ell,t-1}^k)$
        \item for $k \neq B_{\ell,t}$, sample $\bw_{\ell,t}^k \sim q(\bw_{\ell,t}^k \mid \by_t, \bw_{\ell,t-1}^{A_{\ell,t-1}^k}, \bz_t, \bw_{-\ell,t}, \bpsi)$
        \item compute the weights $u_{\ell,t}(\bw_{\ell,1:t}^k)$ and their normalised version, $\tilde{u}_{\ell,t}$, as
        \begin{align*}
            u_{\ell,t}(\bw_{\ell,1:t}^k) & \coloneqq \frac{p(\bw_{\ell,1:t}^k, \by_{1:t} \mid \bz_{1:t}, \bw_{-\ell,1:t}, \bpsi)}{q(\bw_{\ell,1:t}^k \mid \by_{1:t}, \bz_{1:t}, \bw_{\ell,1:t-1}^{A_{t-1}^k}, \bpsi)} \\
            & = \frac{p(\bw_{\ell,1:t}^k \mid \bw_{\ell,1:t-1}, \bpsi) p(\by_{1:t} \mid \bz_{1:t}, \bw_{\ell,1:t}^k, \bw_{-\ell,1:t}, \bpsi)}{q(\bw_{\ell,1:t}^k \mid \by_{1:t}, \bz_{1:t}, \bw_{\ell,1:t-1}^{A_{t-1}^k}, \bpsi)} \\
            \tilde{u}_{\ell,t}^k & \coloneqq \frac{u_{\ell,t}(\bw_{\ell,1:t}^k)}{\sum_{j=1}^N u_{\ell,t}(\bw_{\ell,1:t}^j)},
        \end{align*}
    \end{enumerate}
\end{enumerate}
This Sequential Monte Carlo algorithm yields an approximation of the distribution $p(\bw_{\ell,1:T} \mid \by_{1:T}, \bz_{1:T}, \bw_{-\ell,1:T}, \bpsi)$ given by
\begin{equation*}
    \widehat{p}(\mathrm{d}\bw_{\ell,1:T} \mid \by_{1:T}, \bz_{1:T}, \bw_{-\ell,1:T}, \bpsi) \coloneqq \sum_{k=1}^N \tilde{u}_{\ell,T}^k \, \delta_{\bw_{\ell,1:T}^k}(\mathrm{d}\bw_{\ell,1:T}).
\end{equation*}
In the empirical analysis, we take $q(\cdot \mid \cdot)$ as the prior noncentral gamma transition density and $\mathcal{F}$ as the multinomial distribution (leading to multinomial resampling).
As a result, the previous steps 2. and 3. simplify as follows:
\begin{enumerate}
    \setcounter{enumi}{1}
    \item At time $t=1$, do
    \begin{enumerate}
        \item for $k \neq B_{\ell,1}$, sample $\bw_{\ell,1}^k \sim \mathcal{G}a(\bw_{\ell,1}^k \mid \alpha_\ell, c_1/(1-\beta c_1))$
        \item compute the weights $u_{\ell,1}(\bw_{\ell,1}^k)$ and their normalised version, $\tilde{u}_{\ell,1}$, as
        \begin{align*}
            u_{\ell,1}(\bw_{\ell,1}^k) & = p(\by_1 \mid \bz_1,\bw_{\ell,1}^k,\bw_{-\ell,1},\bpsi),
            &
            \tilde{u}_{\ell,1}^k & = \frac{u_{\ell,1}(\bw_{\ell,1}^k)}{\sum_{j=1}^N u_{\ell,1}(\bw_{\ell,1}^j)}.
        \end{align*}
    \end{enumerate}
    \item For time $t=2,\ldots,T$, do
    \begin{enumerate}
        \item for $k \neq B_{\ell,t}$, sample $A_{\ell,t-1}^k \sim \text{Multin}(A_{\ell,t-1}^k \mid \tilde{u}_{\ell,1}^k,\ldots,\tilde{u}_{\ell,t-1}^k)$
        \item for $k \neq B_{\ell,t}$, sample $\bw_{\ell,t}^k \sim \text{NcGa}(\bw_{\ell,t}^k \mid \alpha_\ell, \beta \bw_{\ell,t-1}^{A_{\ell,t-1}^k}, c_t^{-1})$
        \item compute the weights $u_{\ell,t}(\bw_{\ell,1:t}^k)$ and their normalised version, $\tilde{u}_{\ell,t}$, as
        \begin{align*}
            u_{\ell,t}(\bw_{\ell,1:t}^k) & = p(\by_{1:t} \mid \bz_{1:t}, \bw_{\ell,1:t}^k, \bw_{-\ell,1:t}, \bpsi),
            &
            \tilde{u}_{\ell,t}^k & \coloneqq \frac{u_{\ell,t}(\bw_{\ell,1:t}^k)}{\sum_{j=1}^N u_{\ell,t}(\bw_{\ell,1:t}^j)}.
        \end{align*}
    \end{enumerate}
\end{enumerate}

\subsection{Dependent prior on $c_t$}
Using the shape-rate parametrisation of the Gamma distribution, following \cite{gamerman2013non} we specify the following conditional density for $c_t$ given its past $c_{0:t-1} = \{ c_0,c_!,\dots,c_{t-1} \}$ and the auxiliary variable $\zeta_t \in\Rp$
\begin{align*}
    \zeta_0 \mid c_0 & \sim \mathcal{G}a\big( \zeta_0 \mid \underline{a}_\zeta, \underline{b}_\zeta \big) \\
    \zeta_t & = \omega^{-1} \zeta_{t-1} \upsilon_t \qquad \upsilon_t \mid c_{0:t-1} \sim \mathcal{B}e\big( \upsilon_t \mid \omega \underline{a}_{t-1}, (1-\omega) \underline{b}_{t-1} \big) \\
    c_t \mid \zeta_t, c_{0:t-1} & \sim \mathcal{G}a\big( c_t \mid \underline{a}_c, \zeta_t \big),
\end{align*}
where $\omega\in (0,1)$ tunes the degree of temporal dependence, whereas $\underline{a}_\zeta, \underline{b}_\zeta, \underline{a}_c \in\Rp$ are fixed hyperparameters, and $\underline{a}_{t-1}, \underline{b}_{t-1}$ are defined below.
Integrating out $\upsilon_t$ results in a scaled beta distribution for the transition
{\small
\begin{align*}
    p(\zeta_t \mid \zeta_{t-1}, c_{0:t-1}, \omega) = \frac{\Gamma(\underline{a}_{t-1})}{\Gamma(\omega \underline{a}_{t-1}) \Gamma( (1-\omega)\underline{a}_{t-1})} \Big( \frac{\omega}{\zeta_{t-1}} \Big)^{\omega \underline{a}_{t-1}} \zeta_t^{\omega \underline{a}_{t-1} -1} \Big( 1 - \frac{\omega \zeta_t}{\zeta_{t-1}} \Big)^{(1-\omega)\underline{a}_{t-1}} \I_{\big( 0, \frac{\zeta_{t-1}}{\omega} \big)}(\zeta_t)
\end{align*}}
which is defined on the interval $\big( 0, \frac{\zeta_{t-1}}{\omega} \big)$.

This specification leads to the following closed form filtering distribution:
\begin{align*}
    \zeta_t \mid c_{0:t}, \omega \sim \mathcal{G}a(\zeta_t \mid \underline{a}_t, \underline{b}_t),
\end{align*}
where $\underline{a}_t = \omega \underline{a}_{t-1} + \underline{a}_c$ and $\underline{b}_t = \omega \underline{b}_{t-1} + c_t$. The backward conditional distribution is a shifted Gamma distribution defined on the interval $(\omega \zeta_{t+1}, \infty)$, given by
\begin{align*}
    \zeta_t \mid \zeta_{t+1}, c_{0:t}, \omega \overset{d}{=} \mathcal{G}a(\zeta_t \mid (1-\omega) \underline{a}_t, \underline{b}_t) + \omega \zeta_{t+1},
\end{align*}
or, equivalently, $(\zeta_t -\omega \zeta_{t+1}) \mid \zeta_{t+1}, c_{0:t}, \omega \sim \mathcal{G}a(\zeta_t \mid (1-\omega) \underline{a}_t, \underline{b}_t)$.
Together, these formulas allow to sample the entire path of the auxiliary variable $\boldsymbol{\zeta} = (\zeta_0,\dots,\zeta_T)'$ from the joint distribution $(\boldsymbol{\zeta} \mid \omega, \mathbf{c})$ using a forward filtering backward sampling procedure.

For the parameter tuning the temporal dependence we assume a Beta prior $\omega \sim \mathcal{B}e(\omega \mid \underline{a}_\omega, \underline{b}_\omega)$, leading to the posterior full conditional distribution
\begin{align*}
    p(\omega \mid \boldsymbol{\zeta}) & \propto \omega^{\underline{a}_\omega-1} (1-\omega)^{\underline{b}_\omega-1} \times \prod_{t=1}^T \frac{\Gamma(\underline{a}_{t-1})}{\Gamma(\omega \underline{a}_{t-1}) \Gamma( (1-\omega)\underline{a}_{t-1})} \big( \omega \zeta_t \big)^{\omega \underline{a}_{t-1}} \big( \zeta_{t-1} - \omega \zeta_t \big)^{(1-\omega)\underline{a}_{t-1}-1}
\end{align*}

\newpage

\section{Simulating the SN-M-ARG for continuous $H$}
\label{sec:simulation_continuous}

To better illustrate the role of the main components of the proposed dynamic spatial model, we generated synthetic datasets under a full $2 \times 2 \times 2$ sensitivity grid, varying the temporal dependence parameter $\beta c$, the innovation/intensity parameter $\alpha$, and the kernel bandwidth $\phi$. The remaining quantities were kept fixed across scenarios, with $T=20$, $M=10,000$, $c=0.46 \times 10^3$, observation window $[10,130]\times[10,130]$, and latent spatial domain $[0,140]\times[0,140]$.

The values used in the experiment are reported in Table~\ref{tab:sim_design}. The parameter $\beta c$ controls the persistence of the latent random measure over time. Small values lead to weak temporal dependence and rapidly changing spatial configurations, while large values produce stronger persistence and smoother temporal evolution. The parameter $\alpha$ controls the amount of newly introduced mass at each time point. Hence, increasing $\alpha$ increases the number and/or magnitude of new active components in the latent process, producing richer and more heterogeneous intensity surfaces. Finally, $\phi$ controls the spatial dispersion of observations around their latent locations. Small values of $\phi$ generate highly localized clusters, whereas larger values produce smoother and more diffuse spatial patterns.

\begin{table}[t!h]
\centering
\begin{tabular}{c cc}
\toprule
Parameter & Small value & Large value \\
\midrule
$\beta c$ & $0.12$ & $0.82$ \\
$\alpha$ & $0.02$ & $2.00$ \\
$\phi$   & $0.05$ & $1.00$ \\
\bottomrule
\end{tabular}
\caption{Parameter values used in the additional simulation scenarios.}
\label{tab:sim_design}
\end{table}

\begin{figure}[p]
\centering
\vspace{-10ex}
\renewcommand{\arraystretch}{0.2}
\setlength{\tabcolsep}{6pt}
\begin{tabular}{c c c c}
& & {Small bandwidth ($\phi=0.05$)} & {Large bandwidth ($\phi=1.00$)} \\

\multirow{2}{*}{\begin{rotate}{90} \hspace{-50pt} Small innovation ($\alpha=0.02$) \end{rotate}} \hspace{10pt} & 
\begin{rotate}{90} \hspace{20pt} Small persistence $\beta c$ \end{rotate}
&
\includegraphics[width=0.30\textwidth,trim=0 0 2cm 0,clip]
{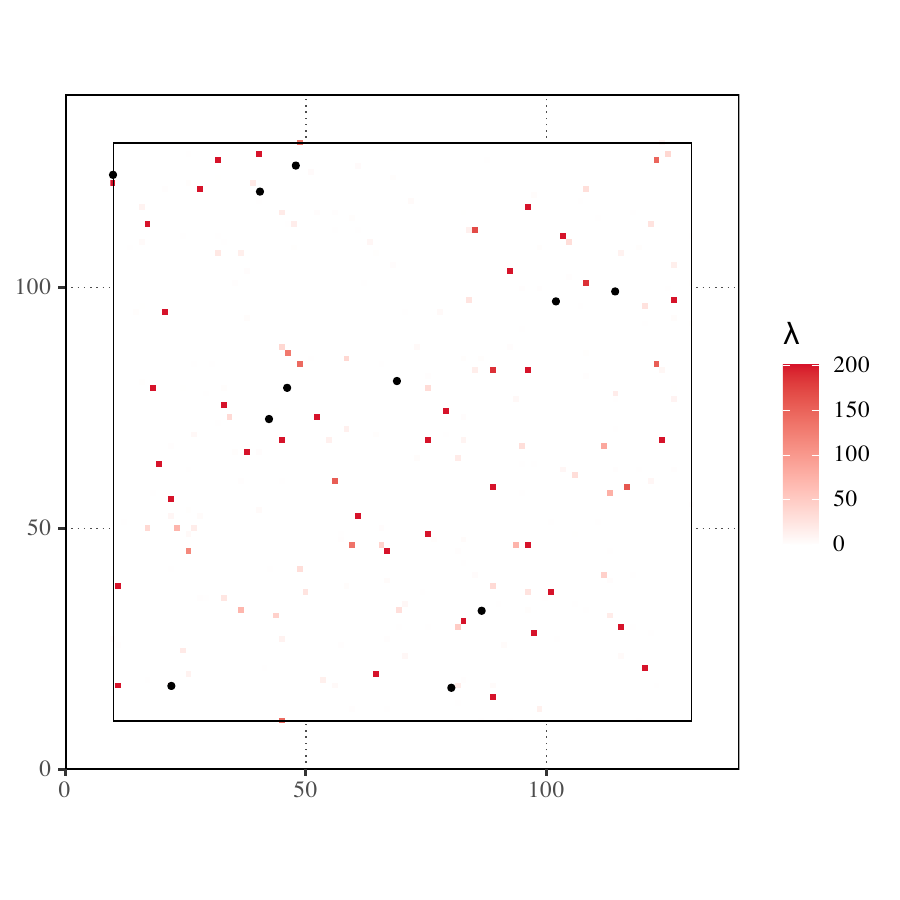}
&
\includegraphics[width=0.30\textwidth,trim=0 0 2cm 0,clip]
{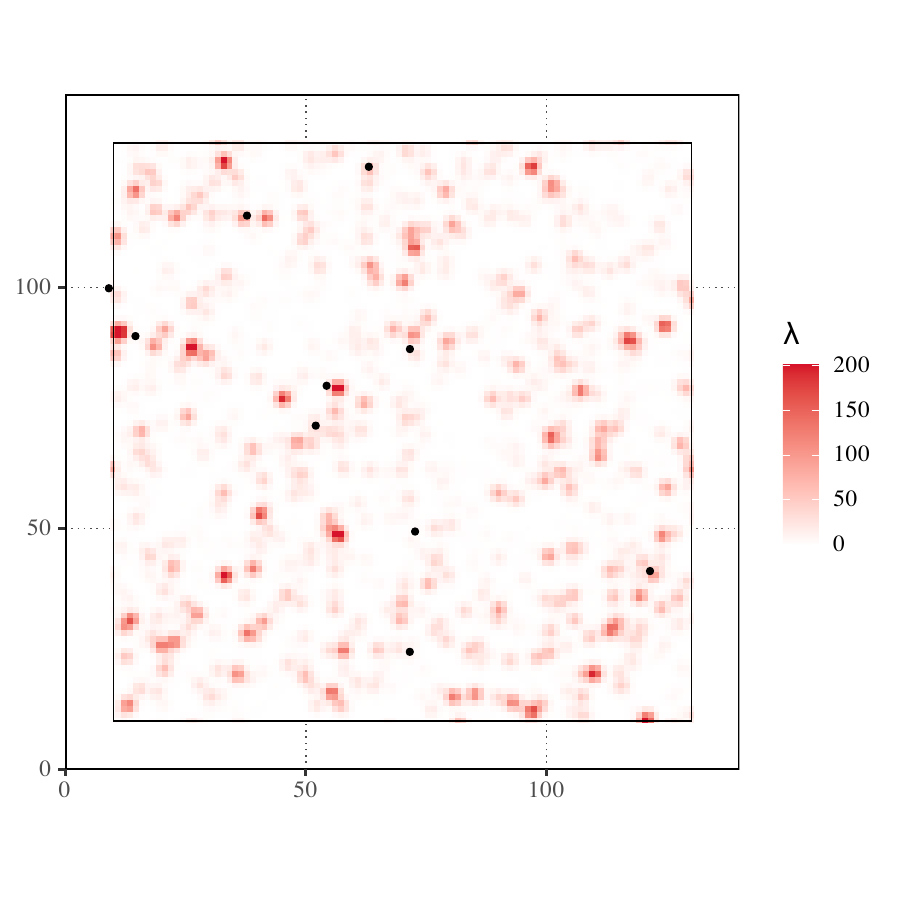}
\\[-10pt]

&
\begin{rotate}{90} \hspace{20pt} Large persistence $\beta c$ \end{rotate}
&
\includegraphics[width=0.30\textwidth,trim=0 0 2cm 0,clip]
{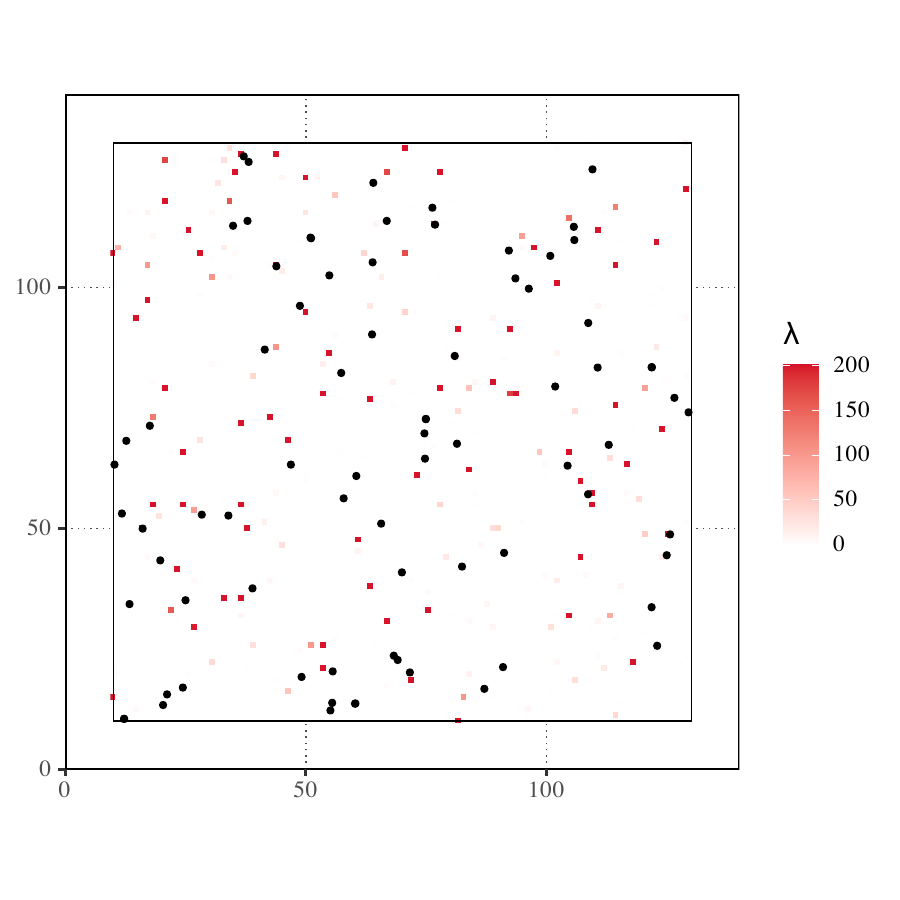}
&
\includegraphics[width=0.30\textwidth,trim=0 0 2cm 0,clip]
{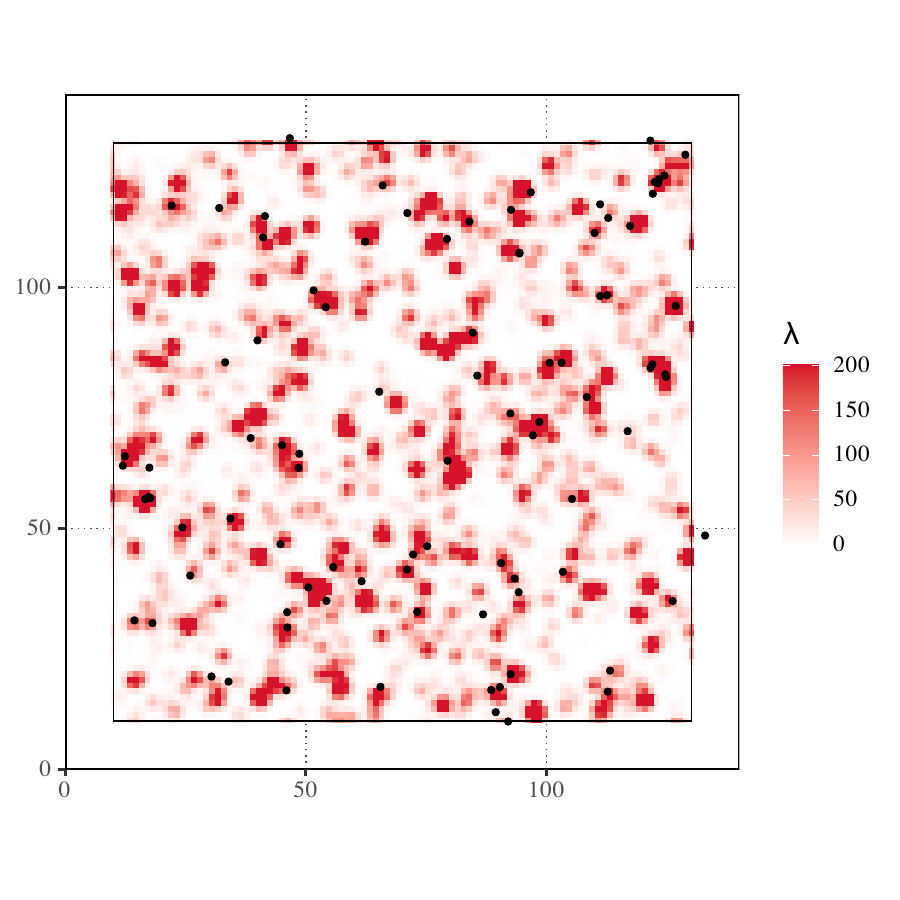}
\\[10pt]

\multirow{2}{*}{\begin{rotate}{90} \hspace{-50pt} Large innovation ($\alpha=2.00$) \end{rotate}} \hspace{10pt} & 
\begin{rotate}{90} \hspace{20pt} Small persistence $\beta c$ \end{rotate}
&
\includegraphics[width=0.30\textwidth,trim=0 0 2cm 0,clip]
{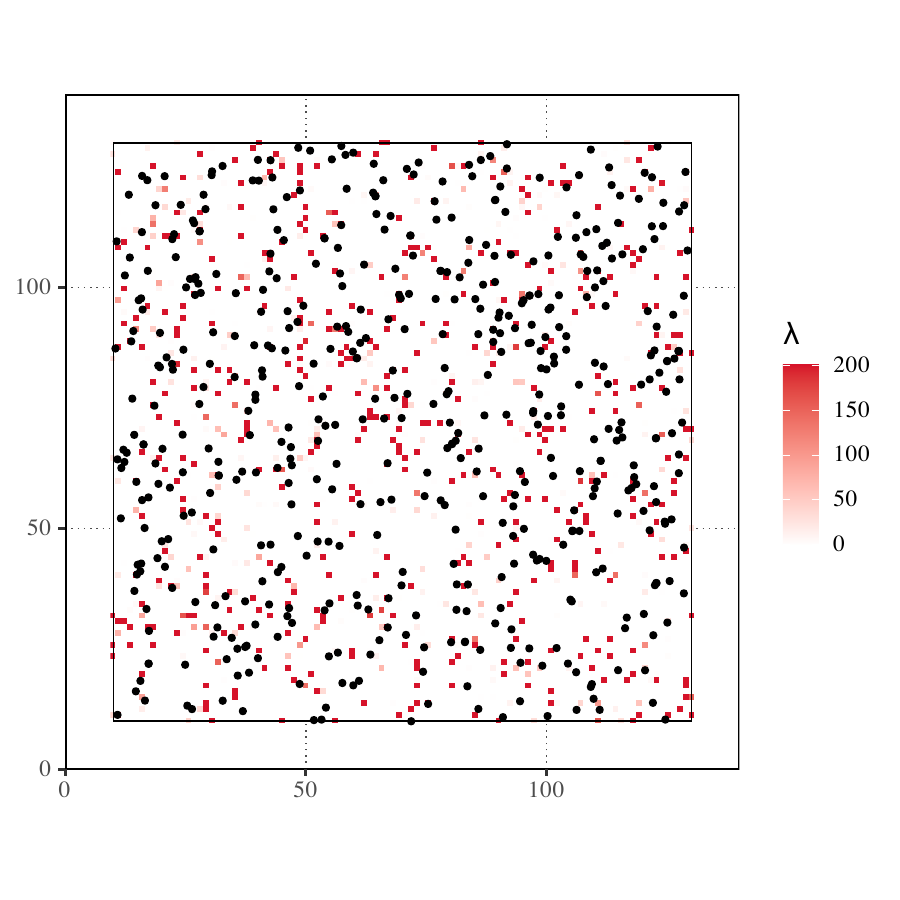}
&
\includegraphics[width=0.30\textwidth,trim=0 0 2cm 0,clip]
{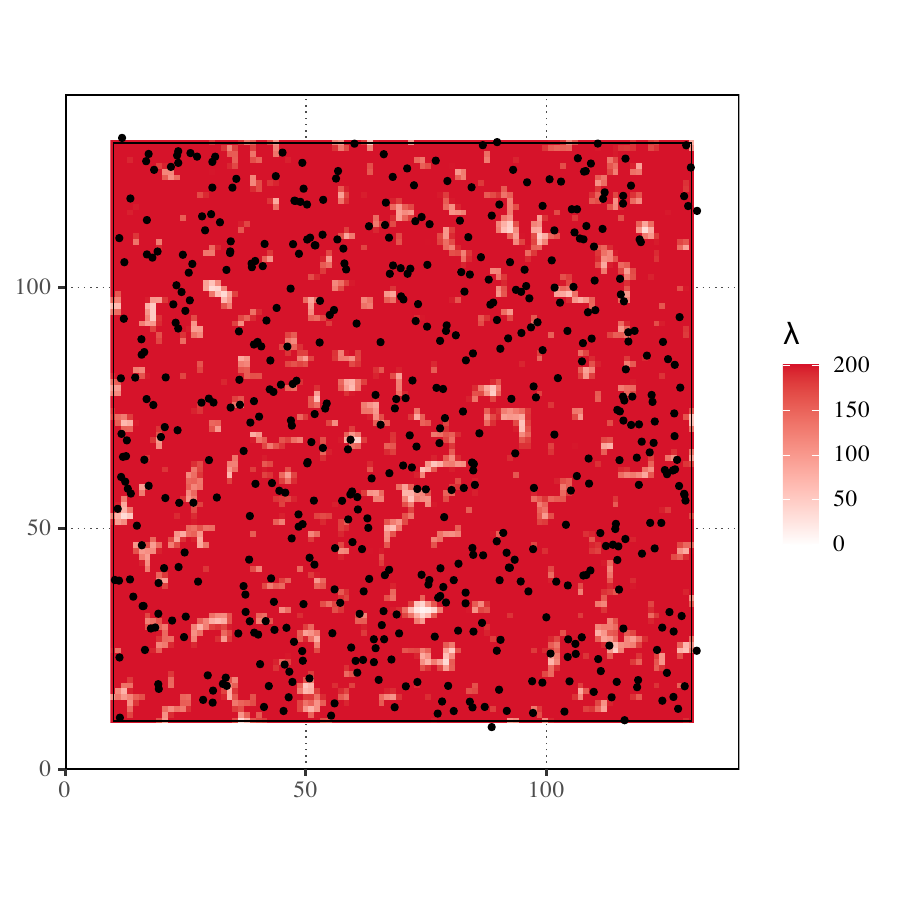}
\\[-10pt]

&
\begin{rotate}{90} \hspace{20pt} Large persistence $\beta c$ \end{rotate}
&
\includegraphics[width=0.30\textwidth,trim=0 0 2cm 0,clip]
{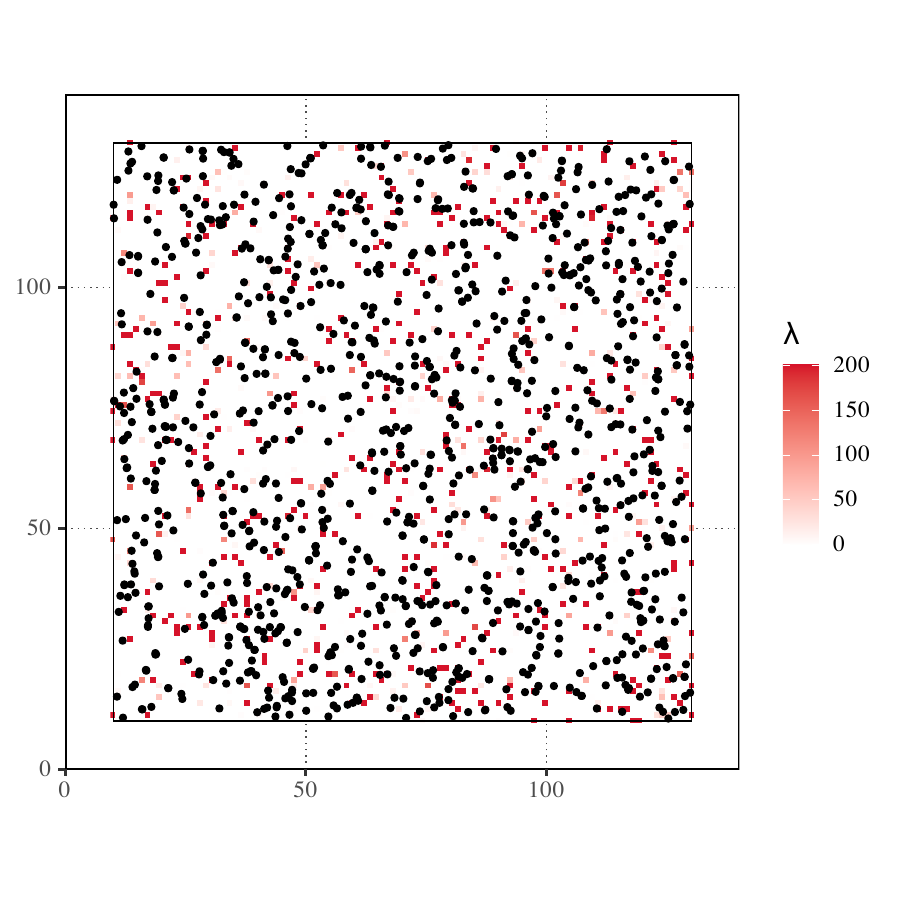}
&
\includegraphics[width=0.30\textwidth,trim=0 0 2cm 0,clip]
{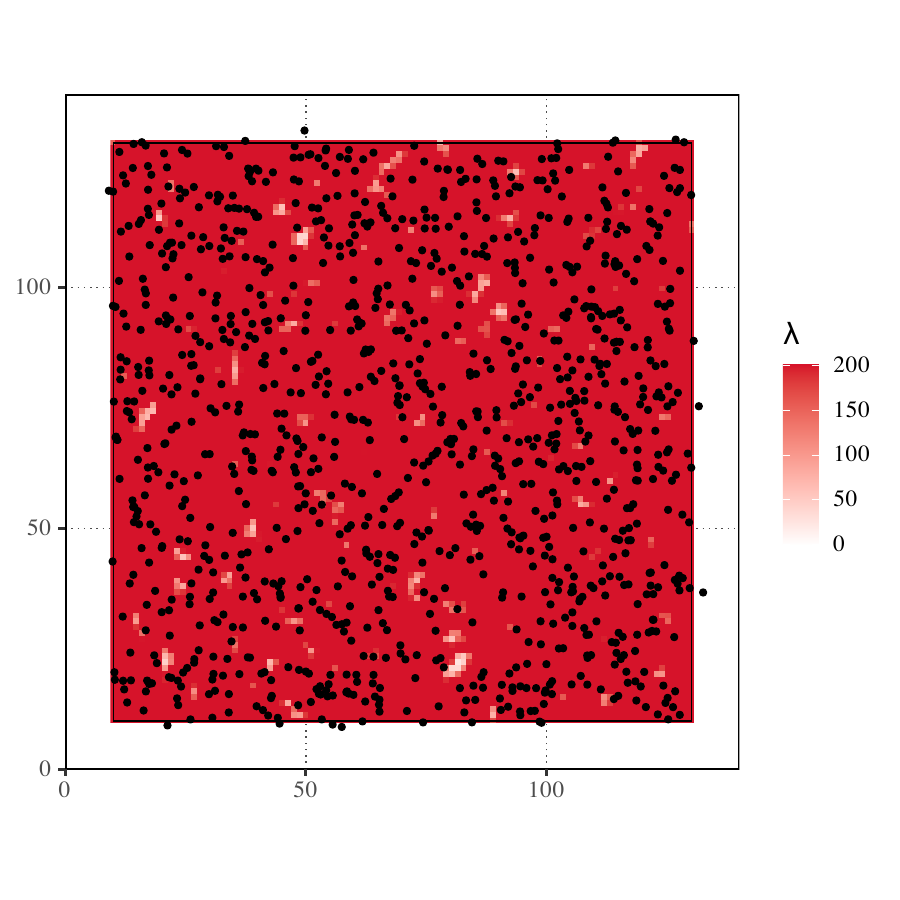}
\end{tabular}
\captionsetup{width=0.90\textwidth}
\caption{Representative simulated intensity surfaces under different parameter configurations. The panels compare the effect of temporal persistence $\beta c$, innovation level $\alpha$, and spatial bandwidth $\phi$. Black dots denote simulated observations.}
\label{fig:sim_components}
\end{figure}

Figure ~\ref{fig:sim_components} reports representative intensity surfaces from selected scenarios. The simulated intensity surfaces confirm the expected behaviour of the model. When $\beta c$ is small, the process exhibits limited temporal memory and the active regions may change substantially from one time point to the next. Conversely, when $\beta c$ is large, latent components are more likely to survive over time, inducing stronger temporal dependence and more persistent spatial hotspots. Increasing $\alpha$ mainly affects the richness of the latent measure: for small $\alpha$, the process is driven by relatively few components, whereas for large $\alpha$ the intensity field becomes more populated and spatially heterogeneous. Finally, increasing $\phi$ smooths the point pattern around the latent atoms, changing the visual appearance from sharply concentrated clusters to broader spatial regions of elevated intensity.

\clearpage

\section{Simulation exercise for discrete $H$}
\label{sec:simulation_exercise}

\subsection{MCMC mixing}
We assess the finite-sample performance of the Bayesian inference procedure through a simulation study based on the isotropic SN-M-ARG model. Data are generated from the hierarchical specification described in Section 4 of the main manuscript, using a Gaussian kernel with bandwidth $\phi=0.06$, autoregressive parameter $\beta=1$, and a time-varying scale process $(c_t)_{t\geq1}$ with a hierarchical Gamma distribution. The deterministic component $\kappa_t$ is assumed either constant or governed by two harmonic covariates. The latent space $\Theta \subseteq [0,1]^2$ is discretised on a $5 \times 5$ regular grid.

\begin{sidewaysfigure}[p]
\centering
\begin{tabular}{ccc}
$\boldsymbol{\alpha}$ boxplot &
$\boldsymbol{c}$ boxplot &
\textbf{$\beta$} \\
\includegraphics[width=.32\textwidth]{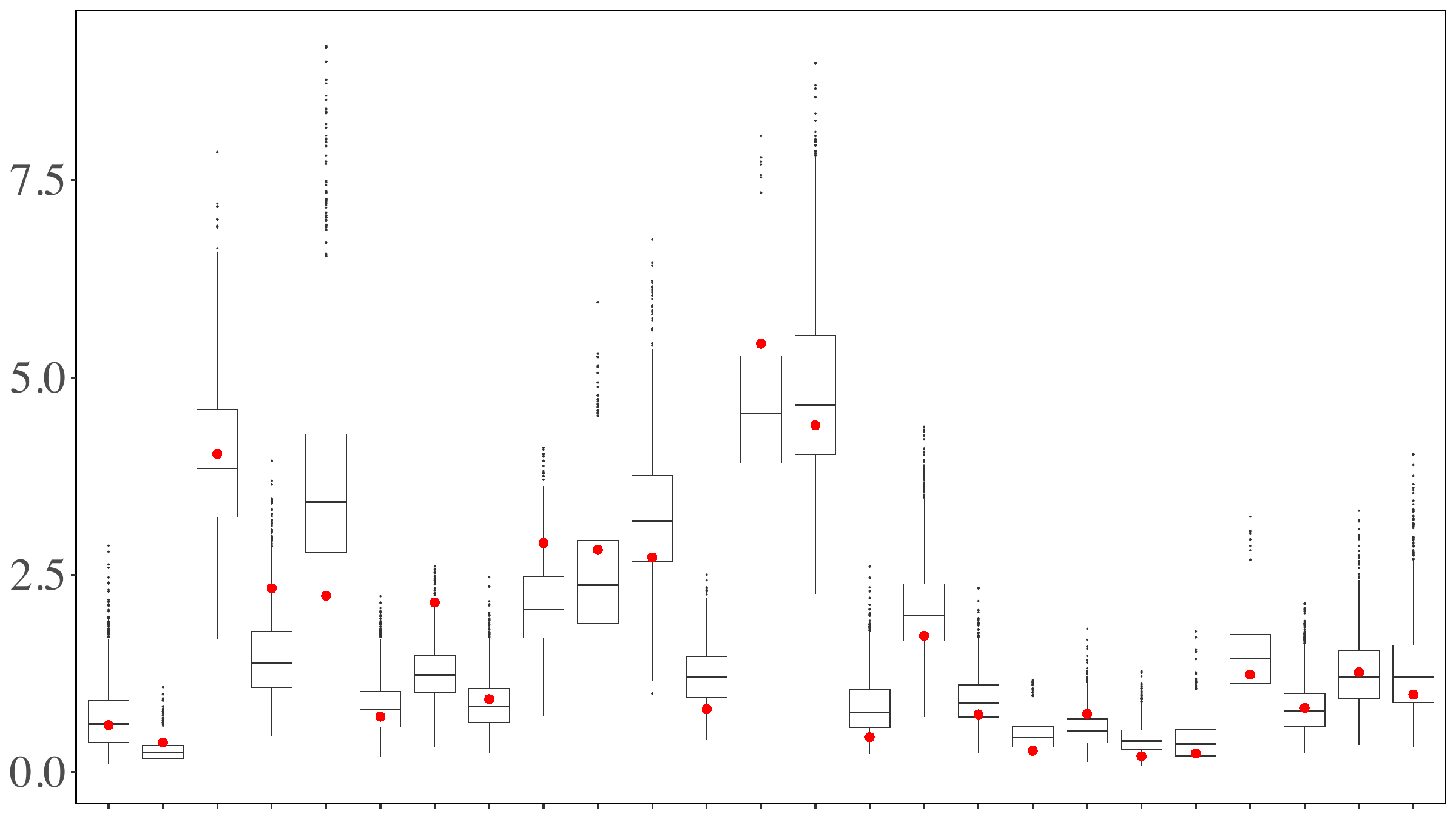} &
\includegraphics[width=.32\textwidth]{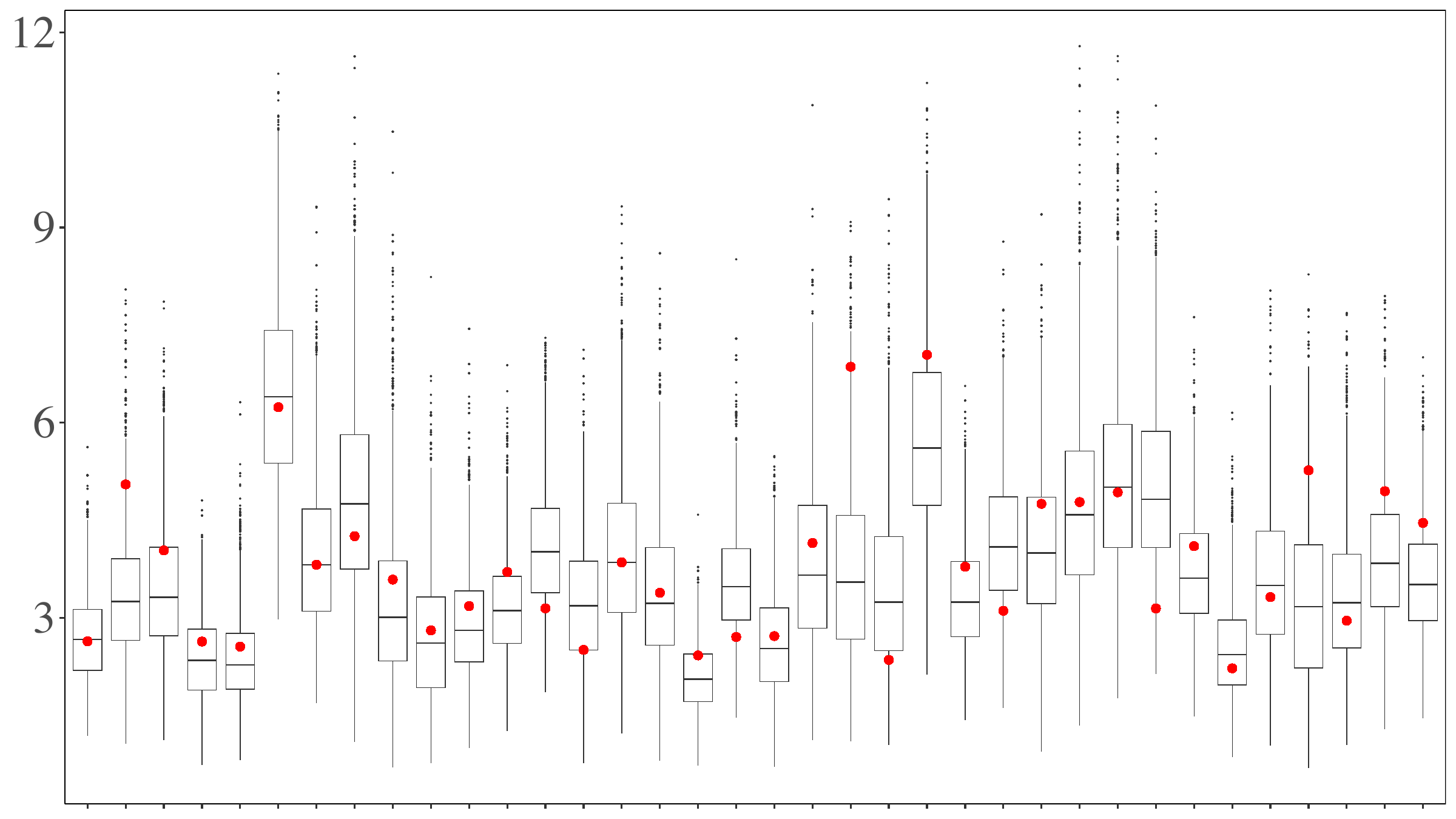} &
\includegraphics[width=.32\textwidth]{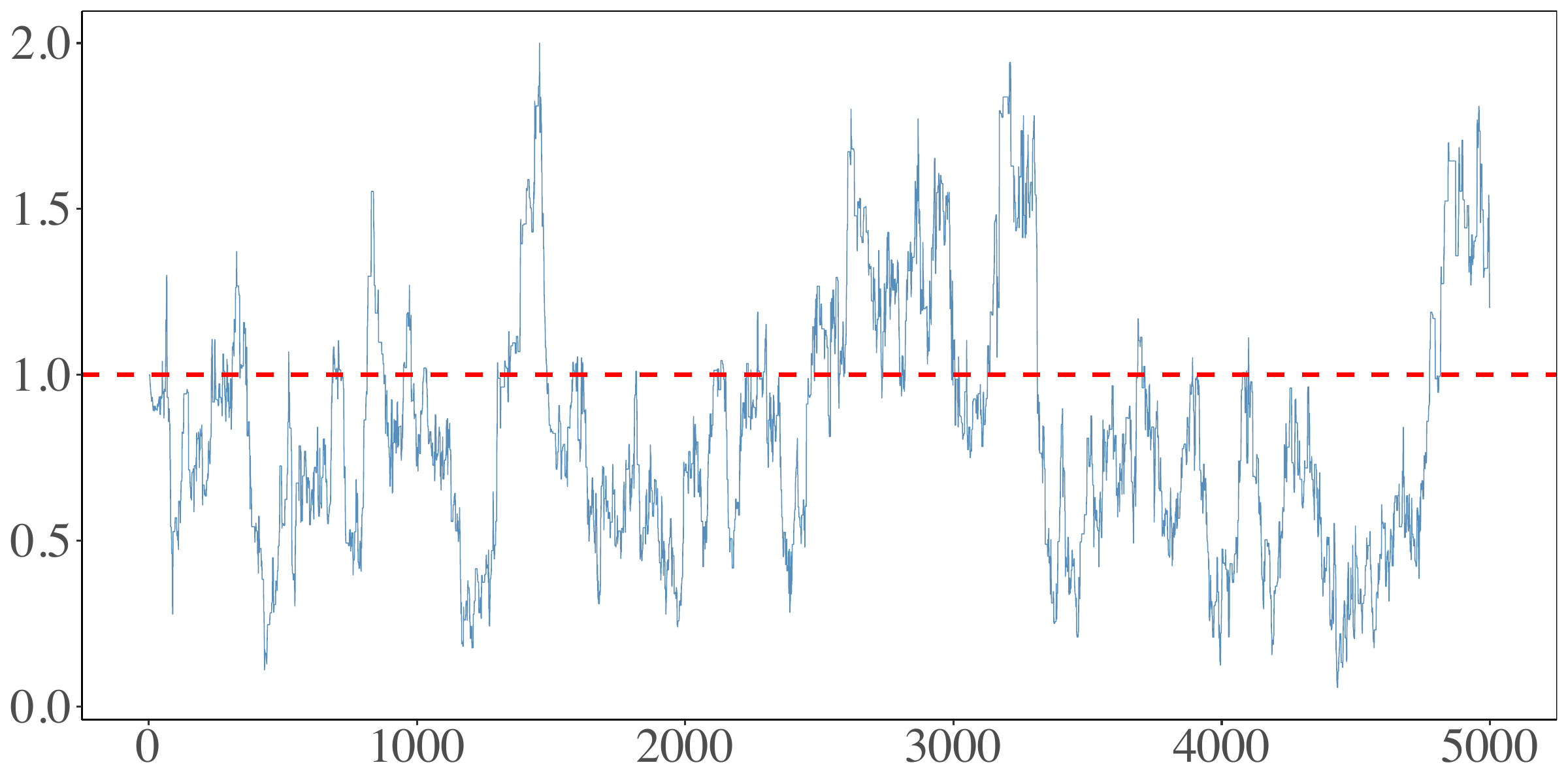}
\\[0.5em]
\textbf{$\phi$} &
\textbf{$\eta_1$} &
\textbf{$\eta_2$} \\
\includegraphics[width=.32\textwidth]{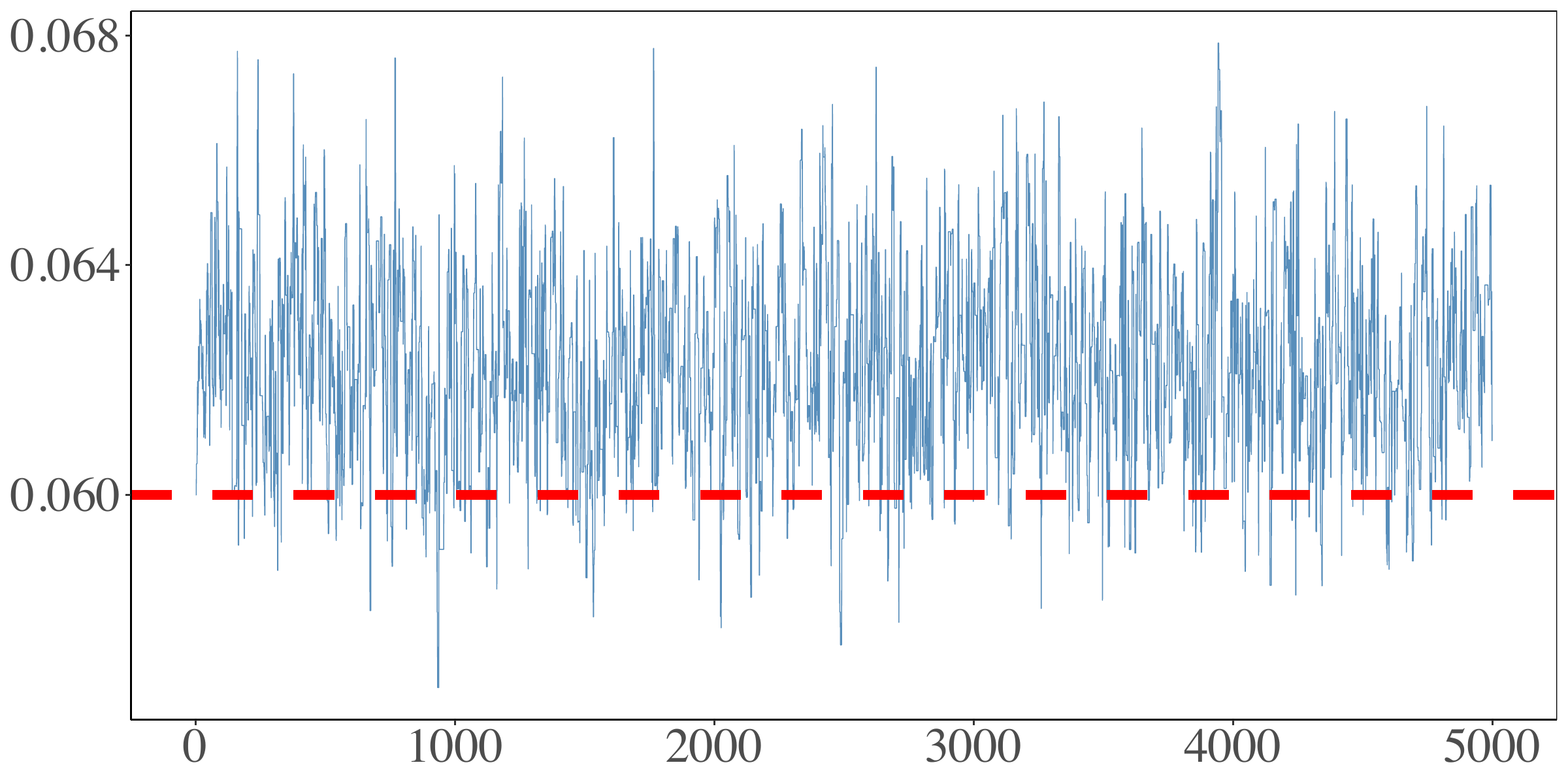} &
\includegraphics[width=.32\textwidth]{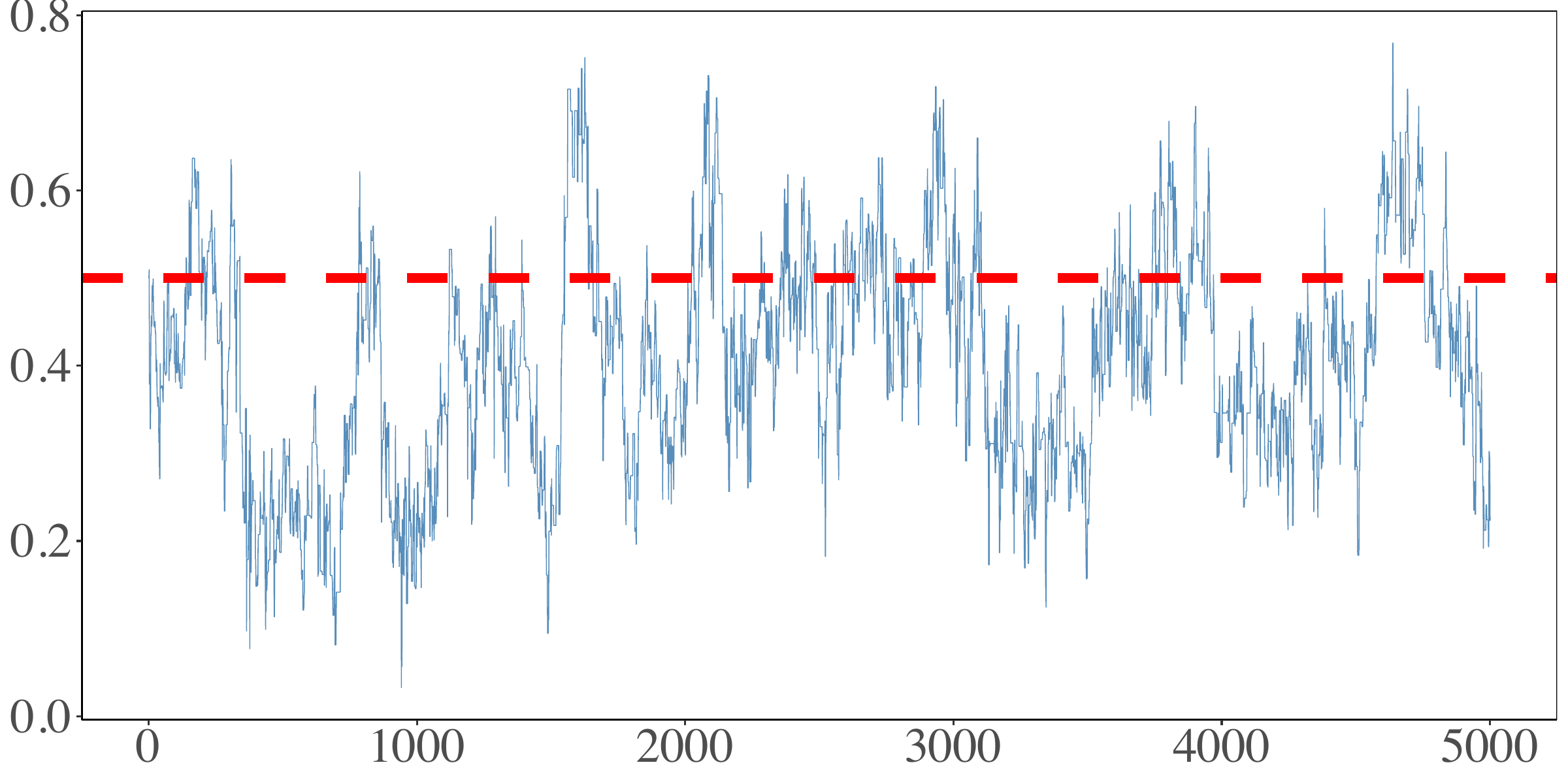} &
\includegraphics[width=.32\textwidth]{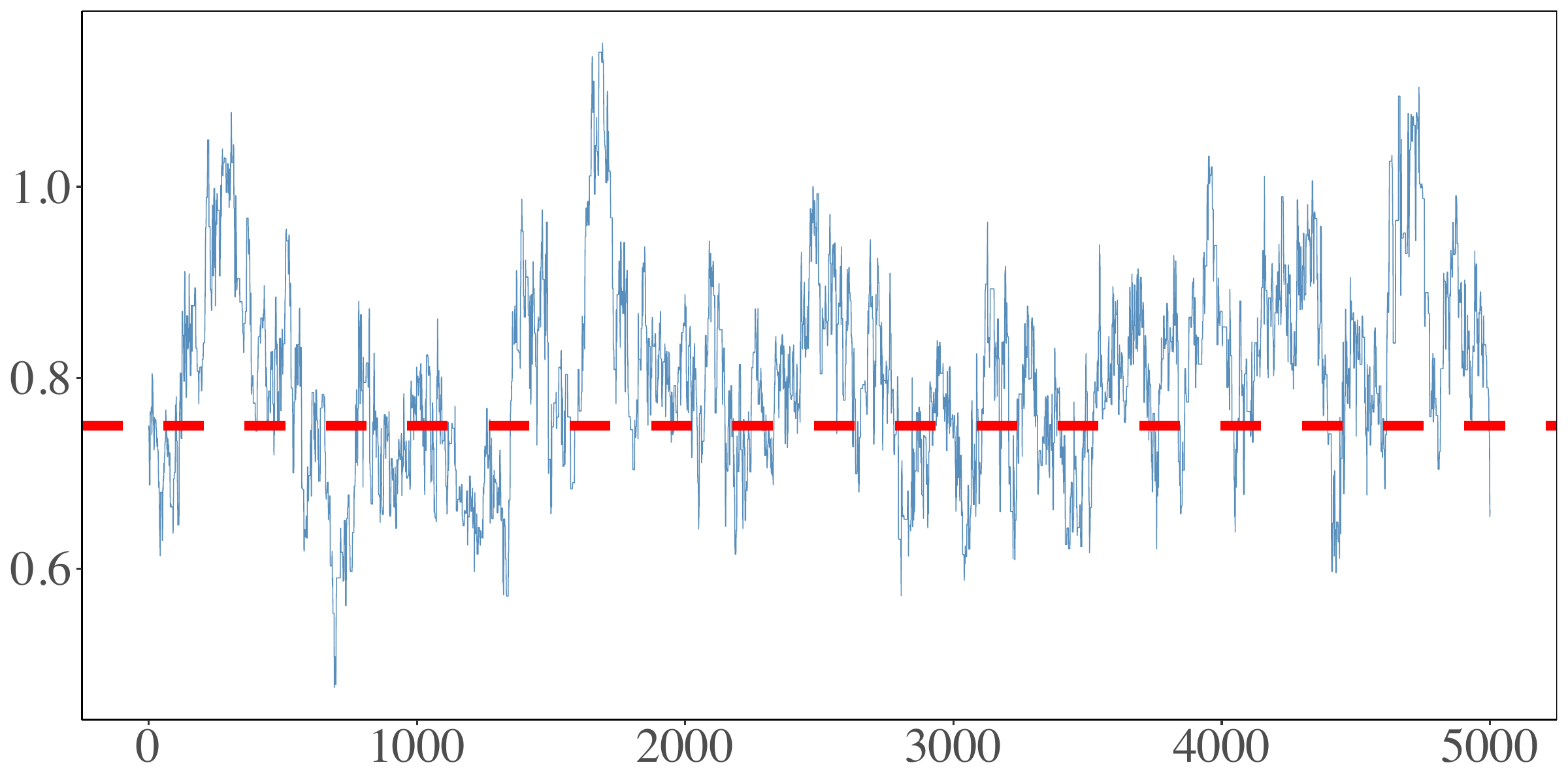}
\\[0.5em]
\textbf{$\sum_j w_{j,1}\kappa_{1}$} &
\textbf{$\sum_j w_{j,6}\kappa_{6}$} &
\textbf{$\sum_j w_{j,12}\kappa_{12}$} \\
\includegraphics[width=.32\textwidth]{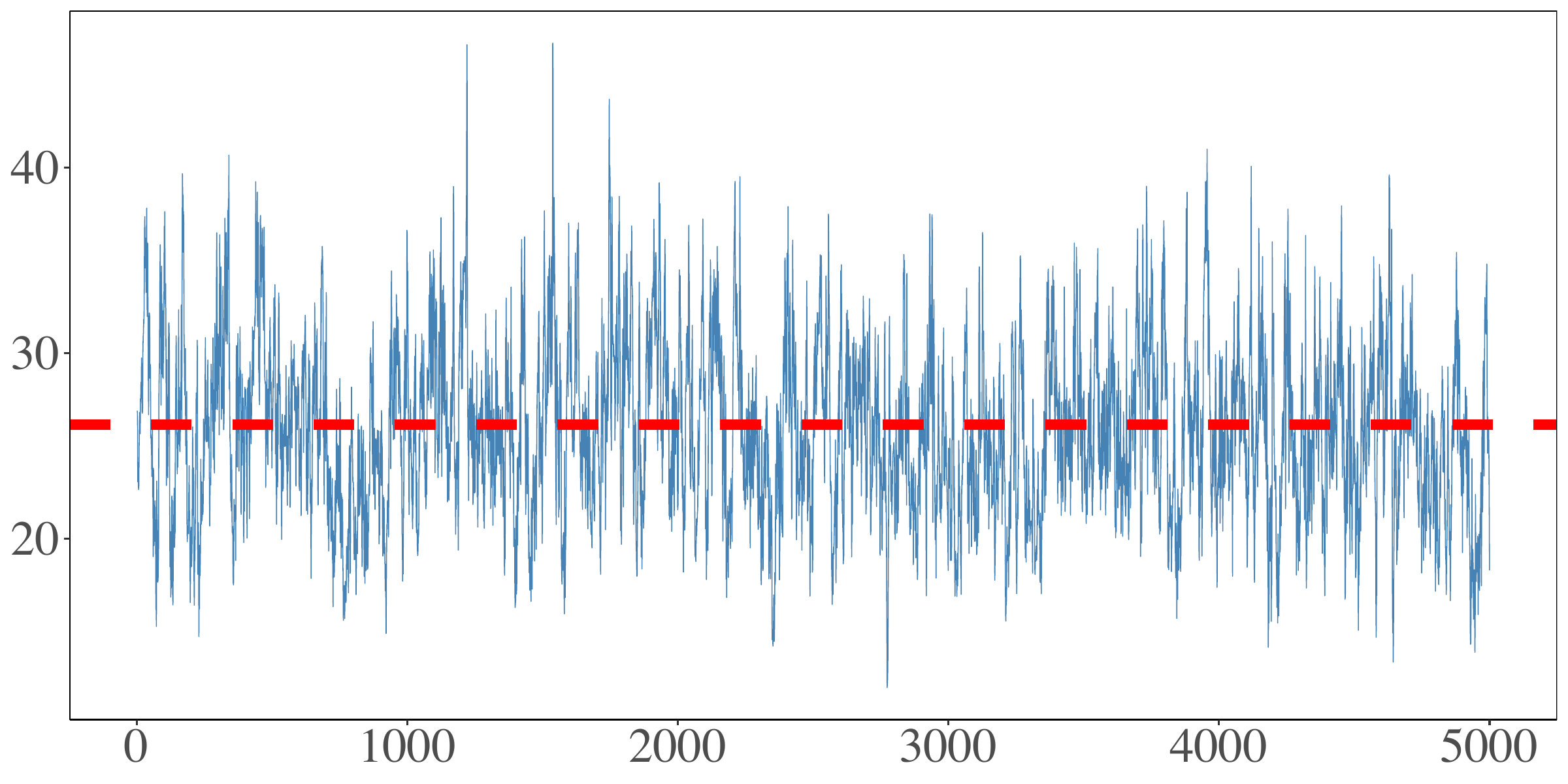} &
\includegraphics[width=.32\textwidth]{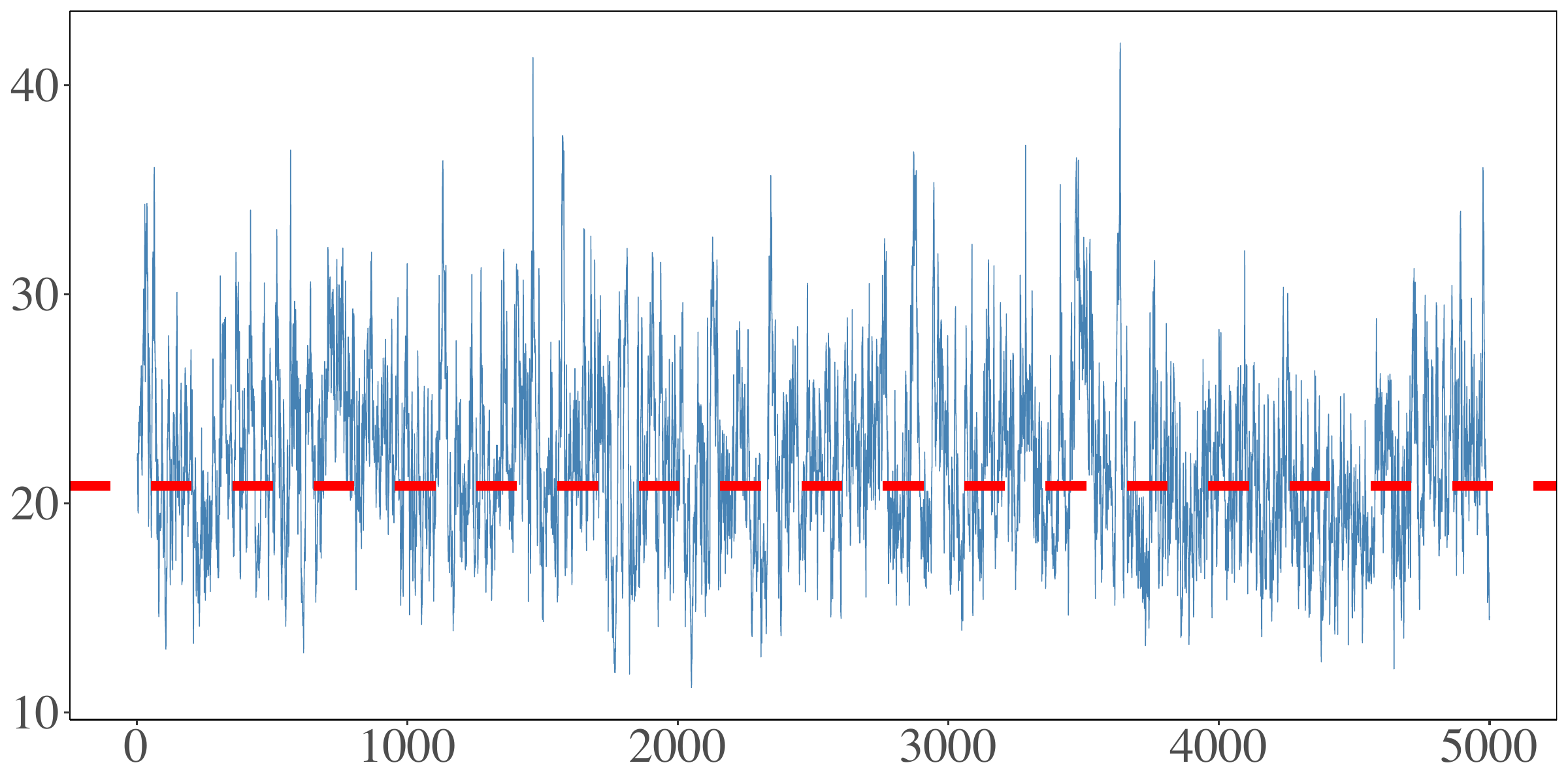} &
\includegraphics[width=.32\textwidth]{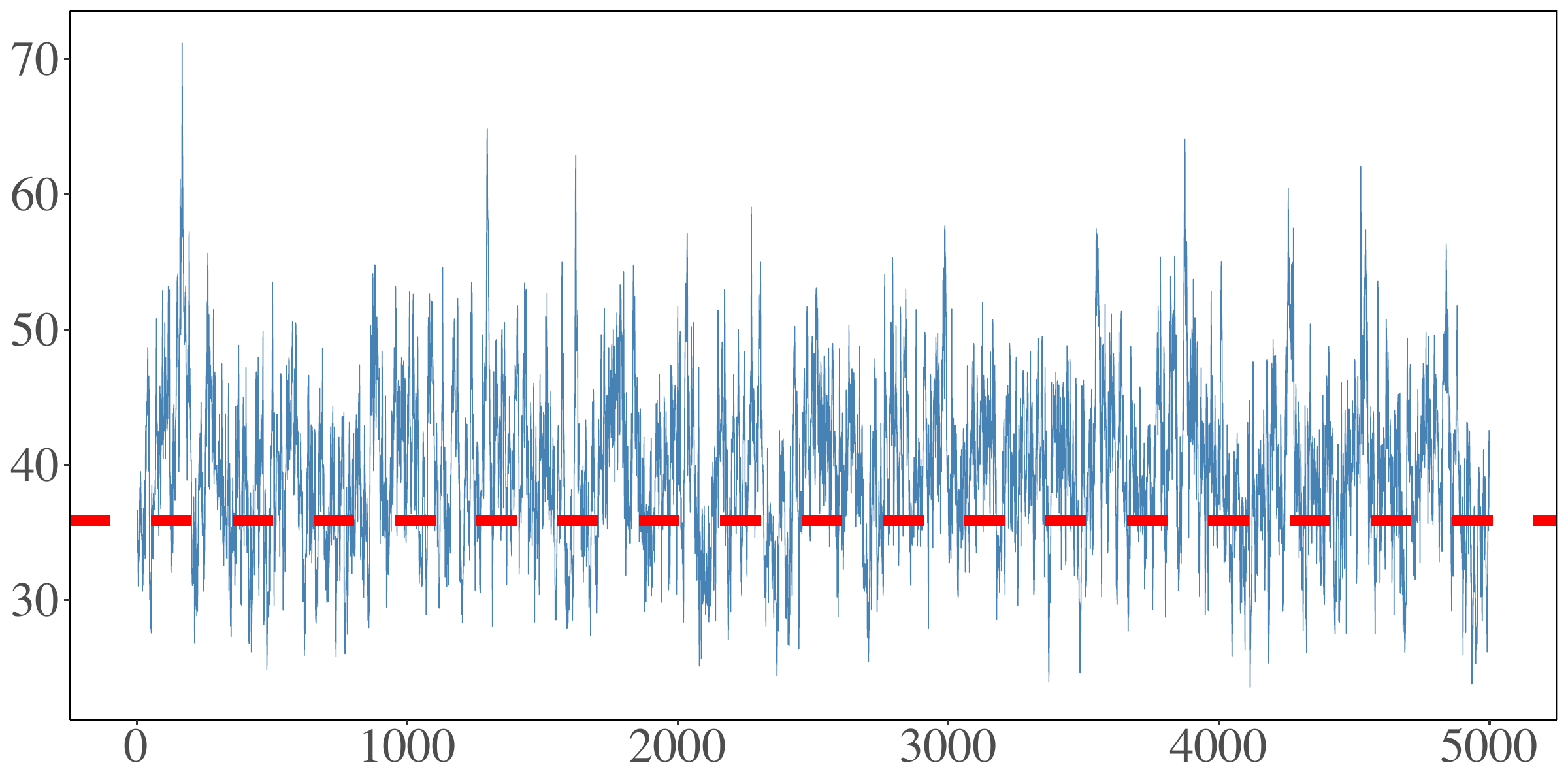}
\end{tabular}
\caption{Boxplots and trace plots for selected model parameters.}
\label{fig:diagnostics_box_trace}
\end{sidewaysfigure}

Figure \ref{fig:diagnostics_box_trace} reports the boxplots and trace plots of the posterior draws for the model specification including covariates. For the latent processes, $w_{j,t}$, we report three representative elements: the remaining ones have a similar pattern. For all parameters, the algorithm correctly retrieves the true value, as indicated by the red dots and dashed red lines.
We remark that the autocorrelation visible in the plots of $\beta,\eta_1,\eta_2$ stems from the fact that the trace plots show all the MCMC iterations, before removing the burn-in and applying thinning.

To further inspect the mixing and estimation accuracy of the proposed MCMC scheme versus alternative algorithms, Table~\ref{tab:ess_mse_covariates} compares the effective sample size (ESS) and mean squared error (MSE) of four sampling schemes: (i) single-move sampler without SMC, (ii) single-move sampler with SMC, (iii) block sampler without SMC for $\boldsymbol{\alpha}$ and $\beta$, and (iv) block sampler with SMC.
Given the large number of $\alpha_j$ and $c_t$ parameters, we report the average ESS and MSE across parameters, denoted $\bar\alpha$ and $\bar c$, respectively.
The block samplers generally achieve higher ESS for several parameters, particularly $\bar c$ and $\eta$, although this pattern is not uniform across all parameters. In terms of MSE, the single-move samplers tend to perform better for $\bar\alpha$, whereas the block samplers yield lower MSE for $\beta$, $\bar c$, and $\sum_j w_{j,t}\kappa_t$. These conclusions remain broadly unchanged after burn-in and thinning.
The computing times for 5,000 MCMC iterations each scheme are 0.9638 minutes (single-move, no SMC), 2.3056 minutes (single-move, SMC), 0.9065 minutes (block, no SMC), and 2.4338 minutes (block, SMC) on a MacBook Air Apple M2. Given its competitive ESS and MSE performance together with the low computational cost, the single-move and block sampler without SMC appears to be the most attractive option for large-scale applications.

\begin{sidewaystable}[p]
\centering
\resizebox{\textwidth}{!}{
\begin{tabular}{l rrrrrr rrrrrr}
\toprule
& \multicolumn{6}{c}{ESS} & \multicolumn{6}{c}{MSE} \\
\cmidrule(lr){2-7} \cmidrule(l){8-13}
Scenario & \multicolumn{1}{c}{$\phi$} & \multicolumn{1}{c}{$\bar\alpha$}
& \multicolumn{1}{c}{$\beta$} & \multicolumn{1}{c}{$\bar c$}
& \multicolumn{1}{c}{$\sum_j w_{j,t}\kappa_t$} & \multicolumn{1}{c}{$\eta$} 
& \multicolumn{1}{c}{$\phi$} & \multicolumn{1}{c}{$\bar \alpha$}
& \multicolumn{1}{c}{$\beta$} & \multicolumn{1}{c}{$\bar c$}
& \multicolumn{1}{c}{$\sum_j w_{j,t}\kappa_t$} & \multicolumn{1}{c}{$\eta$} \\
\midrule
\multicolumn{13}{l}{\textbf{Raw MCMC (5,000 iters)}}\\
\midrule
Single-move, no SMC
& 647.64 & 55.27 & 47.62 & 125.83 & 197.67 & 78.12
& 0.000 & 0.963 & 0.420 & 2.420 & 327.222 & 0.358\\
Single-move, SMC
& 593.53 & 35.12 & 31.72 & 101.84 & 123.88 & 56.49
& 0.000 & 0.744 & 0.295 & 2.319 & 336.074 & 0.416\\
Block, no SMC
& 758.74 & 42.25 & 20.26 & 169.59 & 212.45 & 60.58
& 0.000 & 2.017 & 0.140 & 2.388 & 315.671 & 0.388\\
Block, SMC
& 719.91 & 30.61 & 23.52 & 563.26 & 156.05 & 252.48
& 0.000 & 2.214 & 0.004 & 2.008 & 316.082 & 0.375\\
\addlinespace
\multicolumn{13}{l}{\textbf{Burn-in (1,000 iters)}}\\
\midrule
Single-move, no SMC
& 543.68 & 47.27 & 41.95 & 107.94 & 165.36 & 68.72
& 0.000 & 0.963 & 0.420 & 2.420 & 327.222 & 0.358\\
Single-move, SMC
& 500.63 & 35.78 & 26.70 & 85.80 & 101.37 & 57.39
& 0.000 & 0.744 & 0.295 & 2.319 & 336.074 & 0.416\\
Block, no SMC
& 660.48 & 36.66 & 21.99 & 144.48 & 174.86 & 49.71
& 0.000 & 2.017 & 0.140 & 2.388 & 315.671 & 0.388\\
Block, SMC
& 594.52 & 50.23 & 42.35 & 526.98 & 153.29 & 234.43
& 0.000 & 2.214 & 0.004 & 2.008 & 316.082 & 0.375\\
\addlinespace
\multicolumn{13}{l}{\textbf{Burn-in (1,000 iters) + thinning every 5 iters}}\\
\midrule
Single-move, no SMC
& 519.11 & 43.36 & 44.14 & 86.65 & 156.94 & 60.70
& 0.000 & 0.963 & 0.420 & 2.420 & 327.222 & 0.358\\
Single-move, SMC
& 210.18 & 25.25 & 16.87 & 75.17 & 98.38 & 51.24
& 0.000 & 0.744 & 0.295 & 2.319 & 336.074 & 0.416\\
Block, no SMC
& 563.89 & 33.04 & 22.21 & 129.15 & 165.61 & 35.57
& 0.000 & 2.017 & 0.140 & 2.388 & 315.671 & 0.388\\
Block, SMC
& 638.17 & 21.94 & 12.03 & 371.35 & 113.77 & 184.55
& 0.000 & 2.214 & 0.004 & 2.008 & 316.082 & 0.375\\
\bottomrule
\end{tabular}
}
\captionsetup{width=0.9\textwidth}
\caption{Effective sample size (ESS) and posterior mean squared error (MSE) under the four estimation schemes.}
\label{tab:ess_mse_covariates}
\end{sidewaystable}

\subsection{MCMC sensitivity}
To investigate the impact of model misspecification and scalability with respect to the dimension of the latent state, we consider three grid resolutions: $3 \times 3$, $5 \times 5$, and $7 \times 7$ corresponding to $N^g=9$, $25$, and $49$ latent locations, respectively. To evaluate the effect of the time dimension, we consider three sample lengths, $T=18$,  $T=36$, and $T=54$. Finally, we vary the intensity of the point process through the hyperparameter $\gamma$ governing the prior distribution of the latent weights: smaller values of $\gamma$ generate larger latent weights and consequently larger expected numbers of observations $N_t^y$. 

For each combination of $(T,\gamma)$, we generate independent datasets from the \SNMARG\ model and estimate the posterior distribution of the model with different grid specifications. The quality of the posterior approximation is evaluated using: the MSE and mean absolute error (MAE) of the in-sample and out-of-sample predicted number of counts in the whole area, $[0.0, 1.0]^2$ (subscript ``full'') and in a subregion of dimension $[0.3, 0.7]^2$ (subscript ``inner'').

In all cases, the computing time scales linearly in $T$ and in the number of observations, $N_t^y(A)$, whereas it scales quadratically in the number of points on each side of the regular (square) grid used for $\Theta$. Predictive ability measures are reported in Table \ref{tab:ess_mse_covariates}.  The in-sample accuracy of the sampler over the full area improves as the grid becomes finer, although it is highest under the ($5 \times 5$) configuration, which corresponds to the data-generating process. This may be explained by the fact that, in the inner-area framework, the parameter ($\theta$) adequately covers the boundary regions, whereas the full observation area contains no points outside the domain. In contrast, the out-of-sample forecasting results consistently identify the inner framework as the best-performing approach across all grid resolutions.

\begin{table}[t]
\centering
\footnotesize
\setlength{\tabcolsep}{5pt}
\resizebox{\textwidth}{!}{
\begin{tabular}{ccrrrrrrrrrr}
\toprule
&
&
\multicolumn{5}{c}{(a) In-sample}
&
\multicolumn{5}{c}{(b) Out-of-sample}
\\
\cmidrule(lr){3-7}\cmidrule(lr){8-12}
Setting & Grid &
MSE$_{\rm full}$ &
MAE$_{\rm full}$ &
MSE$_{\rm inner}$ &
MAE$_{\rm inner}$ &
Time &
MSE$_{\rm full}$ &
MAE$_{\rm full}$ &
MSE$_{\rm inner}$ &
MAE$_{\rm inner}$ &
Time
\\
\midrule
\multicolumn{12}{l}{\textbf{Setting 1 ($T=54$)}}\\
\midrule
3x3 & 9  & 2.774 & 1.308 & 2.930 & 1.227 & 0.47 &
1.930 & 1.147 & 0.773 & 0.860 & 0.47 \\
4x4 & 16 & 2.512 & 1.237 & 3.578 & 1.317 & 0.87 &
1.726 & 1.313 & 1.805 & 1.088 & 0.87 \\
5x5 & 25 & 3.927 & 1.558 & 2.234 & 1.074 & 1.43 &
1.395 & 1.159 & 1.248 & 0.984 & 1.43 \\
6x6 & 36 & 4.905 & 1.739 & 2.727 & 1.191 & 2.15 &
2.674 & 1.569 & 1.496 & 1.083 & 2.15 \\
7x7 & 49 & 6.256 & 1.993 & 3.099 & 1.376 & 3.80 &
2.729 & 1.571 & 1.054 & 0.988 & 3.80 \\
\addlinespace

\multicolumn{12}{l}{\textbf{Setting 2 ($T=36$)}}\\
\midrule
3x3 & 9  & 1.267 & 0.980 & 4.171 & 1.325 & 0.43 &
69.801 & 8.112 & 8.369 & 2.862 & 0.43 \\
4x4 & 16 & 1.713 & 1.130 & 5.896 & 1.502 & 1.05 &
67.027 & 7.722 & 7.863 & 2.737 & 1.05 \\
5x5 & 25 & 2.190 & 1.225 & 2.346 & 1.026 & 1.12 &
67.726 & 7.719 & 8.026 & 2.767 & 1.12 \\
6x6 & 36 & 3.374 & 1.571 & 3.511 & 1.279 & 1.68 &
69.826 & 7.856 & 7.752 & 2.754 & 1.68 \\
7x7 & 49 & 6.016 & 2.165 & 3.799 & 1.401 & 2.35 &
68.387 & 7.855 & 8.174 & 2.820 & 2.35 \\
\bottomrule
\end{tabular}}
\caption{Simulation results. Computational cost in minutes (Time) and in-sample (panel a) and out-of-sample (panel b) forecast errors for different grid sizes (Grid), evaluated using the Mean Squared Error (MSE), Mean Absolute Error (MAE) on the full sample (full) and on a subsample (inner).}
\label{tab:simulation_grid_time}
\end{table}

\clearpage

\section{Further details of the empirical application}      \label{Sec:Supp_empirical_appl}

In our application, we consider the global monthly fire location product (MCD14ML), which contains Terra and Aqua MODIS fire pixels in single monthly ASCII files. MODIS Collection 6 NRT Hotspot/Active Fire Detections MCD14ML distributed from NASA FIRMS. Available on-line \url{https://earthdata.nasa.gov/firms}. Fires are detected by the algorithms and swath-level products implemented as part of the Collection 6 land-product reprocessing, starting in May 2015.
The algorithms detect a fire pixel that contains actively burning fires at the time of the satellite overpass \citep{Giglio03, Giglio16}.

In the following, we present some results for models with dry and wet season dummy variables ($j=D$) and with harmonic components ($j=H$) following three specifications: (i) constant scale $\mathcal{M}^{j,c}$; (ii) time-varying scale $\mathcal{M}^{j,v}$; (iii) time-varying scale with monthly seasonal dummy $\mathcal{M}^{j,v12}$.
As a robustness check, we also report in Table~\ref{tab1} the estimated parameters and in-sample performance metrics for a model with an autoregressive scale $\mathcal{M}^{j,AR}$ similar to \cite{gamerman2013non}.

Figures~\ref{fig:Lambda_mean_Dummy} and \ref{fig:Lambda_mean_Harmonic} show the estimates of the fire intensity $\Lambda_t(x)$ for the dummy and harmonic model, respectively. Figures~\ref{fig:Lambda_mean_Dummy_CV} and \ref{fig:Lambda_mean_Harmonic_CV} show the estimates of the iCV of the fire intensity for the dummy and harmonic model, respectively.

\begin{figure}[p]
\centering
\hspace{-20pt}
\captionsetup{width=0.93\linewidth}
\begin{tabular}{ccc}
\multicolumn{3}{c}{\small (a) Global factor $\kappa_t$ with dummy specification} \\
Constant $c$ & iid $c_t$ & Monthly $c_t$ \\
\includegraphics[scale=0.33]{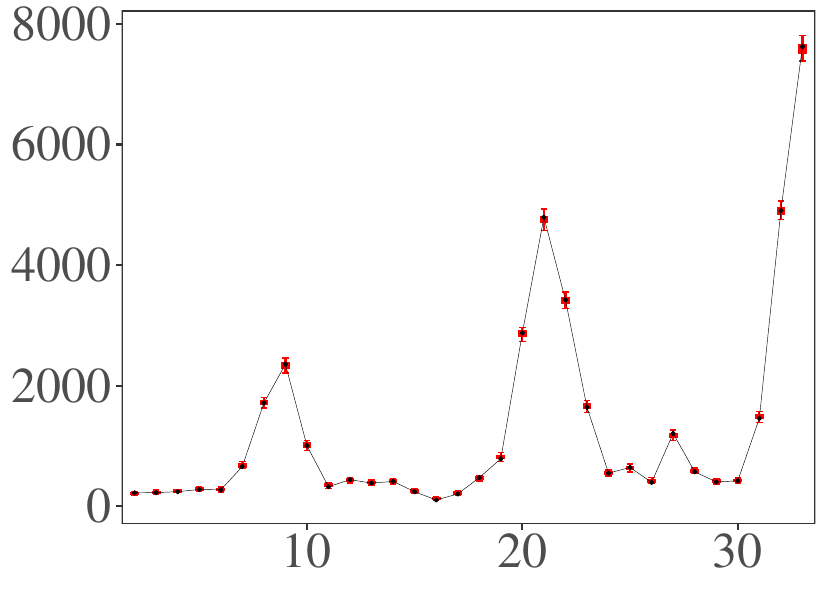}&
\includegraphics[scale=0.33]{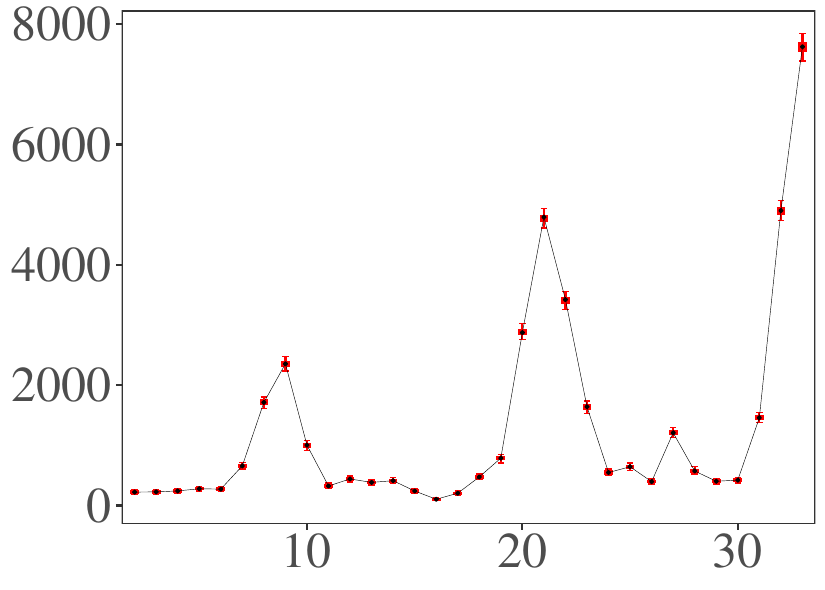}&
\includegraphics[scale=0.33]{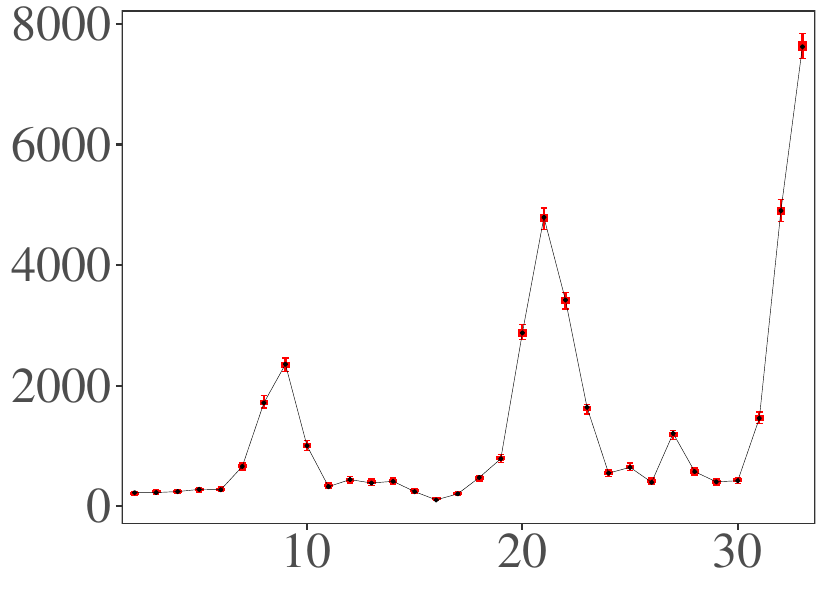}\\
\includegraphics[scale=0.33]{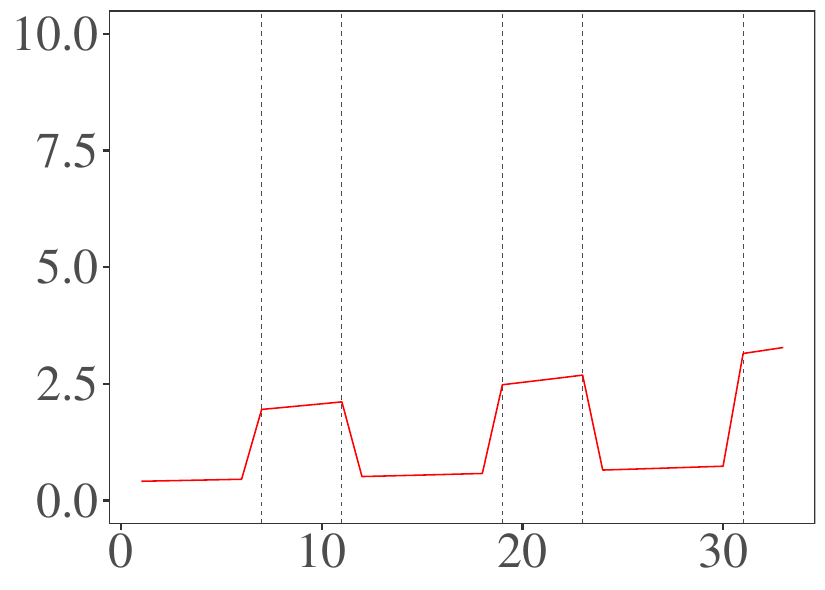}&
\includegraphics[scale=0.33]{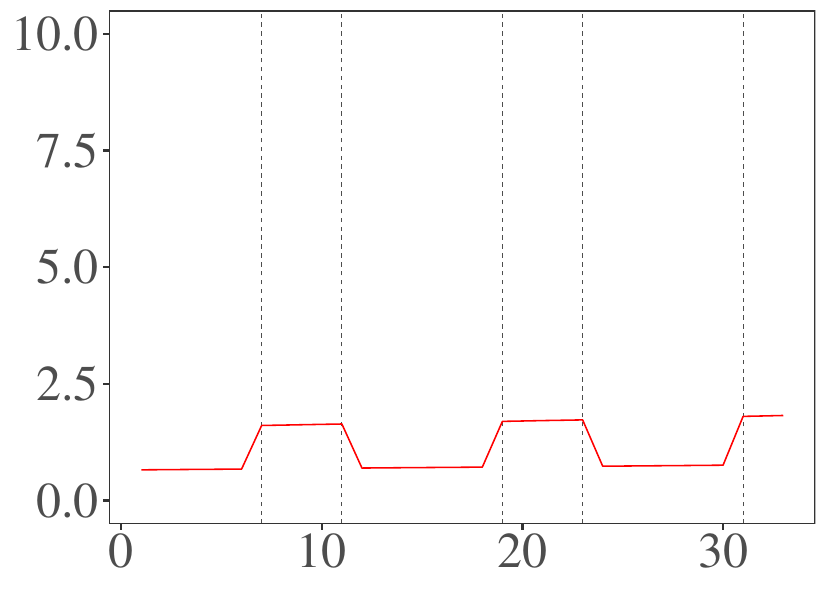}&
\includegraphics[scale=0.33]{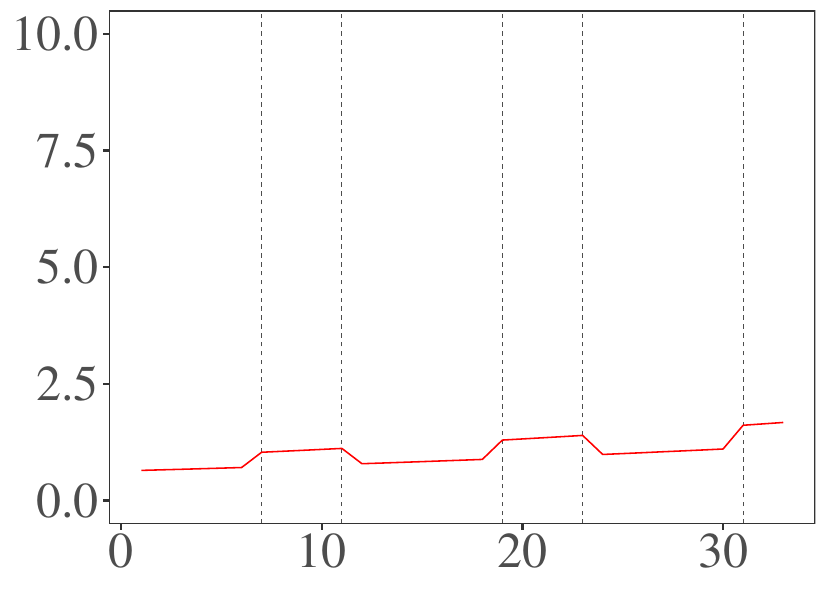}\\
\multicolumn{3}{c}{\small (b) Global factor $\kappa_t$ with harmonic specification}\\
Constant $c$ & iid $c_t$ & Monthly $c_t$ \\
\includegraphics[scale=0.33]{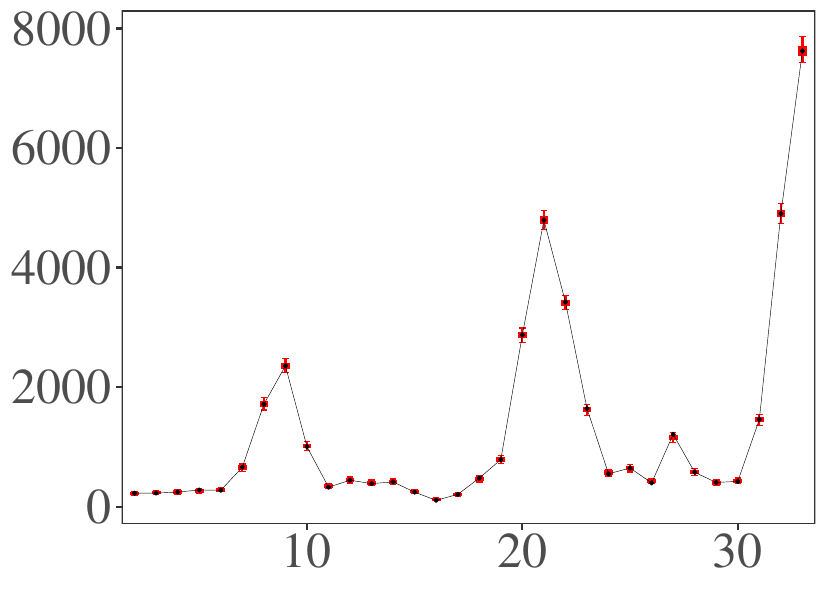}&
\includegraphics[scale=0.33]{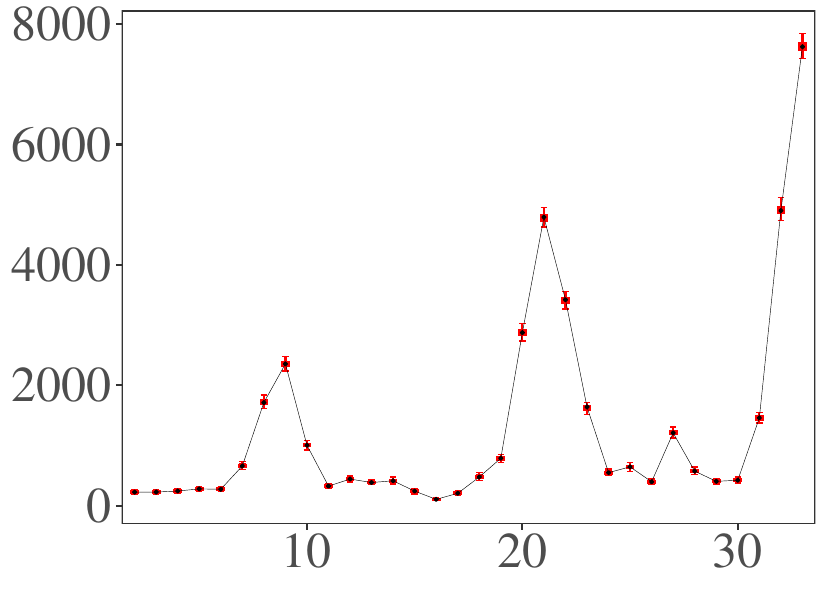}&
\includegraphics[scale=0.33]{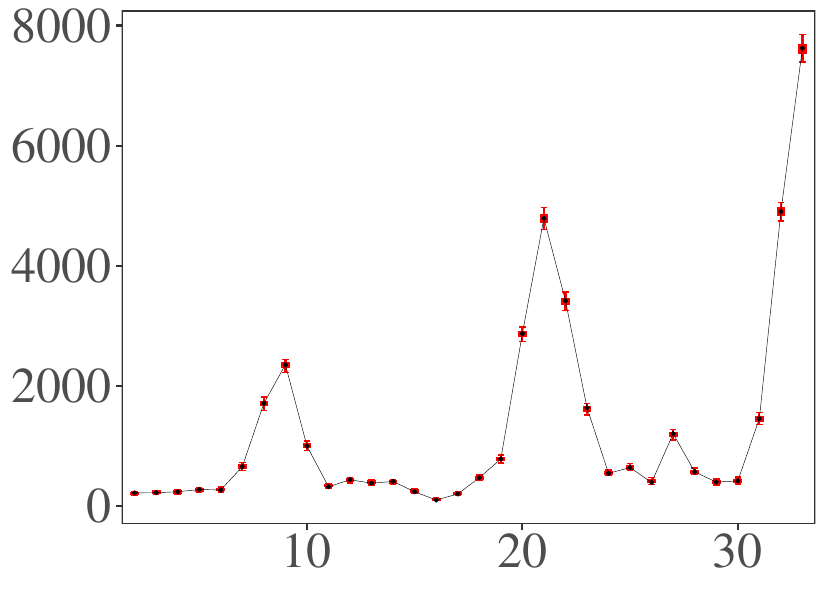}\\
\includegraphics[scale=0.33]{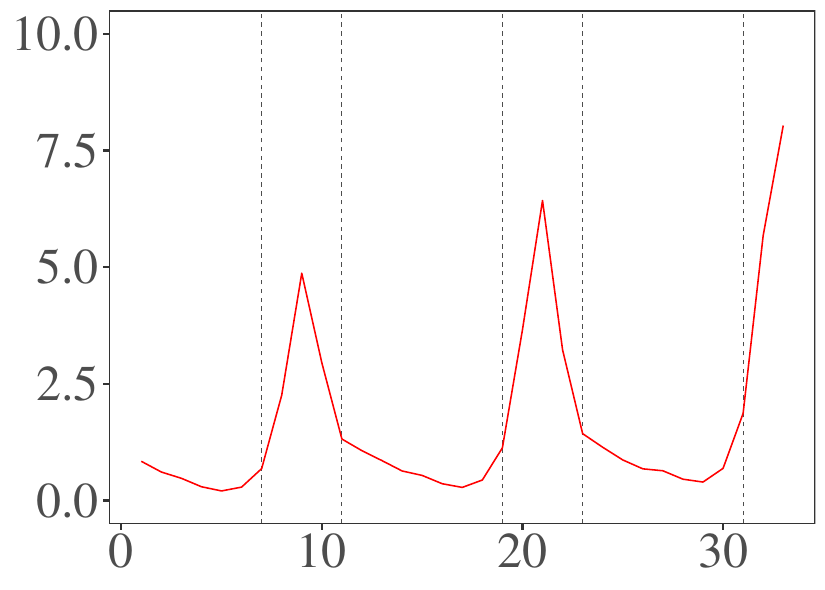}&
\includegraphics[scale=0.33]{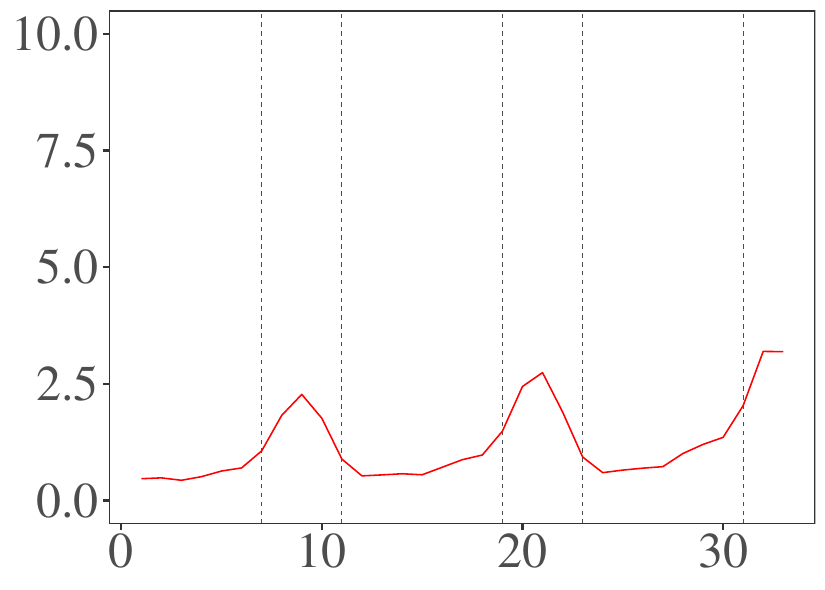}&
\includegraphics[scale=0.33]{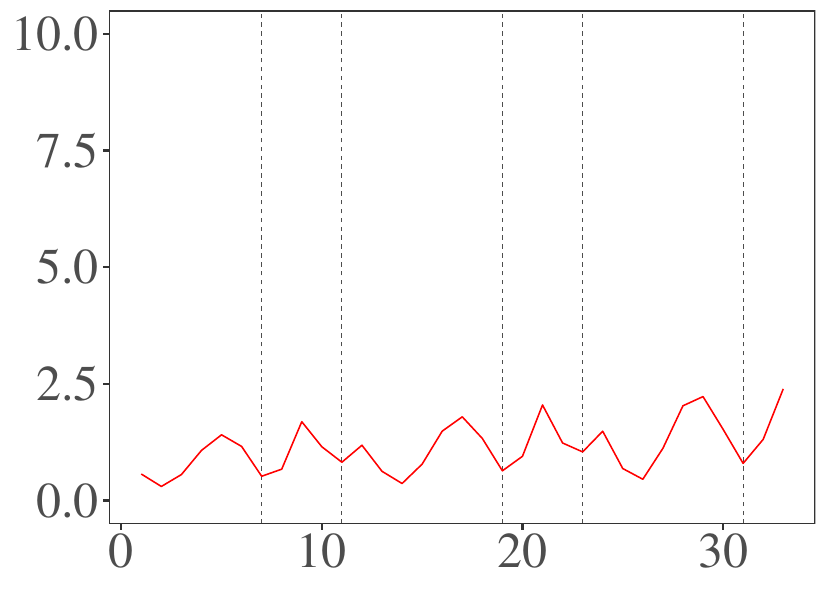}
\end{tabular}
\caption{Fire intensity for the harmonic component (panel a) and the dummy variable (panel b) specifications. In each panel, the global number of fires $N_t^y$ (top,$\bullet$), estimated global fire intensity $\Lambda_t(\Y)$ (top, \textcolor{red}{$\square$}) and the global factor $\kappa_t$ (bottom) at a monthly frequency from February 2018 to September 2020.
Dashed vertical lines denote the start and end of the dry season (July-November).}
\label{fig:LambdaKappaHat}
\end{figure}

To investigate the in-sample performance of the proposed \MARG(1) model, we compare the different scale and covariate specifications already considered in Table 1.
Specifically, for each case, Fig.~\ref{fig:LambdaKappaHat} compares the estimated total latent intensity, $\hat{\Lambda}_t$, with the total number of fires in the dataset. All the models are found to perform well, with a better performance of the harmonic specifications for months characterised by few fires.
Furthermore, we report the estimated global temporal component, $\kappa_t$, which clearly highlights both the trend and the seasonal patterns of the fire counts.
As a robustness check about the findings in 
Fig. 4 of the main article, Fig.~\ref{fig:IQRall} reports the normalised interquantile range of $\Lambda_t(A)$ for different sizes of the sub-regions $A \subset \Y$.
For all the sizes of the sub-regions, the main results are unaltered since there is evidence of higher uncertainty during the dry season (left plots) and, within each season, the uncertainty decreases when the number of observations increases.
Finally, Fig.~\ref{fig:Rtth} shows the value of $\mathscr{R}_{t,t+h}(x,y)$, for different months, locations, and horizons. We find evidence of heterogeneous dependence patterns, both spatially and over time. In fact, areas characterised by clustering features (i.e., $\mathscr{R}_{t,t+h}(x,y) > 1$) and areas with regularity ($\mathscr{R}_{t,t+h}(x,y) < 1$) are found at the same time/horizon around several of the locations considered.
Summarising, this suggests spatial heterogeneity and deviation from the standard Poisson process.

\clearpage
\begin{table}[h!]
\caption{Model estimates and fitting.\label{tab1}}
\def~{\hphantom{0}} 
\centering
\small
\setlength{\tabcolsep}{1pt}
\resizebox{\textwidth}{!}{%
\renewcommand{\arraystretch}{0.8}
\begin{tabular}{@{}ccccccccc@{}}
\toprule
&$\mathcal{M}^{H,c}$&$\mathcal{M}^{H,v}$&$\mathcal{M}^{H,v12}$&$\mathcal{M}^{H,vdep}$&$\mathcal{M}^{D,c}$&$\mathcal{M}^{D,v}$&$\mathcal{M}^{D,v12}$&$\mathcal{M}^{D,vdep}$\\[4pt]
\multicolumn{9}{c}{(a) Parameter estimates}\\[4pt]
$\eta_{TR}$ & 0.019 & 0.018 & 0.018 & -0.003 & 0.020 & 0.005 & 0.018 & -0.002 \\
 & (0.018,0.022) & (0.017,0.019) & (0.017,0.019) & (-0.005,0.000) & (0.016,0.022) & (0.003,0.008) & (0.015,0.020) & (-0.005,-0.000) \\
$\eta_{D,1}$ &  &  &  &  & 1.438 & 0.850 & 0.364 & 0.878 \\
 &  &  &  &  & (1.410,1.482) & (0.784,0.911) & (0.260,0.489) & (0.871,0.886) \\
$\eta_{S,1}$ & -0.929 & -0.697 & -0.209 & -0.759 &  &  &  &  \\
 & (-0.953,-0.898) & (-0.753,-0.646) & (-0.220,-0.198) & (-0.850,-0.707) &  &  &  &  \\
$\eta_{C,1}$ & 0.686 & -0.052 & -0.126 & 0.550 &  &  &  &  \\
 & (0.683,0.688) & (-0.054,-0.049) & (-0.128,-0.123) & (0.541,0.559) &  &  &  &  \\
$\eta_{S,2}$ & 0.139 & -0.025 & -0.443 & -0.112 &  &  &  &  \\
 & (0.130,0.147) & (-0.066,0.052) & (-0.460,-0.420) & (-0.143,-0.089) &  &  &  &  \\
$\eta_{C,2}$ & -0.513 & -0.260 & 0.077 & -0.003 &  &  &  &  \\
 & (-0.543,-0.494) & (-0.321,-0.197) & (0.062,0.089) & (-0.009,0.005) &  &  &  &  \\
$\eta_{S,3}$ & 0.119 & 0.078 & 0.308 & 0.235 &  &  &  &  \\
 & (0.117,0.121) & (0.045,0.103) & (0.307,0.310) & (0.228,0.240) &  &  &  &  \\
$\eta_{C,3}$ & 0.025 & -0.058 & 0.054 & 0.029 &  &  &  &  \\
 & (-0.026,0.050) & (-0.059,-0.055) & (0.052,0.061) & (0.026,0.034) &  &  &  &  \\
$\eta_{S,4}$ & 0.060 & -0.014 & 0.106 & -0.021 &  &  &  &  \\
 & (0.018,0.145) & (-0.031,0.015) & (0.105,0.107) & (-0.027,-0.006) &  &  &  &  \\
$\eta_{C,4}$ & 0.073 & -0.054 & 0.215 & 0.024 &  &  &  &  \\
 & (0.030,0.110) & (-0.058,-0.047) & (0.046,0.304) & (0.010,0.038) &  &  &  &  \\[4pt]
\multicolumn{9}{c}{(b) Model fitting (normalised across models)}\\[4pt]
MSE & 0.240 & 0.032 & 0.121 & 0.021 & 0.350 & 0.043 & 0.142 & 0.051 \\
MAE & 0.183 & 0.064 & 0.146 & 0.061 & 0.250 & 0.063 & 0.163 & 0.071 \\[4pt]
\multicolumn{9}{c}{(c) Model forecasting over horizon $h$ (cumulated and normalised across models)}\\[4pt]
MSE $h = 1$ & 0.0001 & 0.0018 & 0.0000 & 0.1253 & 0.0016 & 0.0028 & 0.0001 & 0.8683 \\
MSE $h = 3$ & 0.0005 & 0.0018 & 0.0000 & 0.1044 & 0.0016 & 0.0028 & 0.0002 & 0.8886 \\
MSE $h = 6$ & 0.0000 & 0.0000 & 0.0000 & 0.2183 & 0.0000 & 0.0000 & 0.0000 & 0.7817 \\[3pt]
MAE h = 1 & 0.0068 & 0.0297 & 0.0037 & 0.2447 & 0.0277 & 0.0363 & 0.0067 & 0.6442 \\
MAE h = 3 & 0.0137 & 0.0304 & 0.0041 & 0.2378 & 0.0291 & 0.0382 & 0.0099 & 0.6369 \\
MAE h = 6 & 0.0007 & 0.0015 & 0.0002 & 0.3880 & 0.0014 & 0.0018 & 0.0005 & 0.6058 \\
\bottomrule
\end{tabular}}
\label{tab:parest}
\begin{tablenotes}
\item \footnotesize Panel (a) reports the posterior mean and $95\%$ posterior credible intervals (in brackets) of the parameters of the global factor $\kappa_t$ across models. Panels (b) and (c) show the in- and out-of-sample model diagnostics.
For the harmonic ($j=H$) and the dry-season dummy ($j=D$) model three specifications are considered: $\mathcal{M}^{j,c}$, constant scale;
$\mathcal{M}^{j,v}$ time-varying scale; $\mathcal{M}^{j,v12}$ time-varying scale with monthly seasonal dummy; $\mathcal{M}^{j,vdep}$ time-dependent scale as in \citet{gamerman2013non}.
For model performance, Mean Square Error (MSE) and Mean Absolute Error (MAE) are computed in-sample and normalised across models.
\end{tablenotes}
\end{table}

\begin{sidewaystable}[p]
\hspace{1.5ex}
\captionsetup{width=0.97\linewidth}
\setlength{\tabcolsep}{5pt}
\begin{tabular}{c c c c c}
 & {\footnotesize June $2020$} & {\footnotesize July $2020$} & {\footnotesize August $2020$} & {\footnotesize September $2020$} \\
\begin{rotate}{90} \hspace{9pt} {\scriptsize Constant scale $c$} \end{rotate} &
\includegraphics[clip,trim=0 0 0 0, scale=0.45]{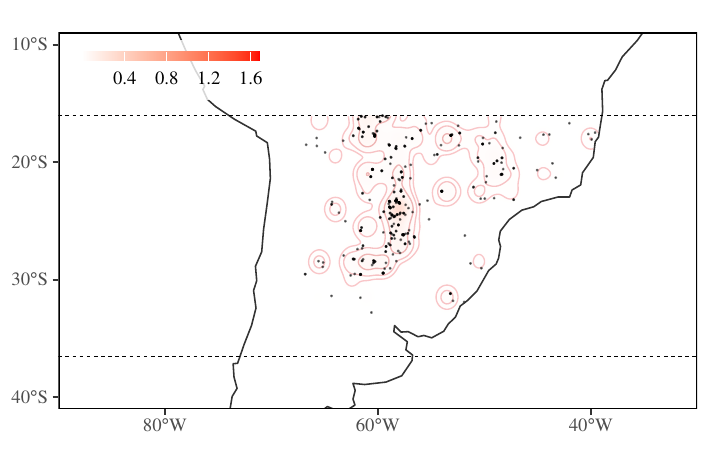}&
\includegraphics[clip,trim=0 0 0 0, scale=0.45]{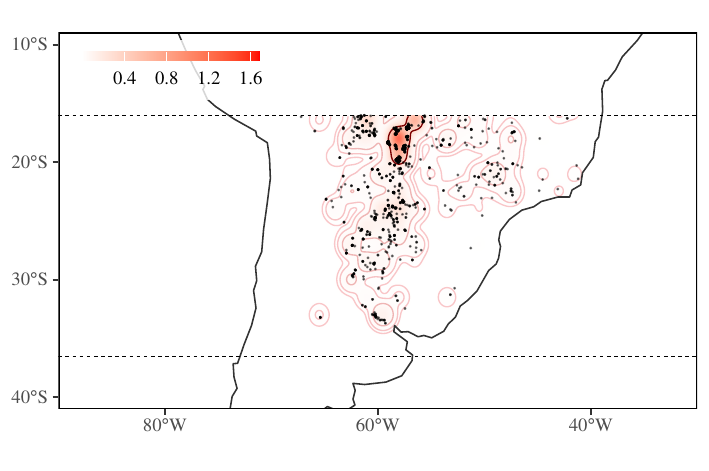}&
\includegraphics[clip,trim=0 0 0 0, scale=0.45]{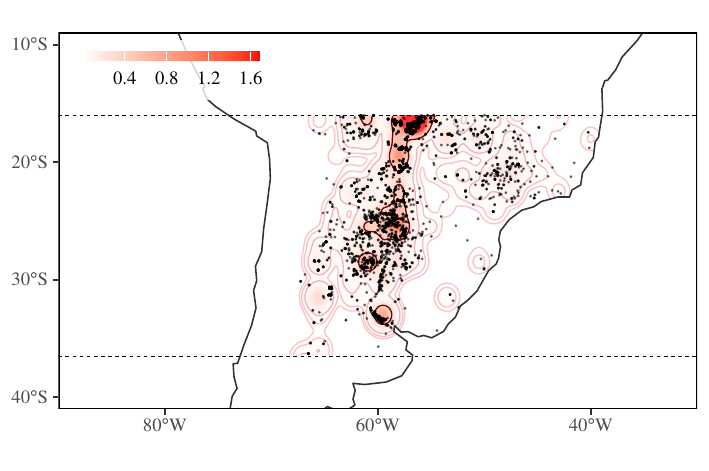}&
\includegraphics[clip,trim=0 0 0 0, scale=0.45]{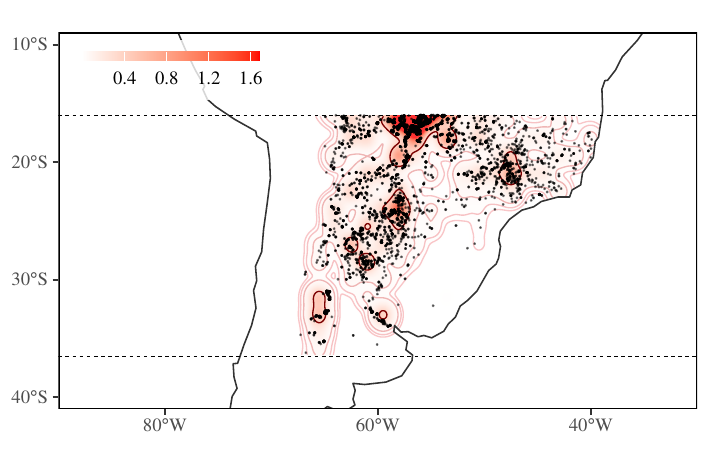} \\
\begin{rotate}{90} \hspace{9pt} {\scriptsize Time-varying scale $c_t$} \end{rotate} &
\includegraphics[clip,trim=0 0 0 0, scale=0.45]{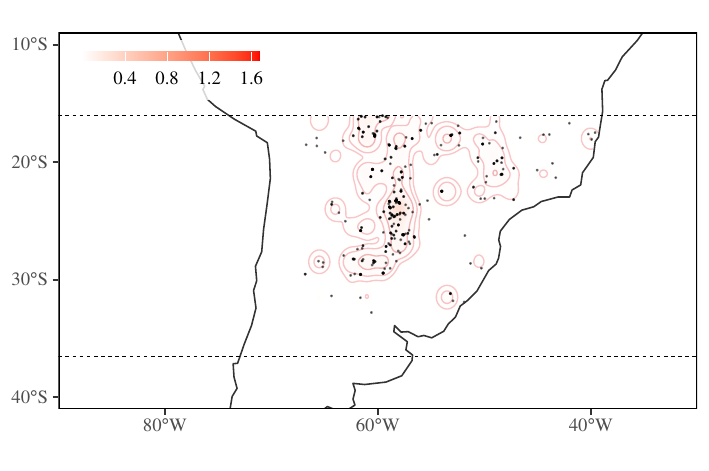}&
\includegraphics[clip,trim=0 0 0 0, scale=0.45]{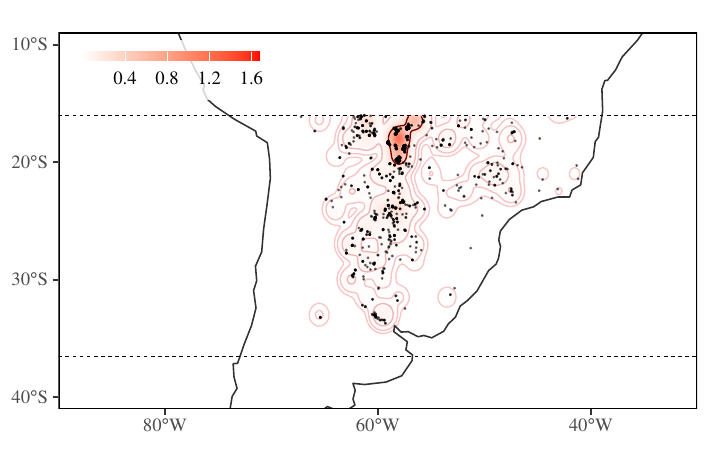}&
\includegraphics[clip,trim=0 0 0 0, scale=0.45]{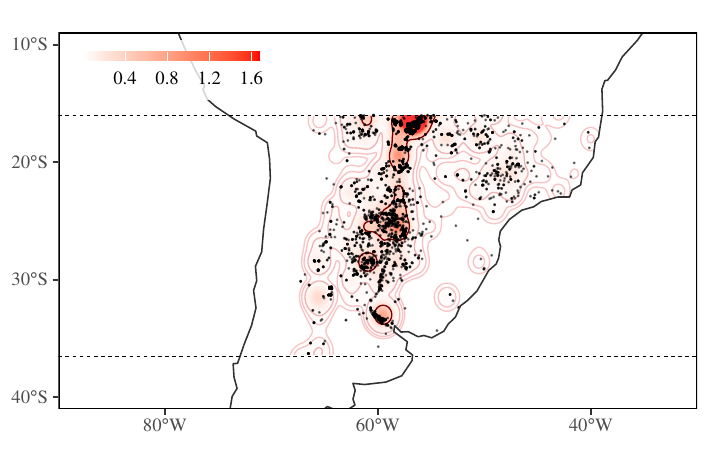}&
\includegraphics[clip,trim=0 0 0 0, scale=0.45]{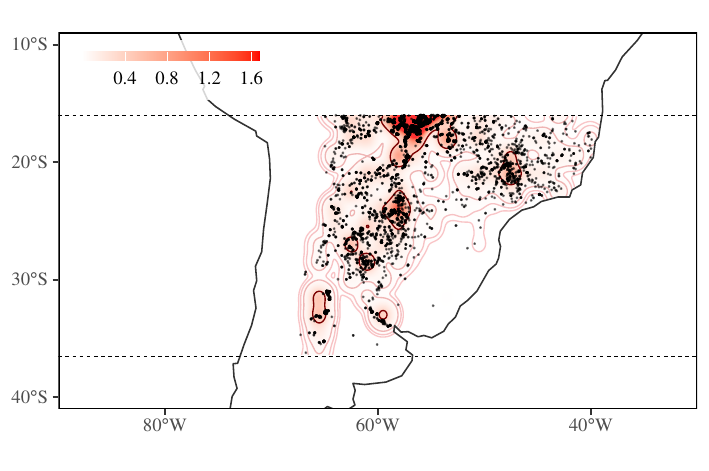} \\
\begin{rotate}{90} \hspace{9pt} {\scriptsize Time-varying scale $c_t(12)$} \end{rotate} &
\includegraphics[clip,trim=0 0 0 0, scale=0.45]{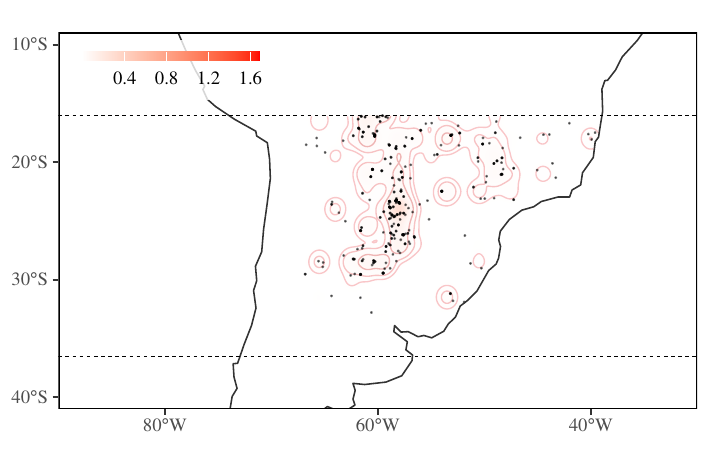}&
\includegraphics[clip,trim=0 0 0 0, scale=0.45]{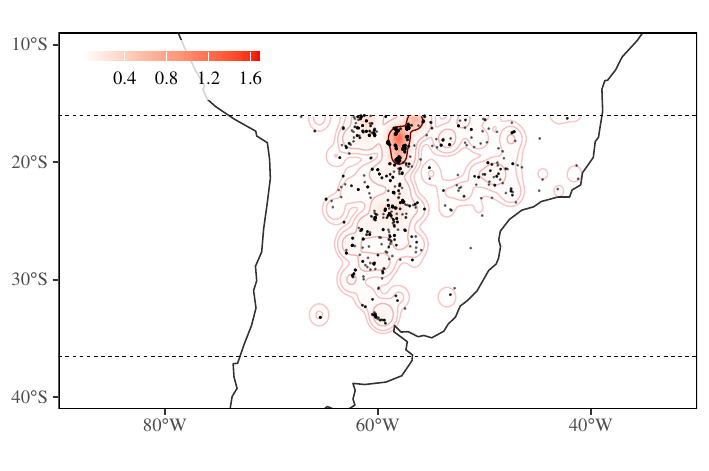}&
\includegraphics[clip,trim=0 0 0 0, scale=0.45]{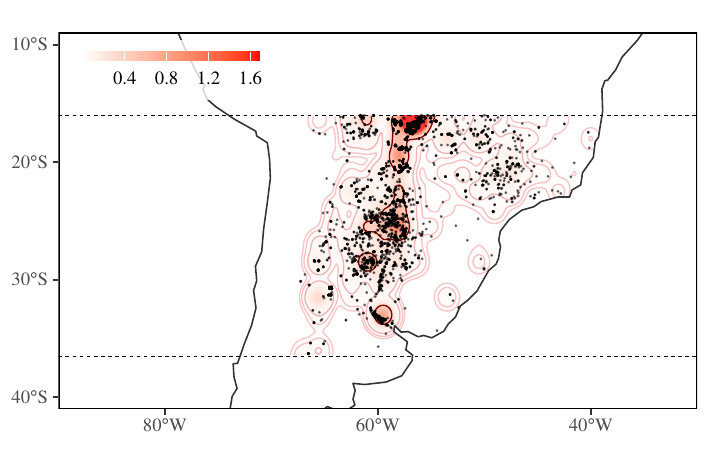}&
\includegraphics[clip,trim=0 0 0 0, scale=0.45]{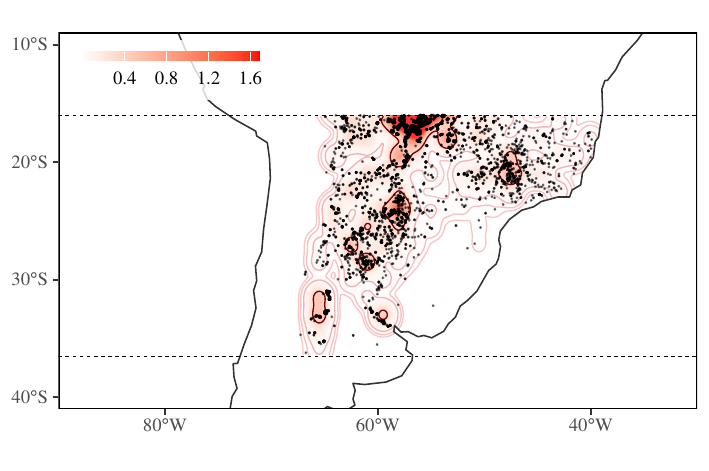} \\
\end{tabular}
\captionof{figure}{Monthly fires (dots) between longitude $80^{o}W$ and $40^{o}W$ and latitude $35^{o}S$ and $10^{o}S$ and estimated fire intensity, $\Lambda_t(x)$ (shaded areas and contour lines) for models with dry and wet seasonal dummy variables and either constant scale $c$ or time-varying scale $c_t$ and  time-varying seasonal scale $c_t(12)$ (different rows).}
\label{fig:Lambda_mean_Dummy}
\end{sidewaystable}

\begin{sidewaystable}[p]
\hspace{1.5ex}
\captionsetup{width=0.97\linewidth}
\setlength{\tabcolsep}{5pt}
\begin{tabular}{c c c c c}
 & {\footnotesize June $2020$} & {\footnotesize July $2020$} & {\footnotesize August $2020$} & {\footnotesize September $2020$} \\
\begin{rotate}{90} \hspace{9pt} {\scriptsize Constant scale $c$} \end{rotate} &
\includegraphics[clip,trim=0 0 0 0, scale=0.45]{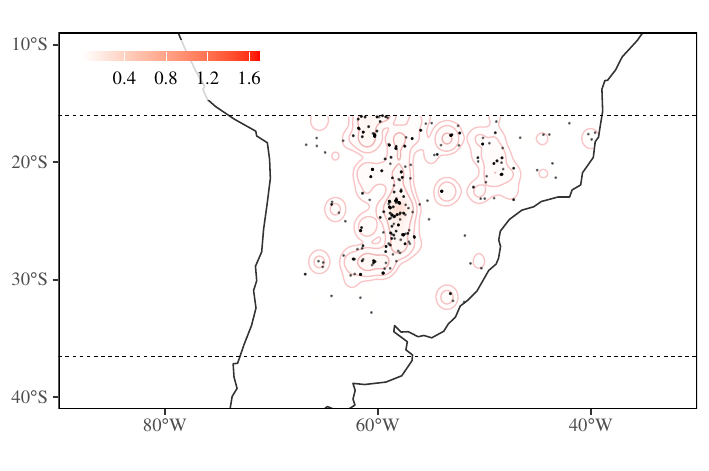}&
\includegraphics[clip,trim=0 0 0 0, scale=0.45]{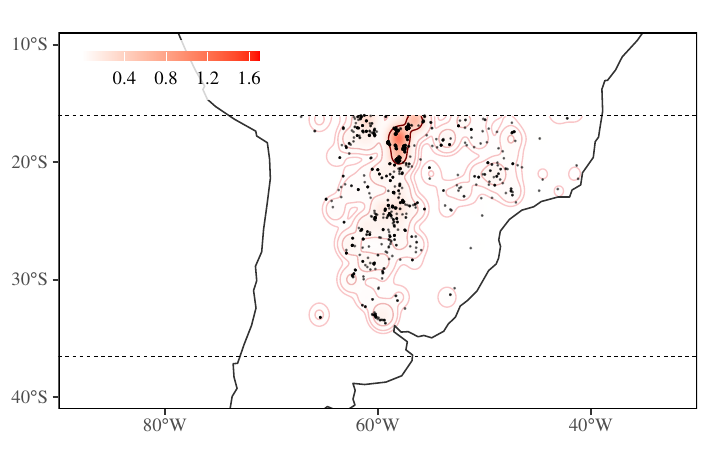}&
\includegraphics[clip,trim=0 0 0 0, scale=0.45]{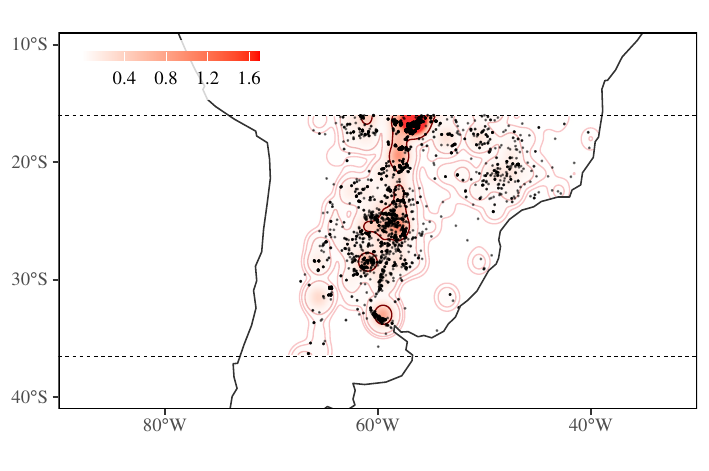}&
\includegraphics[clip,trim=0 0 0 0, scale=0.45]{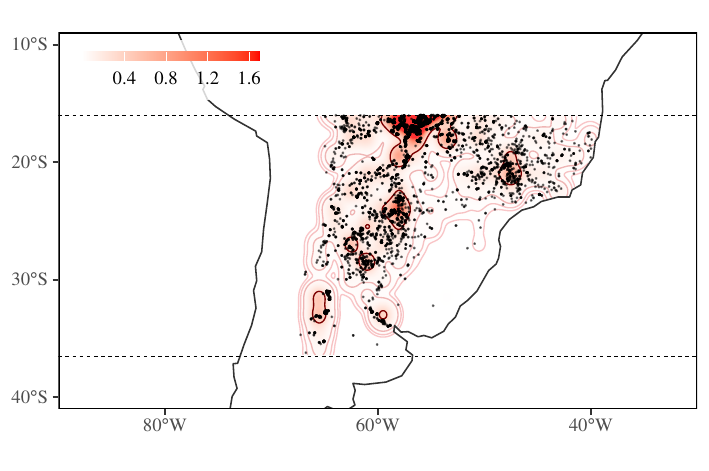} \\
\begin{rotate}{90} \hspace{9pt} {\scriptsize Time-varying scale $c_t$} \end{rotate} &
\includegraphics[clip,trim=0 0 0 0, scale=0.45]{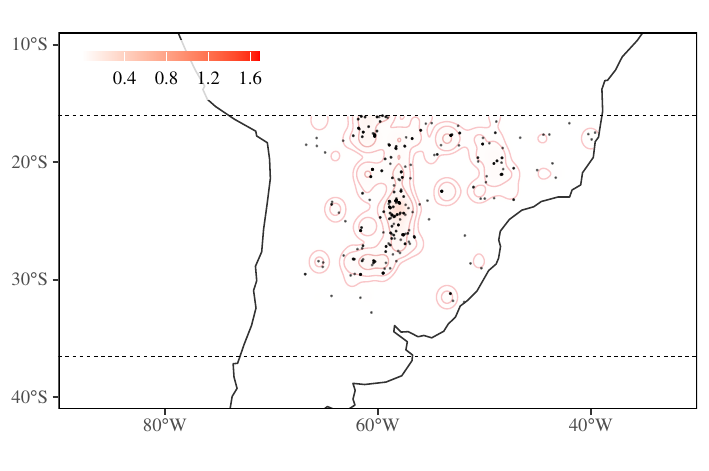}&
\includegraphics[clip,trim=0 0 0 0, scale=0.45]{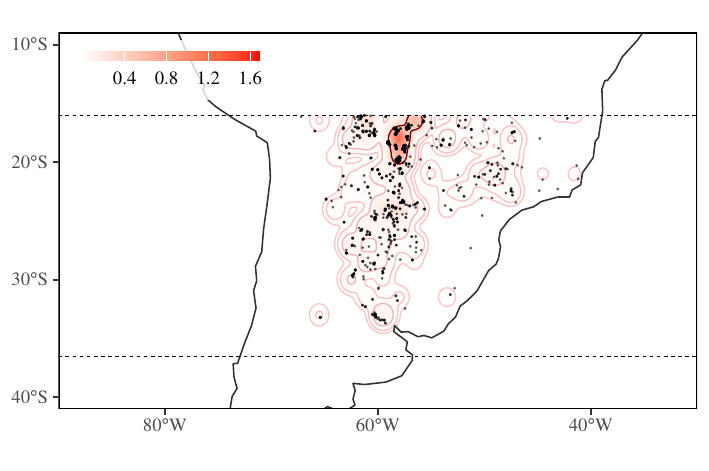}&
\includegraphics[clip,trim=0 0 0 0, scale=0.45]{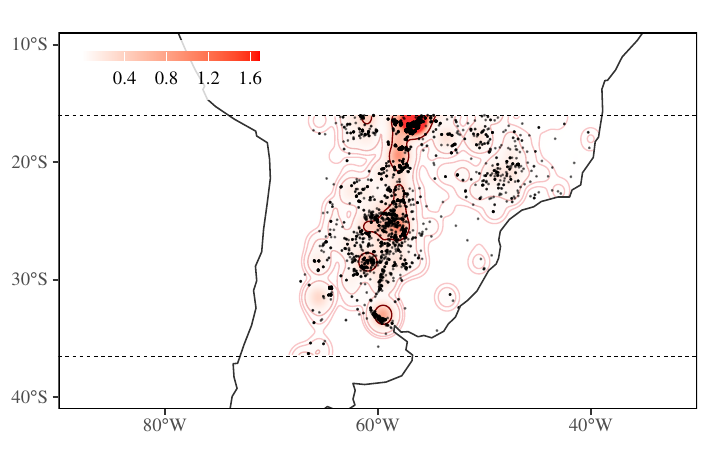}&
\includegraphics[clip,trim=0 0 0 0, scale=0.45]{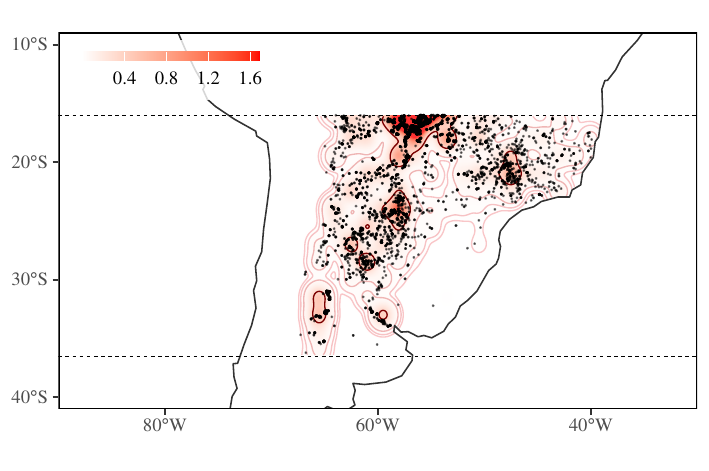} \\
\begin{rotate}{90} \hspace{9pt} {\scriptsize Time-varying scale $c_t(12)$} \end{rotate} &
\includegraphics[clip,trim=0 0 0 0, scale=0.45]{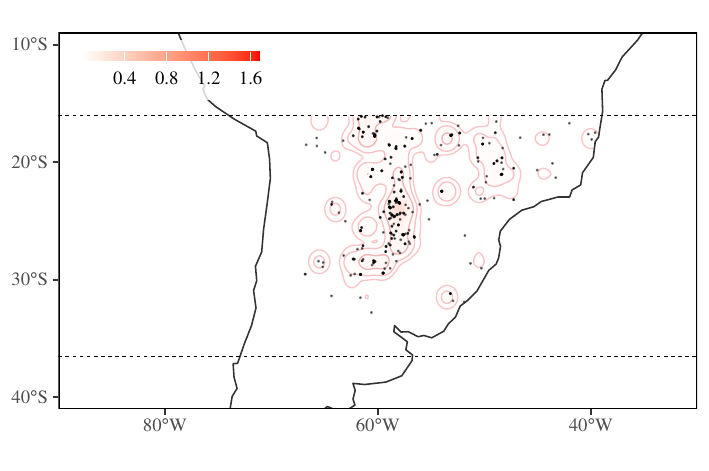}&
\includegraphics[clip,trim=0 0 0 0, scale=0.45]{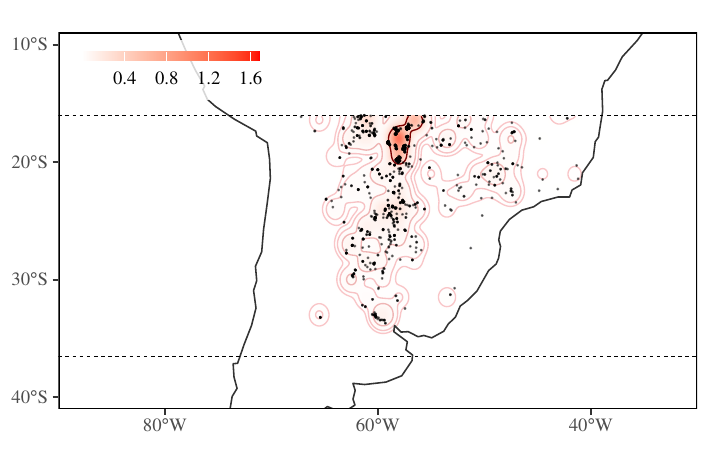}&
\includegraphics[clip,trim=0 0 0 0, scale=0.45]{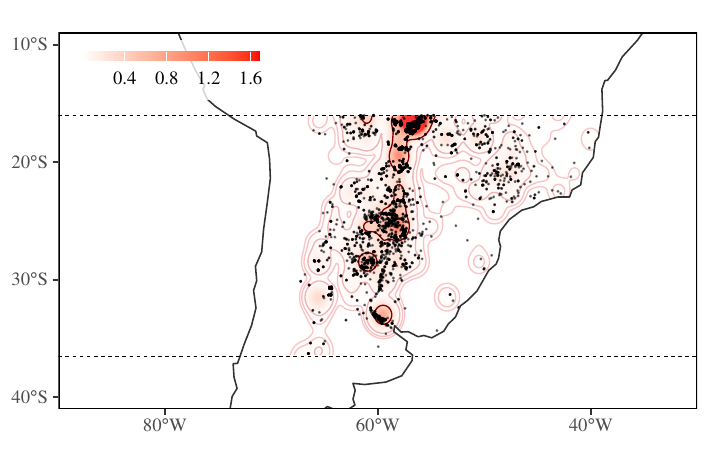}&
\includegraphics[clip,trim=0 0 0 0, scale=0.45]{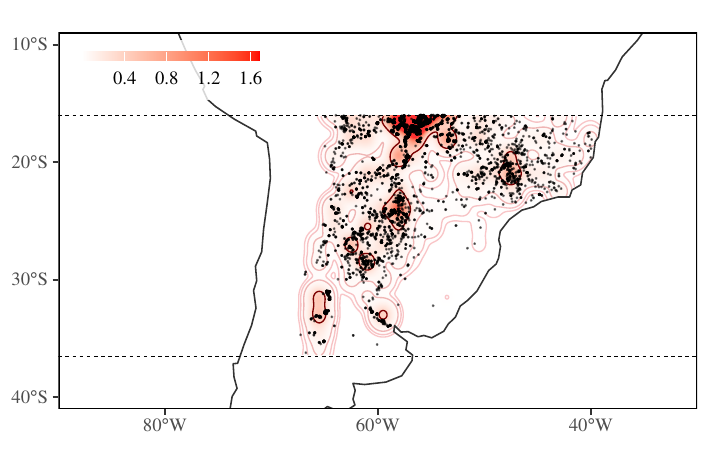} \\
\end{tabular}
\captionof{figure}{Monthly fires (dots) between longitude $80^{o}W$ and $40^{o}W$ and latitude $35^{o}S$ and $10^{o}S$ and estimated fire intensity, $\Lambda_t(x)$ (shaded areas and contour lines) for models with harmonic components and either constant scale $c$ or time-varying scale $c_t$ and  time-varying seasonal scale $c_t(12)$ (different rows).}
\label{fig:Lambda_mean_Harmonic}
\end{sidewaystable}

\begin{sidewaystable}[p]
\hspace{1.5ex}
\captionsetup{width=0.97\linewidth}
\setlength{\tabcolsep}{5pt}
\begin{tabular}{c c c c c}
 & {\footnotesize June $2020$} & {\footnotesize July $2020$} & {\footnotesize August $2020$} & {\footnotesize September $2020$} \\
\begin{rotate}{90} \hspace{9pt} {\scriptsize Constant scale $c$} \end{rotate} &
\includegraphics[clip,trim=0 0 0 0, scale=0.45]{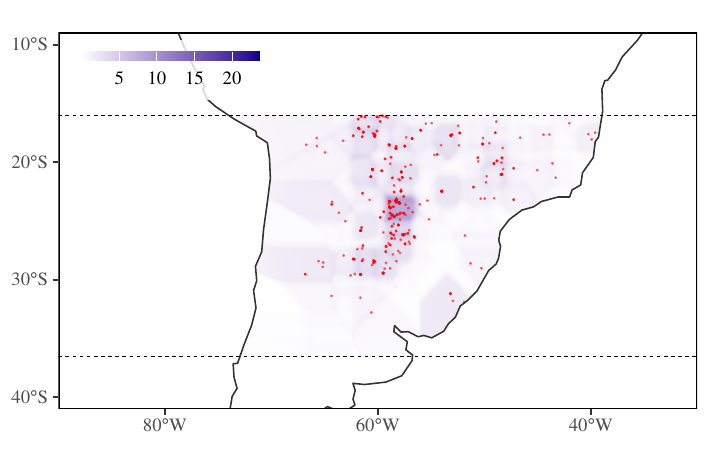}&
\includegraphics[clip,trim=0 0 0 0, scale=0.45]{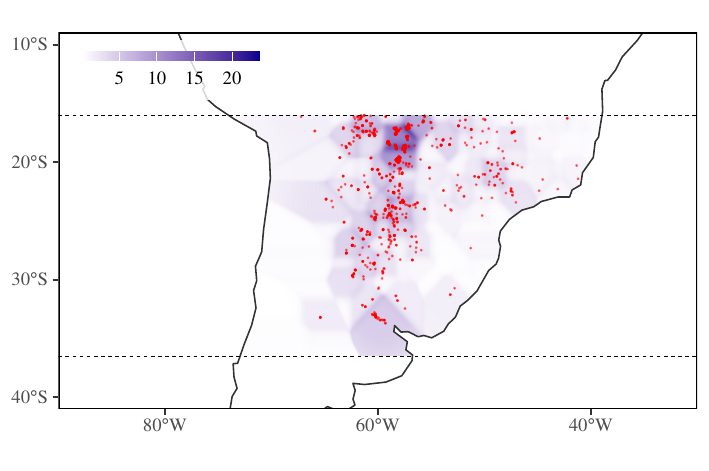}&
\includegraphics[clip,trim=0 0 0 0, scale=0.45]{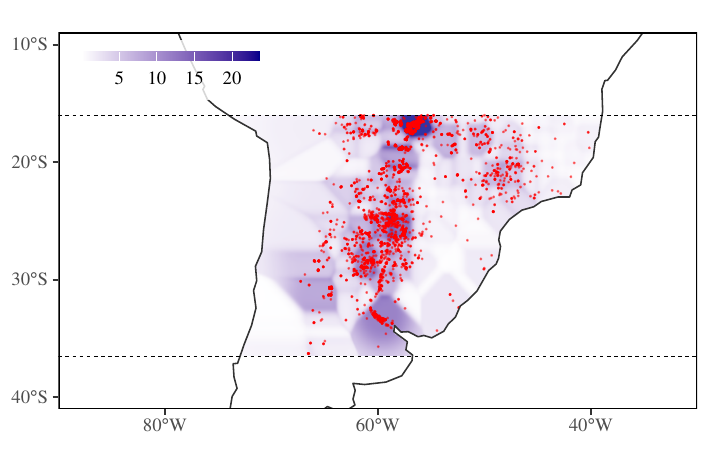}&
\includegraphics[clip,trim=0 0 0 0, scale=0.45]{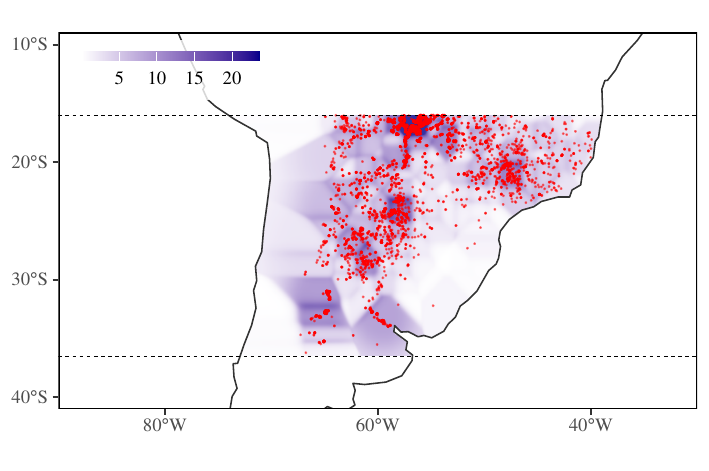} \\
\begin{rotate}{90} \hspace{9pt} {\scriptsize Time-varying scale $c_t$} \end{rotate} &
\includegraphics[clip,trim=0 0 0 0, scale=0.45]{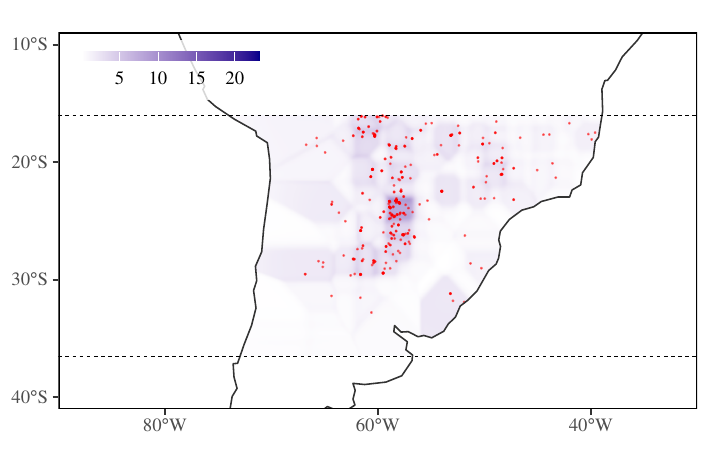}&
\includegraphics[clip,trim=0 0 0 0, scale=0.45]{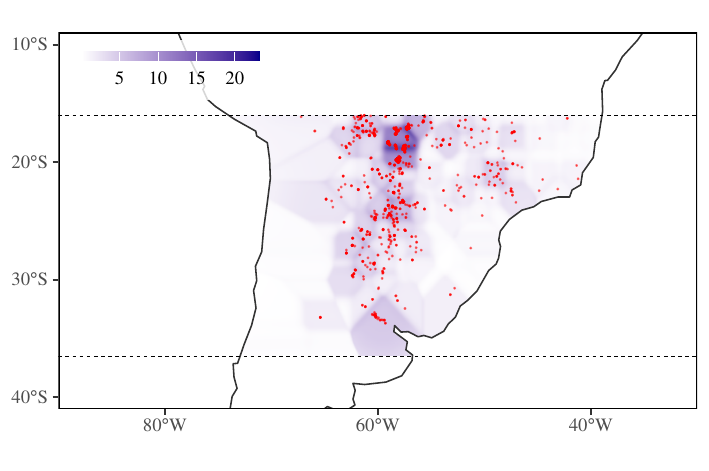}&
\includegraphics[clip,trim=0 0 0 0, scale=0.45]{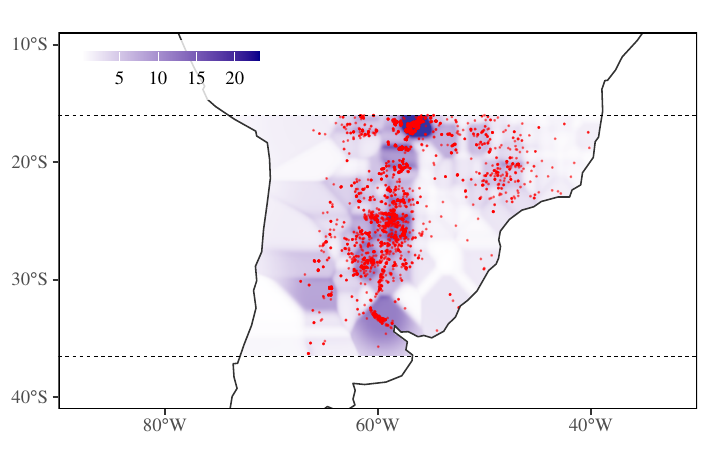}&
\includegraphics[clip,trim=0 0 0 0, scale=0.45]{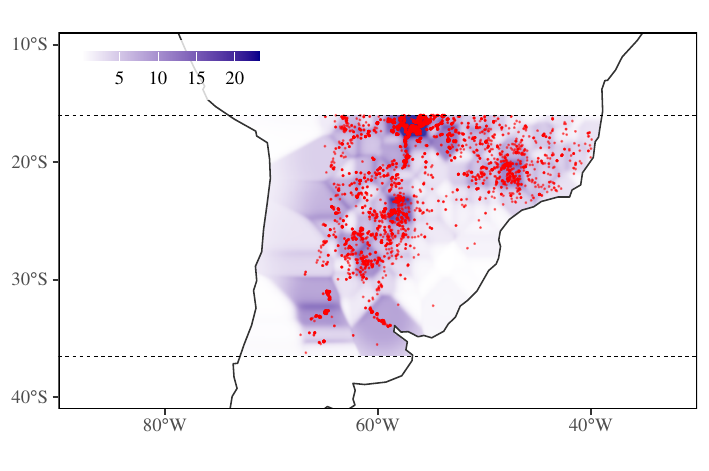} \\
\begin{rotate}{90} \hspace{9pt} {\scriptsize Time-varying scale $c_t(12)$} \end{rotate} &
\includegraphics[clip,trim=0 0 0 0, scale=0.45]{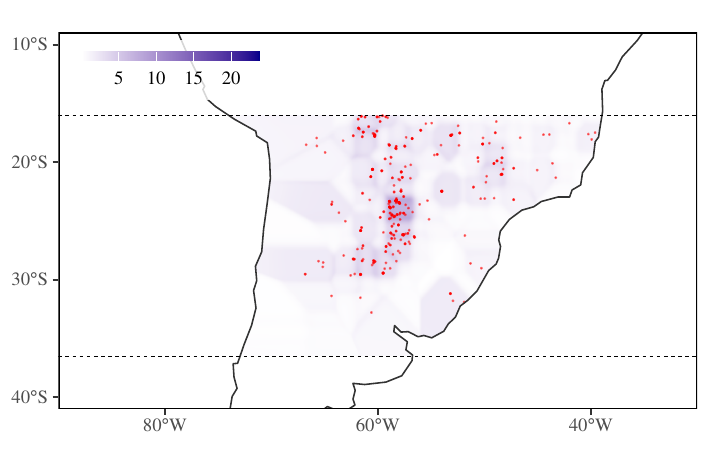}&
\includegraphics[clip,trim=0 0 0 0, scale=0.45]{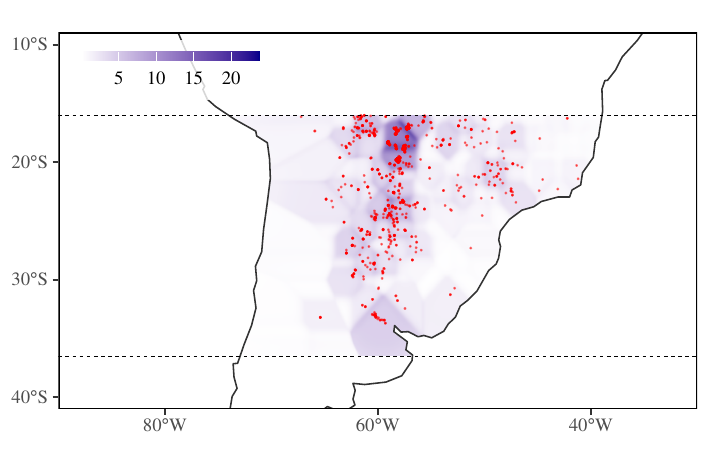}&
\includegraphics[clip,trim=0 0 0 0, scale=0.45]{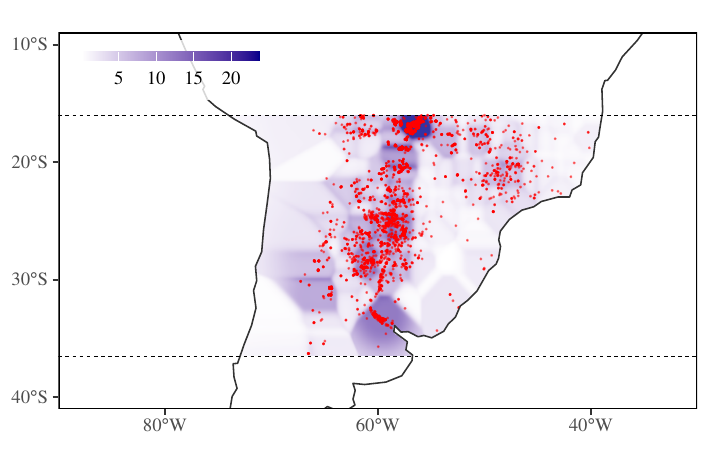}&
\includegraphics[clip,trim=0 0 0 0, scale=0.45]{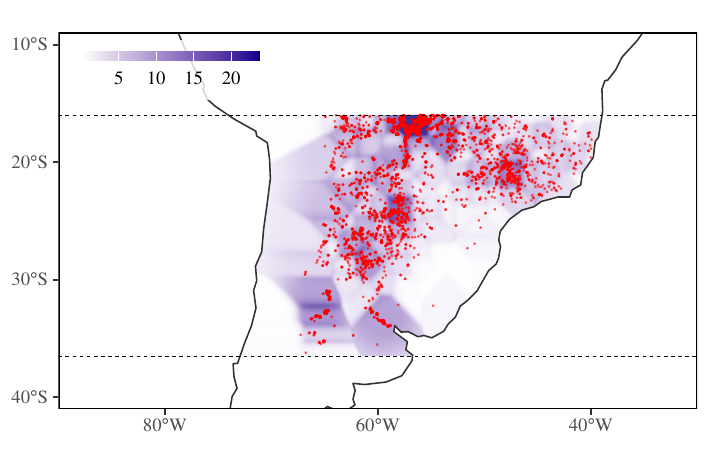} \\
\end{tabular}
\captionof{figure}{Inverse coefficient of variation of the fire intensity, $\Lambda_t(x)$ (shaded areas and contour lines) for models with seasonal dummy variables and either constant scale $c$ or time-varying scale $c_t$ and  time-varying seasonal scale $c_t(12)$ (different rows).}
\label{fig:Lambda_mean_Dummy_CV}
\end{sidewaystable}

\begin{sidewaystable}[p]
\hspace{1.5ex}
\captionsetup{width=0.97\linewidth}
\setlength{\tabcolsep}{5pt}
\begin{tabular}{c c c c c}
 & {\footnotesize June $2020$} & {\footnotesize July $2020$} & {\footnotesize August $2020$} & {\footnotesize September $2020$} \\
\begin{rotate}{90} \hspace{9pt} {\scriptsize Constant scale $c$} \end{rotate} &
\includegraphics[clip,trim=0 0 0 0, scale=0.45]{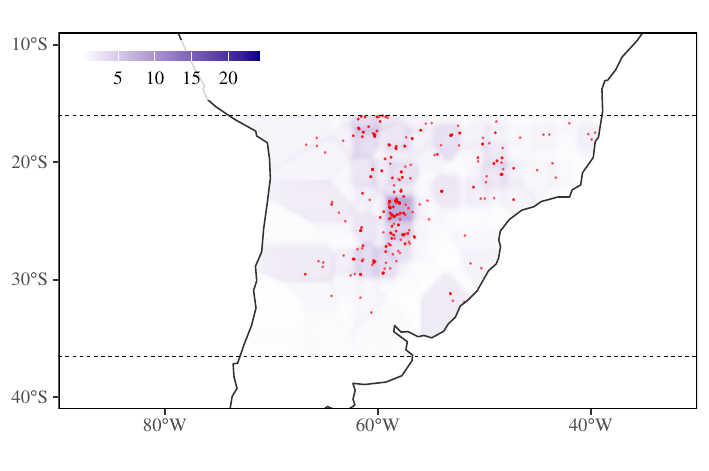}&
\includegraphics[clip,trim=0 0 0 0, scale=0.45]{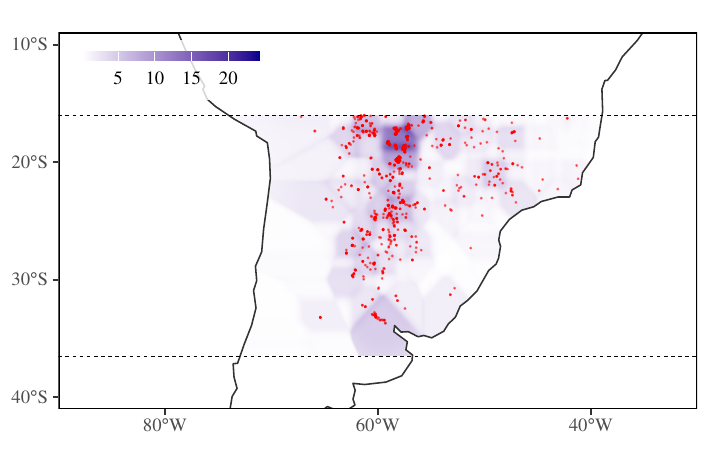}&
\includegraphics[clip,trim=0 0 0 0, scale=0.45]{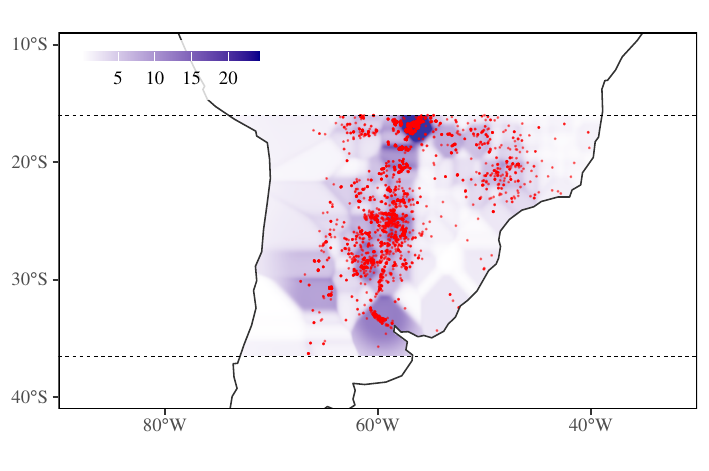}&
\includegraphics[clip,trim=0 0 0 0, scale=0.45]{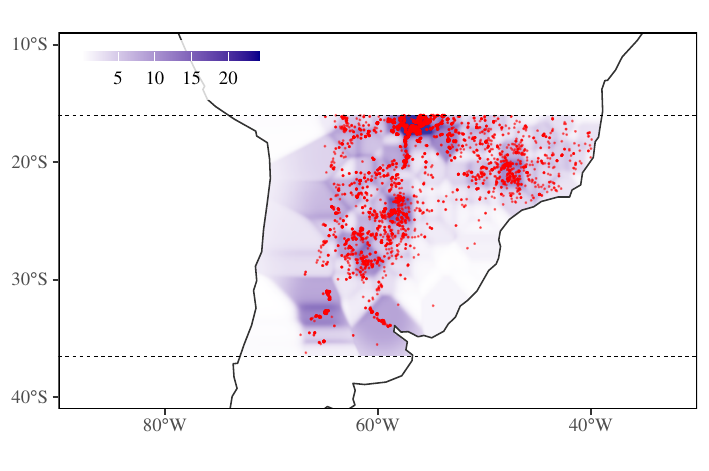} \\
\begin{rotate}{90} \hspace{9pt} {\scriptsize Time-varying scale $c_t$} \end{rotate} &
\includegraphics[clip,trim=0 0 0 0, scale=0.45]{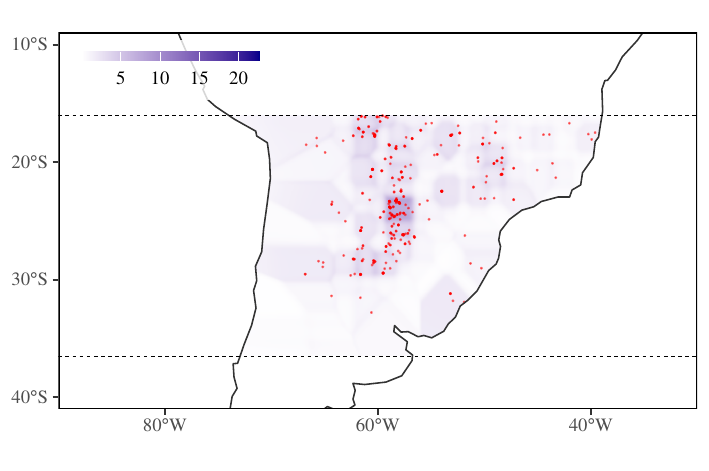}&
\includegraphics[clip,trim=0 0 0 0, scale=0.45]{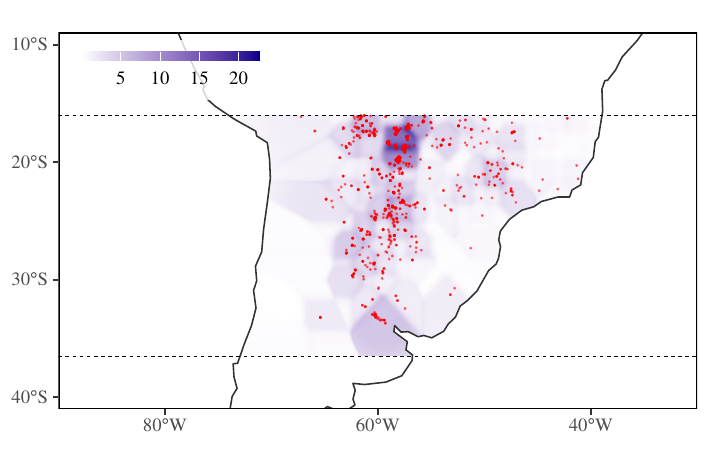}&
\includegraphics[clip,trim=0 0 0 0, scale=0.45]{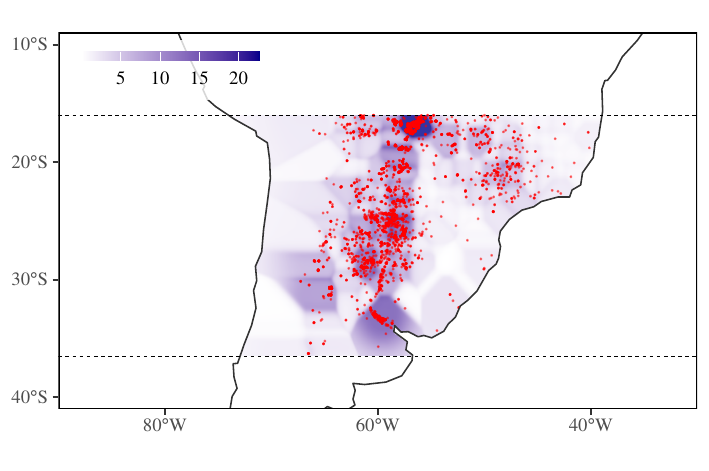}&
\includegraphics[clip,trim=0 0 0 0, scale=0.45]{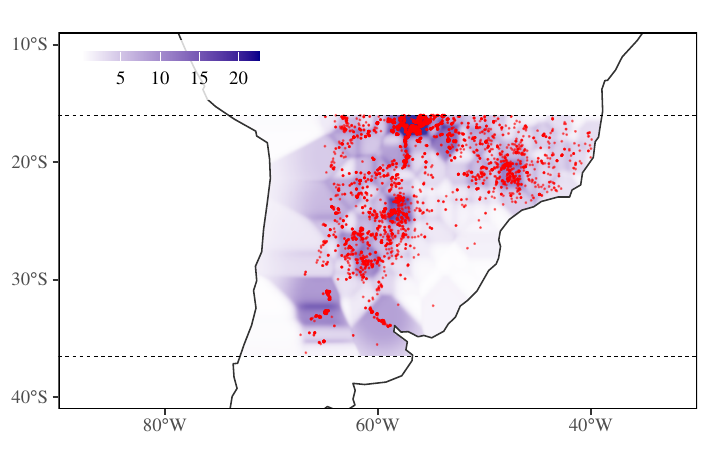} \\
\begin{rotate}{90} \hspace{9pt} {\scriptsize Time-varying scale $c_t(12)$} \end{rotate} &
\includegraphics[clip,trim=0 0 0 0, scale=0.45]{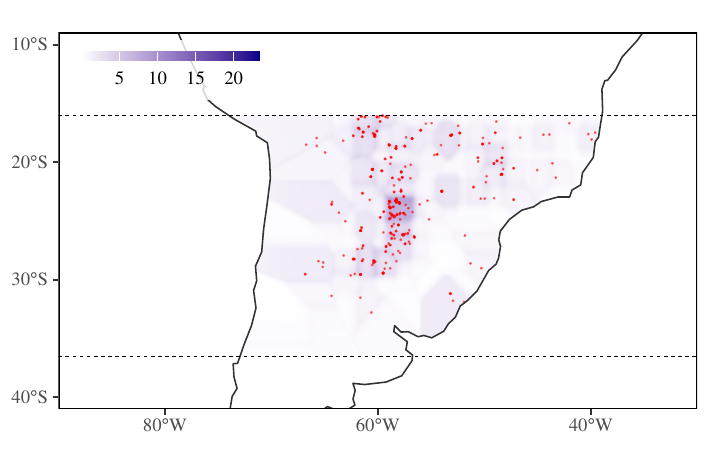}&
\includegraphics[clip,trim=0 0 0 0, scale=0.45]{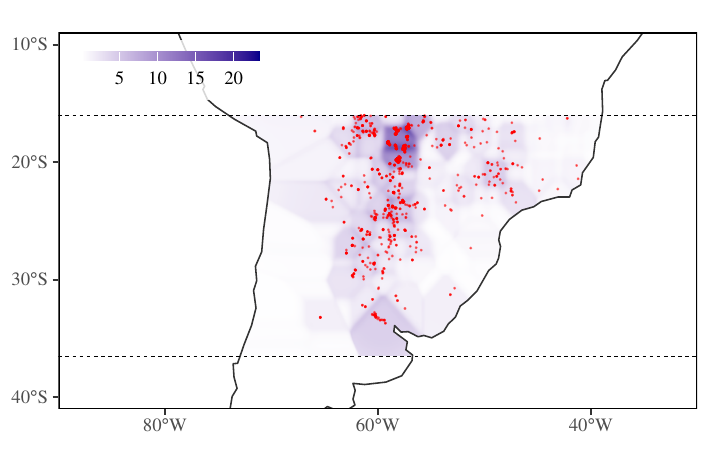}&
\includegraphics[clip,trim=0 0 0 0, scale=0.45]{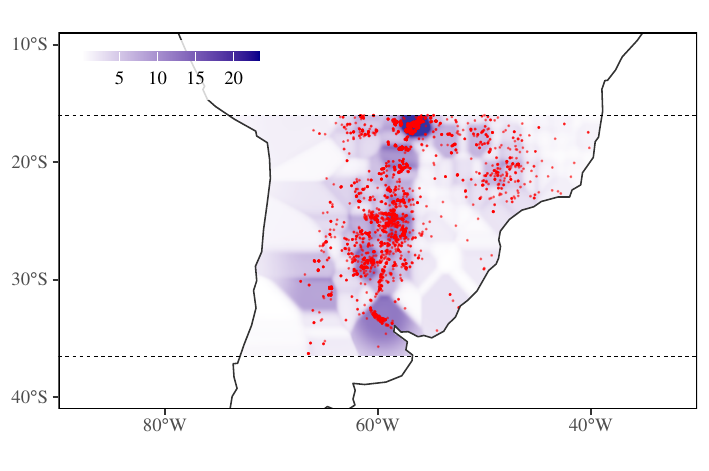}&
\includegraphics[clip,trim=0 0 0 0, scale=0.45]{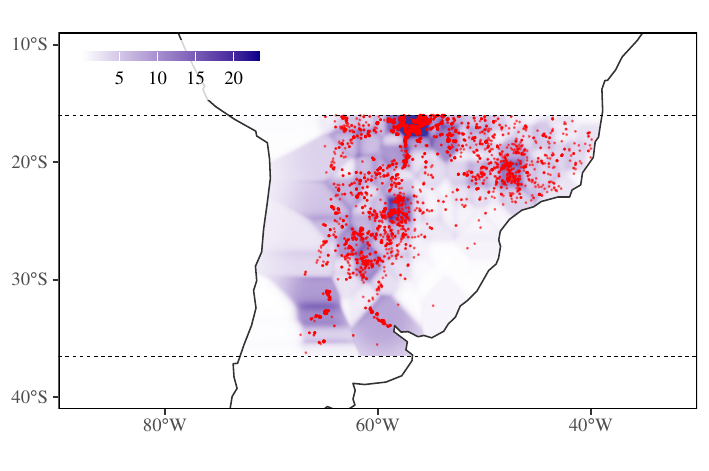} \\
\end{tabular}
\captionof{figure}{Inverse coefficient of variation (iCV) of the fire intensity, $\Lambda_t(x)$ (shaded areas and contour lines) for models with harmonic components and either constant scale $c$ or time-varying scale $c_t$ and time-varying seasonal scale $c_t(12)$ (different rows).}
\label{fig:Lambda_mean_Harmonic_CV}
\end{sidewaystable}


\begin{figure}[p]
\begin{center}
\setlength{\tabcolsep}{10pt}
\begin{tabular}{cc}
$g=0.5^{\circ}$&$g=1^{\circ}$\\
\includegraphics[clip,trim=0 0 0 0, scale=0.45]{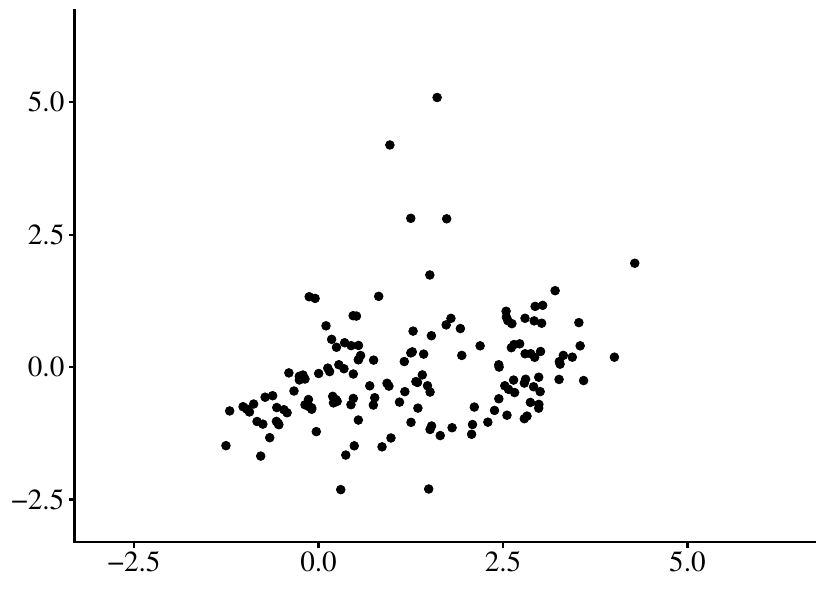}&
\includegraphics[clip,trim=0 0 0 0, scale=0.45]{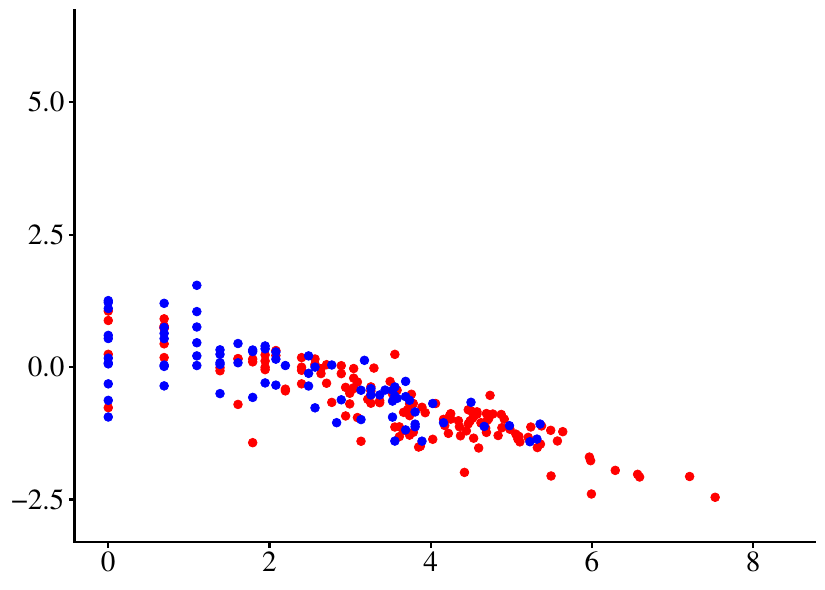}\\
$g=1.5^{\circ}$&$g=2^{\circ}$\\
\includegraphics[clip,trim=0 0 0 0, scale=0.45]{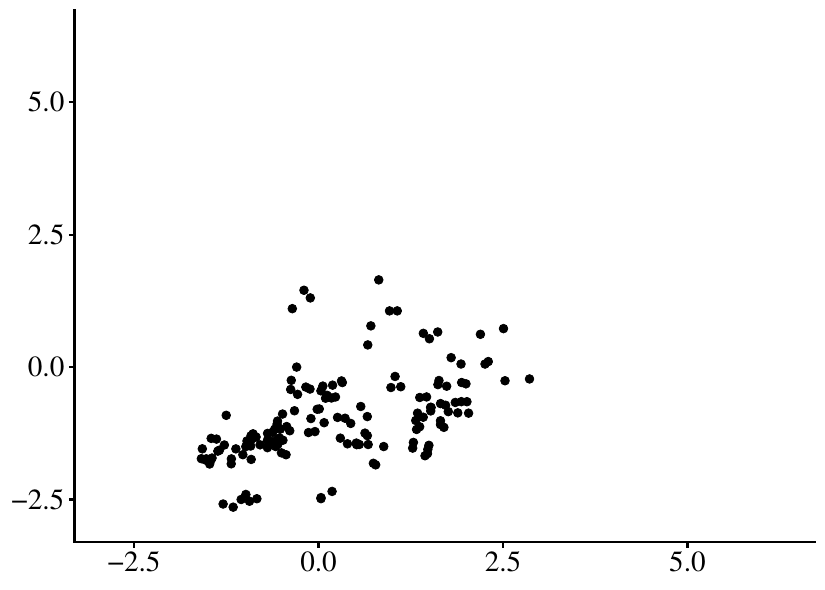}&
\includegraphics[clip,trim=0 0 0 0, scale=0.45]{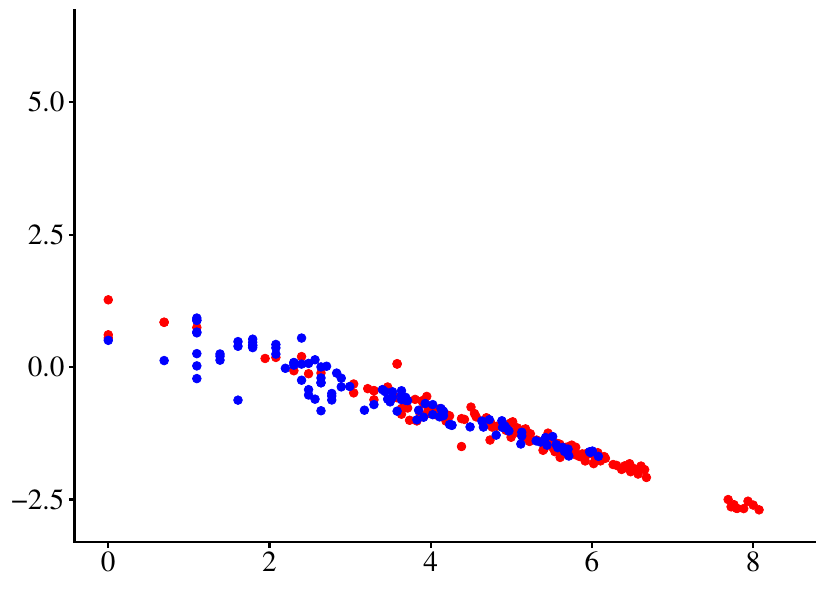}\\
$g=2.5^{\circ}$&$g=3^{\circ}$\\
\includegraphics[clip,trim=0 0 0 0, scale=0.45]{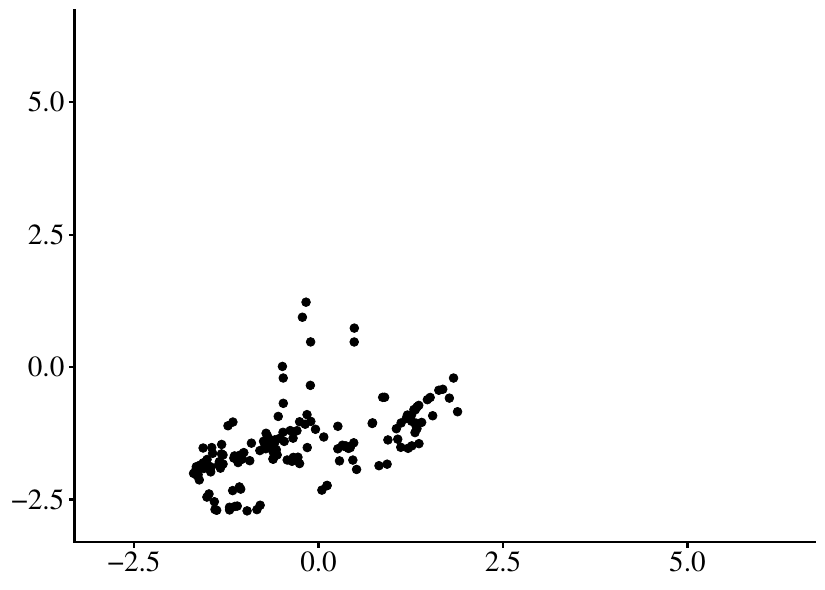}&
\includegraphics[clip,trim=0 0 0 0, scale=0.45]{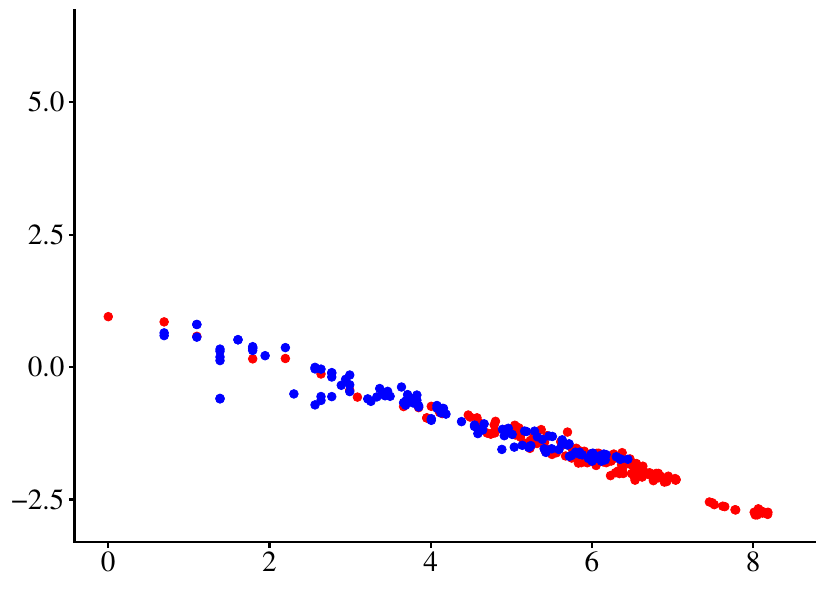}\\
$g=3.5^{\circ}$&$g=4^{\circ}$\\
\includegraphics[clip,trim=0 0 0 0, scale=0.45]{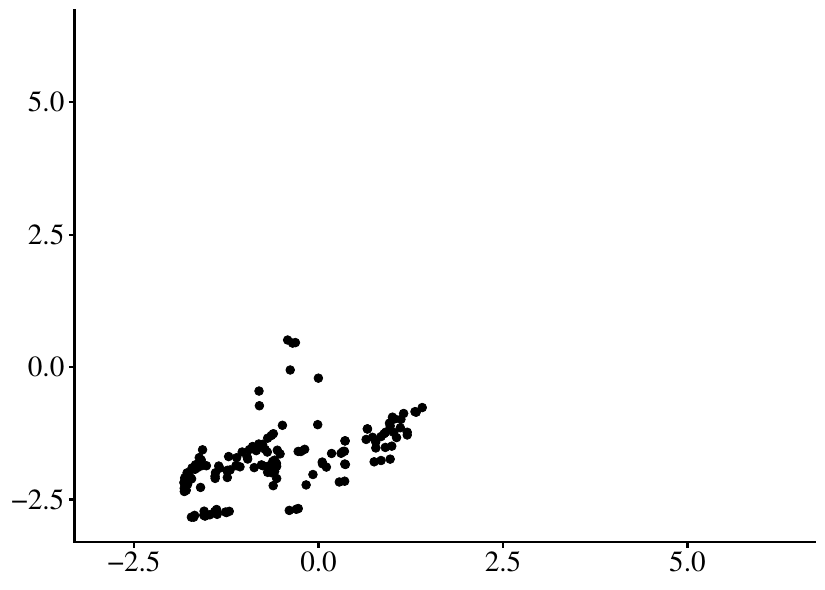}&
\includegraphics[clip,trim=0 0 0 0, scale=0.45]{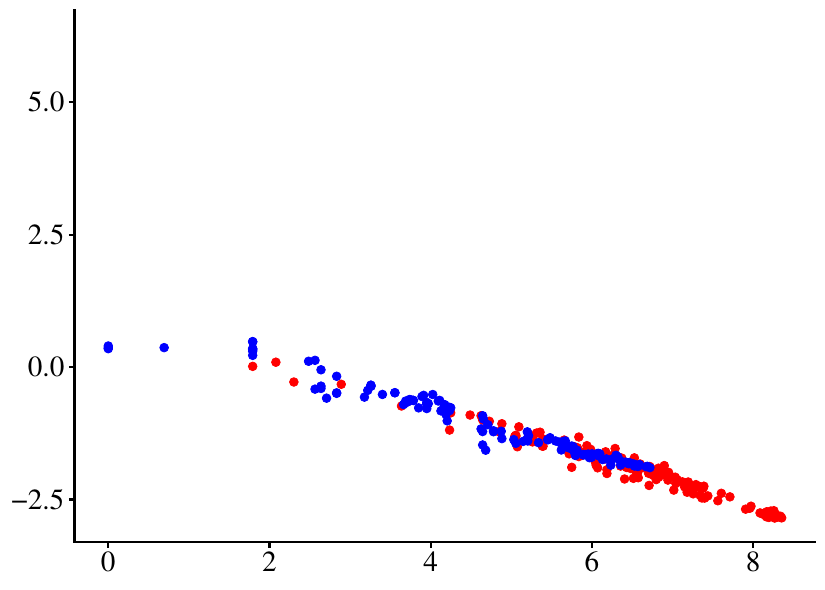}
\end{tabular}
\end{center}
\captionsetup{width=0.9\textwidth}
\caption{Variability of the intensity estimates, measured by the normalised interquantile range, $(Q_{97.5}-Q_{2.5})/Q_{50}$, for 100 different $g\times g$ areas ($\bullet$). Left: Dry season against wet season range.  Right: range against the number of fires in the same areas during dry (\textcolor{red}{$\bullet$}) and wet (\textcolor{blue}{$\bullet$}) seasons.}
\label{fig:IQRall}
\end{figure}


\begin{figure}[p]
\vspace*{-6ex}
\centering
\setlength{\tabcolsep}{10pt}
\begin{tabular}{c c c}
& {\footnotesize Constant scale} & {\footnotesize Time-varying Scale} \\
\begin{rotate}{90} \hspace{36pt} {\footnotesize \parbox{2.5cm}{\centering $t=3$ (March)\\ $h=0$}} \end{rotate} &
\includegraphics[clip,trim=0 0 0 0, scale=0.42]{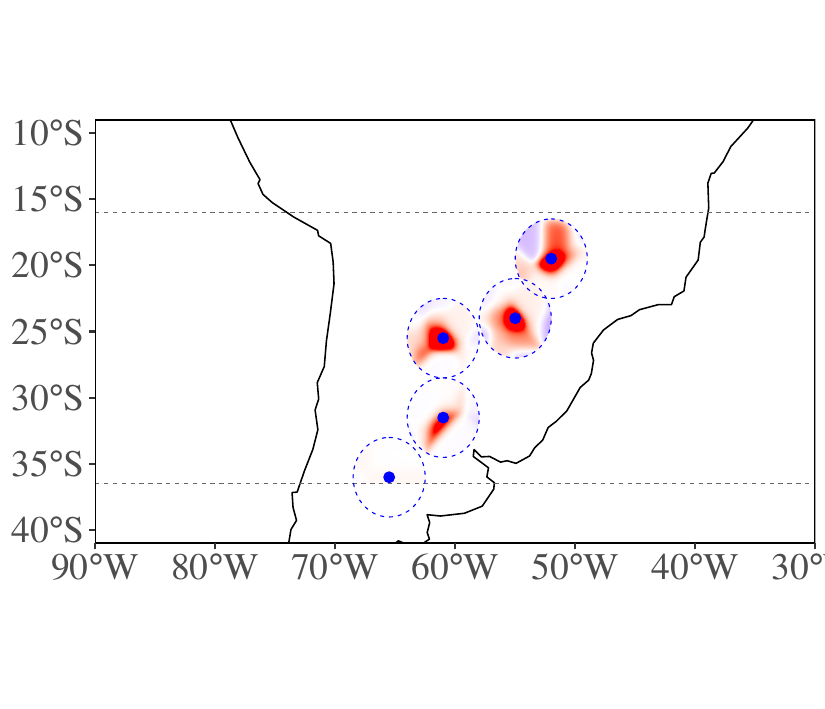} &
\includegraphics[clip,trim=0 0 0 0, scale=0.42]{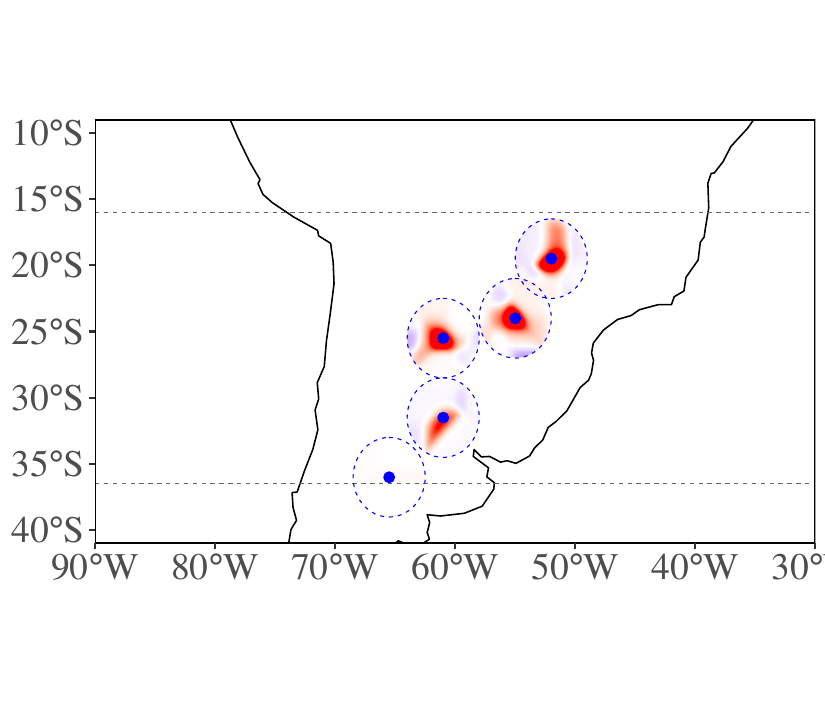} \\
\begin{rotate}{90} \hspace{36pt} {\footnotesize \parbox{2.5cm}{\centering $t=3$ (March)\\ $h=1$}} \end{rotate} &
\includegraphics[clip,trim=0 0 0 0, scale=0.42]{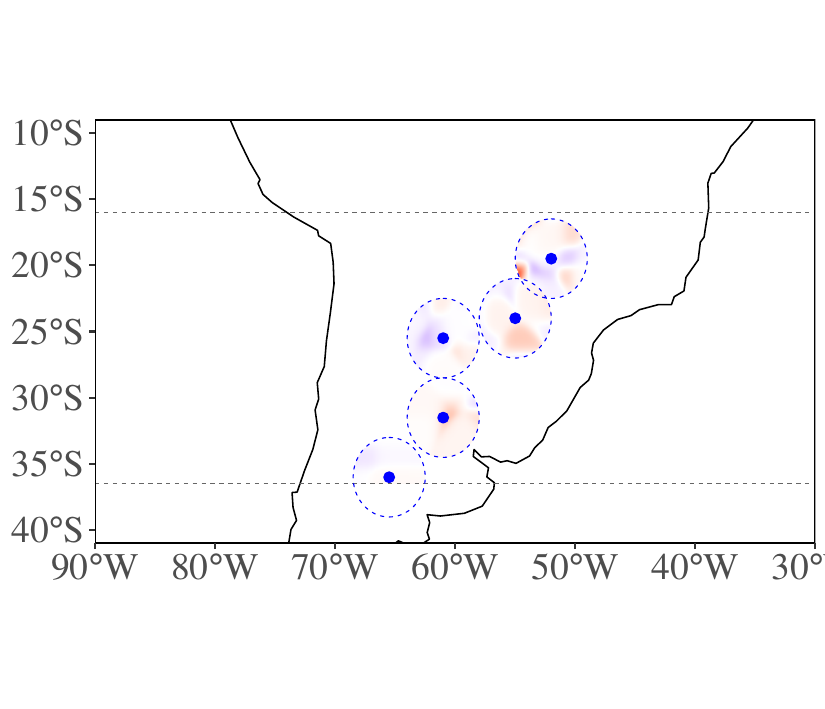} &
\includegraphics[clip,trim=0 0 0 0, scale=0.42]{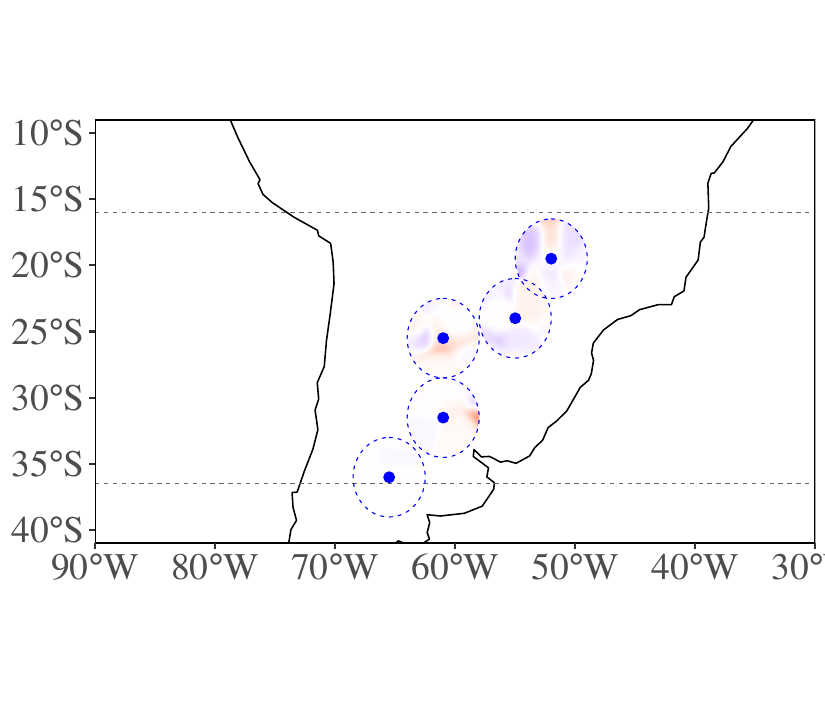} \\
\begin{rotate}{90} \hspace{29pt} {\footnotesize \parbox{2.5cm}{\centering $t=11$ (November)\\ $h=0$}} \end{rotate} &
\includegraphics[clip,trim=0 0 0 0, scale=0.42]{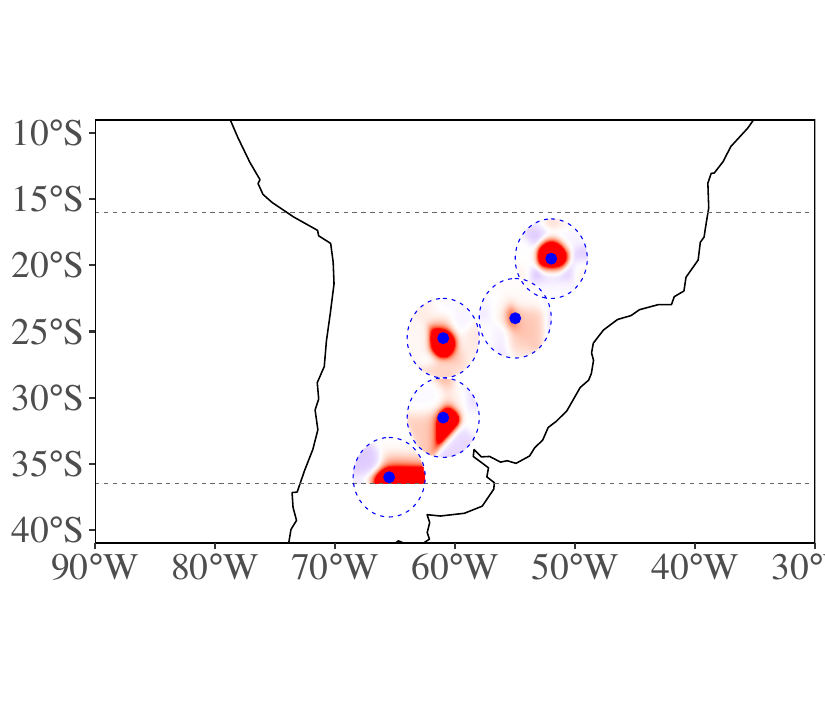} &
\includegraphics[clip,trim=0 0 0 0, scale=0.42]{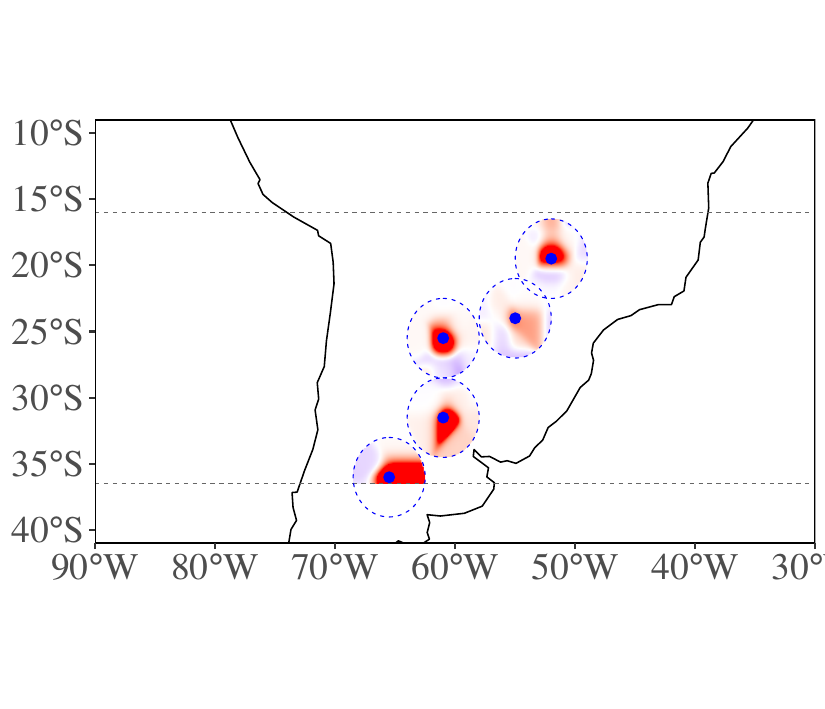} \\
\begin{rotate}{90} \hspace{29pt} {\footnotesize \parbox{2.5cm}{\centering $t=11$ (November)\\ $h=1$}} \end{rotate} &
\includegraphics[clip,trim=0 0 0 0, scale=0.42]{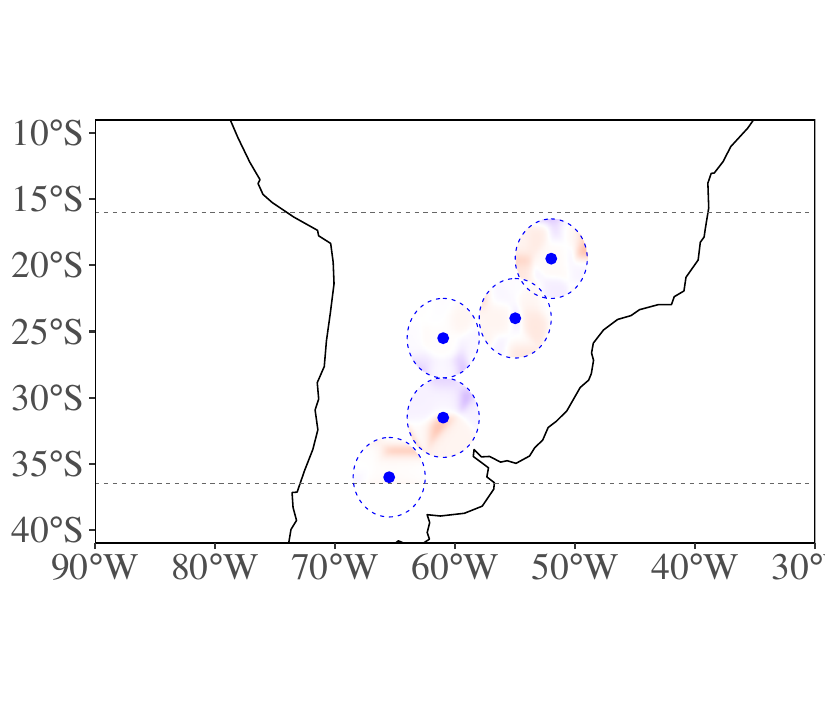} &
\includegraphics[clip,trim=0 0 0 0, scale=0.42]{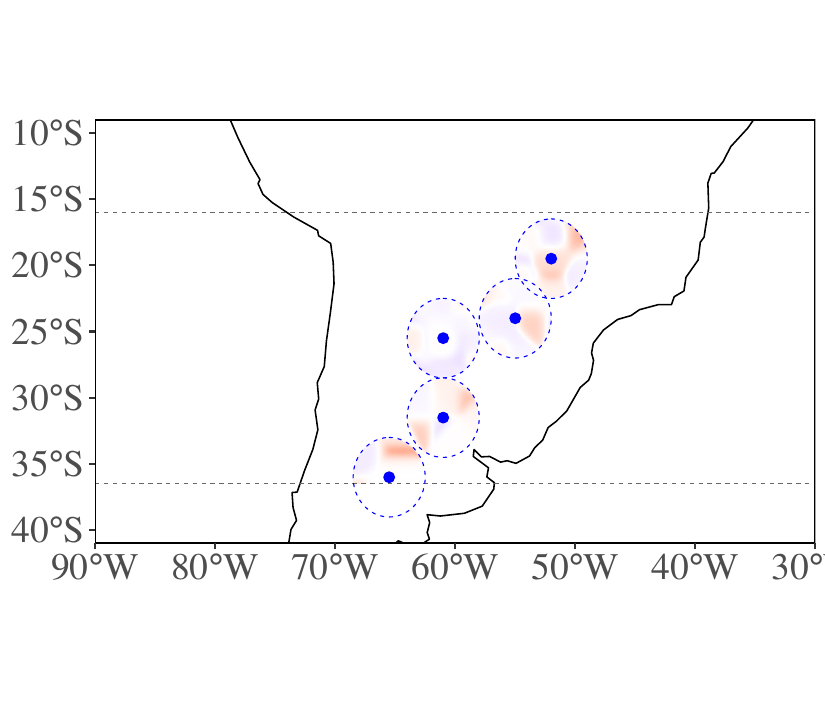}
\end{tabular}
\captionsetup{width=0.9\textwidth}
\captionof{figure}{Pair correlation function $\mathscr{R}_{t,t+h}(x,y)$ for harmonic components models with constant (left) and time-varying (right) scale. In each plot, the function value (colours) for $t=3$ (March) and $t=11$ (November) at selected locations (dots) and alternative locations in a $4^\circ$ degree radius around $x$ (circles). Red shades for $\mathscr{R}_{t,t+h}(x,y)>1$ and blue shades for $\mathscr{R}_{t,t+h}(x,y)\leq 1$.}
\label{fig:Rtth}
\end{figure}

\FloatBarrier

\subsection{Anisotropy and comparison with LGCP}
\label{sec:anisotropy_lgcp}

We further discuss the role of spatial stationarity and isotropy in the proposed framework and compare our model with a log-Gaussian Cox process (LGCP).

In our benchmark specification, the shot-noise kernel is isotropic, hence smoothing depends only on the Euclidean distance between $y$ and $\theta$:
\begin{equation*}
    K_{\phi}(y,\theta) = (\pi\phi^2)^{-1} \exp\big( -\phi^{-2}\|y-\theta\|^2 \big).
\end{equation*}
Spatial stationarity is obtained when the base measure $H$ is translation invariant and the kernel depends only on $y$ and $\theta$. Non-stationarity can instead be introduced through a spatially varying base measure $H$, a non-homogeneous measure $\ell(dy)$, or covariate-dependent components in the deterministic intensity.

To allow for anisotropy, we replace the scalar bandwidth $\phi$ with a positive definite matrix $\Sigma$, and use
\begin{equation*}
    K_{\Sigma}(y,\theta) = \pi^{-1} |\Sigma|^{-1/2} \exp\big( -(y-\theta)^\T \Sigma^{-1}(y-\theta) \big).
\end{equation*}
This specification allows the intensity to be smoothed differently across spatial directions, and the isotropic model is recovered as the special case when $\Sigma = \phi^2 \Id_2$. In our implementation, positive definiteness of $\Sigma$ is enforced (at every iteration) through a Cholesky parametrisation, as described in Section~\ref{sec:posterior}.

This comparison is also useful relative to LGCPs, which are commonly specified through a latent Gaussian process with a stationary and isotropic covariance function. We performed a numerical comparison with the LGCP implemented in the \texttt{R} package \texttt{lgcp} \citep{taylor2013lgcp}. The package follows a three-step estimation strategy: first, the temporal and spatial components of the intensity are estimated separately; second, the parameters of the latent Gaussian process are selected conditionally on these components, using moment-based criteria such as the inhomogeneous pair correlation function or the inhomogeneous $K$ function; finally, the latent Gaussian process is estimated conditionally on the remaining components.

We tested the performance of the two \SNMARG\ specifications, anisotropic and isotropic, against that of the LGCP implementation using the same harmonic covariates and the same $0.1$ degree tessellation grid on the fire location data. For the LGCP benchmark, the fixed spatial component was held constant over the study region and normalized to integrate to 1 over the observation window, while the temporal component was estimated via a Poisson regression using the same trend and harmonic covariates as in our model. The latent Gaussian parameters $(\sigma,\phi)$ were then chosen by minimum-contrast matching of the inhomogeneous pair correlation function under an exponential covariance model, while the temporal parameter $\theta$ was estimated from the empirical autocorrelation of the total monthly counts.

Table~\ref{tab:prediction_assessment} reports the in-sample and three-step-ahead out-of-sample prediction errors for the total number of counts over the study region, together with the computational time required by each method. The results indicate that both \SNMARG\ specifications substantially outperform the LGCP benchmark. In particular, according to the MAE and the Root Mean Squared Logarithmic Error (RMSLE), the isotropic \SNMARG\ model achieves the best performance in both in-sample fitting and out-of-sample prediction. The anisotropic specification also improves considerably upon the LGCP, although it does not match the predictive accuracy of the isotropic model under these two criteria.

A similar pattern is observed when considering the Mean Absolute Percentage Error (MAPE). While the anisotropic \SNMARG\ model attains the lowest out-of-sample MAPE, both \SNMARG\ specifications clearly outperform the LGCP benchmark. Moreover, the computational times reported in Table~\ref{tab:prediction_assessment} highlight a substantial efficiency advantage of the proposed models. The isotropic and anisotropic \SNMARG\ models require approximately 19 and 35 minutes, respectively, whereas the LGCP implementation requires more than two hours.

\begin{table}[t!h]
\centering
\resizebox{\textwidth}{!}{
\begin{tabular}{llrrrr}
\toprule
{\bf Model} & {\bf Sample} & {\bf MAE}  & {\bf RMSLE} & {\bf MAPE}  & {\bf Time} \\
\midrule
SNMARG iso     & in-sample       & 2.310        & 0.00709        & 0.48554       & 19.02         \\
SNMARG aniso   & in-sample       & 3.582        & 0.01018        & 0.74702       & 35.39         \\
LGCP           & in-sample       & 435.886      & 1.01104        & 55.60019      & >120.00        \\ \hline
SNMARG iso     & out-of-sample   & 2272.391     & 0.58951        & 49.39286      & 19.02         \\
SNMARG aniso   & out-of-sample   & 2623.308     & 0.80179        & 45.05701      & 35.39         \\
LGCP           & out-of-sample   & 3060.653     & 0.99655        & 61.78880      & >120.00\\ 
\bottomrule
\end{tabular}}
\caption{Prediction assessment for \SNMARG\ isotropic, \SNMARG\ anisotropic, and LGCP. Time reports the computing time in minutes.}
\label{tab:prediction_assessment}
\end{table}

\begin{sidewaystable}[p]
\centering
\setlength{\tabcolsep}{4pt}
\begin{tabular}{c c c c}
& {\footnotesize \SNMARG\ aniso}
& {\footnotesize \SNMARG\ iso}
& {\footnotesize LGCP} \\
\begin{rotate}{90}\hspace{16pt}{\footnotesize April 2020}\end{rotate}
&
\includegraphics[width=.29\textwidth]{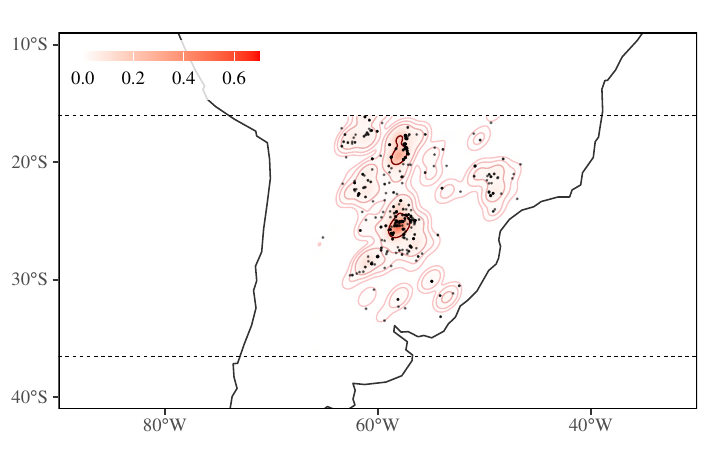}
&
\includegraphics[width=.29\textwidth]{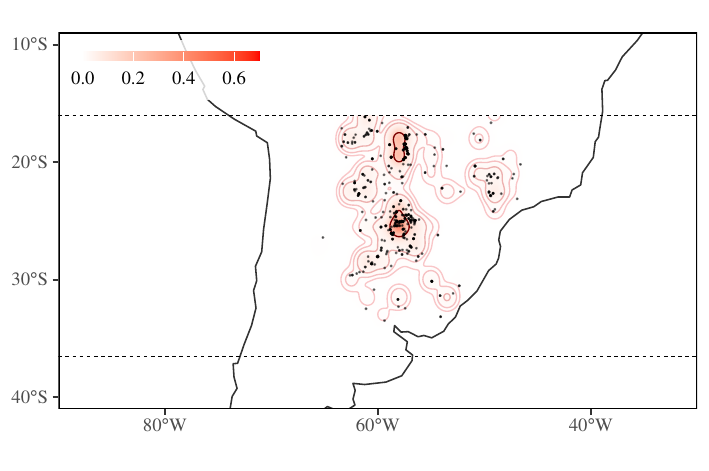}
&
\includegraphics[width=.29\textwidth]{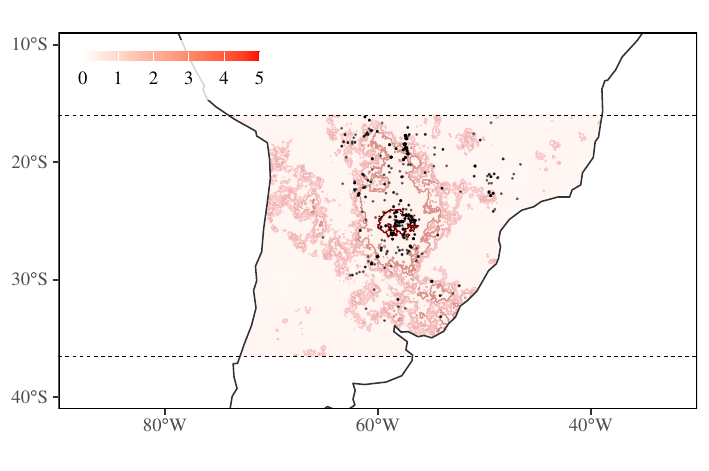}
\\[-0.5em]

\begin{rotate}{90}\hspace{12pt}{\footnotesize May 2020}\end{rotate}
&
\includegraphics[width=.29\textwidth]{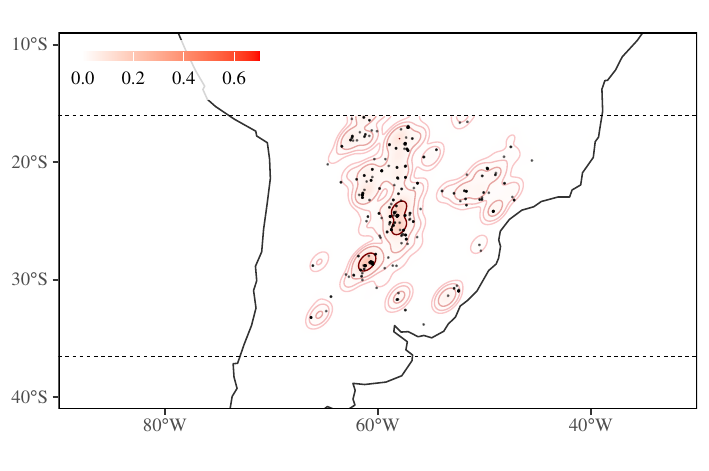}
&
\includegraphics[width=.29\textwidth]{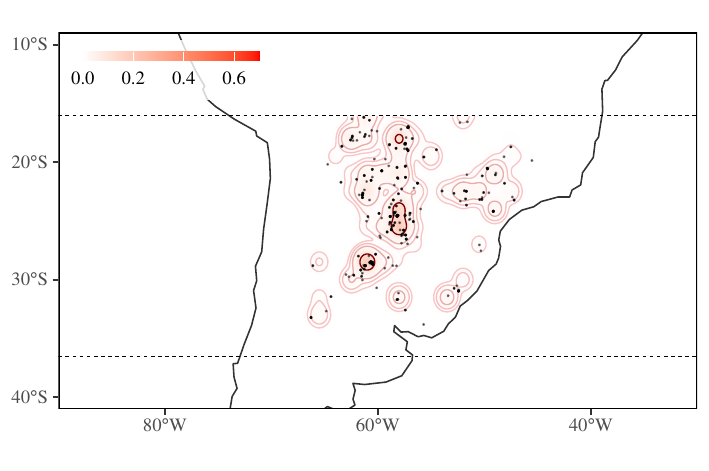}
&
\includegraphics[width=.29\textwidth]{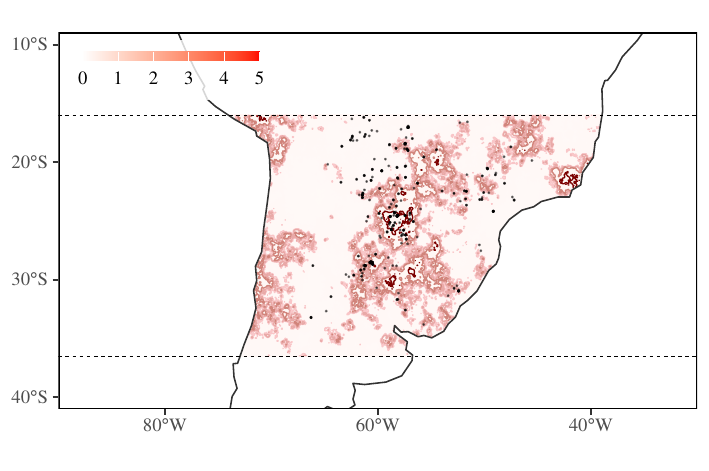}
\\[-0.5em]

\begin{rotate}{90}\hspace{12pt}{\footnotesize June 2020}\end{rotate}
&
\includegraphics[width=.29\textwidth]{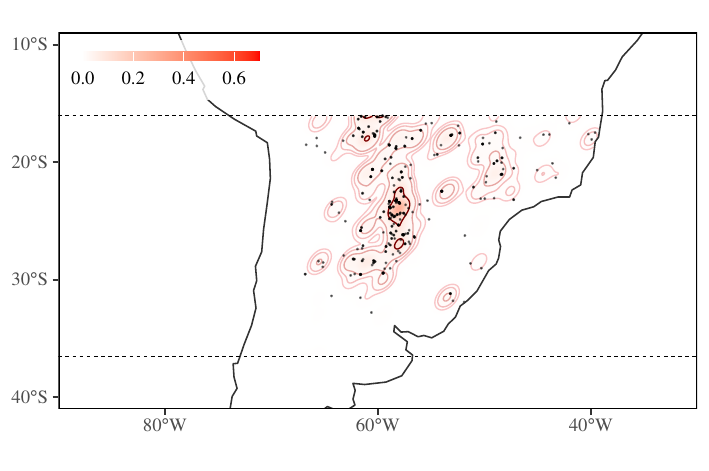}
&
\includegraphics[width=.29\textwidth]{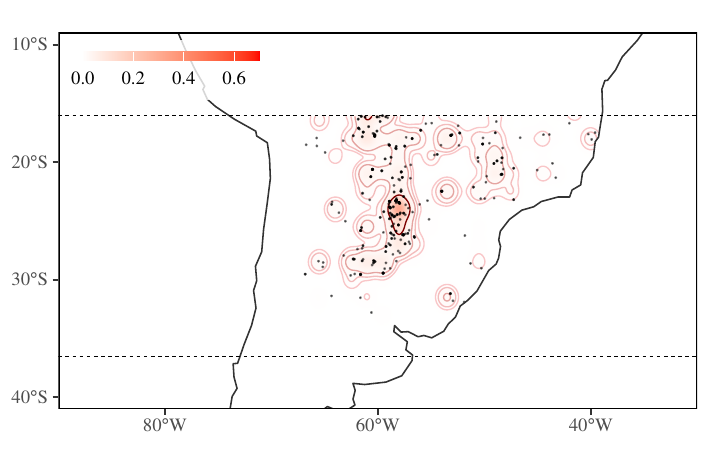}
&
\includegraphics[width=.29\textwidth]{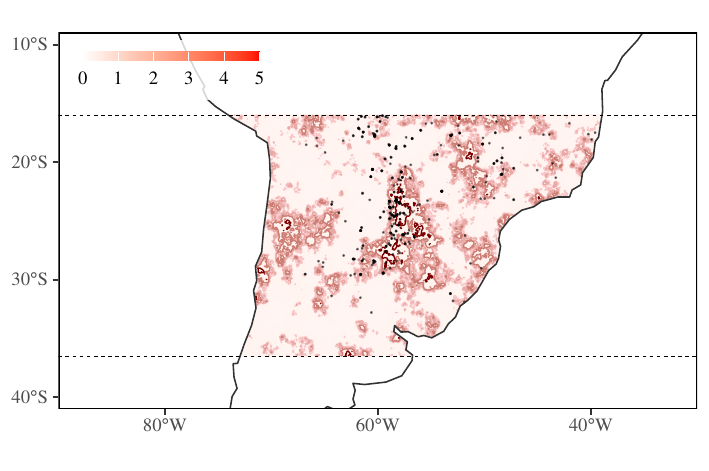}
\end{tabular}
\captionsetup{width=0.90\textwidth}
\caption{Estimated in-sample intensity maps for March--June 2020.
Rows correspond to months and columns correspond to the anisotropic \SNMARG\ model, the isotropic \SNMARG\ model, and the LGCP. The anisotropic specification captures directional spatial dependence and provides a more localized representation of high-intensity regions. Consistent with the results reported in Table~\ref{tab:prediction_assessment}, both \SNMARG\ specifications outperform the LGCP, while the anisotropic model yields the best out-of-sample predictive performance.
}
\label{fig:anisotropy_lgcp_maps}
\end{sidewaystable}

\clearpage

\subsection{Euclidean vs geodesic coordinates}
\label{sec:euclidean_space}

\begin{figure}[p]
\centering
\begin{tabular}{cc}
\multicolumn{2}{c}{July 2020} \\
\includegraphics[width=0.45\textwidth]{figures_new/Figure56/figure_56_lambda_m_FALSE_2_31.pdf} &
\includegraphics[width=0.45\textwidth]{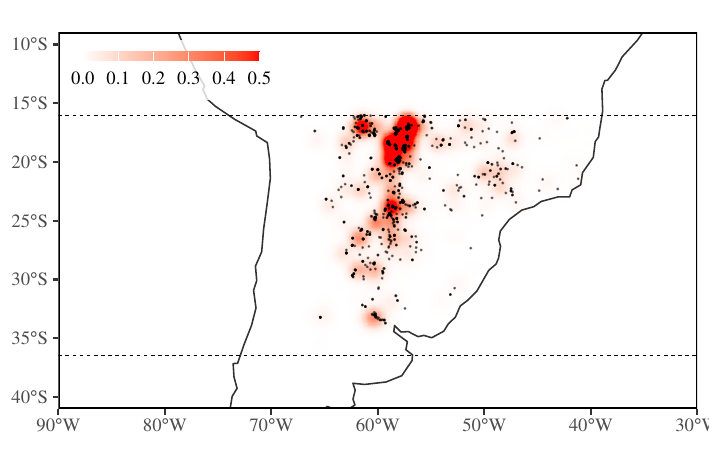} \\[2mm]

\multicolumn{2}{c}{August 2020} \\
\includegraphics[width=0.45\textwidth]{figures_new/Figure56/figure_56_lambda_m_FALSE_2_32.pdf} &
\includegraphics[width=0.45\textwidth]{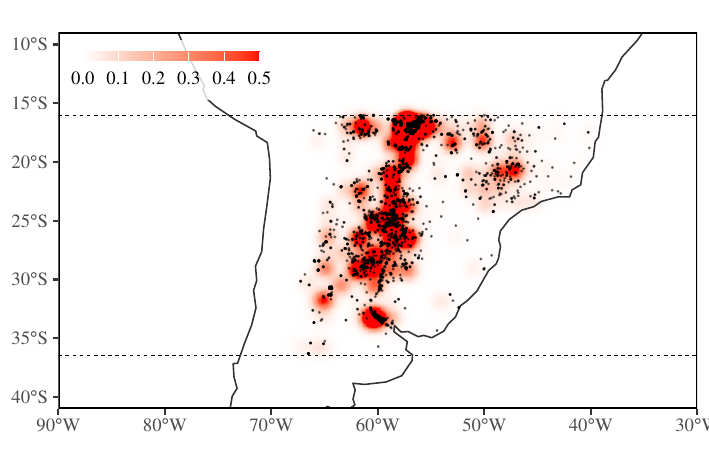} \\[2mm]

\multicolumn{2}{c}{September 2020} \\
\includegraphics[width=0.45\textwidth]{figures_new/Figure56/figure_56_lambda_m_FALSE_2_33.pdf} &
\includegraphics[width=0.45\textwidth]{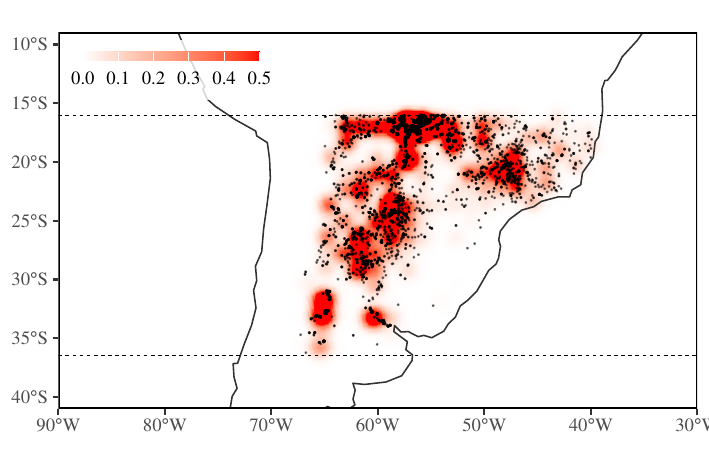}
\end{tabular}
\captionsetup{width=0.90\textwidth}
\caption{Estimated random fields for the last three months of the application using geographical coordinates (top row) and Euclidean coordinates (bottom row). We notice little difference in the estimation of the critical areas.}
\label{fig:geo_euclidean_comparison}
\end{figure}

As a robustness check, we compare the results in the empirical application when working with geodesic coordinates and Euclidean coordinates.
Figure~\ref{fig:geo_euclidean_comparison} reports the plot of the estimated random field in the last four months of the application using geodesic coordinates (top) and Euclidean coordinates (bottom). We notice little difference in the estimation of the critical areas.
To gain further insight into model fit, Table~\ref{tab:geo_euclidean_errors} reports the in-sample MSE and MAE for the total counts. The model based on Euclidean coordinates achieves slightly lower errors than its geodesic-coordinate counterpart, indicating a better overall fit to the observed data. Nevertheless, despite the small differences in the numerical error metrics, the estimated spatial patterns and critical areas remain largely consistent across the two coordinate systems. Consequently, we do not expect the choice between geodesic and Euclidean coordinates to have a substantial impact on the conclusions of the analysis.

\begin{table}[t!h]
\centering
\begin{tabular}{lrr}
\toprule
Coordinates & \multicolumn{1}{c}{MSE} & \multicolumn{1}{c}{MAE} \\
\midrule
Geodesic & 30.501 & 3.831 \\
Euclidean    & 28.787 &  3.700 \\
\bottomrule
\end{tabular}
\caption{In-sample prediction accuracy under geodesic and Euclidean coordinates.}
\label{tab:geo_euclidean_errors}
\end{table}

\end{document}